\documentclass{article}
\usepackage[T1]{fontenc}

\usepackage{mathtools,amsmath,amsthm,amssymb,amsbsy,amsfonts,amscd,mathrsfs}
\usepackage{breqn}
\usepackage{bm}
\usepackage{enumitem}
\usepackage{empheq}
\usepackage{rotating}
\usepackage{ftnxtra}
\usepackage{fnpos}
\usepackage{authblk}
\usepackage[square]{natbib}
\usepackage{tcolorbox}
\usepackage{setspace}
\usepackage{changepage}
\usepackage{booktabs}
\usepackage{multirow}
\usepackage{accents}
\usepackage{tikz-cd} 

\usepackage{hyperref}
\hypersetup{colorlinks=true, linkcolor=blue}
\hypersetup{colorlinks=true,citecolor=blue}
\usepackage{cleveref}

\usepackage{autonum}
\numberwithin{equation}{section}
\theoremstyle{plain}	
 \newtheorem{thm}{Theorem}[section]
 
 \newtheorem{lem}[thm]{Lemma}
 
\theoremstyle{definition}	
 
 \newtheorem{remark}[thm]{Remark}
 \newtheorem{example}[thm]{Example}

\usepackage[text={6.2in,9in},%
	]{geometry}
\usepackage{caption2}

\makeatletter
\newsavebox{\@brx}
\newcommand{\llangle}[1][]{\savebox{\@brx}{\(\m@th{#1\langle}\)}%
  \mathopen{\copy\@brx\mkern2mu\kern-0.9\wd\@brx\usebox{\@brx}}}
\newcommand{\rrangle}[1][]{\savebox{\@brx}{\(\m@th{#1\rangle}\)}%
  \mathclose{\copy\@brx\mkern2mu\kern-0.9\wd\@brx\usebox{\@brx}}}%
\let\oldabs\abs
\def\abs{\@ifstar{\oldabs}{\oldabs*}}
\makeatother

\newcommand{\stilde}[1]{ \skew{3}\tilde{#1} }                               % Skewed tilde for inclined symbols

\newcommand{\incl}{j}

\newcommand{\indx}[1]{\mathfrak{#1}}                                      % Command for alphabetical disl distr index not as a super/subscript
\DeclareRobustCommand{\prsn}[2]{ \accentset{#1}{#2} }                            % Command for disl distr index
\newcommand{\prs}[2]{ \prsn{\indx{#1}}{#2} }                            % Command for gothic disl distr index

\newcommand{\comp}[1]{%                                                          % Command for un-boldsymbol
  \begingroup
    \renewcommand*{\boldsymbol}[1]{##1}%
    #1%
  \endgroup
}

\newcommand{\op}[1]{{\boldsymbol{#1}}}                                            % Operators

\newcommand{\subp}[1]{#1_{ \mspace{-1mu} \scriptscriptstyle{\mathrm P}  }}   % Plastic subscript
\newcommand{\sube}[1]{#1_{ \mspace{-1mu} \scriptscriptstyle{\mathrm E}  }}   % Elastic subscript

\newcommand{\dd}{\boldsymbol\eta}                                                              % dislocation field (no superscript)
\newcommand{\df}{{\boldsymbol\omega}}                                                       % Dislocation form
\newcommand{\bv}{{B}}                                                                       % Burgers director
\newcommand{\lv}{{L}}                                                                         % Line director
\newcommand{\norm}{{N}}                                                                   % Normal director (to the dislocation curve)
\newcommand{\dens}{{\varrho}}                                                                       % Scalar dislocation density
\newcommand{\vel}{{U}}                                                                                  % Dislocation velocity
\newcommand{\var}{{W}}                                                                                 % Dislocation epsilon-velocity
\newcommand{\mvf}{{\boldsymbol{\mu}}}                                                         % Material volume form
\newcommand{\massd}{{\rho}}                                                                         % Mass density
\newcommand{\pk}{{\mathcal{P}}}                                                                         % Mass density

\begin{document}
\bibliographystyle{abbrvnat}

\title{\Large{\textbf{\vspace{-.75in}\\
A Geometric Field Theory of Dislocation Mechanics 
}}}

\author[1]{Fabio Sozio}
\author[2,3]{Arash Yavari\thanks{Corresponding author, e-mail: arash.yavari@ce.gatech.edu}}
\affil[1]{\small \textit{Solid Mechanics Laboratory, \'Ecole Polytechnique, Palaiseau, France}}
\affil[2]{\small \textit{School of Civil and Environmental Engineering, Georgia Institute of Technology, Atlanta, GA 30332, USA}}
\affil[3]{\small \textit{The George W. Woodruff School of Mechanical Engineering, Georgia Institute of Technology, Atlanta, GA 30332, USA}}

\maketitle

%-----------------------------
%-----------------------------
\begin{abstract}
\noindent
In this paper a geometric field theory of dislocation dynamics and finite plasticity in single crystals is formulated. Starting from the multiplicative decomposition of the deformation gradient into elastic and plastic parts, we use Cartan's moving frames to describe the distorted lattice structure via differential $1$-forms. In this theory the primary fields are the dislocation fields, defined as a collection of differential $2$-forms. The defect content of the lattice structure is then determined by the superposition of the dislocation fields.
All these differential forms constitute the internal variables of the system.
The evolution equations for the internal variables are derived starting from the kinematics of the dislocation $2$-forms, which is expressed using the notions of flow and of Lie derivative.
This is then coupled with the rate of change of the lattice structure through Orowan's equation.
The governing equations are derived using a two-potential approach to a variational principle of the Lagrange-d'Alembert type.
As in the nonlinear setting the lattice structure evolves in time,
the dynamics of dislocations on slip systems is formulated by enforcing some constraints in the variational principle.
Using the Lagrange multipliers associated with these constraints, one obtains the forces that the lattice exerts on the dislocation fields in order to keep them gliding on some given crystallographic planes.
Moreover, the geometric formulation allows one to investigate the integrability---and hence the existence---of glide surfaces,  and how the glide motion is affected by it.
Lastly, a linear theory for small dislocation densities is derived, allowing one to identify the nonlinear effects that do not appear in the linearized setting.
\end{abstract}

\begin{description}
\item[Keywords:] Dislocation mechanics, continuum dislocation dynamics, nonlinear elasticity, anelasticity, plasticity, geometric mechanics
\end{description}

\tableofcontents

%%%%%%%%%%%%%%%%%%%%%%%%%%%%%%%%%%%%%%
%%%%%%%%%%%%%%%%%%%%%%%%%%%%%%%%%%%%%%
\section{Introduction} \label{Intro}

The mechanics of plasticity and defects in crystalline solids has a close connection with differential geometry.
Plasticity is a phenomenon that falls under the broader category of anelasticity, which is the study of solids that carry residual stresses.
In particular, anelasticity revolves around the concept of material metric tensor, describing local natural distances in a solid, and distributions of eigenstrains \citep{reissner1931eigenspannungen}.
Therefore, the natural framework for describing plasticity as a source of eigenstrains is Riemannian geometry \citep{eckart1948thermodynamics}, the main predictor of residual stresses being the three-dimensional Riemann curvature tensor.
On the other hand, plasticity can be seen as the study of deformation of a solid in relation to its microstructure, containing more information than the simple change in natural distances considered in anelasticity.
In this case the exterior algebra of differential forms provides a description of the lattice structure and of the line defects associated with it. Differential geometry offers a natural framework for a continuum theory of dislocation plasticity, and of crystallographic defects in general. Although geometric theories for the analysis of equilibrium configurations of distributed defects in nonlinear solids are available in the literature \citep{Gairola1979,Rosakis1988,Zubov1997,acharya2001model,Yavari2012Weyl,yavari2012riemann,yavari2013riemann,yavari2014geometry,Yavari2016,golgoon2018line}, geometric formulations for dislocation dynamics in the nonlinear setting have not been developed systematically to this date.
This is due to its complexity; for instance, when finite deformations are allowed, the lattice structure is time dependent, and therefore crystallographic planes deform into surfaces, and they can even cease to exist.

The mathematical theory of the mechanics of dislocations and disclinations was formulated by Vito Volterra in a series of papers from 1905-1907, which were summarized in \citep{Volterra1907} (for a recent English translation of this paper see \citep{Delphenich2020}).
A few decades later, \citet{taylor1934mechanism}, \citet{orowan1934kristallplastizitat}, and \citet{polanyi1934art} were the first to realize that the motion of dislocations facilitates crystal slip and is the micro-mechanism of plastic deformation in crystals.
The interaction between dislocations and the elastic field was studied by \citet{PeachKoehler1950}, who provided the first expression for what is now commonly known as the Peach-Koehler force.
The notion of dislocation density tensor was introduced by \citet{nye1953some},\footnote{See \citep{Sozio2021Nye} for a recent study of Nye's lattice curvature tensor using Cartan's moving frames.} while the first geometric formulations of plasticity are due to \citet{bilby1955continuous}, \citet{kondo1955geometry}, \citet{kroner1962dislocations}, and of \citet{noll1967materially} and \citet{wang1968geometric}.

More recently, new contributions to the geometric theory of dislocation plasticity have been made. Examples are \citet{clayton2005geometric} who proposed a novel three-term decomposition of the deformation gradient, and  \citet{yavari2012riemann}, who formulated a geometric theory of solids with distributed dislocations using Cartan's moving frames.
Epstein and Segev introduced a geometric framework for discrete dislocations using de Rham's currents \citep{Epstein2014,epstein2014geometric,Epstein2015,epstein2020regular} .
This tool has recently been used in dislocation dynamics by~\citet{starkey2022development}.
\citet{sozio2020riemannian} studied different formats of the governing equations for anelastic solids in both the standard and configurational frameworks. Both the underlying Euclidean structure inherited by the ambient space, and the Riemannian structure induced by the material metric were considered.
It is also worth mentioning that \citet{trzkesowski1997kinematics} was perhaps the first to investigate the issue of the integrability of slip surfaces, and to look at crystals as foliated manifolds.

Aside from the geometric approach, in the past two decades several field dislocation mechanics formulations have been proposed whose focus has been the study of the formation of dislocation patterns and structures at the mesoscale from a continuum perspective.
Examples are the works of \citet{acharya2001model}, \citet{cermelli2001characterization}, and \citet{gurtin2002gradient}.
\citet{sedlavcek2003importance, sedlavcek2007continuum} provided an accurate description of the linear kinematics of dislocations, introducing the concept of virtual motion.
\citet{zhu2013dislocation} investigated the instability of the dislocation motion due to the cross slip of the screw segments.
\citet{xia2015computational, xia2015preliminary} proposed a continuum description as a smeared representation of discrete distributions, in which the occurrence of cross-slip is regulated by a probability function.

Recent years have witnessed the development of many statistical theories of continuum dislocation dynamics.
As opposed to geometrically necessary dislocations, statistically stored dislocations are responsible for strain hardening and cannot be deduced through purely geometric arguments \citep{ashby1970deformation,arsenlis1999crystallographic}.
We should mention the works of \citet{el2000statistical}, \citet{groma2003spatial}, \citet{hochrainer2007three}, \citet{el2007statistical} and \citet{hochrainer2016thermodynamically}.
The existing theories of dislocation dynamics were recently reviewed by \citet{mcdowell2019multiscale},
while some recent developments in plasticity were reviewed by \citet{steigmann2020primer}.

Our goal in this paper is to formulate a geometric theory for nonlinear field dislocation mechanics in single crystals.
In the geometric setting, plastic slip, crystallographic planes, and distributed dislocations are described by differential forms on a Riemannian manifold.
These fields constitute the internal state variables of the model \citep{coleman1967thermodynamics, rice1971inelastic, lubliner1973structure}.
The kinetic equations for the internal variables are derived through a variational approach in the presence of nonholonomic internal constraints.
Variational methods in plasticity have already been used in the works of \citet{hackl1997generalized}, \citet{ortiz1999nonconvex}, \citet{berdichevsky2006continuum}, \citet{junker2014principle}, and as well as in the recent paper by \citet{acharya2021action}.
We should also mention the work of \citet{po2014variational} for variational approaches to the thermodynamics of discrete dislocations.
We will use a two-potential approach \citep{halphen1975materiaux,germain1983continuum}, in which all the constitutive equations can be derived from two functions of the internal variables and their rates.
This is similar to the approach of \citet{ziegler1958attempt}, and \citet{ziegler1987derivation} based on a dissipation function expressing the entropy production.
In particular, we propose a deterministic mesoscale theory, in which the evolution of dislocations is only due to their glide motion, while sources/sinks of dislocations and climb are neglected.
However, nonlocal and micro-inertial effects are included.
The main contributions of this paper can be summarized as follows.

%---------------------
\begin{itemize}\setlength\itemsep{0em}
\item
A metric-free formulation of dislocation fields is presented using differential forms. The incompatibility of the lattice structure is written as the superposition of a number of dislocation fields.
\item
The kinematics of dislocations is formulated using the notion of flow and Lie derivative.
It is shown that Orowan's equation is consistent with the geometric formulation.
\item
The integrability of crystallographic planes in relation to the dislocated lattice is investigated. We discuss the consequences of the non-integrability of the slip planes on the glide of dislocations.
\item
The governing equations are derived variationally, using a two-potential approach to include dissipation.
This allows one to write the kinetic equations for the internal variables without assuming specific forms for the constitutive model.
The only constitutive assumptions that are made are those that guarantee frame indifference, the second law of thermodynamics, etc.
\item
Lagrange multipliers are used to enforce lattice constraints directly in the variational formulation through the methods of nonholonomic mechanics.
\item
A linearized theory in the case of small dislocation densities is derived. We study how the defect content of the lattice structure affects the linearized dynamics of dislocations.
\end{itemize}
%---------------------

This paper is organized as follows.
In~\S\ref{Sec:Lattice} we review nonlinear plasticity, and introduce the concept of distorted lattice structure in the material manifold, given by a frame field representing the underlying crystalline microstructure.
In~\S\ref{Sec:Dislocations} we define decomposable dislocation fields, and discuss some convenient decompositions.
We also study the case of layered dislocation fields, and introduce the notion of integrability of the slip planes.
\S\ref{Sec:Kinematics} is devoted to the kinematic description for the internal variables in terms of some evolution equations.
Dislocation fields are assumed to be convected by a material motion, while the lattice differential forms evolve according to Orowan's equation.
We study the glide motion and its relations with the integrability of slip plane distributions.
In~\S\ref{Sec:Variational} we introduce the variational formulation, using an action principle of the Lagrange-d'Alembert type and a a two-potential approach.
We study the geometric constraints that the lattice puts on the dislocation fields, and their effect on the equations of motion for the dislocation fields.
We also derive the balance of energy.
In~\S\ref{Sec:Problem} we introduce a simplified model for nonlinear dislocation mechanics.
In particular, we assume a purely hyperelastic free energy and derive an expression for the Peach-Koehler force. We also propose a penalty approach to include the effect of the Peierls stress in the dissipation potential.
In~\S\ref{Sec:Linearization} we formulate a linearized theory and look at how the initial lattice structure affects the glide of dislocation fields.
Conclusions are given in \S\ref{Sec:Conclusions}.

%--------------------------------
%--------------------------------
\paragraph{Notation.}
Given a manifold $\mathcal B$, we denote with $T\mathcal B$ the union of all tangent spaces $T_X\mathcal B$ for $X\in\mathcal B$.
Given a diffeomorphism $f$ of manifolds, we indicate with $f_*$ and $f^*$ the pushforward and the pullback operators, respectively.
We denote differential forms and vector and tensor fields using bold letters.
Frames, coframes and all triplets of fields are denoted with Greek symbols and curly brackets, e.g., $\{\boldsymbol f_{\nu}\}$, where $\nu=1,2,3$ is implied.
Coordinate functions are denoted as in $(Z^{\nu})$.
Dislocation fields and associated quantities are indexed using gothic symbols, e.g., the dislocation velocities $\prs{a}{\boldsymbol{\vel}}$.
With an abuse of notation, this might indicate a single field or the whole collection depending on the context.
We use Einstein's summation convention for lattice components and Greek indices, but not for gothic indices.
The symbol $\delta_{\mu\nu}$ as well as $\delta_{AB}$ denotes Kronecker's delta, while $\delta$ without indices is used to denote variations and perturbations.
Pairings of $1$-forms with vectors are denoted as $\langle \boldsymbol \gamma, \boldsymbol V \rangle$, and in components $\gamma_A V^A$.
It extends to tensors of any order.
We also denote with $\langle \boldsymbol A, \boldsymbol B \rangle$ the natural pairing of dual objects $\boldsymbol A$ and $\boldsymbol B$, such as tensor contraction or a form-multivector pairing.\footnote{The pairing of a $k$-form with a $k$-multivector can be seen as a tensor contraction operated on the $3\choose k$ independent index combinations.}
The scalar product associated with a metric $\boldsymbol M$ is denoted by $\llangle, \rrangle_{\boldsymbol M}$, and in components $\llangle \boldsymbol V,\boldsymbol W \rrangle_{\boldsymbol M}= V^A W^B M_{AB}$.
The raising and lowering of indices via a metric is denoted with the musical operators~$^{\sharp}$ and~$^{\flat}$, where the metric used is implied (usually the material metric $\boldsymbol G$).
Given an operator $\op A$, its dual is denoted with $\op A^{\star}$, and is such that $\langle \boldsymbol\gamma, \op A \boldsymbol V \rangle = \langle \op A^{\star} \boldsymbol\gamma, \boldsymbol V \rangle$.
It should not be confused with the adjoint (transpose), that is a metric-dependent notion, i.e., $\llangle \op T \boldsymbol V,  \boldsymbol W \rrangle =\llangle \boldsymbol V, \op T^{\top} \boldsymbol W \rrangle$.
As differential forms are mainly considered in the context of exterior algebra, we use the same symbol $0$ for the zero form. Instead, when treating tensors, such as vectors and operators, a zero tensor is denoted with $\boldsymbol 0$.
The wedge operator $\wedge$ is the exterior product of forms.
$\iota_{\boldsymbol X}$ is the interior product of a form with the vector $\boldsymbol X$, and
$\mathrm d$ denotes the exterior derivative of differential forms.
The derivative of a scalar $f$ along the vector $\boldsymbol V$ is denoted with $\langle \mathrm d f, \boldsymbol V \rangle$.
The advantage of using differential forms is due to the fact that one can reduce the methods of vector calculus to the exterior algebra of differential forms, which is a metric-free description.
This is particularly important in the case of non-Euclidean solids, whose natural distances cannot be represented by the standard metric in $\mathbb R^3$.
However, there are strong analogies between exterior and vector calculi.
For example, the wedge product between two $1$-forms works exactly as a cross product of vectors.
Also, replacing a $2$-form with the axial vector associated with it, its exterior product with a $1$-form becomes similar to a scalar product, while its interior product with a vector works as a cross product.
Moreover, a closed differential $1$-form is the analogue of an irrotational vector field, while a closed $2$-form can be associated with a divergence-free vector field.
We discuss all this in detail in~\S\ref{App:Differential}.

%-----------------------------------------------------------------------------------------
%-----------------------------------------------------------------------------------------
\section{The lattice structure} \label{Sec:Lattice}

In addition to the multiplicative decomposition of the deformation gradient \citep{sadik2017origins}, anelasticity, and in particular plasticity, can be formulated using differential forms \citep{yavari2012riemann}.
This is a natural formulation as it allows the study of dislocation plasticity through the use of exterior algebra. 
In this section we review some concepts of finite plasticity, and provide some insight on the notion of lattice structure.
Starting from the multiplicative decomposition of the deformation gradient, we show that at the continuum scale a crystal can be modeled as a material manifold endowed with a triplet of differential $1$-forms representing the underlying crystalline microstructure, and providing information on the distribution of defects.

%-----------------------------------------------------------------------------------------
%-----------------------------------------------------------------------------------------
\subsection{The multiplicative decomposition of the deformation gradient} \label{Multiplicative}

We work in the framework of continuum mechanics and consider smooth embeddings $\varphi:\mathcal B \to \mathcal S$ representing configurations of a three-dimensional material body $\mathcal B$ in the three-dimensional ambient space $\mathcal S$.
The ambient space is endowed with a Euclidean metric $\boldsymbol g$, expressing the standard scalar product defining distances and angles in the ambient space.
In a continuum theory, crystalline solids carry additional information about the order with which particles are arranged in the discrete lattice, e.g., directions of periodicity, crystallographic symmetries, etc.
We will be referring to this information as \emph{undistorted lattice structure}. 
The fundamental idea in modeling plasticity is that, during motion, the lattice structure does not deform via the macroscopic motion, i.e., the mapping that takes the material points to their current placements. This is based on the fact that plastic slip leaves the crystalline order unaltered, so that the only deformation that the lattice structure undergoes is by definition the elastic one.
More precisely, the deformation gradient $\op F=T\varphi$ is multiplicatively decomposed into plastic and elastic parts as $\op F = \sube{\op{F}} \subp{\op{F}}$, where $\subp{\op{F}}$ is a tensor field on $\mathcal B$, and $\sube{\op{F}}$ is a two-point tensor field, both of type $(1,1)$, i.e., $\subp{\op{F}}(X):T_X\mathcal B \to T_X\mathcal B$, and $\sube{\op{F}}(X):T_X \mathcal B \to T_{\varphi(X)} \mathcal S$ for all points $X\in\mathcal B$.\footnote{For discussions on the reverse decomposition see \citep{Clifton1972,Lubarda1999,YavariSozio2023FeFa}.} Both $\subp{\op{F}}$ and $\sube{\op{F}}$ are assumed invertible and orientation-preserving.

\begin{figure}[tp!]
\centering
\includegraphics[width=.95\textwidth]{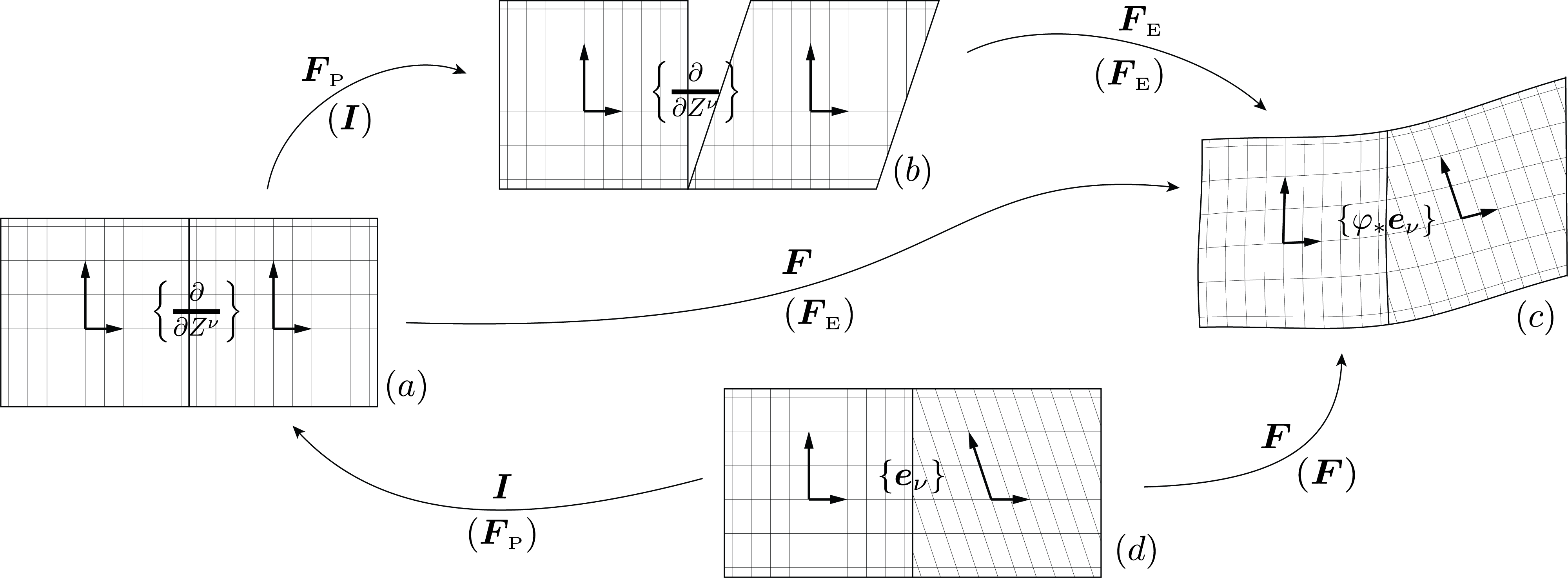}
\vskip 0.1in
\caption{An example of (discontinuous) plastic and elastic deformations.
A parallelogram indicates an infinitesimal piece of material where the deformation can be assumed homogeneous. The lighter grid indicates the lattice structure in every chunk of material.
Each arrow is labeled with a map of material points and, in parentheses, a map for the lattice structure.
The classical multiplicative decomposition of deformation gradient is described by the configurations (a), (b) and (c), while the geometric formalism only uses the configurations (d) and (c).
Note that the lattice structure in both (d) and (c) is incompatible at the interface, whence the presence of defects.
}
\label{fig:plasticity}
\end{figure}

The decomposition $\op F = \sube{\op{F}} \subp{\op{F}}$ is to be interpreted in the following way: starting from an undistorted body, material points plastically deform via $\subp{\op{F}}$ with respect to fixed lattice directions, and then the entire ensemble of ``slipped material points and lattice structure'' is mapped via $\sube{\op{F}}$ to the deformed configuration, see Fig.~\ref{fig:plasticity}.
For our purposes, a periodic lattice structure on $\mathcal B$ can be represented by a Cartesian frame.
Although $\mathcal B$ has not been endowed with any metric yet, Cartesian coordinates $(Z^{\nu})$ can be pulled back from the ambient space together with the standard metric via the use of a reference configuration map. 
More precisely, we take $Z^{\nu} = z^{\nu}\circ\kappa$, where $( z^{\nu} )$ are some Cartesian coordinates on $\mathcal S$, and $\kappa:\mathcal B\to\mathcal S$ is an embedding that fixes a reference configuration for $\mathcal B$.
These coordinates induce a Cartesian frame $\left\lbrace \frac{\partial}{\partial Z^{\nu}} \right\rbrace$ and coframe $\{\mathrm d Z^{\nu}\}$, representing the undistorted lattice structure.
Clearly, $\left\lbrace \frac{\partial}{\partial Z^{\nu}} \right\rbrace$ is orthonormal with respect to the Euclidean metric $\kappa^*\boldsymbol g$ on $\mathcal B$, representing distances in the ideal lattice.
By assumption, the lattice structure is mapped to the deformed configuration via the elastic deformation $\sube{\op{F}}$, to obtain the \emph{deformed lattice structure} $\left\lbrace \sube{\op{F}} \frac{\partial}{\partial Z^{\nu}} \right\rbrace$.

%---------------------
%---------------------
%---------------------
%---------------------

\begin{figure}[tp!]
\centering
\includegraphics[width=.75\textwidth]{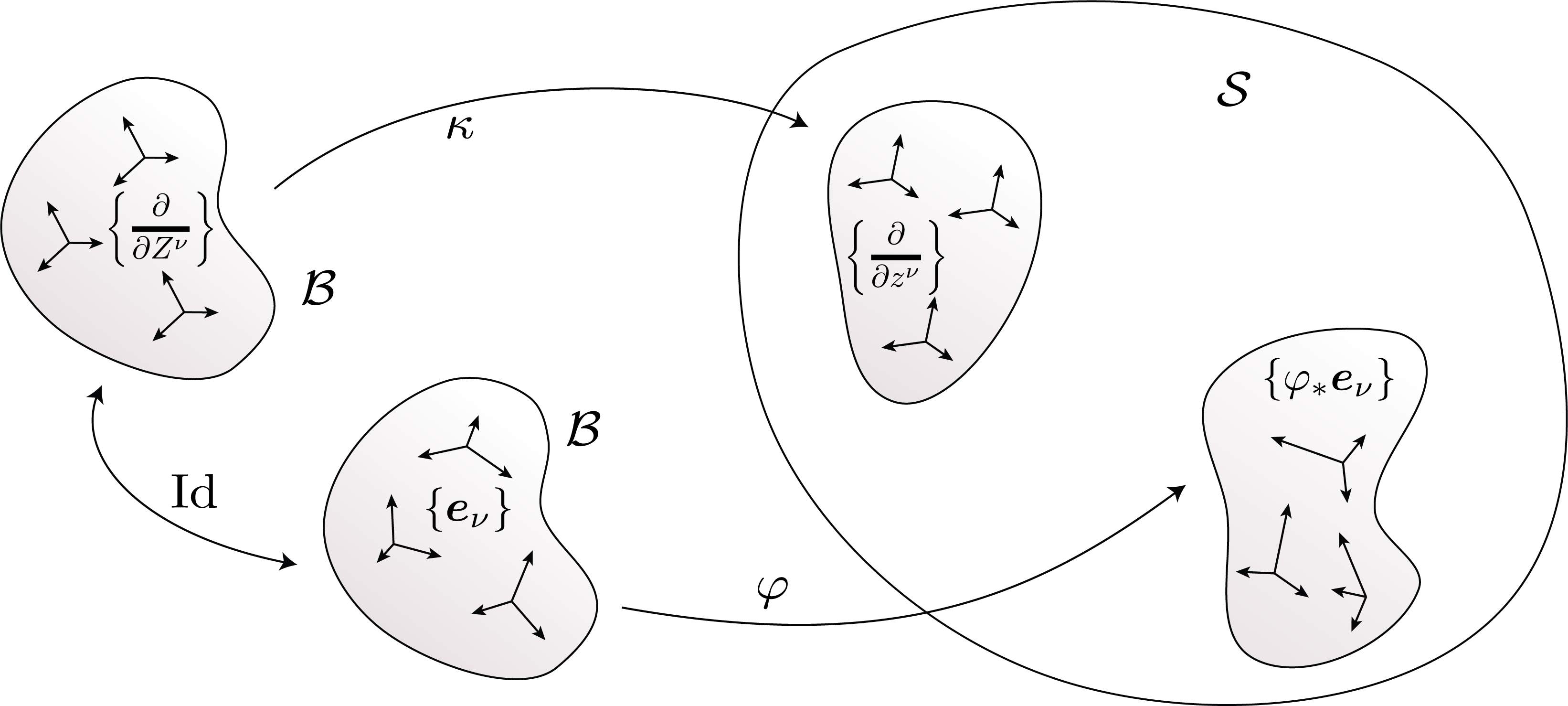}
\vskip 0.1in
\caption{Moving frames representing different lattice structures in a solid. The material manifold $\mathcal B$ is endowed with an undistorted lattice structure represented by the frame $\left\{ \frac{\partial}{\partial Z^{\nu}} \right\}$, which is obtained by pulling back a Cartesian frame $\left\{ \frac{\partial}{\partial z^{\nu}} \right\}$ in $\mathcal S$ via a reference embedding, and with a distorted lattice structure represented by $\{ \boldsymbol e_{\nu} \}$. The deformed lattice structure $\{ \varphi_* \boldsymbol e_{\nu} \}$ on $\varphi(\mathcal B)$ can be thought of as obtained by applying $\sube{\op{F}}$ to $\left\{ \frac{\partial}{\partial Z^{\nu}} \right\}$ or by applying $\op F$ to $\{ \boldsymbol e_{\nu} \}$.}
\label{fig:frames}
\end{figure}

In the geometric approach, the decomposition $\op F = \sube{\op{F}} \subp{\op{F}}$ is seen in the opposite way: the lattice structure is first deformed by $\subp{\op{F}}^{-1}$ with respect to fixed material points to give the distorted lattice structure, and then the ensemble of ``material points and distorted lattice structure'' is mapped to the deformed configuration via a compatible $\op F = T\varphi$, see Fig.~\ref{fig:plasticity}.
The distorted lattice structure is then represented by the so-called lattice frame, a moving frame $\{\boldsymbol e_{\nu}\}$ on $\mathcal B$ defined as
%---------------------
\begin{equation} \label{lattice-frame}
	\boldsymbol e_{\nu} =\varphi^* \left( \sube{\op{F}} \tfrac{\partial}{\partial Z^{\nu}}\right)
	\,,\quad\text{or}\qquad
         \boldsymbol e_{\nu} = \subp{\op{F}}^{-1} \tfrac{\partial}{\partial Z^{\nu} } \,,
\end{equation}
%---------------------
or by the associated lattice coframe $\{\boldsymbol\vartheta^{\nu} \}$, a field of three $1$-forms defined as
%---------------------
\begin{equation} \label{lattice-coframe}
	\boldsymbol \vartheta^{\nu} =\varphi^* \left( \sube{\op{F}}^{\star} \mathrm d Z^{\nu} \right)
	\,,\quad\text{or}\qquad
         \boldsymbol \vartheta^{\nu} = \subp{\op{F}}^{\star} \mathrm d Z^{\nu} \,,
\end{equation}
%---------------------
and such that $\langle\boldsymbol \vartheta^{\nu},\boldsymbol e_{\mu}\rangle=\delta^{\nu}_{\mu}$.
All the different frames defined so far are shown in Fig.~\ref{fig:frames}.
In this geometric approach, the lattice frame $\{\boldsymbol e_{\nu}\}$ is material, in the sense that it is mapped to the deformed lattice frame $\left\lbrace \sube{\op{F}}\frac{\partial}{\partial Z^{\nu}} \right\rbrace=\{\varphi_* \boldsymbol e_{\nu}\}$ via the configuration map $\varphi$ by virtue of~\eqref{lattice-frame}.
In other words, there is no difference between material points and lattice structure in regard to the way they are mapped to~$\mathcal S$.
Eq.~\eqref{lattice-frame} can also be written as a change of frame from Cartesian to lattice frame, viz.
%---------------------
\begin{equation} \label{change-frame}
	\boldsymbol \vartheta^{\nu} = \mathbb{F}^{\nu}{}_{\mu} \mathrm d Z^{\mu}
	\,,\quad
	\boldsymbol e_{\nu} = (\mathbb{F}^{-1})^{\mu}{}_{\nu} \,\tfrac{\partial}{\partial Z^{\mu} } \,,
\end{equation}
%---------------------
where $\mathbb{F}^{\nu}{}_{\mu}$'s are the components of $\subp{\op{F}}$ with respect to both the Cartesian coordinates and the lattice frames:
%---------------------
\begin{equation} \label{active-passive}
	\subp{\op{F}}=\mathbb{F}^{\nu}{}_{\mu} \, \tfrac{\partial}{\partial Z^{\nu} } \otimes \mathrm d Z^{\mu} = \mathbb{F}^{\nu}{}_{\mu} \, \boldsymbol e_{\nu} \otimes \boldsymbol \vartheta^{\mu} \,.
\end{equation}
%---------------------
This means that the lattice coframes carry direct information on $\subp{\op{F}}$.
In order to represent the natural distances in the lattice, a material metric $\boldsymbol G$ on $\mathcal B$ is defined as the one that makes the lattice frame $\{\boldsymbol e_{\nu}\}$ orthonormal, viz.
%---------------------
\begin{equation} \label{material-metric}
	\boldsymbol G = \delta_{\mu\nu} \, \boldsymbol\vartheta^{\mu} \otimes \boldsymbol\vartheta^{\nu}  \,.
\end{equation}
%---------------------
By doing so, vectors with constant components with respect to the frame $\{\boldsymbol e_{\nu}\}$ preserve their length regardless of the evolution of the plastic deformation.
Note that from~\eqref{material-metric}, since the Cartesian frame is orthonormal with respect to $\kappa^*\boldsymbol g $, one has
%---------------------
\begin{equation} \label{material-metric2}
	\llangle\boldsymbol V, \boldsymbol W \rrangle_{\boldsymbol G} = 
	\llangle\subp{\op{F}} \boldsymbol V, \subp{\op{F}}\boldsymbol W \rrangle_{\kappa^*\boldsymbol g}\,,
\end{equation}
%---------------------
for all vectors $\boldsymbol V, \boldsymbol W$.
Therefore, the plastic deformation $\subp{\op{F}}$ is a local isometry from $(T\mathcal B,  \boldsymbol G)$ to $(T\mathcal B, \kappa^* \boldsymbol g)$. 

The volume form $\mvf=\boldsymbol\vartheta^1\wedge\boldsymbol\vartheta^2\wedge\boldsymbol\vartheta^3$ associated with $\boldsymbol G$ is called the material volume form.
The material mass density $\massd$ is a scalar on $\mathcal B$ that defines the mass $3$-form $\boldsymbol{m} = \massd \mvf$.
The Levi-Civita connection associated with $\boldsymbol G$ is denoted with $\nabla$.
In addition to the material metric $\boldsymbol G$ and the Euclidean metric $\kappa^*\boldsymbol g$,
one can define a Riemannian metric on $\mathcal B$ by pulling back the ambient space metric $\boldsymbol g$ via the configuration mapping $\varphi$.
We denote with $\boldsymbol{C}^{\flat}=\varphi^* \boldsymbol g$ this pulled-back metric, while $\op C=(\varphi^* \boldsymbol g)^{\sharp} = \op F^{\top}\op F$ is the right Cauchy-Green strain.

%---------------------
%---------------------
%---------------------
%---------------------

%-----------------------------------------------------------
%-----------------------------------------------------------
\subsection{The distorted lattice} \label{Defects}

The Cartesian frame $\left\lbrace \frac{\partial}{\partial Z^{\nu}} \right\rbrace$ on $\mathcal B$ was introduced as a descriptor of the undistorted lattice structure, and can be viewed as a homogenized representation of the translational symmetry of a periodic lattice.
Translations induced by these vectors are commutative: an ordered sequence of $a$ steps along the coordinate $Z^{\mu}$ followed by $b$ steps along $Z^{\nu}$, gives the same result as the reversed sequence does.
In the deformed lattice structure defined by the moving frame $\{\varphi_* \boldsymbol e_{\nu}\}$ on $\varphi(\mathcal B)$,
the commutativity of translations does not necessarily hold: a translation along $\varphi_* \boldsymbol e_{\mu}$ followed by a translation along $\varphi_* \boldsymbol e_{\nu}$ is not the same operation as in the reversed sequence.\footnote{%
In general, translations along a vector field can be defined as translations along its integral curves.
Given a vector field $\boldsymbol V$, an integral curve $\gamma:\mathbb R\to \mathcal B$ is such that its velocity is $\boldsymbol V \circ \gamma$.}
The difference between the undistorted lattice structure on $\mathcal B$ and the deformed lattice structure on $\varphi(\mathcal B)$ is due to the fact that while the integral curves of $\left\lbrace \frac{\partial}{\partial Z^{\nu}} \right\rbrace$ are also coordinate curves for $( Z^{\nu} )$, forming a grid where the commutativity of translations is clear, this is not the case for the deformed lattice vectors $\{ \varphi_* \boldsymbol e_{\nu} \}$.
This commutative property of translations along the vectors of a frame is equivalent to holonomicity, which is the property of a frame being induced by local coordinates.

Holonomicity is not affected by pullbacks, and therefore, the dislocation content of $\{ \varphi_* \boldsymbol e_{\nu} \}$ is encoded in the anholonomicity of the lattice frame $\{ \boldsymbol e_{\nu} \}$, that we introduced as the descriptor of the distorted lattice structure.
More precisely, a moving frame $\{ \boldsymbol e_{\nu} \}$ is holonomic if there exist local coordinates $( Y^{\nu} )$ such that $\boldsymbol e_{\nu}=\frac{\partial}{\partial Y^{\nu}}$.
This is equivalent to vanishing of the Lie bracket $[\boldsymbol e_{\mu} , \boldsymbol e_{\nu} ]$ for all $\mu,\nu$.\footnote{
Cf. \citep{Sternberg1999, iliev2006handbook,  schouten2013ricci}. In \citep{spivak1970comprehensive} (vol. I, Chapter 5, Theorem 16) it is also shown that $[ \boldsymbol e_{\mu} , \boldsymbol e_{\nu} ]$ represents a second-order approximation to the gaps generated by non-commutative translations along $\{\boldsymbol e_{\nu}\}$.}
This can also be expressed in terms of its coframe as $\boldsymbol\vartheta^{\nu} =\mathrm d Y^{\nu}$, which is equivalent to requiring that the lattice forms be closed.
Since a closed differential form can be seen as locally exact,
the existence of local coordinates $( Y^{\nu} )$ such that $\boldsymbol \vartheta^{\nu}=\mathrm dY^{\nu} $ is guaranteed whenever the lattice forms are closed.
As a matter of fact, invoking~\eqref{der-1-form} one obtains
%---------------------
\begin{equation} \label{exterior-torsion}
	\mathrm d \boldsymbol\vartheta^{\eta} (\boldsymbol e_{\mu},\boldsymbol e_{\nu}) =
	-\langle \boldsymbol\vartheta^{\eta}  , [\boldsymbol e_{\mu},\boldsymbol e_{\nu}] \rangle
	\,.
\end{equation}
%---------------------
At the discrete level, the lack of commutativity of translations along the deformed lattice vectors is due to the presence of dislocations.
Therefore, the $2$-forms $\mathrm d \boldsymbol \vartheta^{\nu}$ are the descriptors for the presence of distributed dislocations in the continuous setting.
In particular, the solid is dislocation-free if and only if the lattice forms are closed.

The presence of distributed dislocations can be detected by calculating the circulation of the lattice coframe along a closed curve~$\gamma$, viz.
%---------------------
\begin{equation} \label{burgers-curve}
	\mathsf{B}^{\nu}(\gamma)=
	\int_{\varphi(\gamma)} \varphi_* \boldsymbol\vartheta^{\nu} =
	\int_{\gamma} \boldsymbol\vartheta^{\nu} =
	\int_{\gamma} \mathbb{F}^{\nu}{}_{\mu} \mathrm d Z^{\mu} \,.
\end{equation}
%---------------------
The scalars $\mathsf{B}^{\nu}(\gamma)$ are usually called the components of the Burgers vector associated with $\gamma$.\footnote{\label{Burger-Note}Technically, $\mathsf{B}^{\nu}(\gamma)$'s are not the components of a vector, as they are not ``attached'' to any point. They are simply three numbers associated with a closed curve, see \citep{sozio2020riemannian}.
See also \citep{ozakin2014affine}.}
When the closed curve $\gamma$ is the only component of the boundary of a surface $\Sigma$,
from Stokes' theorem~\eqref{stokes}$_1$ one can write~\eqref{burgers-curve} as
%---------------------
\begin{equation} \label{burgers-stokes}
	\mathsf{B}^{\nu}(\gamma)=
	\int_{\Sigma} \incl^*\mathrm d \boldsymbol\vartheta^{\nu} \,,
\end{equation}
%---------------------
where $\incl:\Sigma\hookrightarrow\mathcal B$ denotes the inclusion map.

%---------------------
%---------------------
\begin{remark}
In this setting, the presence of defects is a local notion, in the sense that it does not depend on the topology of the body $\mathcal B$.
Global compatibility, i.e., the existence of global coordinates inducing the lattice frame, requires that the lattice forms be not just closed but exact as well, and is therefore related to the topology of the body \citep{Yavari2013,Yavari2020}, whence the notion of topological defects or charges \citep{kupferman2015metric}.
In particular, for a closed form to be exact, one needs vanishing periods on the generators of the first homology group.
\end{remark}
%---------------------
%---------------------

We define the dislocation density as a triplet of vectors $\{\boldsymbol\alpha^{\nu}\}$ given by $\boldsymbol\alpha^{\nu}=\star^{\sharp} \mathrm d \boldsymbol\vartheta^{\nu}$, where $\star^{\sharp}$ is the raised Hodge operator associated to $\boldsymbol G$ defined in~\S\ref{App:Differential}.
Note that since $\mathrm d \mathrm d \boldsymbol\vartheta^{\nu}=0$, from~\eqref{ext-der-hodge} one necessarily has $\operatorname{Div}\boldsymbol\alpha^{\nu}=0$ for $\nu=1,2,3$, where $\operatorname{Div}$ is the divergence operator induced by the material volume form $\mvf$, see~\S\ref{App:Differential}.
From the divergence theorem~\eqref{stokes}$_2$, the Burgers vector~\eqref{burgers-curve} associated with a closed curve $\gamma=\partial\Sigma$ can now be expressed as the flux of the corresponding vector $\boldsymbol\alpha^{\nu}$ across $\Sigma$, viz.
%---------------------
\begin{equation} \label{burgers-surf}
	\mathsf{B}^{\nu}(\gamma)= 
	\int_{\Sigma} \langle \boldsymbol\nu, \boldsymbol\alpha^{\nu} \rangle \, \boldsymbol\varsigma \,,
\end{equation}
%---------------------
where $\boldsymbol{\nu}$ is the normal $1$-form on $\Sigma$, and $\boldsymbol\varsigma$ is the area $2$-form on $\Sigma$, both induced by $\boldsymbol G$, see~\S\ref{App:Differential}.
We should also mention the existence of a Weitzenb\"ock connection $\nabla^{(\mathrm{W})}$ on $\mathcal B$ defined as the connection that parallelizes the lattice frame $\{\boldsymbol{e}_{\nu}\}$.
The Weitzenb\"ock connection acts as the ordinary derivative of the components of a tensor with respect to the lattice frame \citep{sozio2020riemannian}, whence the vanishing of the Weitzenb\"ock derivative of the material metric $\boldsymbol G$.
The torsion $\boldsymbol T$ of the Weitzenb\"ock connection has the expression $\boldsymbol T = \boldsymbol e_{\nu} \otimes \mathrm d \boldsymbol\vartheta^{\nu}$.
The tensorial version of the dislocation density is defined as $\boldsymbol\alpha = \boldsymbol e_{\nu} \otimes \boldsymbol\alpha^{\nu}=\star^{\sharp} \boldsymbol T$, where the raised Hodge operator acts on the lower indices.
Note that, denoting the extension of the divergence operator to double contravariant tensors with $\operatorname{Div}$, one has
%---------------------
\begin{equation}
	\operatorname{Div}\boldsymbol\alpha =
	\operatorname{Div}(\boldsymbol e_{\nu} \otimes \boldsymbol\alpha^{\nu}) =
	\nabla_{\boldsymbol\alpha^{\nu}} \boldsymbol e_{\nu} \,,
\end{equation}
%---------------------
which in general does not vanish \citep{Sozio2021Nye}.

%---------------------
%---------------------
\begin{remark} \label{Rem:same-dislo}
Two different lattice coframes $\{ \boldsymbol\vartheta^{\nu} \}$ and $\{ \stilde{\boldsymbol\vartheta}^{\nu} \}$ can be such that $\mathrm d \boldsymbol\vartheta^{\nu}=\mathrm d\stilde{\boldsymbol\vartheta}^{\nu}$ for all $\nu$.
As a matter of fact, $\mathrm d\stilde{\boldsymbol\vartheta}^{\nu}=\mathrm d \boldsymbol\vartheta^{\nu}$ if and only if $\stilde{\boldsymbol\vartheta}^{\nu} =\boldsymbol\vartheta^{\nu} + \boldsymbol\kappa^{\nu}$, with $\mathrm d \boldsymbol\kappa^{\nu} = 0$.
In other words, a distribution of defects corresponds to a plastic deformation modulo compatible deformations.
We will see that this has implications in the evolution equations for the internal variables.
\end{remark}
%---------------------
%---------------------

%---------------------
%---------------------
\begin{remark} \label{rem:contorted-aeolotropy}
Two different lattice coframes $\{\boldsymbol\vartheta^{\nu}\}$ and $\{\stilde{\boldsymbol\vartheta}^{\nu}\}$ with different incompatibility content, i.e., $\mathrm d \boldsymbol\vartheta^{\nu} \neq \mathrm d\stilde{\boldsymbol\vartheta}^{\nu}$ and $\boldsymbol T \neq \tilde{\boldsymbol T}$, can induce the same material metric $\boldsymbol G$.
When this happens, they are called metric-equivalent or isometric.
It is straightforward to show that isometric coframes are related as $\stilde{\boldsymbol\vartheta}^{\nu} = \mathbb{Q}^{\nu}{}_{\mu }\boldsymbol\vartheta^{\mu}$, where $\mathbb{Q}^{\nu}{}_{\mu}$ is an orthogonal matrix.
Equivalently, the operators $\subp{\op F}$ and $\subp{\tilde{ \op F}}$ defining the two isometric coframes $\{\boldsymbol\vartheta^{\nu}\}$ and $\{\stilde{\boldsymbol\vartheta}^{\nu}\}$ are related as $\subp{\tilde{ \op F}} = \subp{\op F}  \op Q$, where $\op Q$ is a $\boldsymbol G$-orthogonal operator with the following representation:
%---------------------
\begin{equation}
	\op Q = \mathbb{Q}^{\nu}{}_{\mu } \, \boldsymbol e_{\nu} \otimes \boldsymbol\vartheta^{\mu} =
	\mathbb{Q}^{\nu}{}_{\mu } \, \stilde{\boldsymbol e}_{\nu} \otimes \stilde{\boldsymbol\vartheta}^{\mu} \,.
\end{equation}
%---------------------
%
A state of contorted aeolotropy \citep{noll1967materially} is characterized by a lattice coframe $\{\boldsymbol\vartheta^{\nu}\}$ that induces a Euclidean metric $\boldsymbol G$ while $\mathrm d \boldsymbol\vartheta^{\nu}\neq 0$.\footnote{These are also called impotent dislocations \citep{mura1989impotent}, or zero stress dislocations \citep{yavari2012riemann}.}
This means that the body is allowed to locally relax, meaning that there exist local isometric embeddings, i.e., maps $\varphi$ such that $\boldsymbol G = \varphi^*\boldsymbol g$.
Hence, there exist coordinates $Y^{\nu}=z^{\nu}\circ \varphi$ inducing a coframe $\{\mathrm dY^{\nu}\}$ that is isometric to $\{\boldsymbol\vartheta^{\nu}\}$, i.e., such that the material metric can be written as $\boldsymbol G = \delta_{\mu\nu} \, \mathrm dY^{\mu}\otimes  \mathrm dY^{\nu}$.
Thus, the case of contorted aeolotropy is equivalent to the defect-free case modulo non-uniform $\boldsymbol G$-rotations.
\end{remark}
%---------------------
%---------------------

%---------------------
%---------------------
\begin{remark}
The choice of a Cartesian frame to represent the undistorted lattice structure might suggest that the unit cell of the crystal must be cubic.
This is not the case. As a matter of fact, the primitive directions of periodicity in a crystallographic lattice are represented by generic affine coordinates, that might differ from the Cartesian ones.
Affine coordinates $\tilde{Z}^{\nu}$ on $\mathcal{B}$ can be defined by pulling back affine coordinates on $\mathbb R^3$ via a reference mapping $\kappa$ as $\tilde{Z}^{\nu} = \tilde{z}^{\nu} \circ \kappa$.
The change of coordinates from $Z$ to $\tilde{Z}$ is a linear map that can be written as $\mathrm d \tilde{Z}^{\nu} = \mathbb A^{\nu}{}_{\mu} \mathrm d Z^{\mu}$, where $\mathrm d \mathbb A^{\nu}{}_{\mu} = 0$.
Therefore, for a given plastic deformation $\subp{\op F}$, one can define two different distorted lattice structures associated with $\boldsymbol\vartheta^{\nu}= \subp{\op F}^{\star}  \mathrm d Z^{\nu}$ and $\tilde{\boldsymbol\vartheta}^{\nu}= \subp{\op F}^{\star} \mathrm d \tilde{Z}^{\mu}$, that are related as
%---------------------
\begin{equation} \label{cartesian-affine}
	\tilde{\boldsymbol\vartheta}^{\nu}=
	\subp{\op F}^{\star} \mathrm d \tilde{Z}^{\mu} =
	\mathbb A^{\nu}{}_{\mu} \subp{\op F}^{\star}  \mathrm d Z^{\mu} =
	\mathbb A^{\nu}{}_{\mu} \boldsymbol\vartheta^{\mu} \,.
\end{equation}
%---------------------
We want to show that the lattice coframes $\{\boldsymbol\vartheta^{\nu}\}$ and $\{\tilde{\boldsymbol\vartheta}^{\nu}\}$ are equivalent.
First, we note that they induce the same material metric, and hence the same material Riemannian structure on $\mathcal{B}$.
This follows immediately from the fact that~\eqref{material-metric2} depends only on $\subp{\op F}$, which is the same for both lattice structures.
Second, the two coframes induce the same Weitzenb\"ock connection on $\mathcal{B}$.
This can be proved by showing that the two torsions are equal.
Since $\mathrm d \mathbb A^{\nu}{}_{\mu} = 0$, from~\eqref{cartesian-affine} one has $\mathrm d \tilde{\boldsymbol\vartheta}^{\nu} =  \mathbb A^{\nu}{}_{\mu} \mathrm d \boldsymbol\vartheta^{\mu}$, and therefore,
%---------------------
\begin{equation}
	\tilde{\boldsymbol{T}}= 
	\tilde{\boldsymbol e}_{\mu} \otimes \mathrm d \tilde{\boldsymbol\vartheta}^{\nu} = 
	(\mathbb A^{-1})^{\eta}{}_{\nu} \,\boldsymbol e_{\eta} \otimes \mathbb A^{\nu}{}_{\rho}  
	\,\mathrm d \boldsymbol\vartheta^{\rho} = \boldsymbol e_{\nu} \otimes  
	\mathrm d \boldsymbol\vartheta^{\nu} =\boldsymbol{T} \,.
\end{equation}
%---------------------
Thus, the two lattice coframes induce the same Riemannian structure and are associated with the same dislocation content.
Therefore, as was mentioned earlier in this section, for our purposes a periodic lattice structure can be fully represented by a Cartesian frame.
\end{remark}
%---------------------
%---------------------

%%%%%%%%%%%%%%%%%%%%%%%%%%%%%%%%%
%%%%%%%%%%%%%%%%%%%%%%%%%%%%%%%%%
\section{Distributed Dislocations}       \label{Sec:Dislocations}

In the previous section we showed that the dislocation content associated with a field of plastic deformations $\subp{\op{F}}$ is represented by the triplet of $2$-forms $\{\mathrm d \boldsymbol \vartheta^{\nu}\}$, where $\{\boldsymbol \vartheta^{\nu}\}$ is a triplet of $1$-forms associated with the plastic deformation $\subp{\op{F}}$ and representing the distorted lattice structure.
Next we assume the existence of multiple dislocation fields, each one represented by a triplet of differential $2$-forms $\{ \prs{a}{\dd}^{\nu}{} \}$, $\indx{a}=1,2,\hdots,N$, and write the dislocation content $\mathrm d \boldsymbol \vartheta^{\nu}$ as the sum of these $2$-forms, viz.
%---------------------
\begin{equation} \label{gnd-sum}
	\mathrm d \boldsymbol \vartheta^{\nu} = \sum_{\indx{a}=1}^{N} \prs{a}{\dd}^{\nu} \,.
\end{equation}
%---------------------
It should be emphasized that in our formulation the fundamental objects describing the dislocations in a solid are the $2$-forms $\{ \prs{a}{\dd}^{\nu} \}$.
Their kinematics will be discussed in \S\ref{Sec:Kinematics}.
It should also be noted that the present theory is not statistical, so there is no classification of dislocations into geometrically necessary dislocations  and statistically-stored dislocations.\footnote{%
In works such as \citep{arsenlis1999crystallographic,gurtin2002gradient} the dislocation content of plastic slips (the analogue of our $\mathrm d \boldsymbol \vartheta^{\nu}$ or $\boldsymbol \alpha$) is considered to be the fundamental descriptor of the internal state.
The densities of different types of geometrically necessary dislocations are then deduced from the dislocation content on the basis of some extra assumptions.
Statistically-stored dislocations are defined as those distributions that do not contribute to the total dislocation content.
To this extent, our approach is closer to that of \citet{acharya2001model}, and \citet{sedlavcek2003importance, sedlavcek2007continuum}.}
Dislocation fields are simply seen as single-valued smooth fields whose superposition determines the incompatibility of the lattice structure. In this regard, Eq.~\eqref{gnd-sum} represents a link between the internal variables $\boldsymbol \vartheta^{\nu}$ and $\prs{a}{\dd}^{\nu}$.
Morever, Eq.~\eqref{gnd-sum} implies that the sum of all the dislocation fields must be exact for all $\nu$, i.e., for $N$ given $\prs{a}{\dd}^{\nu}$ there must exist a triplet $\{\boldsymbol \vartheta^{\nu}\}$ satisfying~\eqref{gnd-sum}.
For this to hold, it is sufficient (although not necessary) to enforce exactness of each single distribution $\{ \prs{a}{\dd}^{\nu} \}$ for $\indx{a}=1,2,\hdots,N$, i.e., require $\prs{a}{\dd}^{\nu} = \mathrm d \prs{a}{\boldsymbol\epsilon}^{\nu}$ for some triplet of $1$-forms $\{\prs{a}{\boldsymbol\epsilon}^{\nu} \}$.
In the case of a simply-connected body one can simply require that each individual distribution be closed, i.e., $\mathrm d \prs{a}{\dd}^{\nu}= 0$ for all $\indx{a}$ and $\nu$.
These simplifications will be considered later in this section.

%----------------------------------------------------------------------------------------------------
%----------------------------------------------------------------------------------------------------
\subsection{Decomposable dislocation fields} \label{Decomposable}

In the following, for the sake of simplicity we will be omitting---when possible---the gothic index on quantities associated with a particular dislocation field.
A dislocation field $\{ \dd^{\nu}\}$ is said to be \emph{decomposable} if it can be written as
%---------------------
\begin{equation} \label{decomp}
	\dd^{\nu}= \bv^{\nu} \df \,,
\end{equation}
%---------------------
for some triplet of scalar fields $\{ \bv^{\nu}\}$ and a $2$-form $\df$ that we call \emph{dislocation form}.
The $\bv^{\nu}$'s define a vector field $\boldsymbol{\bv} = \bv^{\nu} \boldsymbol e_{\nu}$, that we call \emph{Burgers director}.\footnote{For $\indx{a}=1,2,..,N$, the $\prs{a}{\boldsymbol{\bv}}$'s are fields on $\mathcal B$ and should not be confused with the classical Burgers ``vector'' associated with a curve~\eqref{burgers-curve}, see Footnote~\ref{Burger-Note}. Moreover, while $\mathsf{B}^{\nu}(\gamma)$ refers to the total dislocation content $\mathrm d \boldsymbol\vartheta^{\nu}$, the $\prs{a}{\boldsymbol{\bv}}$'s are associated with the $\indx{a}$-th dislocation field $\prs{a}{\dd}^{\nu}$.}
The integral curves of the vector field $\star^{\sharp} \df$
are called dislocation curves, and are uniquely determined by $\df$.\footnote{%
\label{disl-curves-metric}%
Albeit the raised Hodge operator $\star^{\sharp}$ can be defined with respect to different metric tensors, the dislocation curves are metric-independent.
As a matter of fact, it can be shown that the vector fields obtained from the dislocation $2$-form via $\star^{\sharp}$ differ only by a scalar factor, and hence they define the same integral curves.
In general, a field of $(n-1)$-forms on an $n$-manifold defines a partition of the manifold into a family of curves and the set of points where the form vanishes.
For the definitions of the raised Hodge and interior product see~\S\ref{App:Differential}.}
The scalar field $\dens =\Vert  \star^{\sharp} \df \Vert_{\boldsymbol G} $ is called the scalar dislocation density,
and defines a unit vector $\boldsymbol{\lv}$ as $ \star^{\sharp}\df = \dens \,\boldsymbol{\lv}$, which is called the \emph{dislocation line director}.
From~\eqref{hodge-n-1}, one also has $\df = \dens \, \iota_{ \boldsymbol{\lv}}\mvf$, where $\mvf$ is the volume form introduced in~\S\ref{Multiplicative}.
The Burgers and the dislocation line director span a two-dimensional distribution defined by the $1$-form $\iota_{\boldsymbol{\bv} } \df$, see~\S\ref{App:Frobenius}, that can coincide with a glide plane, see~\S\ref{Layered} and~\S\ref{Glide}.
A dislocation field has a screw character when $\iota_{ \boldsymbol{\bv} }\df =0$, and it has an edge character when $\llangle \boldsymbol{\bv}, \boldsymbol{\lv} \rrangle_{\boldsymbol G} = 0$, which is a metric-dependent condition.
It should be emphasized that an expression of the type~\eqref{decomp} is not unique.
A possible choice for a decomposition consists of using the material metric $\boldsymbol G$ induced by the lattice frame as in~\eqref{material-metric}, and take $\boldsymbol G$-normalized variants of both the Burgers vector density and dislocation form.
In this way, it is possible to write~\eqref{decomp} as
%---------------------
\begin{equation} \label{unit-decomp}
	  \dd^{\nu} =
	  \bv^{\nu} \df
	  \,,\quad\text{such that}\qquad
	  \delta_{\mu\nu} \, \bv^{\mu} \bv^{\nu} = 1
	  \quad\text{or}\qquad
	  \Vert  \boldsymbol{\bv} \Vert_{\boldsymbol G} =1
	  \,.
\end{equation}
%---------------------
Next we consider $N$ decomposable dislocation fields, and refer all the previous quantities to the respective Greek index.
By doing so, one can write~\eqref{gnd-sum} as
%---------------------
\begin{equation} \label{gnd-sum-decomp-1}
	\mathrm d \boldsymbol \vartheta^{\nu} = \sum_{\indx{a}=1}^{N} \prs{a}{\bv}^{\nu} \prs{a}{\df}
	\,.
\end{equation}
%---------------------
Since $\prs{a}{\boldsymbol{\bv}} = \prs{a}{\bv}^{\nu} \boldsymbol e_{\nu}$ and $ \star^{\sharp} \prs{a}{\df} = \prs{a}{\dens} \, \prs{a}{\boldsymbol{\lv} }$ for all $\indx{a}$, one can also obtain the other incompatibility descriptors  $\boldsymbol T = \boldsymbol e_{\nu} \otimes \mathrm d \boldsymbol\vartheta^{\nu}$, $\boldsymbol\alpha^{\nu}=\star^{\sharp} \mathrm d \boldsymbol\vartheta^{\nu}$, and
$\boldsymbol\alpha = \boldsymbol e_{\nu} \otimes \boldsymbol\alpha^{\nu}$, viz.
%---------------------
\begin{equation} \label{gnd-sum-decomp-2}
	\boldsymbol T = \sum_{\indx{a}=1}^N  \prs{a}{\boldsymbol{\bv}} \otimes \prs{a}{\boldsymbol{\df}}
	\,,\quad
	\boldsymbol\alpha^{\nu} = \sum_{\indx{a}=1}^N  \prs{a}{\dens}\,  \prs{a}{\bv}^{\nu} \otimes \prs{a}{\boldsymbol{\lv}}
	\,,\quad
	\boldsymbol\alpha = \sum_{\indx{a}=1}^N  \prs{a}{\dens}\,  \prs{a}{\boldsymbol{\bv}} \otimes \prs{a}{\boldsymbol{\lv}}
	\,.
\end{equation}
%---------------------

We say that a decomposable dislocation field is \emph{distinct} when there exists a decomposition $\dd^{\nu}=\bv^{\nu}\df$ such that $\mathrm d \bv^{\nu} \wedge \df = 0$.
Since $\mathrm d\bv^{\nu} \wedge \df = \dens\,\langle \mathrm d\bv^{\nu},\boldsymbol{\lv} \rangle\,\mvf$ by virtue of~\eqref{wedge-hodge}, this is equivalent to $\langle \mathrm d\bv^{\nu}, \boldsymbol{\lv} \rangle= 0$. 
This means that in the case of distinct dislocation fields the scalar fields $\bv^{\nu}$ are constant along the dislocation curves.
Note that if such a decomposition exists, then one can find infinitely many others simply by rescaling it with a non-vanishing scalar factor that is constant along the dislocation curves.
Therefore, one can always decompose a distinct dislocation field into $\bv^{\nu}\df$ such that i) each $\bv^{\nu}$ has unit $\boldsymbol G$-norm as in~\eqref{unit-decomp}, and ii) each $\bv^{\nu}$ is constant along the dislocation lines.
Let us also note in passing that since the Weitzenb\"ock derivative acts like an ordinary derivate on the components in the lattice frame, if the $\bv^{\nu}$'s are constant along dislocation curves, then the Weitzenb\"ock derivative of $\boldsymbol{\bv} $ along $\lv$ vanishes, i.e., $\nabla^{(\mathrm{W})}_{\boldsymbol{\lv}} \boldsymbol{\bv} =\boldsymbol 0$. 
Finally, a decomposable dislocation field is \emph{uniform} if there exists a decomposition $\bv^{\nu} \df$ such that the scalars $\bv^{\nu}$ are uniform on $\mathcal B$, i.e., such that $\mathrm d \bv^{\nu}=0$.
In other words, the Burgers director field associated with uniform dislocations has the same lattice direction at every point.
This is a common assumption in dislocation dynamics \citep{cermelli2001characterization,gurtin2002gradient,xia2015computational}.
In short, the following classes of dislocation fields have been defined:
%---------------------
\begin{equation}
	\text{Uniform} \subset \text{Distinct} \subset \text{Decomposable} \subset \text{Triplets of $2$-forms} \,.
\end{equation}
%---------------------

%---------------------
\begin{remark}
In the case of a single uniform dislocation field ($N=1$, $\mathrm d \bv^{\nu}=0$),
we assume that for a given closed curve $\gamma=\partial\Sigma$
one can choose a surface $\widetilde\Sigma$ such that i) the Burgers vectors~\eqref{burgers-surf} associated with $\gamma=\partial\Sigma$ and $\widetilde\gamma=\partial\widetilde\Sigma$ are the same, i.e., $\mathsf{B}^{\nu}(\widetilde\gamma)=\mathsf{B}^{\nu}(\gamma)$ (meaning that $\widetilde\Sigma$ crosses the same dislocation curves as $\Sigma$ does), and ii) $\widetilde\Sigma$ is everywhere $\boldsymbol G$-orthogonal to the dislocation lines.
This means that $\boldsymbol{\lv}$ is the unit normal vector on $\widetilde\Sigma$,
and hence the area form $\boldsymbol\varsigma$ on $\widetilde\Sigma$ satisfies $\dens \, \boldsymbol\varsigma = \dens \, \incl^* \iota_{\boldsymbol{\lv}} \mvf = \incl^* \df$, where $\incl: \widetilde\Sigma  \hookrightarrow \mathcal B$ is the inclusion map,
 see~\S\ref{App:Differential}.
Then, one can write the Burgers vector associated with $\gamma$ as
%---------------------
\begin{equation}
	 \mathsf{B}^{\nu}(\gamma)
	 = \int_{\widetilde\Sigma} \bv^{\nu} \, \incl^* \df
	 = \bv^{\nu} \int_{\widetilde\Sigma} \dens \,  \boldsymbol\varsigma \,.
\end{equation}
%---------------------
This shows that $\dens$ represents a Burgers vector density per $\boldsymbol G$-unit area.
Unfortunately, given a family of dislocation curves, such a surface $\widetilde\Sigma$ does not necessarily exist.
The reason for this is that the $1$-form $\star\df$, describing the orientation of the distribution of planes that are normal to the dislocation curves, is not necessarily Frobenius integrable, see~\S\ref{App:Frobenius}.
\end{remark}
%---------------------

%--------------------------------------------------------
%--------------------------------------------------------
\subsection{Closed and exact dislocation fields} \label{Closed-Distributions}

Next we consider the case in which a decomposable dislocation field $\{ \dd^{\nu} \}$ is \emph{closed}, i.e., $\mathrm d \dd^{\nu} = 0$ for all $\nu$.
Let us note in passing that a $2$-form is closed if and only if its corresponding vector obtained through the raised Hodge operator is solenoidal, see~\S\ref{App:Differential}.
In the following lemma the existence of a convenient decomposition for closed decomposable dislocation distributions is discussed.
Recall that distinct dislocation fields were defined as those for which the Burgers director is constant along the dislocation curves.

%-----------------
%-----------------
\begin{lem} \label{lem:closed-decomp}
If a decomposable dislocation field $\{ \dd^{\nu} \}$ is closed, then it is distinct.
Moreover, there exists a decomposition $\dd^{\nu}=\bv^{\nu}\df$ such that the $2$-form $ \df$ is closed and the scalars $\bv^{\nu}$ are constant along the dislocation curves.
In particular, the dislocation field admits a decomposition $\dd^{\nu}=\bv^{\nu}\df$ with $\df$ closed and $\bv^{\nu}$ of unit norm and constant along the dislocation curves.
\end{lem}
%-----------------
\begin{proof}
By assumption $\{ \dd^{\nu} \}$ is decomposable,  i.e., $\dd^{\nu} =\tilde{\bv}^{\nu} \tilde{\df}$ for some $ \tilde{\bv}^{\nu}$ and $\tilde{\df}$.
We look for a scalar $f$ inducing the decomposition $\bv^{\nu}\df$, with $\bv^{\nu}= \tilde{\bv}^{\nu}/f$, $\df =f \tilde{\df} $, and such that $\mathrm d \df= 0$.
One such scalar is $f=\tilde \bv^1$, so that $\df =  \tilde \bv^1 \tilde\df = \tilde \bv^1  \dd^1 / \tilde \bv^1=  \dd^1 $, which is closed by hypothesis.
As for the Burgers director, from~\eqref{ext-der} one obtains
%---------------------
\begin{equation}
	 \mathrm d\bv^{\nu} \wedge \df =
	 \mathrm d (\bv^{\nu} \df ) - 
	 \bv^{\nu}\mathrm d \df = 0 \,,
\end{equation}
%---------------------
as both $\dd^{\nu}$ and $\df$ are closed.
Since $(\mathrm d\bv^{\nu} \wedge \df ) \mvf = \langle \mathrm d\bv^{\nu}, \boldsymbol{\lv} \rangle$, one obtains $\langle \mathrm d\bv^{\nu}, \boldsymbol{\lv} \rangle= 0$, and hence, the scalar fields $\{\bv^{\nu}\}$ are constant along the dislocation curves.
To obtain a Burgers director with unit $\boldsymbol G$-norm it is sufficient to observe that one can replace the decomposition $\bv^{\nu}\df$ with $(\bv^{\nu}/g) (g\,\df)$ for any nowhere vanishing scalar field $g$ that is constant along each dislocation curve, and obtain the same properties derived so far.
Therefore, by setting $g=\Vert\boldsymbol{\bv}\Vert_{\boldsymbol G}$, which is constant along the dislocation curves, one completes the proof.
\end{proof}
%-----------------
%-----------------

Recall that the exactness of the total incompatibility can be enforced by requiring that all dislocation fields in~\eqref{gnd-sum} be exact.
If $\{ \dd^{\nu} \}$ is a triplet of exact forms, then $\dd^{\nu} =\mathrm d \boldsymbol \epsilon^{\nu}$ for some triplet of $1$-forms $\{\boldsymbol \epsilon^{\nu}\}$.
The following lemma is the analogue of Lemma~\ref{lem:closed-decomp} for exact decomposable distributions.

%-----------------
\begin{lem} \label{lem:exact-decomp}
If a decomposable distribution $\{ \dd^{\nu} \}$ is exact, then there exists a decomposition $\dd^{\nu}=\bv^{\nu}\mathrm d \boldsymbol\epsilon$, for some $1$-form $\boldsymbol\epsilon$, with $\bv^{\nu}$ constant along the dislocation curves.
\end{lem}%
\begin{proof}
By assumption $\{ \dd^{\nu} \}$ is decomposable, i.e., $\dd^{\nu} =\widetilde{\bv}^{\nu} \widetilde{\df}$ for some $ \widetilde{\bv}^{\nu}$ and $\widetilde{\df}$, and exact, i.e., $ \dd^{\,\nu}=\mathrm d \widetilde{\boldsymbol\epsilon}^{\,\nu}$ for some $1$-forms $\widetilde{\boldsymbol\epsilon}^{\,\nu}$.
We look for a scalar $f$ inducing the decomposition $\bv^{\nu}\df$, with $\bv^{\nu}= \widetilde{\bv}^{\nu}/f$, $\df = f\widetilde{\df}$, and such that there exists a $1$-form $\boldsymbol\epsilon$ for which $\df = \mathrm d \boldsymbol\epsilon$.
Choosing $f=\widetilde{\bv}^1$, one has $\df = \dd^1$, which is exact by hypothesis.
Therefore, one obtains $\boldsymbol\epsilon=\widetilde{\boldsymbol\epsilon}^{\,1}$,  and the proof is complete.
The Burgers director is constant along the dislocation curves by Lemma~\ref{lem:closed-decomp} as exactness implies closedness.
\end{proof}
%-----------------

By virtue of Lemmas~\ref{lem:closed-decomp} and~\ref{lem:exact-decomp},
in the remaining of the paper we will assume decompositions with unit norm as in~\eqref{unit-decomp}.
Unfortunately, in the case of exact dislocation fields it is not possible to write a decomposition of the type $\bv^{\nu}\mathrm d \boldsymbol\epsilon $ with unit $\bv^{\nu}$ without further assumptions on $\boldsymbol\epsilon$.\footnote{%
Starting from a decomposition $\bv^{\nu}\mathrm d \boldsymbol\epsilon $ as in Lemma~\ref{lem:exact-decomp},
in order to obtain unit $\bv^{\nu}$ one would need $\mathrm d \Vert\boldsymbol{\bv}\Vert \wedge \boldsymbol\epsilon = 0$, which in general does not hold.
%(they would also be constant along the dislocation lines as exactness implies closedness).
Geometrically, this means that the $1$-form $\boldsymbol\epsilon$ (serving as a dislocation potential) must define a plane distribution that is tangent to the level surfaces of $\Vert\boldsymbol{\bv}\Vert $.
On the other hand, by Lemma~\ref{lem:closed-decomp} $\Vert\boldsymbol{\bv}\Vert $ is constant along the dislocation curves defined by $\df=\mathrm d \boldsymbol\epsilon$.
This means that in order to have  $\bv^{\nu}  \mathrm d \boldsymbol\epsilon = \mathrm d ( \bv^{\nu} \boldsymbol\epsilon)$ the dislocation curves need to be tangent to the plane distribution defined by $\boldsymbol\epsilon$.
This can be expressed by the requirement $\mathrm d \boldsymbol\epsilon \wedge \boldsymbol\epsilon =0$, which is the Frobenius integrability condition for $\boldsymbol\epsilon$.
Therefore, one has $\dd^{\nu}=\bv^{\nu}\mathrm d \boldsymbol\epsilon$ with unit $\bv^{\nu}$ if and only if $\boldsymbol\epsilon$ is Frobenius integrable.}
Lastly, in the case of uniform dislocation fields all these issues become trivial, as the uniform scalars $\bv^{\nu}$ do not alter the derivatives.
It should also be noticed that if the body does not contain any cavities, then closed $2$-forms are exact as well.
The following is an example of a hollow sphere with a closed dislocation field that is not exact.

%---------------------
\begin{example}[A non-exact closed dislocation field] \label{Ex: exact}
Let $\mathcal B$ be a thick hollow sphere that in spherical coordinates $(R,\Phi,\Theta)$ is the set $\mathcal B=\{(R,\Phi,\Theta) \vert ~ R_i \leq R \leq R_o\}$ for $0<R_i<R_o$.
Let us consider the decomposable distribution $\dd^{\nu}=\bv^{\nu}\df$, with $\bv^{\nu}$ constant scalar fields, and $\df=\sin\Phi \,\mathrm d \Phi \wedge \mathrm d \Theta$, which is well-defined in $\mathcal B$.
The dislocation curves for this distribution are straight radial lines.
It is straightforward to show that $\df$, and hence $\dd^{\nu}$, is closed but not exact, i.e., there does not exist any triplet $\{ \boldsymbol\vartheta^{\nu} \}$ such that $\mathrm d \boldsymbol\vartheta^{\nu} = \dd^{\nu}$.
This shows that radial dislocation lines with uniform Burgers director are not realizable by any plastic deformation field.
\end{example}
%---------------------

%----------------------------------------------------------------------------------
%----------------------------------------------------------------------------------
\subsection{Slip planes and layered dislocation fields}     \label{Layered}

A crystallographic plane is defined up to nonzero factors by three scalars $\{ \pi_{\nu} \}$.
These scalars can be seen as the lattice components of a $1$-form $\boldsymbol \pi = \pi_{\nu} \boldsymbol\vartheta^{\nu}$,
and provide a representation of a distribution $\Pi$ of planes in the dislocated structure on $\mathcal B$, defined as
%---------------------
\begin{equation}
	\Pi = \bigcup_{X\in\mathcal B}
	\Big\lbrace \boldsymbol V_X \in T_X \mathcal B ~\vert~  \langle\boldsymbol \pi_X, 
	\boldsymbol V_X\rangle=0 \Big\rbrace \,,
\end{equation}
%---------------------
see also~\S\ref{App:Frobenius}.
It should be noted that all $1$-forms differing by a nonzero factor provide an equivalent representation of the same distribution.
For example, the Miller index of a crystallographic plane is obtained by choosing the three smallest integers.
Instead, we assume that the triplet $\{ \pi_{\nu} \}$ has unit norm with respect to the metric $\delta_{\mu\nu}$, i.e., that the $1$-form $\boldsymbol \pi$ has unit $\boldsymbol G$-norm.
By doing so, a distribution is defined by a unique differential $1$-form up to a sign.
In a single crystal, the scalars $\pi_{\nu}$ associated with a lattice plane do not change from point to point, whence $\mathrm d \pi_{\nu}=0$.
This is true regardless of the plastic deformation.
This property can also be expressed by the vanishing of the lattice Weitzenb\"ock derivative of $\boldsymbol\pi$, see~\S\ref{Defects}.
However, it should be noted that in the presence of dislocations the exterior derivative $\mathrm d \boldsymbol\pi = \pi_{\nu}\,\mathrm d \boldsymbol\vartheta^{\nu}$ does not vanish, in general, as $\mathrm d \boldsymbol\vartheta^{\nu}\neq 0$.
Moreover, since $\boldsymbol\pi$ has unit $\boldsymbol G$-norm everywhere, the covariant derivative $\nabla \boldsymbol\pi$ is a $2$-form representing how fast the plane $\Pi_X$ rotates as $X$ changes (in Remark~\ref{integr-surfaces} we show that in the integrable case its projection on $\Pi$ gives the second fundamental form).
In short, one has $\nabla^{\text{(W)}} \boldsymbol\pi =\boldsymbol 0$, $\nabla \boldsymbol\pi \neq \boldsymbol 0$, and $\mathrm d \boldsymbol\pi \neq 0$.

Next we look at the special class of dislocation fields that are layered on stacks of slip planes.
This is a common assumption in continuous dislocation dynamics \citep{acharya2001model, sedlavcek2003importance, xia2015computational}, yet it is violated by those distributions that account for climbed dislocations and prismatic loops.
In the case of layered decomposable dislocation fields, the slip $1$-form $\boldsymbol \pi$ can be used to constrain both the Burgers director and the dislocation line director to lie on the slip plane, viz.
%---------------------
\begin{equation} \label{slip-constr}
	\pi_{\nu}\bv^{\nu}=0
	\,,\qquad
	\boldsymbol \pi \wedge \df = 0 \,.
\end{equation}
%---------------------
The conditions~\eqref{slip-constr} are equivalent to $\langle \boldsymbol \pi,\boldsymbol{\bv} \rangle = 0$, and $\langle \boldsymbol \pi,\boldsymbol{\lv} \rangle=0$; they can also be written as $\pi_{\mu}\dd^{\mu}=0$, and $\boldsymbol \pi \wedge \dd^{\nu} = 0$ for all $\nu$.
A dislocation field satisfying~\eqref{slip-constr} is said \emph{layered} (or strongly layered, in order to distinguish it from weakly layered dislocation fields defined in \S\ref{Glide}).
A summary of all the internal variables and their classification is shown in Table~\ref{Table-1}.
Note also that when~\eqref{slip-constr} are satisfied, one has $\iota_{\boldsymbol{\bv}}\df = f \boldsymbol\pi$ for some scalar $f$, with $f=0$ when the dislocation field has a screw character or when it vanishes.
Denoting with $\boldsymbol{\norm}$ the unit vector in $\Pi$ normal to the dislocation curves (and such that the $\boldsymbol G$-orthonormal frame $\{\boldsymbol{\lv},\boldsymbol{\norm},\boldsymbol\pi^{\sharp}\}$ induces the same orientation as $\mvf$),
one has
%---------------------
\begin{equation} \label{norm-dens}
	\iota_{\boldsymbol{\norm}}\df = \dens \, \boldsymbol\pi \,.
\end{equation}
%---------------------

%----------------
%----------------
\begin{table}[tp!]
\centering
\setstretch{1.4} \small
\begin{tabular}{lrlrl}
\toprule
\multicolumn{5}{l}{Plastic deformation: $\subp{\op{F}}$} \\
\multicolumn{5}{l}{Lattice frame and coframe: $\boldsymbol e_{\nu} = \subp{\op{F}}^{-1} \,\frac{\partial}{\partial Z^{\nu}}$ and $\boldsymbol \vartheta^{\nu} = \subp{\op{F}}^{\star} \,\mathrm d Z^{\nu}$ } \\
\multicolumn{5}{l}{Total dislocation content: $\mathrm d \boldsymbol \vartheta^{\nu}  \,,~~ \boldsymbol{T} = \boldsymbol e_{\nu} \otimes \mathrm d \boldsymbol \vartheta^{\nu}  \,,~~ \boldsymbol\alpha^{\nu} = \star^{\sharp} \mathrm d \boldsymbol \vartheta^{\nu}  \,,~~ \boldsymbol{\alpha} = \boldsymbol e_{\nu} \otimes \boldsymbol\alpha^{\nu}  \,,~~  [\boldsymbol e_{\mu},\boldsymbol e_{\nu}] $} \\
\addlinespace[.1in] \midrule \addlinespace[.075in]
\multicolumn{5}{l}{Dislocation fields: $~\prs{1}\dd^{\nu}, \prs{2}\dd^{\nu}, \hdots , \prsn{N}\dd^{\nu}$ ~~($N$ triplets of $2$-forms) } \\
\hspace{0in} & Decomposable: &  $\dd^{\nu}=\bv^{\nu} \df $ & \multicolumn{2}{l}{ $\Longrightarrow ~~ \prs{a}{\boldsymbol{\bv}} = \prs{a}{\bv}^{\nu} \boldsymbol e_{\nu}$  and $\prs{a}{\dens} \, \prs{a}{\boldsymbol{\lv}} =\star^{\sharp} \prs{a}{\df}$ } \\
 & Distinct: & \multicolumn{3}{l}{  $\exists \,\{b^{\nu}\}~\text{such that}~\mathrm d b^{\nu}\wedge\df=0$ } \\
 & Uniform: & \multicolumn{3}{l}{ $\exists \,\{b^{\nu}\}~\text{such that}~\mathrm d b^{\nu}=0$ } \\
 & Closed: &  $\mathrm d \dd^{\nu} = 0$ & Decomposable \& Closed & $\Longrightarrow$~~Distinct \\
\addlinespace[-.05in] & & &  & $\Longrightarrow ~~ \dd^{\nu}=\bv^{\nu} \df$ with $\Vert \boldsymbol{\bv} \Vert_{\boldsymbol G}=1$ and $\mathrm d\df=0$ \\
 & Exact: &  $\dd^{\nu} = \mathrm d \boldsymbol\epsilon^{\nu}$ & \multicolumn{1}{l}{$\Longrightarrow$~~Closed } & \\
 & & & Decomposable \& Exact & $\Longrightarrow$~~$\mathrm d \dd^{\nu}=\bv^{\nu} \df$ with $\df=\mathrm d \boldsymbol\epsilon$
\\
 & Layered: & \multicolumn{3}{l}{$\pi_{\nu}\bv^{\nu}=0\,,~~\boldsymbol \pi \wedge \df = 0$ with $\Pi = \bigcup_{X\in\mathcal B}
	\Big\lbrace \boldsymbol Y_X \in T_X \mathcal B ~\vert~  \langle\boldsymbol \pi_X, \boldsymbol Y_X\rangle=0 \Big\rbrace$} \\
\addlinespace[.1in] \midrule \addlinespace[.075in]
\multicolumn{5}{l}{The lattice structure and the dislocation fields are related as: $~\mathrm d \boldsymbol \vartheta^{\nu} = \sum_{\indx{a}=1}^{N} \prs{a}{\dd}^{\nu} $ } \\
\addlinespace[.1in] & Decomp. case: & \multicolumn{3}{l}{ $
 	\mathrm d \boldsymbol \vartheta^{\nu} = \sum_{\indx{a}=1}^{N} \prs{a}{\bv}^{\nu} \prs{a}{\df}^{\nu}
	\,,~~
	\boldsymbol T = \sum_{\indx{a}=1}^N  \prs{a}{\boldsymbol{\bv}} \otimes \prs{a}{\boldsymbol{\df}}
	\,,~~
	\boldsymbol\alpha^{\nu} = \sum_{\indx{a}=1}^N  \prs{a}{\dens} \, \prs{a}{\bv}^{\nu} \otimes \prs{a}{\boldsymbol{\lv}}
	\,,~~
	\boldsymbol\alpha = \sum_{\indx{a}=1}^N  \prs{a}{\dens} \, \prs{a}{\boldsymbol{\bv}} \otimes \prs{a}{\boldsymbol{\lv}}
	$} \\
\addlinespace[.05in] \bottomrule
\end{tabular}
\caption{Summary of the internal variables and their characterization.}
\label{Table-1}
\end{table}
%----------------
%----------------

As we have just seen, the slip $1$-form $\boldsymbol \pi$ depends on the distorted lattice structure.
So one may ask how the presence of plastic slip affects the geometry of the distribution $\Pi$.
In this regard, \citet{el2000statistical} and \citet{el2007statistical} suggested that the glide planes can deform into 3D surfaces by the effect of finite plastic deformations. 
But what happens when the plastic deformation is not compatible?
Can one still define slip surfaces?
\citet{trzkesowski1997kinematics} addressed this question in the case of a single dislocation field, which as we will see is a trivial case.
In our analysis, we make no assumption on the number of smooth dislocation fields.

\begin{figure}[tp]
\centering
\includegraphics[width=.9\textwidth]{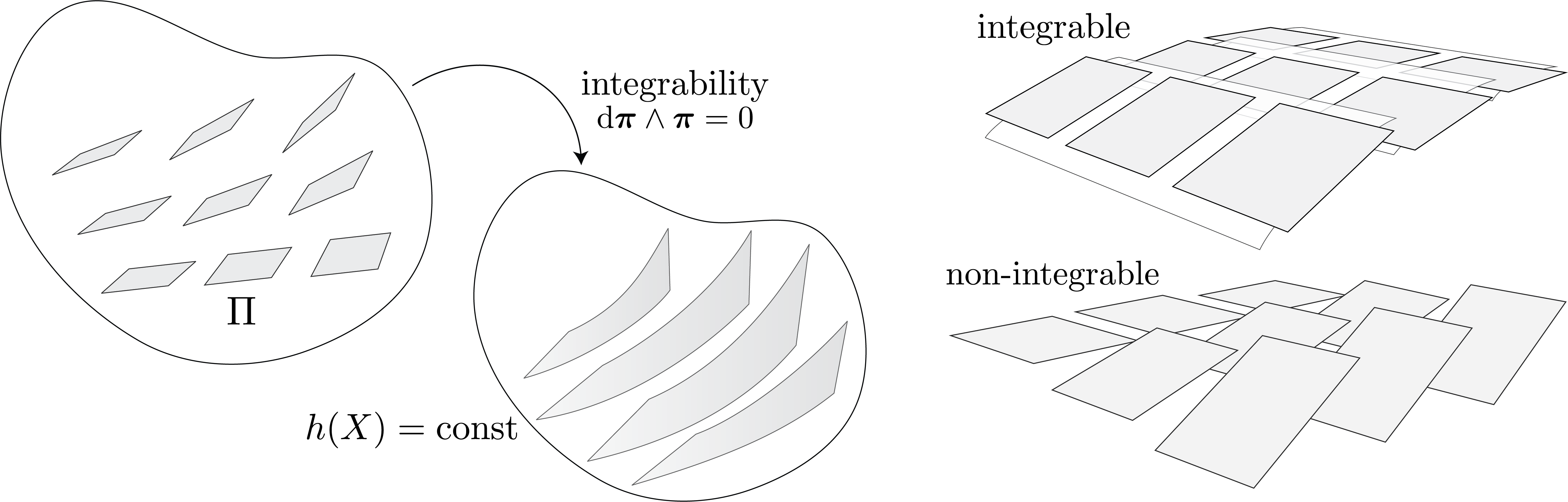}
\vskip 0.1in
\caption{Frobenius integrability of crystallographic plane distributions.
Left: A plane distribution $\Pi$ is defined by a $1$-form $\boldsymbol \pi = \pi_{\nu}\boldsymbol\vartheta^{\nu}$. In the integrable case, the lattice surfaces can be thought as level sets of the scalar function $h$, i.e., $h(X)=\mathrm{const}$, with $\mathrm d h = f \boldsymbol\pi$ for some scalar field $f\neq 0$.
Right: Integrable (top) plane distribution of the type $\boldsymbol\pi = KZ^1 \,\mathrm d Z^1 + \mathrm d Z^3$, and non-integrable (bottom) plane distribution of the type $\boldsymbol\pi = KZ^2 \,\mathrm d Z^1 + \mathrm d Z^3$.}
\label{fig:slips}
\end{figure}

The integrability condition for a plane distribution $\Pi$ defined by the $1$-form $\boldsymbol \pi$ is given by $\mathrm d \boldsymbol \pi \wedge \boldsymbol \pi = 0$ as a consequence of the Frobenius theorem, see \S\ref{App:Frobenius}. Therefore, since $\boldsymbol \pi = \pi_{\nu}\boldsymbol\vartheta^{\nu}$, one obtains the following condition:
%---------------------
\begin{equation}
	\mathrm d \! \left( \pi_{\mu} \,\boldsymbol \vartheta^{\mu} \right) \wedge \pi_{\nu}\, \boldsymbol \vartheta^{\nu} =
	\pi_{\nu}\, \mathrm d \pi_{\mu} \wedge \boldsymbol \vartheta^{\mu} \wedge \boldsymbol \vartheta^{\nu} +
	\pi_{\mu} \,\pi_{\nu}\, \mathrm d \boldsymbol \vartheta^{\mu} \wedge \boldsymbol \vartheta^{\nu} =0 \,.
\end{equation}
%---------------------
Under the assumption of a single crystal, i.e., $\mathrm d \pi_{\nu}=0$, one obtains the integrability condition of the plane distribution $\Pi$ in terms of the lattice forms, viz.
%---------------------
\begin{equation} \label{F-integr}
	\pi_{\mu}\, \pi_{\nu} \,
	 \mathrm d \boldsymbol \vartheta^{\mu} \wedge \boldsymbol \vartheta^{\nu} =0 \,.
\end{equation}
%---------------------
In the case of no net defect content, i.e., when $\mathrm d \boldsymbol \vartheta^{\mu}= 0$, the condition~\eqref{F-integr} is automatically satisfied, and the existence of slip surfaces is guaranteed (see Fig.~\ref{fig:slips}).
This means that a compatible plastic slip deforms the lattice planes without dismantling them. 
Note that from~\eqref{wedge-hodge} one obtains $\mathrm d \boldsymbol \vartheta^{\mu} \wedge \boldsymbol \vartheta^{\nu} = \langle  \boldsymbol \vartheta^{\nu},  \boldsymbol \alpha^{\mu} \rangle \,  \mvf= \alpha^{\mu\nu} \mvf$, where $ \boldsymbol \alpha^{\mu}$ and $ \boldsymbol \alpha$ denote the two variants of the dislocation density tensor defined in~\S\ref{Sec:Lattice}.
Hence, Eq.~\eqref{F-integr} can be written as $\pi_{\mu} \,\pi_{\nu} \,\alpha^{\mu\nu} = \langle \boldsymbol\pi \otimes \boldsymbol\pi ,  \boldsymbol\alpha \rangle=0$.
Next we introduce the integrability object $\mathfrak I_{\Pi}$ associated with the plane distribution $\Pi$ as the scalar field $\mathfrak I_{\Pi}$ defined by $\mathrm d \boldsymbol\pi \wedge \boldsymbol\pi= \mathfrak I_{\Pi} \, \mvf$.
This allows one to write the Frobenius integrability condition simply as $\mathfrak I_{\Pi}=0$.
Note that from~\eqref{hodge-star} one also has $\mathfrak I_{\Pi} = \star(\mathrm d \boldsymbol\pi \wedge \boldsymbol\pi )$, and from what we just showed, $\mathfrak I_{\Pi} = \langle \boldsymbol\pi \otimes \boldsymbol\pi ,  \boldsymbol\alpha \rangle$.
Therefore, invoking the expression~\eqref{gnd-sum-decomp-2}$_3$ for the dislocation density tensor $\boldsymbol\alpha$, one can write the following identities:
%---------------------
\begin{equation} \label{int-object}
	\mathfrak I_{\Pi} =
	\star(\mathrm d \boldsymbol\pi \wedge \boldsymbol\pi ) =
	\langle \boldsymbol\pi \otimes \boldsymbol\pi ,  \boldsymbol\alpha \rangle 
	=\sum_{\indx{a}=1}^N \prs{a}{\dens} \, \langle \boldsymbol\pi, \prs{a}{\boldsymbol{\bv}} \rangle   
	\langle\boldsymbol\pi, \prs{a}{\boldsymbol{\lv}} \rangle \,.
\end{equation}
%--------------------
It should be noticed that $\mathfrak I_{\Pi}$ is invariant under changes of sign---and hence of orientation---of $\boldsymbol\pi$.
The following result relates the integrability of a plane distribution with the dislocation fields that generate the incompatibility of the lattice structure.

%---------------------
%---------------------

%---------------------
%---------------------
\begin{lem} \label{Lemma:int}
The integrability of a plane distribution is controlled by only those dislocation fields whose Burgers director does not belong to the plane distribution.
In particular, layered dislocation fields do not affect the integrability of the plane distribution on which they lie.
\end{lem}
%---------------------
\begin{proof}
This result is an immediate consequence of~\eqref{int-object}, from which the integrability condition~\eqref{F-integr} can be expanded to read
%---------------------
\begin{equation}
	\sum_{\indx{a}=1}^N 
	\prs{a}{\dens} \,
	\langle \boldsymbol\pi, \prs{a}{\boldsymbol{\bv}} \rangle   \langle\boldsymbol\pi, \prs{a}{\boldsymbol{\lv}} \rangle
	= 0 \,,
\end{equation}
%---------------------
where the terms for which $\langle \boldsymbol\pi, \prs{a}{\boldsymbol{\bv}} \rangle=0$ do not contribute to the sum.
In particular, by virtue of~\eqref{slip-constr}$_1$, only the dislocation fields that do not lie on $\Pi$ contribute to the sum.
\end{proof}
%---------------------
%---------------------

As a consequence of Lemma~\ref{Lemma:int}, in the case of a single layered dislocation field, the integrability of its slip plane distribution is automatically guaranteed. This agrees with a result obtained by \citet{trzkesowski1997kinematics}, who studied single dislocation fields, and showed that distributions of slip planes are always integrable.
%

%---------------------
%---------------------
\begin{remark} \label{integr-surfaces}
One can look at the Frobenius integrability of plane distributions in the light of the geometry of surfaces.
First we assume the existence of a crystallographic surface, endowed with the geometry inherited from $\boldsymbol G$.
For this surface the $1$-form $\boldsymbol\pi$ is the unit normal $1$-form, and hence we define the second fundamental form as $\boldsymbol{I\!I}(\boldsymbol V, \boldsymbol W) = -\langle \nabla_{\boldsymbol V} \boldsymbol\pi , \boldsymbol W \rangle$, for all tangent vectors $\boldsymbol V, \boldsymbol W$.
As was mentioned earlier, the single crystal assumption $\mathrm d \pi_{\nu}=0$ implies that the Weitzenb\"ock derivative of $\boldsymbol\pi$ vanishes, and so one can express the second fundamental form in terms of the contorsion tensor $\boldsymbol K$ as\footnote{%
The contorsion tensor $\boldsymbol K$ is defined as the difference between the Weitzenb\"ock connection and the Levi-Civita connection associated to $\boldsymbol G$ \citep{yavari2012riemann, sozio2020riemannian}.}
%---------------------
\begin{equation} 
	\boldsymbol{I\!I}(\boldsymbol V, \boldsymbol W) =
	\langle \boldsymbol\pi , \boldsymbol K(\boldsymbol V, \boldsymbol W)\rangle \,.
\end{equation}
%--------------------
Note that the second fundamental form of a surface is symmetric by construction.
Therefore, as the anti-symmetric part of the contorsion tensor is the torsion tensor $\boldsymbol T$ defined in~\S\ref{Defects}, the symmetry requirement is equivalent to
%---------------------
\begin{equation} \label{int-surface}
	\langle \boldsymbol\pi , \boldsymbol T(\boldsymbol V, \boldsymbol W) \rangle = 0  \,.
\end{equation}
%--------------------
Note that $T^D{}_{BC}\,V^B W^C=\alpha^{DA}\, \comp\mvf_{ABC}\, V^B W^C$,
whereas for tangent vectors one has $\comp\mvf_{ABC}\, V^B W^C = f \pi_A$, for some scalar $f$.
Therefore, \eqref{int-surface} is equivalent to $\pi_D \,\pi_A\, \alpha^{DA} =0$, and hence we have recovered the necessity of~\eqref{F-integr} in the form $\mathfrak I_{\Pi}=0$.
\end{remark}
%---------------------
%---------------------

%---------------------
\begin{figure}[tp]
\centering
\includegraphics[width=.95\textwidth]{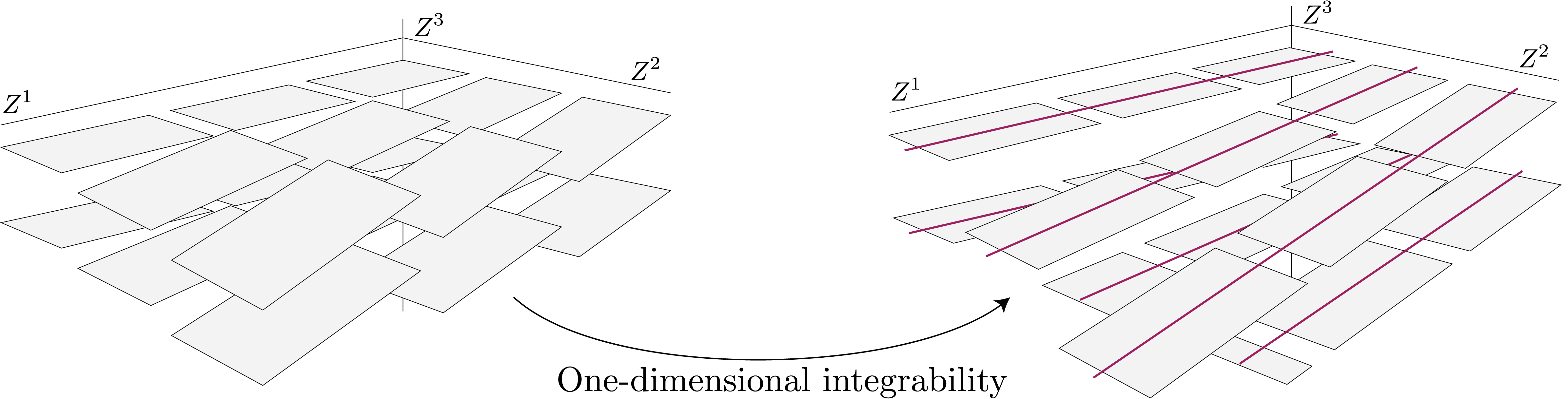}
\vskip 0.2in
\caption{Dislocations on a non-integrable plane distribution. Left: Non-integrable plane distribution $\Pi$ of the type $\boldsymbol\pi = f(Z^2) \, \mathrm d Z^1 +\mathrm d Z^3$ as in Example~\ref{Ex:nonintegrable}.
Right: The dislocation form $\df = \mathrm d Z^2 \wedge \mathrm d Z^3 - f(Z^2) \, \mathrm d Z^1 \wedge \mathrm d Z^2$ layered on $\Pi$.
Although there exists no family of surfaces (foliation) that are tangent to $\Pi$, one can still define dislocation fields that are layered on~$\Pi$. This comes from the fact that one-dimensional distributions are always integrable.}
\label{fig:exint}
\end{figure}
%---------------------

%---------------------
\begin{example}[Dislocations on a non-integrable plane distribution] \label{Ex:nonintegrable}
Setting Cartesian coordinates $( Z^{\nu} )$ on $\mathcal B$, we consider a plane distribution $\Pi$ defined by $\boldsymbol\pi = f (Z^2) \, \mathrm d Z^1 +\mathrm d Z^3$, that we assume is of unit norm with respect to an unspecified material metric $\boldsymbol G$.
Then, the integrability object reads
%---------------------
\begin{equation}
	\mathrm d \boldsymbol\pi \wedge \boldsymbol\pi = -  f'(Z^2) \, \mathrm d Z^1 \wedge \mathrm d Z^2 \wedge \mathrm d Z^3 \,.
\end{equation}
%---------------------
Thus, for non-constant functions $f$ there exists no foliation of $\mathcal B$ into slip surfaces tangent to $\Pi$, see Fig.~\ref{fig:slips}.
Nonetheless, it is still possible to define a (decomposable) dislocation field layered on $\Pi$, for example by taking the dislocation form
%---------------------
\begin{equation}
	\df = \mathrm d Z^2 \wedge \mathrm d Z^3 - f(Z^2) \, \mathrm d Z^1 \wedge \mathrm d Z^2 \,,
\end{equation}
%---------------------
associated with a line director of components $(1,0,-f(Z^2))$.
To see if the dislocation curves locally lie on the slip plane distribution, one simply checks~\eqref{slip-constr}${}_2$, i.e., $\df\wedge \boldsymbol\pi = 0$.
Moreover, it should be noted that $\df$ is closed as $\mathrm d f \wedge \mathrm d Z^1 \wedge \mathrm d Z^2 = 0$.
Therefore, although a non-integrable plane distribution does not admit the existence of surfaces that are tangent to it, it is still possible to define dislocation fields that are layered on it (see Fig.\ref{fig:exint}).
In other words, the non-integrability of a plane distribution does not affect the possibility of stacking dislocations on it.
In~\S\ref{Sec:Kinematics}, we will see that what is affected by non-integrability is the glide motion of dislocations.
\end{example}
%---------------------

\begin{figure}[tp!]
\centering
\includegraphics[width=.95\textwidth]{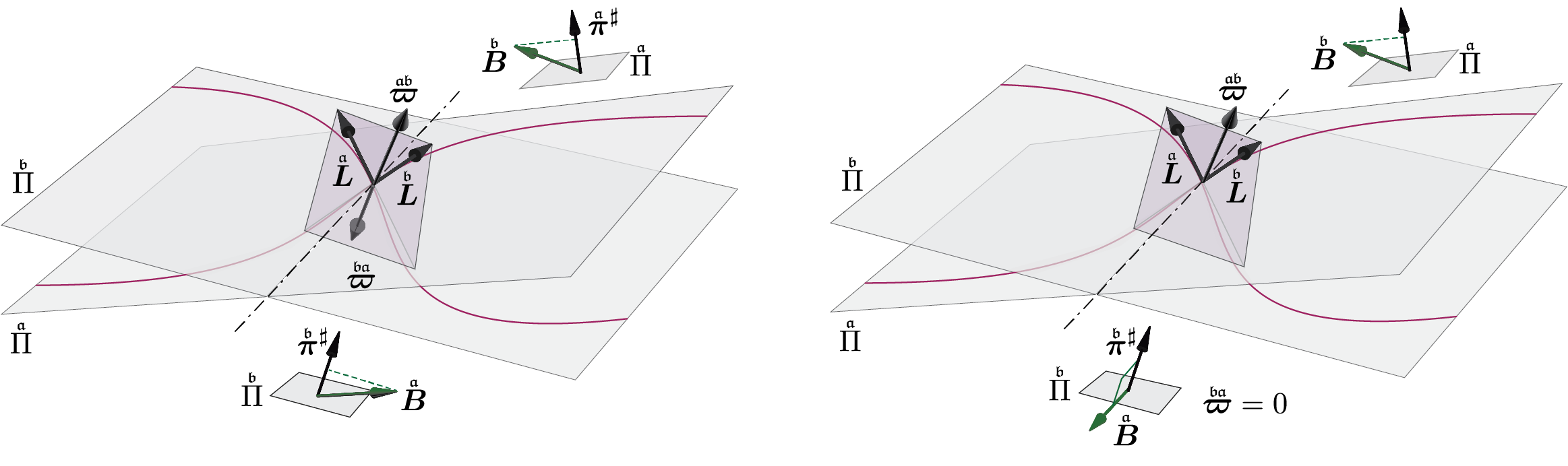}
\vskip 0.2in
\caption{The slip coupling $1$-forms
$\prs{a\!b}{\boldsymbol\varpi}$ and $\prs{b\!a}{\boldsymbol\varpi}$ operate as a triple product with $ \prs{a}{\boldsymbol{\lv}}$ and $ \prs{b}{\boldsymbol{\lv}}$ multiplied $\langle \prs{a}{\boldsymbol\pi} , \prs{b}{\boldsymbol{\bv} } \rangle$ and $\langle \prs{b}{\boldsymbol\pi} , \prs{a}{\boldsymbol{\bv} } \rangle$, respectively.
Left: Both $\langle \prs{a}{\boldsymbol\pi} , \prs{b}{\boldsymbol{\bv} } \rangle$ and $\langle \prs{b}{\boldsymbol\pi} , \prs{a}{\boldsymbol{\bv} } \rangle$---and hence $\prs{a\!b}{\boldsymbol\varpi}$ and $\prs{b\!a}{\boldsymbol\varpi}$---are nonzero.
Right: $\prs{a\!b}{\boldsymbol\varpi}$ is nonzero, but the Burgers director $\prs{a}{\boldsymbol{\bv} }$ lies on $\prs{b}{\Pi}$, and hence $\prs{b\!a}{\boldsymbol\varpi}$ vanishes.
Note that in both cases the slip coupling forms $\prs{a\!b}{\boldsymbol\varpi}$ and $\prs{b\!a}{\boldsymbol\varpi}$ define the plane spanned by $\prs{a}{\boldsymbol{\lv}}$ and $ \prs{b}{\boldsymbol{\lv}}$ (hence they are represented as arrows normal to it).}
\label{fig:coupling}
\end{figure}

Lastly, for every pair of dislocation fields $\indx{a}$ and $\indx{b}$, we define a $1$-form~$\prs{a\!b}{\boldsymbol\varpi}$ as
%---------------------
\begin{equation} \label{cross1}
	\langle \prs{a\!b}{\boldsymbol\varpi} , \boldsymbol V \rangle \, \mvf =
	\langle \prs{a}{\boldsymbol\pi} , \prs{b}{\boldsymbol{\bv} } \rangle \,\,
	\prs{a}{\df} \wedge \iota_{\boldsymbol V } \prs{b}{\df}
	= \prs{a}{\pi}_{\nu} \,	\prs{a}{\df} \wedge \iota_{\boldsymbol V } \prs{b}{\dd}^{\nu}\,,
\end{equation}
%---------------------
for any vector $\boldsymbol X$.
The $1$-form $\prs{a\!b}{\boldsymbol\varpi}$ defines the plane that is locally spanned by the dislocation line directors $ \prs{a}{\boldsymbol{\lv}}$ and $\prs{b}{\boldsymbol{\lv}}$.
Hence, one has $\prs{a\!b}{\boldsymbol\varpi}=f  \prs{b\!a}{\boldsymbol\varpi}$ for some scalar $f$, and $\prs{a\!a}{\boldsymbol\varpi}=0$.
Note that one also has $\langle\prs{a\!b}{\boldsymbol\varpi}, \boldsymbol V \rangle= \prs{a}{\dens} \prs{b}{\dens} \,\langle \prs{a}{\boldsymbol\pi} , \prs{b}{\boldsymbol{\bv} } \rangle  \, \mvf( \boldsymbol V ,\prs{a}{\boldsymbol{\lv}}, \prs{b}{\boldsymbol{\lv}})$, showing that $\prs{a\!b}{\boldsymbol\varpi}$ represents some kind of a triple product with $ \prs{a}{\boldsymbol{\lv}}$ and $ \prs{b}{\boldsymbol{\lv}}$.
The $1$-form $\prs{a\!b}{\boldsymbol\varpi}$ contains information about the component of the $\indx{b}$-th Burgers director normal to the $\indx{a}$-th plane, and is proportional to both scalar dislocation densities. In particular, it is maximum when the $\indx{b}$-th Burgers director is normal to the $\indx{a}$-th plane (see Fig.~\ref{fig:coupling}).
It should be noticed from Lemma~\ref{Lemma:int}, that it is the normal component of the Burgers director with respect to a plane distribution that has an effect on the integrability of the plane distribution.
Hence, $\prs{a\!b}{\boldsymbol\varpi}$ carries information on the influence of the $\indx{b}$-th dislocation field on the integrability of the $\indx{a}$-th plane distribution.
In~\S\ref{Sec:Kinematics}, $\prs{a\!b}{\boldsymbol\varpi}$ will be used to express a condition for dislocation glide, while
in~\S\ref{Sec:Variational} we will show that $\prs{a\!b}{\boldsymbol\varpi}$ is involved in coupling mechanisms between slip systems.
For this reason, we call $\prs{a\!b}{\boldsymbol\varpi}$ the \emph{slip coupling} from $\indx{a}$ to $\indx{b}$.

%---------------------
%---------------------
%---------------------
%---------------------

%%%%%%%%%%%%%%%%%%%%%%%%%%%%%%%%%%%%%%%%%%
%%%%%%%%%%%%%%%%%%%%%%%%%%%%%%%%%%%%%%%%%%
\section{Kinematics}    \label{Sec:Kinematics}

In the previous sections we introduced the lattice $1$-forms and the dislocation field $2$-forms as the internal variables of our geometric model.
These are time-dependent objects, i.e., material fields $\boldsymbol\vartheta^{\nu} (X,t)$ and $\prs{a}{\dd}^{\nu} (X,t) $
depending on an extra independent variable $t\in \mathbb R$.
This section is devoted to the kinematics of the internal variables, which is derived in terms of some evolution equations \citep{hochrainer2007three}.
These equations do not involve the external variables, 
meaning that the evolutions of both the lattice structure and the dislocation fields do not explicitly depend on the spatial configuration and deformations. Later on in the paper we will see that the internal and external variables are coupled via the kinetic equations \citep{rice1971inelastic, lubliner1973structure}.
In particular, each dislocation field will be assumed to be convected by a material motion, representing the movement of the dislocations in the crystal. Under this assumption, the evolution of dislocations is formulated in a geometric setting using the notion of flow, that we review in~\S\ref{App:Flows}.
The evolution of the moving frame is then derived in terms of the dislocation variables, and Orowan's equation is introduced \citep{orowan1940problems}.
We also look at the case in which dislocations are forced to glide on the plane distribution that they are layered on, and study how the lack of integrability of a plane distribution affects the glide motion.
First we review some measures of the rate of deformation in dislocation plasticity.

%-----------------------------------------------------------------
%-----------------------------------------------------------------
\subsection{Rates of deformation}     \label{Rates}

We denote the partial time derivative with $\partial_t$.
In classical multiplicative plasticity, the rate of change of plastic deformation is defined as the following tensors of type $(1,1)$:
%---------------------
\begin{equation} \label{starting-L}
	\subp{\mathbf{L}} = \subp{\op{F}}^{-1} \, \partial_t \subp{\op F}
	\,,\qquad
	\subp{\mathbf{l}} = \partial_t \subp{\op F} \,\subp{\op{F}}^{-1}
	\,,
\end{equation}
%---------------------
where $\subp{\mathbf{L}}$ is referred to the undeformed configuration, while $\subp{\mathbf{l}}$ is referred to the ``intermediate configuration''.\footnote{%
For further discussions on the intermediate configurations see \citep{soare2014plasticity}, and \citep{goodbrake2020mathematical}, and \citep{YavariSozio2023FeFa}.}
The tensor $\subp{\mathbf{L}}$ defined in~\eqref{starting-L} can be used to describe the rate of change of the lattice structure via~\eqref{lattice-frame} and~\eqref{lattice-coframe}, viz.
%---------------------
\begin{equation} \label{frame-rate2}
	\partial_t \boldsymbol \vartheta^{\nu} = (\partial_t \subp{\op{F}}^{\star}) \mathrm d Z^{\nu} 
	= (\partial_t \subp{\op{F}}^{\star}) \subp{\op{F}}^{-\star}  \boldsymbol \vartheta^{\nu} 
	= (\subp{\op{F}}^{-1} \partial_t \subp{\op{F}})^{\star} \, \boldsymbol \vartheta^{\nu} 
	= \subp{\mathbf{L}}^{\star} \boldsymbol\vartheta^{\nu} \,, 
\end{equation}
%---------------------
and
%---------------------
\begin{equation} \label{frame-rate3}
	\partial_t \boldsymbol e_{\nu} = (\partial_t \subp{\op{F}}^{-1}) \tfrac{\partial}{\partial Z^{\nu} }
	=(\partial_t \subp{\op{F}}^{-1}) \subp{\op{F}} \boldsymbol e_{\nu} 
	= -  \subp{\op{F}}^{-1} \, (\partial_t \subp{\op F}) \boldsymbol e_{\nu} 
	= -\subp{\mathbf{L}} \boldsymbol e_{\nu}
	\,.
\end{equation}
%---------------------
This implies that $\subp{\mathbf{L}} =\boldsymbol e_{\nu} \otimes  \partial_t \boldsymbol\vartheta^{\nu} =- \partial_t \boldsymbol e_{\nu} \otimes  \boldsymbol\vartheta^{\nu}$.
An equivalent description consists of using the change of frame approach of Eq.~\eqref{change-frame}, and letting $\mathbb L^{\nu}{}_{\mu} =  \partial_t \mathbb{F}^{\nu}{}_{\eta}  \, (\mathbb{F}^{-1})^{\eta}{}_{\mu}$, so that one obtains
%---------------------
\begin{equation} \label{frame-rate3}
	\partial_t\boldsymbol\vartheta^{\nu} = \mathbb L^{\nu}{}_{\mu}\, \boldsymbol\vartheta^{\mu}	\,,\quad
	\partial_t \boldsymbol e_{\nu} = -\mathbb L^{\mu}{}_{\nu} \,\boldsymbol e_{\mu}
	\,.
\end{equation}
%---------------------
In this way, $\subp{\mathbf{L}}$ and $\mathbb L^{\mu}{}_{\nu}$ are related as
%---------------------
\begin{equation} \label{L-components}
	\subp{\mathbf{L}} = \mathbb L^{\mu}{}_{\nu} \, \boldsymbol e_{\mu} \otimes  \boldsymbol\vartheta^{\nu} \,.
\end{equation}
%---------------------
%
The rate of change of all the quantities derived from the lattice structure can then be expressed using either~\eqref{frame-rate2} or~\eqref{frame-rate3}. For the time derivative of the material metric~\eqref{material-metric}, one can simply write $\partial_t\boldsymbol G=\subp{\mathbf{L}}^{\star} \boldsymbol G + \boldsymbol G \subp{\mathbf{L}}$.
The rate of change of the material volume $\upsilon$ is defined as $\partial_t \mvf=\upsilon  \mvf $ and can be written as
%---------------------
\begin{equation} \label{volume-rate}
	\upsilon =
	\frac{1}{2} \langle \partial_t \boldsymbol G , \boldsymbol G^{\sharp} \rangle =
	\langle \partial_t\boldsymbol\vartheta^{\nu} , \boldsymbol e_{\nu} \rangle =
	\operatorname{tr} \subp{\mathbf{L}}   \,.
\end{equation}
%---------------------
As the mass form $\boldsymbol{m} = \massd \mvf$ is assumed constant in time by virtue of mass conservation, one obtains the rate of change of the mass density as $\partial_t\massd = -\upsilon\massd$.

The rate of change of plastic deformation can be used to express the evolution of the incompatibility of the lattice structure.
From~\eqref{frame-rate3}, since $\mathrm d$ and $\partial_t$ commute, one obtains 
%---------------------
\begin{equation} \label{rate-incomp-L}
	\mathrm d \! \left( \partial_t \boldsymbol\vartheta^{\nu} \right) 
	= \mathrm d \mathbb L^{\nu}{}_{\mu} \wedge \boldsymbol \vartheta^{\mu} 
	+\mathbb L^{\nu}{}_{\mu} \mathrm d \boldsymbol \vartheta^{\mu} \,,
\end{equation}
%---------------------
that can be used to calculate the rate of change of the torsion tensor $\boldsymbol T= \boldsymbol e_{\nu} \otimes \mathrm d \boldsymbol\vartheta^{\nu}$ associated with the Weitzenb\"ock connection $\nabla^{(\mathrm{W})}$.
Invoking~\eqref{frame-rate3} and~\eqref{L-components}, and using the fact that the Weitzenb\"ock derivative of a tensor is equivalent to the ordinary derivative of the components of the tensor with respect to the lattice frame \citep{sozio2020riemannian}, from~\eqref{rate-incomp-L} one obtains\footnote{See also \citep{cleja2007material} for the time-derivative of the Weitzenb\"ock connection.}
%---------------------
\begin{equation}
\begin{split}
	 \partial_t \boldsymbol{T} &=
	 \partial_t \boldsymbol e_{\nu} \otimes \mathrm d \boldsymbol\vartheta^{\nu} + 
	 \boldsymbol e_{\nu} \otimes \partial_t \mathrm d \boldsymbol\vartheta^{\nu} \\
	 &= - \mathbb L^{\mu}{}_{\nu} \boldsymbol e_{\mu} \otimes \mathrm d \boldsymbol\vartheta^{\nu} 
	 +  \boldsymbol e_{\nu} \otimes \left( \mathrm d \mathbb L^{\nu}{}_{\mu} 
	 \wedge \boldsymbol \vartheta^{\mu}  \right) 
	 +  \boldsymbol e_{\nu} \otimes 
	 \left( \mathbb L^{\nu}{}_{\mu} \mathrm d \boldsymbol \vartheta^{\mu} \right) \\
	 &= \boldsymbol e_{\nu} \otimes \left( \mathrm d \mathbb L^{\nu}{}_{\mu} 
	 \wedge \boldsymbol \vartheta^{\mu}  \right) \\
	 &= \nabla^{(\mathrm{W})}_{\boldsymbol e_{\mu}} \subp{\mathbf{L}} \wedge \boldsymbol \vartheta^{\mu}  \\
	 &= \mathsf{alt} ( \nabla^{(\mathrm{W})} \subp{\mathbf{L}} ) \,,
\end{split}
\end{equation}
%---------------------
where the operator $\mathsf{alt}$ acts on the lower indices.\footnote{The operator $\mathsf{alt}$ maps a tensor $\boldsymbol A$ of order $(0,k)$ to the $k$-form with components $(\mathsf{alt}\,\boldsymbol A)_{J_1 J_2 \hdots, J_k} = \frac{1}{k!}\sum_{\sigma} \mathsf{sign} (\sigma) A_{\sigma(J_1 J_2 \hdots J_k)}$, where $\sigma$ denotes a permutation of the indices $J_1, J_2 , \hdots,  J_k$.}
Similarly, the rate of change of $\boldsymbol\alpha^{\nu}= \star^{\sharp} \mathrm d \boldsymbol\vartheta^{\nu}$ is written as
$\partial_t\boldsymbol\alpha^{\nu} = \upsilon \boldsymbol\alpha^{\nu} + \star^{\sharp} \mathrm d ( \partial_t \boldsymbol\vartheta^{\nu} )$,
where the rate of change of the material volume form shows up because of the raised Hodge operator.
As for $\boldsymbol\alpha= \boldsymbol e_{\nu}\otimes \boldsymbol\alpha^{\nu}$,
denoting with $ \mvf^{\sharp}$ the multivector obtained by raising the indices of the volume form, i.e., $\mvf^{\sharp}=\boldsymbol e_1 \wedge \boldsymbol e_2 \wedge \boldsymbol e_3$, one has\footnote{Cf. \citep{berdichevsky2006continuum}.}
%---------------------
\begin{equation}
	 \partial_t \boldsymbol\alpha =
	 \upsilon \boldsymbol\alpha +
	 \mvf^{\sharp} : \nabla^{(\mathrm{W})} \subp{\mathbf{L}} \,,
	 \end{equation}
%---------------------
with components $(\partial_t\alpha)^{AB}=\upsilon\alpha^{AB}+\mu^{AHK} \, \nabla^{(\mathrm{W})}_H  \, \mathbb L^B{}_K$.
Next we consider the configuration mapping and the deformations associated with it.
A spatial motion is a one-parameter family of embeddings $\varphi: \mathcal B \times \mathbb R \to \mathcal S$, and the deformation gradient is a time-dependent two-point tensor $\op F(X,t)=T_X\varphi_t$.
The velocity vector $ \boldsymbol V $ is defined as the velocity of the orbits of $\varphi_X: t\mapsto\varphi_{t}(X)$ for fixed $X$,
while, for every $t$, one can define the vector field $\boldsymbol v (x,t)=\boldsymbol{V} ( \varphi^{-1}_t(x), t)$ on $\varphi_t(\mathcal B)$ as the instantaneous velocity field.
The velocity field can be used to write the rate of change of the pulled-back metric as $\partial_t \boldsymbol C^{\flat} = \partial_t (\varphi^* \boldsymbol g) = \varphi^* \mathfrak L_{\boldsymbol v} \boldsymbol g$ \citep{marsden1983mathematical}.
It should be noted that while the expressions $\partial_t \subp{\op F}$ and $\partial_t \boldsymbol C^{\flat}$ are well-defined, $\partial_t \op F$ and $\partial_t \sube{\op F}$ are not.
The reason for this is that $\op F$ and $\sube{\op F}$ are two-point tensors, and hence their base point in the ambient space $\mathcal S$ moves along the trajectory of the motion when $t$ changes.
In order to define the time derivative of the two-point tensors $\sube{\op F}$ and $\op F$, one needs to identify tangent spaces at different points of $\mathcal S$, e.g., via a connection in $\mathcal S$.\footnote{See \citep{yavari2016nonlinear} and references therein for discussions on covariant time derivatives.}
With this in mind, one can define
%---------------------
\begin{equation} \label{L-LE}
	\sube{\mathbf{L}} = \sube{\op{F}}^{-1} \,\partial_t\sube{\op F}
	\,,\qquad
	\sube{\mathbf{l}} = \partial_t\sube{\op F}\, \sube{\op{F}}^{-1}
	\,,\qquad
	\mathbf{L} = \op F^{-1} \,\partial_t \op F
	\,,\qquad
	\mathbf{l} = \partial_t\op F \, \op F^{-1}
	\,.
\end{equation}
%---------------------
If one chooses the connection $\nabla^{(\boldsymbol g)}$, then $\mathbf{l}= \nabla^{(\boldsymbol g)}\boldsymbol v$, and $\mathbf{L} = \varphi^* \nabla^{(\boldsymbol g)}\boldsymbol v$.
Moreover, one has $\partial_t \boldsymbol C^{\flat} = 2\,\mathsf{sym} \,\boldsymbol{\mathbf{L}}^{\flat}$, and $ \mathfrak L_{\boldsymbol v} \boldsymbol g = 2\,\mathsf{sym} \,\boldsymbol{\mathrm l}^{\flat}$.
The different rates of deformation are then related as
%---------------------
\begin{equation} \label{all-L}
	\mathbf{L} = \subp{\op{F}}^{-1} \sube{\mathbf{L}}\, \subp{\op{F}} + \subp{\mathbf{L}}	\,,\quad
	\mathbf{l} = \sube{\mathbf{l}} + \sube{\op{F}} \,\subp{\mathbf{l}} \,\sube{\op{F}}^{-1}
	 \,.
\end{equation}
%---------------------

%--------------------------------------------------------------
%--------------------------------------------------------------
\subsection{Evolution of dislocation fields} \label{Evolution}

%---------------------
\begin{figure}
\centering
\includegraphics[width=.75\textwidth]{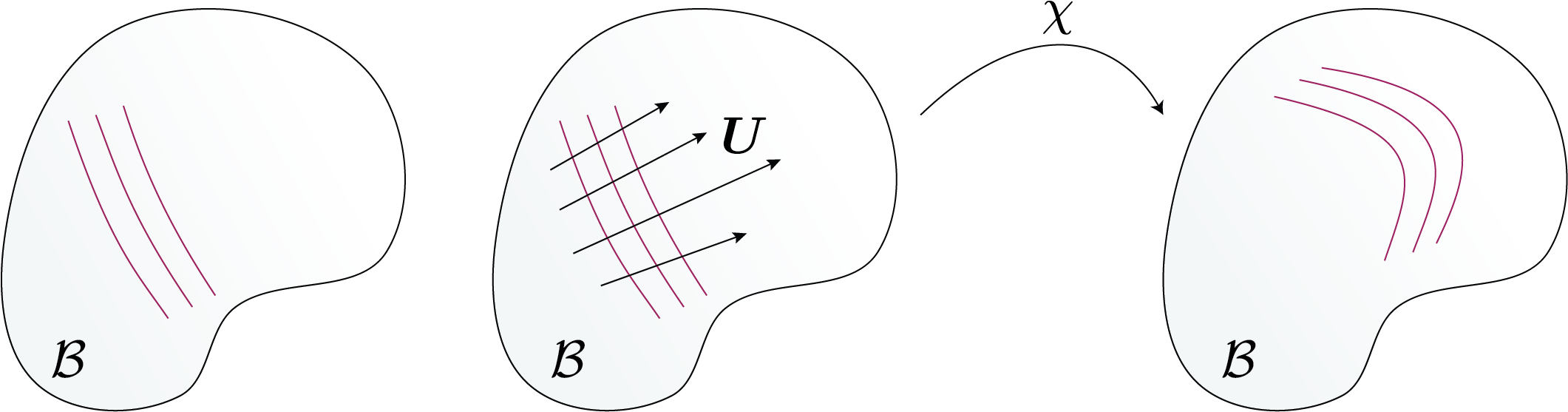}
\vskip 0.1in
\caption{A sketch of the convection of a distribution of dislocations in the material manifold.
Left: A family of dislocation curves.
Center: The material motion velocity field convects the dislocation curves.
Right: The new shape and position of the family of dislocation curves after some time.}
\label{fig:convection}
\end{figure}
%---------------------

We assume that a dislocation field $\{\dd^{\nu}\}$ is convected by a smooth material motion $\chi: \mathcal B \times \mathbb R \to \mathcal B$.\footnote{%
We call it material motion as it occurs at the level of the material body and represents the evolution of internal variables associated with defects, independently of the spatial motion.
However, no migration or diffusion of material within the solid is considered in this model.}
This means that we assume the following evolution equation:
%---------------------
\begin{equation}\label{conv-disl}
	\dd^{\nu}_t = \chi_{t}{}_* \dd^{\nu}_0  \,,
\end{equation}
%---------------------
for all the $2$-forms $\dd^{\nu}$ in the triplet. Eq.~\eqref{conv-disl} can also be written using relative motions as $\dd^{\nu}_t = (\chi^s_{t})_* \, \dd^{\nu}_s $, see~\S\ref{App:Flows}.
Note that $\dd^{\nu}$ is completely determined at any time by the material motion $\chi$ and the initial condition $ \dd^{\nu}_0$.
We denote with $\boldsymbol{\vel}$ the time-dependent velocity field associated with $\chi$.
From~\eqref{conv-disl} the non-autonomous Lie derivative $\mathsf L_{\boldsymbol{\vel}} \dd^{\nu}$ vanishes, and hence from~\eqref{evolution} one obtains an evolution equation in the rate form as
%---------------------
\begin{equation}\label{rate-Lie}
	\partial_t\dd^{\nu} =
	- \mathfrak L_{\boldsymbol{\vel}} \dd^{\nu} \,.
\end{equation}
%---------------------
The motion of a dislocation field as observed in the ambient space is given, for a fixed $X\in\mathcal B$, by the map $t \mapsto \varphi(\chi(X,t),t)$.
Therefore, the velocity with which dislocations travel in the deformed lattice defined on $\varphi(\mathcal B)$ is written as $ \op F \boldsymbol{\vel} + \boldsymbol v$.

If the dislocation field is decomposable, one can obtain time-dependent decompositions $\bv^{\nu}\df$ such that both the Burgers director and the dislocation density $2$-form are convected by the same material motion $\chi$.
This can be easily achieved by convecting the decomposition at the initial time, viz.
%---------------------
\begin{equation} \label{evol-decomp}
	\bv^{\nu}_t = \bv^{\nu}_0 \circ \chi_t^{-1} \,,\qquad
	\df_t = \chi_{t}{}_* \df_0 \,,
\end{equation}
%---------------------
and in rate form
%---------------------
\begin{equation} \label{rate-decomp}
	\partial_t \bv^{\nu}
	= -\langle \mathrm d\bv^{\nu}, \boldsymbol{\vel}\rangle
	\,,\qquad
	\partial_t \df
	= - \mathfrak L_{\boldsymbol{\vel}} \df
	\,.
\end{equation}
%---------------------
It should be noted that in the normalized decomposition~\eqref{unit-decomp}  the scalars $\bv^{\nu}$ are convected, meaning that~\eqref{rate-decomp}$_1$ does not alter the property $\Vert \boldsymbol{\bv} \Vert_{\boldsymbol G}=1$.
However, the director $\boldsymbol{\bv} = \bv^{\nu}\boldsymbol e_{\nu}$ is not convected,
as the lattice forms $\boldsymbol\vartheta^{\nu}$ follow a different evolution equation, see~\S\ref{Orowan}.
For the same reason, the material metric $\boldsymbol G$ and the scalar dislocation density $\dens$ are not convected either.
Moreover, a dislocation field that is initially closed remains closed at all times during convection, as the exterior derivative and pushforward commute, or equivalently, exterior derivative and Lie derivative commute, see~\S\ref{Orowan}.
For the same reason, by convecting the decomposition of Lemma~\ref{lem:closed-decomp} one obtains a decomposition that satisfies the same properties at all times.
The following lemma clarifies the geometric meaning of convected decompositions.

%---------------------
%---------------------
\begin{lem} \label{Lem:dislo-curves}
The dislocation curves associated with a convected $\df$ are convected by the same map.
\end{lem}
%---------------------
\begin{proof}
We fix two times, $0$ and $t$, and denote with $\star^{\sharp}_t$ the raised Hodge operator induced by $\boldsymbol G_t$, and with $\star^{\sharp}_0$ the one induced by $\boldsymbol G_0$.
Then, the vector $\star^{\sharp}_t \, \df_t$ is tangent to the dislocation curves for $\df_t$, while the vector $\chi_{t}{}_* \star^{\sharp}_0 \, \df_0$ is tangent to the convected dislocation curves.
In order to prove the lemma, one needs to show that the two vectors are parallel, i.e.,  $(\chi_{t}{}_* \circ \star^{\sharp}_0 ) \, \df_0 = f \star^{\sharp}_t  \df_t$ for some scalar $f\neq0$.
One can write
%---------------------
\begin{equation}
	(\chi_{t}{}_* \circ \star^{\sharp}_0 ) \, \boldsymbol{\df}_0 =
	(\chi_{t}{}_* \circ \star^{\sharp}_0 \circ \chi_{t}{}^* )\, \boldsymbol{\df}_t \,.
\end{equation}
%---------------------
Note that $\chi_{t}{}_* \circ \star^{\sharp}_0 \circ \chi_{t}{}^*$ is the raised Hodge operator induced by $\chi_{t}{}_* \boldsymbol G_0$.
As in dimension three all $3$-forms differ by a nonzero multiplicative factor, say $K$, one finds that $f=K$.
\end{proof}
%---------------------
%---------------------

Since $\bv^{\nu}$ and the dislocation curves are convected by the same map $\chi$, distinct dislocation fields remain distinct, see~\S\ref{Decomposable}.
In the case of uniform dislocation distributions one has $\mathrm d\bv^{\nu}=0$, and hence from~\eqref{rate-decomp} the rate of change of the Burgers director vanishes.
This means that uniform dislocation distributions stay uniform and the Burgers vector density is constant in time.
The contribution of the dislocation field $\dd^{\nu}$ to the Burgers vector associated with the boundary of a convected surface $\Sigma_t = \chi_t (\Sigma_0)$ is conserved.
As a matter of fact, recalling~\eqref{burgers-stokes}, one can write
%---------------------
\begin{equation}
	 \int_{\Sigma_t} \incl^*  \dd^{\nu}_t
	 = \int_{\chi_t(\Sigma_0)} \incl^*  \left(  \chi_t{}_* \dd^{\nu}_0\right)
	 = \int_{\chi_t(\Sigma_0)} \chi_t{}_* \left( \incl^* \dd^{\nu}_0 \right)
	 =\int_{\Sigma_0} \incl^*\dd^{\nu}_0
	 \,,
\end{equation}
%---------------------
as $\incl^*\circ \chi_t{}_* = \chi_t{}_* \circ \incl^*$ when the surface is convected.\footnote{%
The maps $\chi_t$ and $\incl$ commute. More precisely, $\incl_{\Sigma_t} \,\circ\, \chi_t = \chi_t\vert_{\Sigma_0} \, \circ \, \incl_{\Sigma_0}$, as for $X_0\in\Sigma_0$ one has $(\incl_{\Sigma_t} \,\circ\, \chi_t) (X_0) = (\chi_t\vert_{\Sigma_0} \, \circ \, \incl_{\Sigma_0} )(X_0)=\chi_t (X_0)$.}
Hence, the integral is time independent.

%---------------------
%---------------------
\begin{remark} \label{rem:motion-boundary}
Similar to the spatial motion $\varphi$, a material motion for a dislocation field has been defined as a family of diffeomorphisms $\chi: \mathcal B \times \mathbb R \to \mathcal B$.
By doing so, the dislocation velocity $\boldsymbol{\vel}$ associated with $\chi$ must be tangent to the boundary $\partial \mathcal B$, as diffeomorphisms map the boundary of a manifold to itself.
This means that dislocations can neither emerge on the boundary of the crystal nor enter from the outside.
This is a restrictive assumption, as grain boundaries play a crucial role in plasticity acting as sources, absorbers and barriers for dislocations.
In order to allow a non-tangent $\boldsymbol{\vel}$ on the boundary we consider the following construction:
for each point $X$ in the interior $\mathring{\mathcal B}$ and for each time $t$ we take a sufficiently small time interval $[t-\tau(X,t),t+\tau(X,t)]$ such that the integral curve of $\boldsymbol{\vel}(X,t)$ that passes through $X$ at time $t$ does not intersect the boundary $\partial \mathcal B$.
Next, one defines the subbody $\mathcal P_{t,u} = \{ X\in\mathring{\mathcal B} \vert u \leq \tau(X,t) \}$, i.e., the set of points having a well-defined trajectory during the interval $[t-u,t+u]$.
Then, the motion $\chi^t_s: \mathcal P_t \to \mathcal B $ for $s \in [t-u,t+u]$ is well-defined.\footnote{See \S\ref{App:Flows} for the notation of flows and material motions.}
In other words, by relaxing the restriction of tangent $\boldsymbol{\vel}$, one can still define a material motion for smaller time intervals and subbodies.
For the sake of simplicity, and with an abuse of notation, we will still be referring to a material motion as a single well-defined map $\chi$.
\end{remark}
%---------------------
%---------------------

Next we assume that each dislocation field $\{ \prs{a}{\dd}^{\nu} \}$ moves across the solid with velocity $\prs{a}{\boldsymbol{\vel}}$, associated with the material motion $\prs{a}{\chi}$, $\indx{a}=1,2,\hdots,N$.
Note that the rate of change of the dislocation fields is related to the rate of change of the lattice forms via~\eqref{gnd-sum}.
Then, the evolution equation~\eqref{rate-Lie} applied to each dislocation field allows one to write
%---------------------
\begin{equation} \label{rate-incomp}
	\mathrm d \! \left( \partial_t  \boldsymbol\vartheta^{\nu} \right) =
	\partial_t \left( \mathrm d \boldsymbol\vartheta^{\nu} \right) =
	\partial_t \sum_{\indx{a}=1}^{N} \, \prs{a}{\dd}^{\nu} =
	\sum_{\indx{a}=1}^{N} \, \partial_t \prs{a}{\dd}^{\nu} =
	- \sum_{\indx{a}=1}^{N}  \, \mathfrak L_{\prs{a}{\boldsymbol{\vel}} } \prs{a}{\dd}^{\nu} \,,
\end{equation}
%---------------------
as the derivatives $\mathrm d$ and $\partial_t$ commute.
After integration, one obtains
%---------------------
\begin{equation} \label{kine-lattice}
	\partial_t \boldsymbol\vartheta^{\nu} = 
	\boldsymbol\gamma^{\nu} + \boldsymbol\beta^{\nu} \,,
	\quad\text{with}\qquad
	\mathrm d \boldsymbol\gamma^{\nu} = - \sum_{\indx{a}=1}^{N}  \, \mathfrak L_{\prs{a}{\boldsymbol{\vel}} } \prs{a}{\dd}^{\nu} \,,
\end{equation}
%---------------------
for some triplet of closed $1$-forms $\{\boldsymbol\beta^{\nu} \}$, and the triplet of $1$-forms $\{\boldsymbol\gamma^{\nu}\}$ carrying the whole incompatible content of the rate of plastic deformation.
Eq.~\eqref{kine-lattice} is an evolution equation for the lattice coframe,
and is based on the fact that the dislocation fields contribute to the defect content according to~\eqref{gnd-sum}, and that they evolve as prescribed by~\eqref{rate-Lie}.
It should be noted that the evolution of the lattice coframe is then determined modulo closed $1$-forms,
see Remark~\ref{Rem:same-dislo}.
This in turn means that the mechanics of a solid with moving distributed dislocations cannot be fully described by~\eqref{gnd-sum} and~\eqref{rate-Lie} alone.\footnote{%
One can split $\partial_t \boldsymbol\vartheta^{\nu}$ in a unique way by using the Helmholtz decomposition induced by $\boldsymbol G$, i.e., by choosing $\boldsymbol\gamma^{\nu}$ such that $(\star\mathrm d\star) \boldsymbol\gamma^{\nu}=0$.
Although in a non-rate form, \citet{wenzelburger1998kinematic} suggested that if one uses the Helmholtz decomposition, then $\boldsymbol\gamma^{\nu}$ is the part that is associated to plastic deformations, while $\mathrm d c^{\nu}$ represents an elastic deformation. This is incorrect mainly because the elastic deformation is incompatible as well. Moreover, there is no physical basis for which the Helmholtz decomposition should provide any information on how to eliminate the indeterminacy of $\partial_t \boldsymbol\vartheta^{\nu}$.}
Therefore, in order to fix the indeterminacy of the integration forms $\boldsymbol\beta^{\nu}$ in~\eqref{kine-lattice}, a closure model based on the underlying physics is needed, e.g., Orowan's equation that is discussed next.

%-------------------------------------------------------------
%-------------------------------------------------------------
\subsection{Orowan's equation} \label{Orowan}

Next we consider a closed dislocation field $\{\dd^{\nu}\}$, for which $\mathrm d\dd^{\nu}= 0$ for all $\nu$, as in~\S\ref{Closed-Distributions}.
As was mentioned earlier, when a dislocation field is convected according to~\eqref{conv-disl}, if it is closed at a particular time, then it is closed at all times.
This is due to the fact that exterior derivative and pushforward commute, and hence $\mathrm d\dd^{\nu}_t = (\mathrm d \circ (\chi^s_t)_*) \dd^{\nu}_s = ( (\chi^s_t)_* \circ \mathrm d ) \dd^{\nu}_s$.
Moreover, by virtue of Cartan's formula~\eqref{Cartan}, the evolution equation~\eqref{rate-Lie} becomes
%---------------------
\begin{equation}\label{rate-Cart}
	\partial_t \dd^{\nu} = - \mathrm d ( \iota_{\boldsymbol{\vel}}\dd^{\nu} ) \,,
\end{equation}
%---------------------
and for a decomposable dislocation field~\eqref{rate-decomp}$_2$ reads
%---------------------
\begin{equation}\label{rate-Cart-decomp}
	\partial_t \df  = - \mathrm d ( \iota_{\boldsymbol{\vel}} \df ) \,.
\end{equation}
%---------------------
Both expressions~\eqref{rate-Lie} and~\eqref{rate-Cart} were derived by \citet{hochrainer2007three}, although this was done in the context of a linear theory and for a single dislocation field. 
The kinematics of closed dislocation fields has the following property.

%---------------------
%---------------------
\begin{lem} \label{Lem:equiv-evol}
The evolution equation of closed distributions of dislocations is defined up to material motions in the direction of the dislocation curves.
\end{lem}
\begin{proof}
By virtue of~\eqref{rate-Lie}, it is sufficient to show that $\mathfrak L_{\widetilde{\boldsymbol{\vel}}} \dd^{\nu} = \mathfrak L_{\boldsymbol{\vel}} \dd^{\nu}$ for any $\widetilde{\boldsymbol{\vel}} = \boldsymbol{\vel} + f \boldsymbol{\lv}$, with $f$ a scalar field.
By the linearity of the Lie derivative one has
%---------------------
\begin{equation}
	\mathfrak L_{\widetilde{\boldsymbol{\vel}}} \dd^{\nu} =
	\mathfrak L_{\boldsymbol{\vel}} \dd^{\nu} +  \mathfrak L_{f \boldsymbol{\lv}} \dd^{\nu} \,.
\end{equation}
%---------------------
On the other hand, Cartan's formula~\eqref{Cartan} allows us to write
%---------------------
\begin{equation}
	\mathfrak L_{f\boldsymbol{\lv}} \dd^{\nu} =
	\mathrm d ( \iota_{f \boldsymbol{\lv}} \dd^{\nu} ) +
	\iota_{f \boldsymbol{\lv}} ( \mathrm d\dd^{\nu} ) =
	\mathrm d f \wedge \iota_{\boldsymbol{\lv}} \dd^{\nu}  +
	f\mathrm d ( \iota_{\boldsymbol{\lv}} \dd^{\nu} ) +
	f \iota_{\boldsymbol{\lv}} ( \mathrm d\dd^{\nu} ) \,,
\end{equation}
%---------------------
where use was made of~\eqref{ext-der} and of the linearity of the interior product.
Note that $\mathrm d\dd^{\nu}=0$ by hypothesis, and $\iota_{\boldsymbol{\lv}} \dd^{\nu} = \bv^{\nu} \iota_{\boldsymbol{\lv}} \df=0$ as $\boldsymbol{\lv}$ and $\star^{\sharp} \df$ are parallel.
Therefore, $\mathfrak L_{\widetilde{\boldsymbol{\vel}}} \dd^{\nu} = \mathfrak L_{\boldsymbol{\vel}} \dd^{\nu}$, which proves the lemma.
\end{proof}
%---------------------
%---------------------

It should be emphasized that Lemma~\ref{Lem:equiv-evol} holds only under the hypothesis of closed dislocation fields.
The reason for this is the fact that the Burgers director of a non-closed dislocation field is not necessarily constant along the dislocation curves. This means that the convection along the dislocation curves of a non-closed dislocation field changes the dislocation field itself by translating the non-uniform Burgers director along the dislocation curves.

As was mentioned in \S\ref{Evolution}, the evolution equation leaves an indeterminacy in the lattice forms, whence the need of a closure model coming from the physics of the problem.
In the case of closed dislocations, this extra information is provided by Orowan's equation.
Note that, under the assumption of closed dislocation fields,~\eqref{rate-Cart} allows one to write~\eqref{rate-incomp} as
%---------------------
\begin{equation} \label{evol-defect}
	\mathrm d \! \left(\partial_t \boldsymbol\vartheta^{\nu} \right) = 
	- \sum_{\indx{a}=1}^{N} \, \mathrm d \! \left(  \iota_{\prs{a}{\boldsymbol{\vel}} } \prs{a}{\dd}^{\nu} \right) = 
	- \mathrm d   \sum_{\indx{a}=1}^{N}  \iota_{\prs{a}{\boldsymbol{\vel}} } \prs{a}{\dd}^{\nu} \, ,
\end{equation}
%---------------------
where the linearity of the exterior derivative was used.
Hence, after integration one obtains
%---------------------
\begin{equation} \label{gnd-evol}
	\partial_t \boldsymbol\vartheta^{\nu} = 
	- \sum_{\indx{a}=1}^{N} \, \iota_{\prs{a}{\boldsymbol{\vel}} } \prs{a}{\dd}^{\nu}
	+ \boldsymbol\beta^{\nu} \,,
\end{equation}
%---------------------
where $\{ \boldsymbol\beta^{\nu}\}$ are three arbitrary closed $1$-forms as in~\eqref{kine-lattice}.
Orowan's equation is then obtained by choosing $\boldsymbol\beta^{\nu}=\boldsymbol 0$, i.e.,
%---------------------
\begin{equation} \label{orowan}
	\partial_t \boldsymbol\vartheta^{\nu} =
	- \sum_{\indx{a}=1}^{N} \, \iota_{\prs{a}{\boldsymbol{\vel}}} \prs{a}{\dd}^{\nu} \,.
\end{equation}
%---------------------
Note that, similar to $\boldsymbol\eta^{\nu}$, 
the forms $\boldsymbol\vartheta^{\nu}$ are completely determined at any time by the material motion $\chi$ and the initial condition $ \boldsymbol\vartheta^{\nu}_0$.
From~\eqref{orowan} it is clear that the lattice coframe $\{\boldsymbol\vartheta^{\nu}\}$---and hence the material metric $\boldsymbol G$---is not convected.
Assuming decomposable dislocation fields $\prs{a}{\boldsymbol\eta}^{\nu} =  \prs{a}{\bv}^{\nu} \prs{a}{\df}$, and recalling $\prs{a}{\df} = \prs{a}{\dens}\,\iota_{\prs{a}{\boldsymbol{\lv}} } \mvf$ from~\S\ref{Decomposable}, Orowan's equation~\eqref{orowan} can be rewritten as
%---------------------
\begin{equation} 
	\partial_t \boldsymbol\vartheta^{\nu} 
	=\sum_{\indx{a}=1}^{N}-\prs{a}{\bv}^{\nu} \,  \iota_{ \prs{a}{\boldsymbol{\vel}} } \prs{a}{\df }
	=\sum_{\indx{a}=1}^{N}-\prs{a}{\dens}\,\prs{a}{\bv}^{\nu}\, \iota_{ \prs{a}{\boldsymbol{\vel}}} \iota_{\prs{a}{\boldsymbol{\lv}} }  \mvf
	=\sum_{\indx{a}=1}^{N} \prs{a}{\dens}\, \prs{a}{\bv}^{\nu}\, \iota_{\prs{a}{\boldsymbol{\lv}} } \iota_{ \prs{a}{\boldsymbol{\vel}} } \mvf 
	=\sum_{\indx{a}=1}^{N}  \prs{a}{\dens}\, \prs{a}{\bv}^{\nu} \mvf ( \prs{a}{\boldsymbol{\vel}} , \prs{a}{\boldsymbol{\lv}}  ) \,,
\end{equation}
%---------------------
%%---------------------
where $\mvf ( \prs{a}{\boldsymbol{\vel}} , \prs{a}{\boldsymbol{\lv}}  ) = \iota_{\prs{a}{\boldsymbol{\lv}} } \iota_{ \prs{a}{\boldsymbol{\vel}} } \mvf = \mu_{ABC}\, \prs{a}{\vel}^A  \prs{a}{\lv}^B  \mathrm d X^C$.
As for the rate of change of volume, from~\eqref{volume-rate} one obtains
%---------------------
\begin{equation} \label{orowan-volume}
	\upsilon =
	\sum_{\indx{a}=1}^{N}  \prs{a}{\dens}\,
	 \mvf ( \prs{a}{\boldsymbol{\vel}} , \prs{a}{\boldsymbol{\lv}} , \prs{a}{\boldsymbol{\bv}} ) \,,
\end{equation}
%---------------------
where $\mvf ( \prs{a}{\boldsymbol{\vel}} , \prs{a}{\boldsymbol{\lv}}  , \prs{a}{\boldsymbol{\bv}} )  = \mu_{ABC}\, \prs{a}{\vel}^A  \prs{a}{\lv}^B \prs{a}{\bv}^C$ denotes a triple product in $(\mathcal B, \boldsymbol G)$.

%---------------------
%---------------------
\begin{remark} \label{rem:isochoric}
Let us now look at the contribution to the rate of change of the lattice coframe $\{\boldsymbol\vartheta^{\nu}\}$ of a single decomposable dislocation field $\{\dd^{\nu}\}$ convected by $\boldsymbol{\vel}$.
From~\eqref{orowan} and~\eqref{orowan-volume} one writes
%---------------------
\begin{equation} \label{orowan-single}
	\partial_t \boldsymbol\vartheta^{\nu} =
	-\iota_{\boldsymbol{\vel}}\dd^{\nu}=
	\dens \,\bv ^{\nu} \mvf ( \boldsymbol{\vel},\boldsymbol{\lv} )
	\,,\qquad
	\upsilon = \dens \,
	 \mvf ( \boldsymbol{\vel},\boldsymbol{\lv} , \boldsymbol{\bv} ) \,.
\end{equation}
%---------------------
It should be noticed that the lattice coframe is not convected even in the case of a single dislocation field.
Eq.~\eqref{orowan-single}${}_1$ is the common form of Orowan's equation \citep{sedlavcek2003importance}, from which one concludes that if the material motion velocity is tangent to the dislocation lines there is no plastic deformation.
This is formalized in Lemma~\ref{Lem:equiv-orowan}.
Moreover, by the criterion of linear independence, from~\eqref{orowan-single}${}_2$ one concludes that the plastic slip is isochoric if and only if $\boldsymbol{\bv}$, $\boldsymbol{\vel}$ and $\boldsymbol{\lv}$ are coplanar, e.g., in the case of glide motion of dislocations, see \S\ref{Glide}.
\end{remark}
%---------------------
%---------------------

%---------------------
%---------------------
\begin{lem} \label{Lem:equiv-orowan}
In the case of decomposable dislocation fields, Orowan's equation is invariant under superimposed material motions in the direction of the dislocation curves.
\end{lem}
\begin{proof}
It is sufficient to show that $\iota_{ \widetilde{\boldsymbol{\vel}} } \dd^{\nu}=\iota_{\boldsymbol{\vel}} \dd^{\nu}$ for any $\widetilde{\boldsymbol{\vel}} = \boldsymbol{\vel} + f \boldsymbol{\lv}$, where $f$ is an arbitrary scalar field. By linearity of the interior product one can write
%---------------------
\begin{equation}
	\iota_{ \widetilde{\boldsymbol{\vel}} } \dd^{\nu} =
	\bv^{\nu} \iota_{ \widetilde{\boldsymbol{\vel}} } \df=
	\bv^{\nu} \left(
	\iota_{\boldsymbol{\vel}} \df+
	f \, \iota_{\boldsymbol{\lv}} \df
	\right) \,.
\end{equation}
%---------------------
Since $\iota_{\boldsymbol{\lv}} \df = \iota_{\boldsymbol{\lv}} \iota_{\boldsymbol{\lv}} \mvf = 0$, Orowan's equation~\eqref{orowan-single} is unaltered.
\end{proof}
%---------------------
%---------------------

%---------------------
%---------------------
\begin{example} \label{Ex:orowan}
We look at the motion of a closed dislocation field and at the change of lattice structure induced by the effect of Orowan's equation.
We fix Cartesian coordinates $( Z^{\nu} )$ on $\mathcal B$, and consider the following material motion:
%---------------------
\begin{equation} \label{straight-vel}
	\chi: (Z^1,Z^1,Z^3)\mapsto(Z^1+\vel t,Z^1,Z^3) \,,
\end{equation}
%---------------------
for a constant scalar $\vel$, associated with the velocity $\boldsymbol{\vel} = \vel \tfrac{\partial}{\partial Z^1}$.
We also consider the following closed dislocation field:
%---------------------
\begin{equation} \label{straight-disl}
	\dd^{\nu} = \dens \bv^{\nu} \mathrm d Z^1 \wedge \mathrm d Z^2 \,,
\end{equation}
%---------------------
with constant $\dens$ and $\bv^{\nu}$.
Eqs.~\eqref{straight-vel} and~\eqref{straight-disl} represent a forest of uniformly distributed straight dislocations moving sidewise.\footnote{In this example the dislocation curves are straight in the sense of Cartesian coordinate chart representation, while a coordinate-free notion of straightness and curvature should rely on a metric, such as $\boldsymbol G$.}
It should be noted that such a motion leaves the dislocation field unchanged. As a matter of fact, the evolution equation gives
%---------------------
\begin{equation} \label{straight-evo}
	\partial_t \dd^{\nu} =
	-\mathrm d \iota_{\boldsymbol{\vel}} \dd^{\nu} =
	-\mathrm d \! \left( \vel \bv^{\nu} \mathrm d Z^2 \right) =
	0 \,,
\end{equation}
%---------------------
which agrees with the time independence of $\dd^{\nu}$, while Orowan's equation~\eqref{orowan-single} reads
%---------------------
\begin{equation} \label{straight-oro}
	\partial_t \boldsymbol \vartheta^{\nu} =
	- \iota_{\boldsymbol{\vel}} \dd^{\nu} =
	- \vel\bv^{\nu} \mathrm d Z^2 \,.
\end{equation}
%---------------------
It should be noticed how in this case the constant dislocation field $\dd^{\nu}$ induces a change in the lattice forms.
However, by assuming a vanishing dislocation velocity, i.e., $\boldsymbol{\vel}=\boldsymbol 0$ instead of $\boldsymbol{\vel} = \vel \tfrac{\partial}{\partial Z^1}$, one obtains the same time-independent $\dd^{\nu}$, while Orowan's equation leaves the lattice unchanged, i.e., $\partial_t \boldsymbol \vartheta^{\nu}=0$.
In other words, as was discussed in~\S\ref{Evolution}, the evolution of the lattice structure cannot be deduced exclusively from the evolution of the dislocation fields; instead, it explicitly depends on the material motion.
This shows that Orowan's equation cannot be deduced from the kinematics of the dislocation fields, as any triplet of closed $1$-forms instead of~\eqref{straight-oro} would be compatible with it. 
Assuming $\bv^{\nu}=(-1 ,0,0)$, one obtains the classic forest of straight edge dislocations of sign ``$\bot$'' moving towards right, so that~\eqref{straight-oro} gives $\partial_t \mathbb{F}^1{}_2 = \vel\dens$ as the only non-vanishing component of $\partial_t\subp{\op F}$.
\end{example}
%---------------------
%---------------------

%----------------------------------------------------
%----------------------------------------------------
\subsection{Glide motion} \label{Glide}

Let $\Pi$ be a plane distribution represented by a $1$-form $\boldsymbol\pi$ of unit $\boldsymbol G$-norm, as in \S\ref{Layered}.
A glide motion is defined by a dislocation velocity that locally lies on a plane distribution.
In particular, a dislocation motion $\chi$ with velocity $\boldsymbol{\vel}$ is a glide motion along $\Pi$ if it satisfies
%---------------------
\begin{equation}\label{glide}
	\langle \boldsymbol \pi,\boldsymbol{\vel} \rangle=0 \,.
\end{equation}
%---------------------
One may ask if the condition~\eqref{glide} is appropriate for an evolving lattice structure, i.e., whether it fully takes into account the fact that $\boldsymbol \pi$ is evolving during the glide of dislocations.
What suggests the need for a correction is the fact that, in the case of a particle constrained to move on an evolving surface defined by the normalized $1$-form $\boldsymbol \pi$, the velocity $\boldsymbol V$ of the particle satisfies $\langle \boldsymbol \pi , \boldsymbol V \rangle=S$, with $S$ being the velocity of the moving surface in the normal direction.
Similarly, we can look at our case as a short dislocation segment gliding on a small portion of a surface that can be approximated by a planar surface at a sufficiently close distance.
However, the quantity $S$ can be set to zero as
the evolution of a plane distribution constitutes a mere change in the local orientation of the crystallographic planes
and does not contribute directly to the motion of the dislocations.

In the following we consider the glide motion of layered dislocation fields, i.e., those whose Burgers and dislocation line directors are in $\Pi$, and hence satisfy~\eqref{slip-constr}.
Note that since $\boldsymbol{\lv}$ and $\boldsymbol{\vel}$ span the slip plane given by $\boldsymbol\pi$, the $1$-forms $\iota_{\boldsymbol{\vel}}\df$ and $\boldsymbol\pi$ define the same plane distribution.\footnote{%
This can also be shown by using the relation~\eqref{contr} and invoking~\eqref{slip-constr}${}_2$ and~\eqref{glide}: $\boldsymbol\pi  \wedge \iota_{\boldsymbol{\vel}}\df = \iota_{\boldsymbol{\vel}}(\boldsymbol\pi \wedge \df) -\langle \boldsymbol\pi, \boldsymbol{\vel} \rangle \, \df  =0$.
}
In particular, we set
%---------------------
\begin{equation} \label{sliprate}
	 \iota_{\boldsymbol{\vel}} \df = -\gamma \boldsymbol\pi \,.
\end{equation}
%---------------------
The scalar $\gamma$ is the rate of plastic slip associated with $\Pi$.
Moreover, since $\langle \boldsymbol\pi ,\boldsymbol{\vel} \rangle=0$, one can write $\boldsymbol{\vel} = \llangle \boldsymbol{\vel} , \boldsymbol{\lv} \rrangle_{\boldsymbol G} \,\boldsymbol{\lv} + \llangle \boldsymbol{\vel} ,\boldsymbol{\norm} \rrangle_{\boldsymbol G} \, \boldsymbol{\norm} $.
Therefore, settting $\vel^{\bot} = \llangle \boldsymbol{\vel}  , \boldsymbol{\norm} \rrangle_{\boldsymbol G}$, from~\eqref{hodge-n-1} and~\eqref{norm-dens} one obtains
%---------------------
\begin{equation} \label{lUspanPi}
	\iota_{\boldsymbol{\vel}} \df =
	\iota_{\boldsymbol{\vel}} \iota_{\dens \, \boldsymbol{\lv}} \mvf =
	\dens \, \llangle \boldsymbol{\vel} , \boldsymbol{\lv} \rrangle_{\boldsymbol G} \,
	\iota_{\boldsymbol{\lv}} \iota_{\boldsymbol{\lv}} \mvf +
	\dens \,  \llangle \boldsymbol{\vel} ,\boldsymbol{\norm} \rrangle_{\boldsymbol G} \,
	\iota_{\boldsymbol{\norm}} \iota_{\boldsymbol{\lv}} \mvf =
	\dens \, \vel^{\bot} \boldsymbol\pi
	\,,
\end{equation}
%---------------------
and hence, $\gamma = -\dens\, \vel^{\bot}$. Eq.~\eqref{lUspanPi} can also be written as $\mvf ( \boldsymbol{\lv}, \boldsymbol{\vel} ) = \iota_{\boldsymbol{\vel}} \iota_{\boldsymbol{\lv}} \mvf = \vel^{\bot} \boldsymbol\pi$.
The evolution equation~\eqref{rate-Cart} can be specialized to closed layered dislocation fields as
%---------------------
\begin{equation}
	\partial_t \dd^{\nu} = \mathrm d ( \gamma \bv^{\nu} \boldsymbol \pi ) \,.
\end{equation}
%---------------------
Hence, the defect content of the overall rate of change of the lattice forms~\eqref{evol-defect} reads
%---------------------
\begin{equation}
	\mathrm d \! \left( \partial_t\boldsymbol\vartheta^{\nu} \right) = 
	\sum_{\indx{a}=1}^{N}  \mathrm d ( \prs{a}{\gamma} \prs{a}{\bv}^{\nu}  \prs{a}{\boldsymbol \pi}) \,,
\end{equation}
%---------------------
where $\prs{a}{\boldsymbol\pi}$ is the slip $1$-form associated with the $\indx{a}$-th dislocation field.
Assuming that the closed layered dislocation fields obey Orowan's equation,~\eqref{orowan} is simplified to read
%---------------------
\begin{equation} \label{orowan-slip}
	\partial_t \boldsymbol\vartheta^{\nu} =
	\sum_{\indx{a}=1}^{N} \, \prs{a}{\gamma} \, \prs{a}{\bv}^{\nu} \, \prs{a}{\boldsymbol\pi} \,,
	\quad~\text{or}\qquad
	\subp{\mathbf{L}} =
	\sum_{\indx{a}=1}^{N} \, \prs{a}{\gamma} \, \prs{a}{\boldsymbol{\bv}} \otimes \prs{a}{\boldsymbol\pi} \,.
\end{equation}
%---------------------
Note that from~\eqref{glide} one concludes that $\boldsymbol{\vel}$, $\boldsymbol{\lv}$ and $\boldsymbol{\bv}$ are coplanar, and hence Orowan's equation implies that the rate of change of volume vanishes by virtue of~\eqref{orowan-volume}.
For this reason, the glide of dislocations is said to be a conservative motion.\footnote{In the dislocation dynamics literature, a conservative motion refers to a dislocation motion inducing a plastic deformation that is volume preserving \citep{nabarro1952mathematical}.}
A summary of the evolution equations for the internal variables is shown in Table~\ref{Tab:ev-eq}.

%---------------------
%---------------------
\begin{remark} \label{rmk:slip-rate}
In the absence of changes of phase one can assume lattice characteristics that are constant in time, and hence
$\partial_t \pi_{\nu} = 0$ for all $\nu$.
Then, plane $1$-forms evolve with time according to $\partial_t \boldsymbol\pi =  \pi_{\nu} \partial_t \boldsymbol\vartheta^{\nu}$.
However, they are not convected objects, as the $1$-forms $\partial_t \boldsymbol\vartheta^{\nu}$ follow Orowan's equation.
Therefore, when they exist, slip surfaces are not convected by the material dislocation velocity $\boldsymbol{\vel}$ either.
In particular, Eq.~\eqref{orowan-slip}$_1$ implies that the glide of a dislocation field on its own slip plane does not affect the evolution of the slip plane itself.
\end{remark}
%---------------------
%---------------------

%----------------
%----------------
\begin{table}[t] 
\centering
\setstretch{1.4} \small
\begin{tabular}{rccc}
\toprule
 & Dislocation fields & Defect content & Lattice coframe \\
\midrule \addlinespace[.05in] 
General: & $\partial_t \dd^{\nu} = - \mathfrak L_{\boldsymbol{\vel}} \dd^{\nu} $ & $\mathrm d \! \left( \partial_t \boldsymbol\vartheta^{\nu} \right)= - \sum_{\indx{a}=1}^{N}  \, \mathfrak L_{\prs{a}{\boldsymbol{\vel}} } \prs{a}{\dd}^{\nu}$ &
$\partial_t \boldsymbol\vartheta^{\nu} = \boldsymbol\gamma^{\nu} + \mathrm d c^{\nu}$ \\
Closed: & $\partial_t \dd^{\nu} = - \mathrm d  \iota_{\boldsymbol{\vel}}\dd^{\nu} $ & $\mathrm d \! \left( \partial_t \boldsymbol\vartheta^{\nu} \right)= - \mathrm d   \sum_{\indx{a}=1}^{N}  \iota_{\prs{a}{\boldsymbol{\vel}} } \prs{a}{\dd}^{\nu} $ & 
$\partial_t \boldsymbol\vartheta^{\nu} = - \sum_{\indx{a}=1}^{N} \, \iota_{\prs{a}{\boldsymbol{\vel}} } \prs{a}{\dd}^{\nu} + \boldsymbol\kappa^{\nu} $ \\
\addlinespace[.05in] \cmidrule{4-4} \addlinespace[.05in] 
& & & Orowan's equation: $\boldsymbol\kappa^{\nu} = 0$ \\
& & &  $\partial_t \boldsymbol\vartheta^{\nu} = - \sum_{\indx{a}=1}^{N} \, \iota_{\prs{a}{\boldsymbol{\vel}} } \prs{a}{\dd}^{\nu} $ \\
\addlinespace[.05in] \cmidrule{4-4}\addlinespace[.05in] 
Layered: & $\partial_t \dd^{\nu} = \mathrm d ( \gamma \bv^{\nu} \boldsymbol \pi ) $ & $\mathrm d \! \left( \partial_t \boldsymbol\vartheta^{\nu} \right) = \sum_{\indx{a}=1}^{N}  \mathrm d \! \left( \prs{a}{ \gamma} \prs{a}{\bv}^{\nu}  \prs{a}{ \boldsymbol \pi } \right)$ & $\partial_t \boldsymbol\vartheta^{\nu} = \sum_{\indx{a}=1}^{N} \, \prs{a}{\gamma} \prs{a}{\bv}^{\nu} \prs{a}{ \boldsymbol\pi } $ \\
\addlinespace[.05in]  \bottomrule
\end{tabular}
%\vskip -0.1in
\caption{Summary of the evolution equations for the internal variables.}
\label{Tab:ev-eq}
\end{table}
%----------------
%----------------

Contrary to intuition, the glide condition is not enough to guarantee that layered dislocation fields remain layered.
As a matter of fact, as was pointed out in Remark~\ref{rmk:slip-rate}, the dislocation curves and the plane distributions evolve according to different mechanisms.
On the other hand, in \S\ref{Layered} we showed that a plane distribution is not necessarily integrable, in the sense that the lattice can be so warped that not only do the crystallographic planes bow, but they may not even exist.
The issue we address in the following lemma is what happens to the glide motion in the case of a time-dependent non-integrable plane distribution.

%---------------------
%---------------------
\begin{lem} \label{Lem:glide}
Let $\Pi$ be a plane distribution on $\mathcal B$ defined by the unit $1$-form $\boldsymbol\pi$.
Let $\df$ be a dislocation $2$-form initially layered on $\Pi$, i.e., such that $\boldsymbol\pi\wedge\df =0$ at $t=0$.
Let us assume that $\df$ is only allowed to glide on $\Pi$, i.e., $\langle \boldsymbol\pi , \boldsymbol{\vel} \rangle =0$ at all times $t$.
Then, the dislocation field remains layered on $\Pi$ at all times $t\geq 0$ if and only if
%---------------------
\begin{equation} \label{integr-bal}
	\partial_t \boldsymbol\pi \wedge \df =
	\dens \, \vel^{\bot}  \, \mathfrak I_{\Pi} \, \mvf
	\,,
\end{equation}
%---------------------
where $\mathfrak I_{\Pi}$ is the integrability object associated with $\Pi$ and $\gamma$ is the rate of plastic slip.
\end{lem}
%---------------------
\begin{proof}
Recalling the definition of the non-autonomous Lie derivative (see~\S\ref{App:Flows}),
first we notice that $\mathsf L_{\boldsymbol{\vel}} (\boldsymbol\pi\wedge\df)=0$ implies that the $3$-form $ \chi^t_s{}^* (\boldsymbol\pi_s \wedge\df_s )$ is constant for all $t$ and $s$.
Therefore, under the assumption $\boldsymbol\pi\wedge\df=0$ at time $t=0$, $\mathsf L_{\boldsymbol{\vel}} (\boldsymbol\pi\wedge\df)=0$ is necessary and sufficient for $\boldsymbol\pi\wedge\df=0$ at all times.
Next, we calculate $\mathsf L_{\boldsymbol{\vel}} (\boldsymbol\pi\wedge\df)$. From the evolution equation~\eqref{rate-decomp}${}_2$ one has $\mathsf L_{\boldsymbol{\vel}} \df =0$, and hence one can write
%---------------------
\begin{equation}
	\mathsf L_{\boldsymbol{\vel}} (\boldsymbol\pi\wedge\df) =
	\mathsf L_{\boldsymbol{\vel}} \boldsymbol\pi\wedge\df =
	\partial_t  \boldsymbol\pi \wedge \df+ \mathfrak L_{\boldsymbol{\vel}}\boldsymbol\pi \wedge \df \,.
\end{equation}
%---------------------
The second term can now be rewritten using Cartan's formula as
%---------------------
\begin{equation}
	\mathfrak L_{\boldsymbol{\vel}}\boldsymbol\pi \wedge \df =
	\left( \mathrm d  \iota_{\boldsymbol{\vel}}\boldsymbol\pi  +  \iota_{\boldsymbol{\vel}}  \mathrm d \boldsymbol\pi  \right) \wedge \df =
	\iota_{\boldsymbol{\vel}}  \mathrm d \boldsymbol\pi \wedge \df \,,
\end{equation}
%---------------------
where use was made of the glide condition~\eqref{glide}, i.e., $ \iota_{\boldsymbol{\vel}}\boldsymbol\pi = \langle \boldsymbol\pi , \boldsymbol{\vel} \rangle =0$.
Invoking~\eqref{contr} for the interior product, one can write $\iota_{\boldsymbol{\vel}}  \mathrm d \boldsymbol\pi \wedge \df =- \mathrm d \boldsymbol\pi \wedge \iota_{\boldsymbol{\vel}}  \df$, as $\mathrm d \boldsymbol\pi \wedge \df $ is a $4$-form, and hence it vanishes in dimension three.
Therefore, one is left with
%---------------------
\begin{equation} \label{non-aut-layer}
	\mathsf L_{\boldsymbol{\vel}} (\boldsymbol\pi\wedge\df) =
	\partial_t \boldsymbol\pi \wedge \df - \mathrm d \boldsymbol\pi \wedge \iota_{\boldsymbol{\vel}} \df \,.
\end{equation}
%---------------------
Recalling~\eqref{sliprate}, and the definition of the integrability object $\mathfrak I_{\Pi}$ in~\eqref{int-object},
one can write
%---------------------
\begin{equation}
	\mathrm d \boldsymbol\pi \wedge \iota_{\boldsymbol{\vel}} \df = 
	-\gamma \, \mathrm d \boldsymbol\pi \wedge \boldsymbol\pi =
	-\gamma \, \mathfrak I_{\Pi} \, \mvf =
	\dens \, \vel^{\bot} \, \mathfrak I_{\Pi} \, \mvf \,,
\end{equation}
%---------------------
where $\mvf$ is the material volume form.
Hence, the proof is complete.
\end{proof}
%---------------------
%---------------------

\begin{figure}
\centering
\includegraphics[width=.45\textwidth]{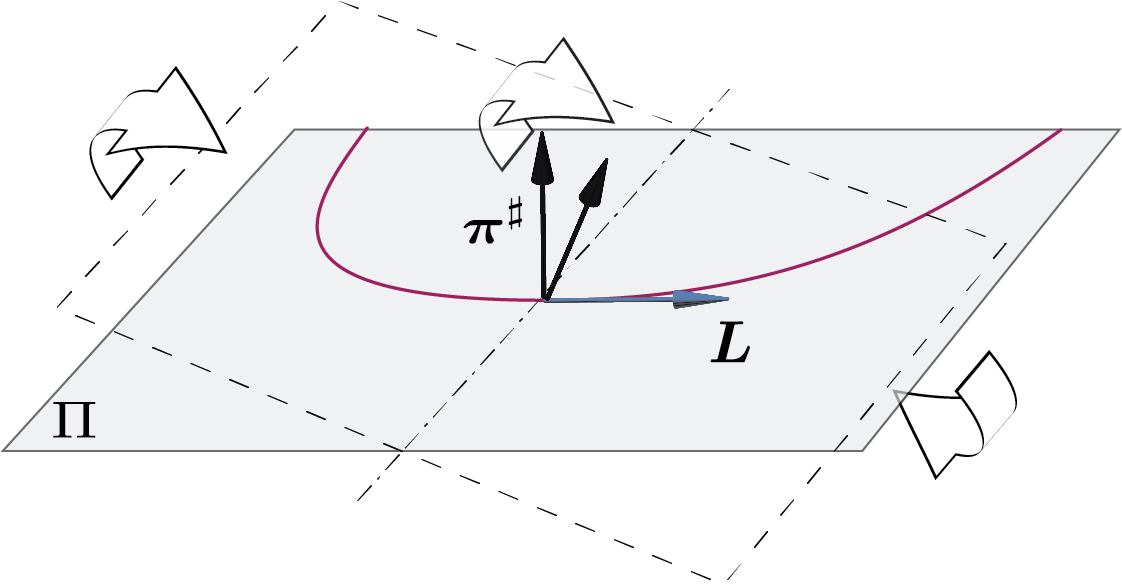}
\vskip 0.1in
\caption{Glide motion of bundles of dislocations.
For each non-integrable slip dislocation field, the non-integrability is compensated by a rotation of the slip system in the direction of the dislocation line director.
}
\label{fig:glide}
\end{figure}

In other words, having glide does not guarantee that initially layered dislocation fields remain layered at all times.
Eq.~\eqref{integr-bal} shows that this is due to two mechanisms:
i) the lattice structure changes with time, and hence so does the plane distribution $\Pi$;
ii) the non-integrability of the slip distribution.
It should also be noticed that Eq.~\eqref{integr-bal} can be written as $\langle \partial_t \boldsymbol\pi , \boldsymbol{\lv}\rangle = -\vel^{\bot}  \mathfrak I_{\Pi} $.
Therefore, in the case of layered dislocations on non-integrable plane distributions, in order to accommodate the glide motion, the slip plane must rotate towards the dislocation line director at a rate that is proportional to the glide velocity and to the non-integrability content of the slip distribution, see Fig.~\ref{fig:glide}.

%---------------------
%---------------------

%---------------------
%---------------------
\begin{remark} \label{rmk:single-rate}
In the case of a single dislocation field following Orowan's equation~\eqref{orowan-single}, one has $\partial_t \boldsymbol\pi = - \pi_{\nu} \bv^{\nu} \iota_{\boldsymbol{\vel}}\df=0$ by virtue of~\eqref{slip-constr}${}_1$, and hence the slip surfaces are time independent.
If $\Pi$ is non-integrable, from~\eqref{integr-bal} one obtains $\partial_t \boldsymbol\pi=0$, and hence $\vel^{\bot}=0$.
This means that single dislocation fields layered on non-integrable slip planes cannot glide.
In other words, non-integrable slip planes behave as anchors for distributions of dislocations.
\end{remark}
%---------------------
%---------------------

Next we consider the entire ensemble of $N$ dislocation fields and their associated slip systems.
We focus on what happens to the $\indx a$-th glide motion.
Notice that $\prs{a}{\boldsymbol\pi} = \prs{a}{\pi}_{\nu}\boldsymbol\vartheta^{\nu}$, with $\prs{a}{\pi}_{\nu}$ uniform in space and constant in time.  By assuming closed dislocation fields obeying Orowan's equation~\eqref{orowan}, one writes the left-hand side of~\eqref{integr-bal} as
%---------------------
\begin{equation} 
	\partial_t \prs{a}{\boldsymbol\pi} \wedge \prs{a}{\df} =
	- \sum_{\indx{b}=1}^{N}
	\prs{a}{\pi}_{\nu} \prs{b}{\bv}^{\nu} \,
	\iota_{\prs{b}{\boldsymbol{\vel}} } \prs{b}{\df} \,
	\wedge \prs{a}{\df} \,.
\end{equation}
%---------------------
Thus, recalling the definition~\eqref{cross1} of the non-Schmid forms, Eq.~\eqref{integr-bal} becomes
%---------------------
\begin{equation} \label{integr-bal7}
	\sum_{\indx{b}=1}^{N}
	\langle \prs{a\!b}{\boldsymbol\varpi} , \prs{b}{\boldsymbol{\vel}} \rangle 
	+\prs{a}{\dens} \, \prs{a}{\vel}^{\bot} \, \mathfrak I_{\prs{a}{\pi}} 
	=0
	\,,\qquad
	\indx{a} = 1,2,\hdots,N \,.
\end{equation}
%---------------------
From~\eqref{non-aut-layer}, one should note that the right-hand side of~\eqref{integr-bal} can also be written in the form $\mathrm d \boldsymbol\pi \wedge \iota_{\boldsymbol{\vel}} \df$, which invoking~\eqref{gnd-sum} becomes
%---------------------
\begin{equation}
	\mathrm d \prs{a}{\boldsymbol\pi} \wedge \iota_{ \prs{a}{\boldsymbol{\vel}} } \prs{a}{\df} =
	- \sum_{\indx{b}=1}^{N}
	\prs{a}{\pi}_{\nu} \prs{b}{\bv}^{\nu} \,
	\iota_{\prs{b}{\boldsymbol{\vel}} } \prs{b}{\df} \,
	\wedge \prs{a}{\df} \,,
\end{equation}
%---------------------
and hence, using~\eqref{cross1} again, one can recast~\eqref{integr-bal7} as
%---------------------
\begin{equation} \label{integr-bal6}
	\sum_{\indx{b}=1}^{N}
	\langle \prs{a\!b}{\boldsymbol\varpi} , \prs{b}{\boldsymbol{\vel}} \rangle 
	=
	\Big\langle \sum_{\indx{b}=1}^{N} \prs{a\!b}{\boldsymbol\varpi} ~,\, \prs{a}{\boldsymbol{\vel}} \Big\rangle 
	\,,\qquad
	\indx{a} = 1,2,\hdots,N \,.
\end{equation}
%---------------------

%---------------------
%---------------------
\begin{example}[A single dislocation field on a non-integrable slip distribution] \label{example-oneslip}
We consider a dislocation field on a non-integrable plane distribution.
In Remark~\ref{rmk:single-rate} we showed that glide on a stationary non-integrable plane distribution is not allowed.
However, according to Lemma~\ref{Lem:glide}, the glide motion can be unlocked when the orientation of the plane distribution changes in time by the effect of the evolution of the lattice structure.
This can be caused by the glide of other dislocation fields (as in Example~\ref{example-twoslips}), or by any other type of anelastic process.
As an example, let us take the following dislocation form:
%---------------------
\begin{equation} \label{df-example-oneslip}
	\df (Z^2,t)= a(Z^2,t) \, \mathrm d Z^2 \wedge \mathrm d Z^3 - c(Z^2,t) \, \mathrm d Z^1 \wedge \mathrm d Z^2 \,,
\end{equation}
%---------------------
that we assume is convected by the material velocity $\boldsymbol{\vel} = \vel(Z^2,t) \frac{\partial}{\partial Z^2}$.
Note that the conditions $\partial_t a =-\frac{\partial}{\partial Z^2}(a\vel)$ and $\partial_t c=-\frac{\partial}{\partial Z^2}(c\vel)$ must hold in order to satisfy the evolution equation~\eqref{rate-Cart-decomp}.
We also take the following $1$-form:
%---------------------
\begin{equation}
	\boldsymbol\pi (Z^2,t) =
	c(Z^2,t) \, \mathrm d Z^1 + a(Z^2,t) \, \mathrm d Z^3  \,,
\end{equation}
%---------------------
defining a non-integrable plane distribution $\Pi$ as
%---------------------
\begin{equation}
	\mathrm d\boldsymbol\pi\wedge\boldsymbol\pi 
	= \left(c\, \frac{\partial a}{\partial Z^2}-a\, \frac{\partial c}{\partial Z^2} \right) 
	\mathrm d Z^1 \wedge \mathrm d Z^2 \wedge \mathrm d Z^3 \,.
\end{equation}
%---------------------
In particular, referring the integrability object to a Euclidean volume form, one has $\mathfrak{I}_{\Pi}=c\, \frac{\partial a}{\partial Z^2} -a\, \frac{\partial c}{\partial Z^2}$.
However, the time dependency of $\boldsymbol\pi$ allows the plane distribution to accommodate the dislocation glide
as both the layer condition~\eqref{slip-constr}$_2$ 
and the glide condition~\eqref{glide} are satisfied at all times.
In particular, one obtains the rate of change of $\boldsymbol\pi$ in the direction of the dislocation line director as
%---------------------
\begin{equation}
\begin{split}
	\partial_t \boldsymbol\pi\wedge\df 
	= (a\, \partial_t c - c\, \partial_t a ) \,\mathrm d Z^1 \wedge \mathrm d Z^2 \wedge \mathrm d Z^3 
	= \left(c\, \frac{\partial a}{\partial Z^2} -a\, \frac{\partial c}{\partial Z^2} \right)  
	\vel  \,\mathrm d Z^1 \wedge \mathrm d Z^2 \wedge \mathrm d Z^3\,,
\end{split}
\end{equation}
%---------------------
where use was made of the conditions $\partial_t a =-\frac{\partial }{\partial Z^2}( a\vel)$, and $\partial_t c=-\frac{\partial }{\partial Z^2}(c \vel)$.
Hence, we have obtained $\partial_t \boldsymbol\pi\wedge\df=\vel \, \mathrm d\boldsymbol\pi\wedge\boldsymbol\pi$, which is consistent with Lemma~\ref{Lem:glide}.
\end{example}
%---------------------
%---------------------

%---------------------
\begin{figure}[tp]
\centering
\includegraphics[width=\textwidth]{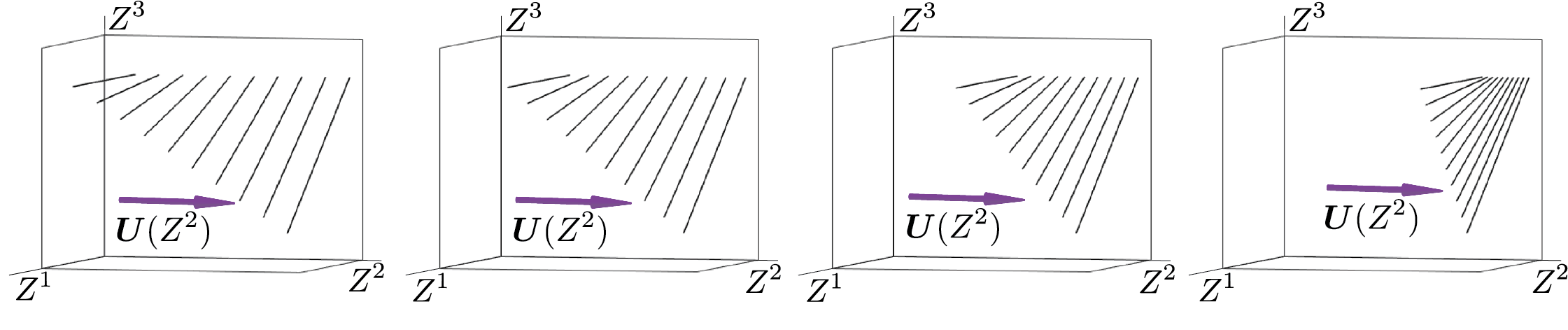}
\vskip 0.0in
\caption{Glide of a dislocation field on a time-dependent non-integrable plane distribution at four different times.
The dislocations move towards the right side of the sample box.
At each point, the orientation of the glide plane needs to change in time in order to accommodate new dislocations coming with different slopes.
Another effect of the time dependence of the plane distribution, combined with a non-uniform velocity, is the evolution of the ruled surface generated by the dislocation lines.}
\label{fig:nonint-glide}
\end{figure}
%---------------------

%---------------------
\begin{example}[Two slip distributions of dislocations.] \label{example-twoslips}
In this example, we consider the case of two closed dislocation fields on two different glide planes.
The condition~\eqref{integr-bal6} is written as
%---------------------
\begin{equation}
	\langle \prs{1\!2}{\boldsymbol\varpi} , \prs{2}{\boldsymbol{\vel}} \rangle 
	=
	\langle \prs{1\!2}{\boldsymbol\varpi} , \prs{1}{\boldsymbol{\vel}} \rangle 
	\,,\qquad
	\langle \prs{2\!1}{\boldsymbol\varpi} , \prs{1}{\boldsymbol{\vel}} \rangle 
	=
	\langle \prs{2\!1}{\boldsymbol\varpi} , \prs{2}{\boldsymbol{\vel}} \rangle 
	\,.
\end{equation}
%---------------------
Since $\prs{1\!2}{\boldsymbol\varpi}=f\, \prs{2\!1}{\boldsymbol\varpi}$ for some scalar $f$, the two equations are dependent. Therefore, one obtains the following single condition for the gliding motion:
%---------------------
\begin{equation} \label{integr-bal-ex}
	\langle \prs{1\!2}{\boldsymbol\varpi} , \prs{1}{\boldsymbol{\vel}} \rangle 
	=
	\langle \prs{1\!2}{\boldsymbol\varpi} , \prs{2}{\boldsymbol{\vel}} \rangle 
	\,.
\end{equation}
%---------------------
As was mentioned earlier, the $1$-form $ \prs{1\!2}{\boldsymbol\varpi} $ generates the distribution of planes spanned by $\prs{1}{\boldsymbol{\lv}}$ and $\prs{2}{\boldsymbol{\lv}}$, see Fig.~\ref{fig:coupling}.
Therefore, the out-of-plane components of the two velocities need to be equal.
Unlike a single dislocation field on a non-integrable slip distribution described in Example~\ref{example-oneslip}, in this case, when the two dislocation velocities satisfy~\eqref{integr-bal-ex}, glide is allowed.
\end{example}
%--------------------

\begin{figure}[tp]
\centering
\includegraphics[width=\textwidth]{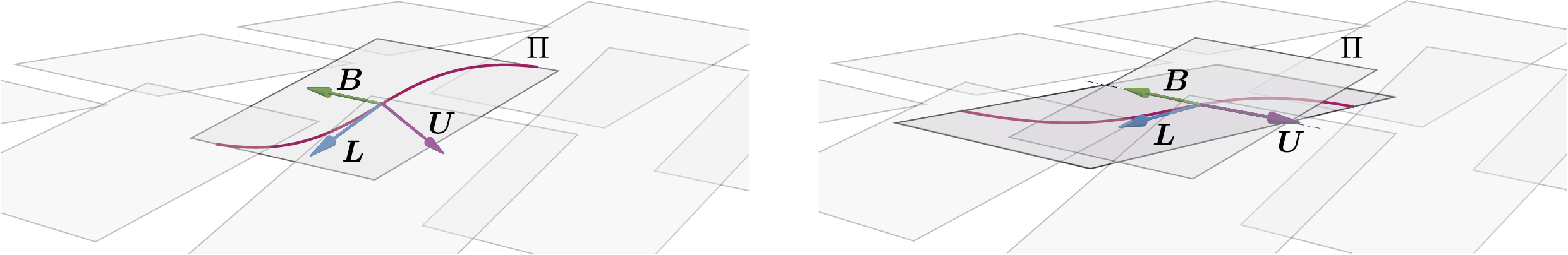}
\vskip 0.2in
\caption{Different layer conditions for gliding dislocations. Left: Strongly layered dislocation fields, for which both $\boldsymbol{\bv}$ and $\boldsymbol{\lv}$---as well as $\boldsymbol{\vel}$---belong to $\Pi$, i.e., $\langle \boldsymbol{\pi} , \boldsymbol{\bv} \rangle = \langle \boldsymbol{\pi} , \boldsymbol{\lv} \rangle = \langle \boldsymbol{\pi} , \boldsymbol{\vel} \rangle = 0$. Right: Weakly layered dislocation fields, satisfying $\langle \boldsymbol{\pi} , \boldsymbol{\bv} \rangle = \langle \boldsymbol{\pi} , \boldsymbol{\vel} \rangle = 0$, and $\iota_{ \boldsymbol{\vel} } \iota_{ \boldsymbol{\bv} } \df = 0$ (conservative motion condition), whence $\boldsymbol{\vel} \parallel \boldsymbol{\bv} $.}
\label{fig:weak}
\end{figure}

Lastly, we look at the glide of dislocation fields that are not strongly layered on their slip plane distribution.
We do this by dropping the layer condition~\eqref{constr-recap}$_2$ on the line director, while keeping the layer condition~\eqref{constr-recap}$_1$ on the Burgers director valid.
By allowing $\boldsymbol\pi \wedge \df \neq 0$, Eq.~\eqref{sliprate} is no longer satisfied.
In other words, when the dislocation curves do not locally lie on the slip plane,
the $1$-forms $\boldsymbol\pi$ and  $\iota_{\boldsymbol{\vel}}\df$ define two different plane distributions.
Hence, instead of~\eqref{slip-constr}$_2$ we introduce the following condition:
%---------------------
\begin{equation} \label{weak-condition}
	\iota_{ \boldsymbol{\vel} } \iota_{ \boldsymbol{\bv} } \df = 0 \,,
\end{equation}
%---------------------
enforcing the glide on the plane that is spanned by the Burgers and the line directors.
A dislocation field satisfying~\eqref{slip-constr}$_1$ and~\eqref{weak-condition} is said to be \emph{weakly layered}.
In other words, we are requiring that $\boldsymbol{\vel}$ belongs to the two planes defined by $\boldsymbol \pi$ and $\iota_{ \boldsymbol{\bv} }  \df $, see Fig.~\ref{fig:weak}.
It should be noted that $\iota_{ \boldsymbol{\vel} } \iota_{ \boldsymbol{\bv} } \df = 0 $ is equivalent to $\mvf( \boldsymbol{\lv} , \boldsymbol{\bv} , \boldsymbol{\vel} )= 0 $, and hence, recalling~\eqref{orowan-single}, Eq.~\eqref{weak-condition} enforces the dislocation motion to preserve the material volume via Orowan's equation, i.e., to be conservative.
Assuming $\langle \boldsymbol \pi , \boldsymbol{\vel} \rangle=0$, Eq.~\eqref{weak-condition} holds if and only if the dislocation field falls into one of the following disjoint categories:
%---------------------
\begin{enumerate}  \setlength\itemsep{0em}
\item Dislocation fields with a screw character, i.e, for which $\iota_{ \boldsymbol{\bv}} \df=0$, see~\S\ref{Decomposable};
\item Dislocation fields with $\iota_{ \boldsymbol{\bv}} \df \neq0$ (edge or mixed) that are layered on $\boldsymbol\pi$, as $\iota_{ \boldsymbol{\bv} } \df = \boldsymbol\pi $;
\item Dislocation fields with $\iota_{ \boldsymbol{\bv}} \df \neq0$ (edge or mixed) that are not layered on $\boldsymbol\pi$ but for which $\boldsymbol{\vel}$ and $\boldsymbol{\bv}$ are parallel.
\end{enumerate}
%---------------------
The previous result is the analogue of Lemma~\ref{Lem:glide} for the condition~\eqref{weak-condition}.
The proof is straightforward and consists of simply listing the cases in which~\eqref{weak-condition} is satisfied (leaving out the case $\iota_{ \boldsymbol{\vel} } \df = 0$).
In other words $\iota_{ \boldsymbol{\vel} } \iota_{ \boldsymbol{\bv} } \df = 0$ includes the case $\boldsymbol\pi \wedge \df = 0$, and hence, is a weaker condition.
Moreover, similar to~\eqref{lUspanPi}, we set
%---------------------
\begin{equation}
	 \iota_{\boldsymbol{\vel}} \df  = \dens \, \widetilde{\vel}^{\bot} \widetilde{\boldsymbol\pi}   \,,
\end{equation}
%---------------------
where $\widetilde{\vel}^{\bot} = \llangle \boldsymbol{\vel}  , \widetilde{\boldsymbol{\norm}} \rrangle_{\boldsymbol G} $, and $\iota_{\widetilde{\boldsymbol{\norm}}} \df  = \dens \, \widetilde{\boldsymbol\pi}$.
Note that $\widetilde{\boldsymbol\pi} \neq\boldsymbol\pi$ as $\df\wedge \boldsymbol\pi \neq 0$.
Similar to~\eqref{sliprate} $\iota_{\boldsymbol{\vel}} \df = -\widetilde\gamma \widetilde{\boldsymbol\pi}$, one has $\widetilde\gamma = -\dens \, \widetilde{\vel}^{\bot}$,
while~\eqref{orowan-slip} becomes
%---------------------
\begin{equation}
	\partial_t \boldsymbol\vartheta^{\nu} =
	\sum_{\indx{a}=1}^{N} \, \prs{a}{\widetilde\gamma} \, \prs{a}{\bv}^{\nu} \, \prs{a}{\widetilde{\boldsymbol\pi}} \,,
	\quad\text{or}\qquad
	\subp{\mathbf{L}} =
	\sum_{\indx{a}=1}^{N} \, \prs{a}{\widetilde\gamma} \, \prs{a}{\boldsymbol{\bv}} 
	\otimes \prs{a}{\widetilde{\boldsymbol\pi}} \,.
\end{equation}
%---------------------

%---------------------
%---------------------

%%%%%%%%%%%%%%%%%%%%%%%%%%
%%%%%%%%%%%%%%%%%%%%%%%%%%
\section{The variational formulation} \label{Sec:Variational}

In this section we discuss the dynamics of dislocation fields in single crystals.
We start with a variational principle of the Lagrange-d'Alembert type for non-conservative processes as in \citep{marsden2013introduction} to obtain the governing equations as the Euler-Lagrange equations associated with both spatial and material variations.
While spatial variations provide the standard balance of linear momentum, material variations give the kinetic equations.
These are expressed in terms of generalized material forces and represent a balance of microstructural actions.
To this extent, the present theory falls within the model-building framework for the mechanics of complex materials. 
We use a two-potential approach, in which the free energy is a function of the strain and of the internal variables, and the dissipation potential depends on the internal variables and on their rates.
Micro-inertial effects associated with the motion of dislocations are taken into account, as well as some geometric constraints that the lattice exerts on the dislocation fields and their motions in order to keep them gliding on assigned lattice planes.
The balance of energy is also derived.

%------------------------------------------------------------------------------
%------------------------------------------------------------------------------
\subsection{The action principle}    \label{Variational}

In~\S\ref{Sec:Kinematics} it was shown that, assuming Orowan's equation, the kinematics of both the dislocation fields $\prs{a}{\dd}^{\nu}$ and the lattice forms $\boldsymbol\vartheta^{\nu}$ is determined by the $N$ material motions $\prs{a}{\chi}$  via~\eqref{rate-Cart} and~\eqref{orowan}, with the associated initial conditions.
Therefore, the generalized configuration space $\mathcal{C}$ for a single crystal is made of $(N+1)$-tuples of the type $\left( \varphi, \prs{1}{\chi}, \hdots ,\prsn{N}{\chi} \,\right)$,
expressing placements of points of~$\mathcal B$ in the ambient space~$\mathcal S$, as well as those of the dislocation fields in the body~$\mathcal B$.
As the degrees of freedom of the system are given by maps of~$\mathcal B$, the configuration space $\mathcal{C}$ is infinite dimensional.
As was mentioned earlier, our goal is to derive the governing equations of the system as the Euler-Lagrange equations associated with a variational principle.
Therefore, we fix a time interval $[t_1 , t_2]$, and look at paths $c:[t_1,t_2]\to\mathcal{C}$ in the configuration space.
We define an action functional $\mathsf{A}$ on the space of paths, viz.
%---------------------
\begin{equation} \label{action}
	\mathsf{A} ( c ) =
	\int_{t_1}^{t_2}  \int_{\mathcal B} \mathscr{L} \, \boldsymbol{m} \, \mathrm dt \,,
\end{equation}
%---------------------
where the scalar field $\mathscr{L}$ is the Lagrangian density per unit mass.
We consider variations of the path $c$, i.e., one-parameter families of homotopic curves $c(t,\epsilon)$  in $\mathcal{C}$ with fixed end points,
and formally denote the first variation operator with $\delta = \frac{\mathrm d}{\mathrm d\epsilon}\vert_{\epsilon=0}$.
In particular, variations of the generalized configurations $(\varphi,\prs{1}\chi, \hdots, \prsn{N}\chi)$ are one-parameter families of motions $\varphi_{t,\epsilon}$ and $\prs{a}\chi_{t,\epsilon}$. 
It should be noted that the variation $\delta\varphi$ is a vector in the ambient space, whereas $\delta\prs{a}{\chi}$ is a vector in the material manifold.
\S\ref{Spatial} and~\S\ref{Material} are devoted to the formalization of the concepts of spatial and material variations, respectively.

We assume the existence of a free energy density per unit mass as a function $\mathscr{F}$ of the pulled-back metric $\boldsymbol{C}^{\flat}=\varphi^*\boldsymbol g$, the lattice forms $ \boldsymbol\vartheta^{\nu}  $, and the dislocation fields $ \prs{a}{\dd}^{\nu} $.
In short, $\mathscr{F}=\mathscr{F}\left(\boldsymbol{C}^{\flat},  \boldsymbol\vartheta^{\nu}  , \prs{a}{\dd}^{\nu}  \right)$.
The total free energy is then defined as the integral $\int_{\mathcal B}\mathscr{F} \, \boldsymbol{m}$, where $\boldsymbol{m}=\massd \mvf$ is the mass $3$-form, with $\massd$ being the material mass density.
The kinetic energy density per unit mass $\mathscr{T}$ can be written as the sum of a spatial part $\mathscr{T}_S = \frac{1}{2}\Vert\boldsymbol V\Vert^2_{\boldsymbol g}$ associated with the motion $\varphi$ in the ambient space,
plus a material part $\mathscr{T}_M (  \boldsymbol\vartheta^{\nu}  , \prs{a}{\dd}^{\nu} , \prs{a}{\boldsymbol{\vel}} )$ associated with the dislocation motions.
This micro-kinetic energy $\mathscr{T}_M$ is responsible for the presence of inertial terms in the kinetic equations, i.e., the equations of motion for the dislocation fields \citep{eshelby1953equation,hirth1998forces}.
Then, one defines the Lagrangian density as $\mathscr{L} = \mathscr{T} - \mathscr{F} = \mathscr{T}_S+ \mathscr{T}_M - \mathscr{F}$, or with explicit reference to its arguments, as
%---------------------
\begin{equation} \label{lag-state-var}
	\mathscr{L} (\boldsymbol{C}^{\flat}, \boldsymbol V, \boldsymbol\vartheta^{\nu}  , \prs{a}{\dd}^{\nu}  ,  
	\prs{a}{\boldsymbol{\vel}} )
	=\mathscr{T}_S ( \boldsymbol V ) +
	\mathscr{T}_M (  \boldsymbol\vartheta^{\nu}  , \prs{a}{\dd}^{\nu}  ,  \prs{a}{\boldsymbol{\vel}} ) -
	\mathscr{F} (\boldsymbol{C}^{\flat},  \boldsymbol\vartheta^{\nu}  , \prs{a}{\dd}^{\nu} ) \,.
\end{equation}
%---------------------

We assume a Lagrange-d'Alembert type variational principle, stating that the generalized configurations at $\epsilon=0$ satisfy the following identity:
%---------------------
\begin{equation} \label{lag-dal}
	\delta \mathsf{A} ~+~
	\sum_{\indx{a}=1}^N \int_{t_1}^{t_2} \int_{\mathcal B} \langle \prs{a}{\boldsymbol\tau} , \delta  \prs{a}{\chi} \rangle \, \mvf \, \mathrm dt +
	\int_{t_1}^{t_2} \int_{\partial \mathcal B} \langle \mathbf{t} , \delta\varphi \rangle \, \boldsymbol{\varsigma} \,\mathrm dt+
	\int_{t_1}^{t_2} \int_{\mathcal B} \langle \mathbf{b} , \delta\varphi \rangle \, \boldsymbol{m}\, \mathrm dt = 0 \,,
\end{equation}
%---------------------
for all variation fields $\delta\varphi$ and $\delta  \prs{a}{\chi}$, $\indx{a}=1,2,\hdots,N$.
The $1$-forms $\prs{a}{\boldsymbol\tau}$, $\mathbf{t}$ and $\mathbf{b}$ in~\eqref{lag-dal} have the following interpretations.
%---------------------
\begin{itemize} \setlength\itemsep{0em}
	\item $\prs{a}{\boldsymbol\tau}$ is a field of material $1$-forms on $\mathcal B$ representing the drag force exerted by the lattice on the motion of the $\indx{a}$-th dislocation field in the material manifold, $\indx{a} = 1,2,\hdots,N$. This means that glide is an irreversible process that dissipates energy.
	\item $\mathbf{t}$ is a field of spatial $1$-forms on the boundary $\partial \varphi(\mathcal B)$ representing the contact forces (tractions) referred to the material area element.
	\item $\mathbf{b}$ is a field of spatial $1$-forms on $\varphi(\mathcal B)$ representing a body force density per unit mass.
\end{itemize}
%---------------------
We also assume the existence of a dissipation potential $\mathscr{D}$,\footnote{%
See \citep{germain1983continuum}, \citep{goldstein2002classical}, and \citep{kumar2016two} for discussions on the dissipation potential in the context of thermomechanics, classical mechanics, and viscoelasticity. For a similar approach, see \citet{ziegler1958attempt} and \citet{ziegler1987derivation} for discussions on the dissipation function.}
which is a function of the internal variables and a convex function of the dislocation velocities $\prs{a}{\boldsymbol{\vel}} $, such that the drag force on the motion of the $\indx{a}$-th dislocation field can be written as
%---------------------
\begin{equation} \label{drag}
	\prs{a}{\boldsymbol\tau} = - \massd \, \frac{\partial \mathscr{D}}{\partial \prs{a}{\boldsymbol{\vel}} } \,.
\end{equation}
%---------------------
%
It should be noted that the body $\mathcal B$ is assumed fixed,\footnote{%
This is the case for most problems in anelasticity, unless some accretion process, such as growth by addition of material or by diffusion/migration of nutrients, is involved  \citep{sozio2019nonlinear}.}
and so is the material mass form $\boldsymbol{m}$, as it does not depend on any of the internal variables by virtue of mass conservation \citep{sozio2020riemannian}.
Therefore, the variation operator $\delta$ in the first term of~\eqref{lag-dal} can be brought inside the integral~\eqref{action}, which is one of the advantages of working with densities per unit mass. This means that our variational approach consists of calculating the quantity $\delta \mathscr{L}$ associated with both the spatial and material variations $\delta\varphi$ and  $\delta\prs{a}{\chi}$.

%---------------------------------------------------------------
%---------------------------------------------------------------
\subsection{Spatial variations}       \label{Spatial}

Spatial variations can be seen as perturbations of the configuration mapping $\varphi$, while the material motions $\prs{1}\chi,\hdots,\prsn{N}\chi$ are kept unperturbed.
They allow one to obtain the standard balance of linear momentum and traction boundary conditions.
We consider one-parameter families of motions $\varphi_{t,\epsilon}$ during a time interval $[t_1,t_2]$, such that for $\epsilon=0$ one recovers the motion $\varphi_t$ that satisfies the Lagrange-d'Alembert principle, and such that all the trajectories agree at their endpoints, i.e., $\varphi_{t_1,\epsilon}$ and $\varphi_{t_2,\epsilon}$ do not depend on $\epsilon$.
We denote with $\delta\varphi$ the vector field tangent to the curves $\epsilon\mapsto\varphi_{t,\epsilon}(X)$ in $\mathcal S$, for $X$ and $t$ fixed, and evaluated at $\epsilon=0$.
In geometric terms, setting $\varphi_{t,X} : \epsilon\mapsto\varphi_{t,\epsilon}(X)$ for $X$ and $t$ fixed, $\delta\varphi$ can be expressed as
%---------------------
\begin{equation} \label{lag-spatial}
	\delta \varphi = (\varphi_{t,X})_* \left( \tfrac{\partial}{\partial \epsilon} \right) \Big\vert_{\epsilon=0} \,.
\end{equation}
%---------------------
%
Note that one has $\delta\varphi(t_1) = \delta\varphi(t_2) = \boldsymbol 0$ as all the trajectories agree at their endpoints.
As was mentioned earlier, we denote $\delta = \frac{\mathrm d}{\mathrm d\epsilon}\vert_{\epsilon=0}$.\footnote{%
Unlike scalar fields, the first variation of a tensor field is well-defined when the base point does not change as $\epsilon$ varies.
For example, while $\delta \boldsymbol{C}^{\flat}$ is well-defined, the expression $\delta \boldsymbol V$ is not, as by changing $\epsilon$, one also changes the trajectory. In order to define $\delta \boldsymbol V$ one needs to identify tangent spaces at different points of $\mathcal S$, e.g., via a connection.}
It should be noted that the micro-kinetic energy $\mathscr{T}_M$ in~\eqref{lag-state-var} was assumed independent of both $\boldsymbol V$ and $\boldsymbol{C}^{\flat}$, and hence has vanishing spatial variation.
Following a procedure that is well-known in solid mechanics, one obtains the balance of linear momentum
in terms of the first Piola-Kirchhoff stress tensor \citep{simo1988hamiltonian, sozio2020riemannian}.

Let us next introduce and discuss stress tensors.
First, as is customary in hyperelasticity, we define the second Piola-Kirchhoff stress as the following symmetric tensor of the type $(2,0)$:
%---------------------
\begin{equation} \label{second-pk}
	\boldsymbol S = 2 \massd \frac{\partial \mathscr{F}}{\partial \boldsymbol{C}^{\flat} } \,, 
\end{equation}
%---------------------
which is conjugate to variations of the pulled-back metric $\boldsymbol{C}^{\flat}$.
For every point $X\in\mathcal B$ one can look at the tensor $\boldsymbol S_X$ as an operator $\boldsymbol S_X:T_X^*\mathcal B\to T_X\mathcal B$ acting on a $1$-form $\boldsymbol \nu$ to give a vector $\boldsymbol S (\boldsymbol \nu)$ with components $S^{AB}\nu_B$.
With the same notation, the first Piola-Kirchhoff stress $\boldsymbol P$ is defined as the two-point tensor of the type $(2,0)$ that satisfies
%---------------------
\begin{equation} \label{P-S}
	\boldsymbol P (\boldsymbol\nu) =
	\op F \boldsymbol S (\boldsymbol\nu)  \,,
\end{equation}
%---------------------
for any $1$-form $\boldsymbol \nu$.
This is sometimes written as $\boldsymbol P=\op F \boldsymbol S$.
Hence, at every point $X\in\mathcal B$ the first Piola-Kirchhoff stress can be seen as an operator $\boldsymbol P_X:T_X^*\mathcal B\to T_{\varphi(X)}\varphi(\mathcal B)$ acting on a material $1$-form $\boldsymbol \nu$ to give a spatial vector $\boldsymbol P (\boldsymbol \nu)$ with components $P^{aB}\nu_B$.
One can also define $(1,1)$ variants of the second Piola-Kirchhoff stress by lowering one of the two indices (either one by symmetry) via the material metric $\boldsymbol G$ to obtain a tensor $\boldsymbol{S}^{\flat}$, or via the pulled-back metric $\boldsymbol{C}^{\flat}=\varphi^*\boldsymbol g$ to obtain a tensor $\op M$, viz.
%---------------------
\begin{equation} \label{mixed-second-mandel}
	\langle \boldsymbol\nu , \boldsymbol{S}^{\flat} \boldsymbol Y \rangle =
	\llangle \boldsymbol S (\boldsymbol\nu) , \boldsymbol Y \rrangle_{\boldsymbol G}
	\,,\qquad
	\langle \boldsymbol\nu , \op M \boldsymbol Y \rangle =
	\llangle \boldsymbol S (\boldsymbol\nu) , \boldsymbol Y \rrangle_{\boldsymbol{C}^{\flat}} \,,
\end{equation}
%---------------------
for any $1$-form $\boldsymbol \nu$ and vector $\boldsymbol Y$.
From the symmetry of $\boldsymbol{S}$, it follows that $\boldsymbol{S}^{\flat}$ and $\op M$ are self-adjoint with respect to $\boldsymbol G$ and $\boldsymbol{C}^{\flat}$, respectively \citep{epstein1990energy}.
The tensor $\op M$ is called the Mandel stress.
The $(1,1)$ variant of the first Piola-Kirchhoff stress is obtained by lowering the spatial index via the ambient metric $\boldsymbol g$, viz.
%---------------------
\begin{equation} \label{P-op}
	\langle  \boldsymbol\nu , \op P\boldsymbol y \rangle =
	\llangle \boldsymbol P (\boldsymbol\nu) , \boldsymbol y \rrangle_{\boldsymbol g} \,,
\end{equation}
%---------------------
for any $1$-form $\boldsymbol \nu$ and vector $\boldsymbol y$.
This can be seen as a map $T\varphi(\mathcal B) \to T\mathcal B$, or, in the dual version $\op P^{\star}$, as a map $T^*\mathcal B \to T^*\varphi(\mathcal B)$.
Note that since $\boldsymbol{C}^{\flat} = \varphi^* \boldsymbol g$ one can also define the Mandel stress as $\op M = \op P \op F$.
The balance of linear momentum in terms of the mixed first Piola-Kirchhoff tensor $\op P$ is expressed as
%---------------------
\begin{equation} \label{EL-spatial}
	\operatorname{Div} \op P + \massd \boldsymbol \beta = \massd \boldsymbol A^{\flat} \,,
\end{equation}
%---------------------
%
with the boundary condition
%---------------------
\begin{equation} \label{EL-spatial-boundary}
	\op P^{\star} \boldsymbol \nu = \mathbf{t}
	\quad\text{on}~\partial\mathcal B \,.
\end{equation}
%---------------------
It should be noted that when assigning the $1$-form $\mathbf{t}$ on $\mathcal B$, Eq.~\eqref{EL-spatial-boundary} represents a traction boundary condition.
However, when the deformation map (displacement) is assigned on part or on the entire boundary $\partial \mathcal B$, the $1$-form $\mathbf{t}$ becomes a Lagrange multiplier associated with the constraint $\varphi = \overline\varphi$.

%---------------------
%---------------------
\begin{remark} \label{Rem:Mandel-int}
The Mandel stress can be interpreted in the light of work conjugacy.
First, from~\S\ref{Rates} one should recall that the time derivative of the pulled-back metric can be written as $ \partial_t \boldsymbol{C}^{\flat} = \partial_t \varphi^* \boldsymbol g = 2\,\mathsf{sym} \boldsymbol{\mathrm l}^{\flat} $.
Then, from the definition of Mandel stress~\eqref{mixed-second-mandel}$_2$, one obtains
%---------------------
\begin{equation} \label{rate-C-mandel}
	\frac{1}{2} \langle  \boldsymbol S , \partial_t \boldsymbol{C}^{\flat} \rangle =
	\langle  \boldsymbol S , \varphi^* \mathsf{sym} \boldsymbol{\mathrm l}^{\flat} \rangle =
	\langle  \boldsymbol S , \varphi^* \boldsymbol{\mathrm l}^{\flat} \rangle =
	\langle  \op M , \mathbf{L} \rangle \,,
\end{equation}
%---------------------
where use was made of $\mathbf{L} = \varphi^*\boldsymbol{\mathrm l}\,$.
This means that the Mandel stress is work conjugate to the total rate of deformation $\mathbf{L} = \varphi^* \nabla^{(\boldsymbol g)}\boldsymbol v$.
See also Remark~\ref{Rem:Mandel-elastic} for an interpretation of Mandel stress under the assumption of purely hyperelastic free energy.
The Mandel stress can also be interpreted in the light of Cauchy's stress theorem.
As a matter of fact, from $\op M = \op P \op F$ one can write \eqref{EL-spatial-boundary} as $\mathbf{t}= \op F^{-\star} \op M^{\star} \boldsymbol\nu$, and hence, $\varphi^*\mathbf{t}= \op M^{\star} \boldsymbol\nu$.
In other words, the Mandel stress represents the pullback of contact forces, and therefore, it does work on pullbacks of displacements.
This is a consequence of the fact that $\op M$ is defined by lowering an index of $\boldsymbol S$ using the pulled-back metric $\boldsymbol{C}^{\flat}=\varphi^*\boldsymbol g$.
On the other hand, the tensor $\boldsymbol{S}^{\flat}$ was obtained through the material metric $\boldsymbol G$, and hence it cannot be paired with spatial displacements.
\end{remark}
%---------------------
%---------------------

%------------------------------------------------------------------
%------------------------------------------------------------------
\subsection{Material variations}       \label{Material}

Next we consider variations of the dislocation motions $\prs{1}\chi,\hdots,\prsn{N}\chi$.
The goal is to obtain the kinetic equations for the internal variables as the Euler-Lagrange equations associated with variations of the material motions.
We refer to a generic Lagrangian as in~\eqref{lag-state-var}, and the results will be used to derive the governing equations for more specific cases in~\S\ref{Sec:Problem}.
We focus on the $\indx{a}$-th dislocation material motion while keeping all the others unperturbed.
In particular, we define a two-parameter family of diffeomorphisms as a map $\prs{a}{\chi}: \mathcal B \times [t_1,t_2] \times [-\epsilon_o, \epsilon_o] \to \mathcal B$ \citep{bruni2003two}.\footnote{%
As was discussed in Remark~\ref{rem:motion-boundary}, although a material motion during the time interval $[t_1,t_2]$ might be defined for a subbody of $\mathcal B$ (or might not even be defined at all), given a material dislocation velocity one can locally define sequences of well-defined material motions on smaller intervals, so that the same results hold as if one assumes a well-defined global material motion on the whole time interval $[t_1,t_2]$.
In the case of two-parameter families of diffeomorphisms, this argument extends to the parameter $\epsilon$.}
The family $\prs{a}{\chi}_{t,\epsilon}$ is such that at $\epsilon=0$ one recovers the material motion $\prs{a}{\chi}_t$ that satisfies the Lagrange-d'Alembert principle, and such that all the trajectories agree at their endpoints, i.e., $\prs{a}{\chi}_{t_1,\epsilon}=\chi_{t_1}$ and $\prs{a}{\chi}_{t_2,\epsilon}=\chi_{t_2}$ for all $\epsilon$.
Let $\prs{a}{\chi} _{\epsilon,X}: t\mapsto\prs{a}{\chi}_{t,\epsilon}(X)$ for $X$ and $\epsilon$ fixed, and
$\prs{a}{\chi} _{t,X}: \epsilon\mapsto\prs{a}{\chi}_{t,\epsilon}(X)$ for $X$ and $t$ fixed.
Then, we define the vectors
%---------------------
\begin{equation}
	\prs{a}{ \boldsymbol u } = \prs{a}{\chi}_{\epsilon,X}{}_* \left( \partial_t \right)\,,\qquad
	\prs{a}{ \boldsymbol w } = \prs{a}{\chi} _{t,X}{}_* \left( \tfrac{\partial}{\partial \epsilon} \right)\,.
\end{equation}
%---------------------
Notice that $\prs{a}{\boldsymbol{u}}$ is the velocity of the orbits of $\prs{a}{\chi} _{\epsilon,X}$, and $\prs{a}{\boldsymbol{w}}$ is the $\epsilon$-velocity of the orbits of $\prs{a}{\chi} _{t,X}$.
Therefore, for every pair $(t,\epsilon)$ the following vector fields are defined on $\mathcal B$:
%---------------------
\begin{equation} \label{u-w-def}
	\prs{a}{\boldsymbol{\vel}} (X,t,\epsilon) = \prs{a}{\boldsymbol{u}} (\prs{a}{\chi}_{t,\epsilon}^{-1}(X),t,\epsilon)
	\,,\qquad
	\prs{a}{\boldsymbol{\var}} (X,t,\epsilon)= \prs{a}{\boldsymbol{w}} (\prs{a}{\chi}_{t,\epsilon}^{-1}(X),t,\epsilon) \,.
\end{equation}
%---------------------
%

For a fixed $X$, the velocities $\prs{a}{\boldsymbol{u}}_X$ and $\prs{a}{\boldsymbol{w}}_X$ can be considered as vector fields on the surface $\mathcal O_X\subset \mathcal B$ made of points $\prs{a}{\chi}_{t,\epsilon}(X)$ spanned by $t$ and $\epsilon$ in the respective intervals (we assume $\mathcal O_X$ to be regular).
By doing so, $\prs{a}{\boldsymbol{u}}$ and $\prs{a}{\boldsymbol{w}}$ are coordinate vectors $(t,\epsilon)$ on $\mathcal O_X$, and therefore they satisfy $[\prs{a}{\boldsymbol{u}}, \prs{a}{\boldsymbol{w}}]=\boldsymbol 0$ on $\mathcal O_X$.
In components, this reads $\partial_{\epsilon}  \prs{a}{u}^A = \partial_t \prs{a}{w}^A$, and is a consequence of the fact that the second derivatives of a smooth $\prs{a}{\chi}^A$ commute.
On the other hand, for the ``Eulerian fields'' $\prs{a}{\boldsymbol{\vel}}$ and $\prs{a}{\boldsymbol{\var}}$ on $\mathcal B$, in general one has $\partial_{\epsilon}  \prs{a}{\boldsymbol{\vel}} \neq \partial_t \prs{a}{\boldsymbol{\var}}$, and $[\prs{a}{\boldsymbol{\vel}}, \prs{a}{\boldsymbol{\var}}]\neq\boldsymbol 0$,
as shown in the following lemma, where we omit the index $\indx{a}$ for the sake of simplicity.\footnote{%
The identity shown in Lemma~\ref{Lem:2-fam-comm} is not original as it can be found in~\citep{marsden2013introduction} in the context of incompressible fluids and Lin constraints.
However, to our knowledge it has not been used in the context of dislocation plasticity.}

%---------------------
%---------------------
\begin{lem} \label{Lem:2-fam-comm}
Let $\boldsymbol{\vel}$ and $\boldsymbol{\var}$ be the generators of the two-parameter family of diffeomorphisms $\chi_{t,\epsilon}$ on $\mathcal B$. Then,
$\partial_{\epsilon} \boldsymbol{\vel} - \partial_t \boldsymbol{\var} = [\boldsymbol{\vel}, \boldsymbol{\var}]$.
\end{lem}
%---------------------
\begin{proof}
We prove the result using components.
The derivative of the $t$-velocity $\boldsymbol{v}$ along the $\epsilon$-orbit is written as
%---------------------
\begin{equation}
	\partial_{\epsilon}  u^A (t , \epsilon) =
	\partial_{\epsilon}  \vel^A (\chi(X, t ,\epsilon), t , \epsilon) =
	\vel^A,_B \var^B + \partial_{\epsilon}  \vel^A \,,
\end{equation}
%---------------------
while the derivative of the $\epsilon$-velocity $\boldsymbol{w}$ along the $t$-orbit reads
%---------------------
\begin{equation}
	\partial_t w^A (t , \epsilon) =
	\partial_t \var^A (\chi(X, t ,\epsilon), t , \epsilon) =
	\var^A,_B \vel^B + \partial_{t} \var^A \,.
\end{equation}
%---------------------
As was discussed earlier, $\partial_{t} w^A=\partial_{\epsilon} v^A$.
Therefore, one obtains
%---------------------
\begin{equation}
	\partial_{\epsilon}  \vel^A - \partial_{t} \var^A =
	\vel^B \var^A,_B  - \var^B \vel^A,_B \,,
\end{equation}
%---------------------
where the right-hand side is the component representation of $[\boldsymbol{\vel},\boldsymbol{\var}]$.
\end{proof}
%---------------------
%---------------------

The vector field $\delta\prs{a}{\chi} (X,t)= \prs{a}{\boldsymbol{\var}}(X,t,0)$ is called the first variation of $\prs{a}{\chi}$.
From Lemma~\ref{Lem:2-fam-comm}, evaluating at $\epsilon=0$, one obtains
the first variation of the $\indx{a}$-th dislocation material velocity $\prs{a}{\boldsymbol{\vel}}_{\epsilon}$ as
%---------------------
\begin{equation} \label{vel-variation}
	\delta  \prs{a}{\boldsymbol{\vel}} = \partial_t \delta \prs{a}{\chi} + [\prs{a}{\boldsymbol{\vel}} , \delta \prs{a}{\chi} ] \,,
\end{equation}
%---------------------
while the variations of all the other velocities $\prs{b}{\boldsymbol{\vel}}_{\epsilon}$ associated with the unperturbed distributions, i.e., for $\indx{b}\neq\indx{a}$, are zero.

As the kinematics of both $\prs{a}{\dd}^{\nu}$ and $\boldsymbol\vartheta^{\nu}$ is fully determined by the material motions $\prs{a}{\chi}$, the introduction of the two-parameter family of diffeomorphisms $\prs{a}{\chi}_{t,\epsilon}$ allows one to derive the evolution of the internal variables along the $\epsilon$ parameter.
In particular, the dislocation fields $\prs{a}{\dd}^{\nu}_{\epsilon}$ follow~\eqref{conv-disl}, and hence one can write
%---------------------
\begin{equation}\label{var-conv}
	\prs{a}{\dd}^{\nu}_{t,\epsilon} = (\prs{a}{\chi}_{t,\epsilon})_*  \prs{a}{\dd}_{t,0}  \,.
\end{equation}
%---------------------
By differentiating~\eqref{var-conv} with respect to $\epsilon$ one obtains the $\epsilon$-evolution equation of a dislocation field as
%---------------------
\begin{equation} \label{evol-variation}
	\partial_{\epsilon}  \prs{a}{\dd}^{\nu} = -\mathfrak L_{\prs{a}{\boldsymbol{\var}} } \prs{a}{\dd}^{\nu} \,,
\end{equation}
%---------------------
which is the analogue of~\eqref{rate-Lie}.
This can be evaluated at $\epsilon=0$ to give the first variation of a dislocation field, viz. 
%---------------------
\begin{equation} \label{evol-variation}
	\delta  \prs{a}{\dd}^{\nu} =
	-\mathfrak L_{\delta \prs{a}{\chi} } \prs{a}{\dd}^{\nu} \,.
\end{equation}
%---------------------
In the case of closed dislocation fields one has
%---------------------
\begin{equation} \label{evol-variation}
	\delta  \prs{a}{\dd}^{\nu}  = -\mathrm d \iota_{\delta \prs{a}{\chi} } \prs{a}{\dd}^{\nu} \,.
\end{equation}
%---------------------
The variations of all the distributions $\prs{b}{\dd}^{\nu}_{\epsilon}$ vanish for $\indx{b}\neq\indx{a}$.
Next we derive an expression for the variation of the lattice frame $\boldsymbol \vartheta^{\nu}_{\epsilon}$ when the dislocation field is closed and Orowan's equation~\eqref{orowan} is assumed.
This result relies on the hypothesis that the first variations $\delta\prs{a}{\chi}$ are coplanar with the dislocation curves and dislocation velocity of the respective dislocation field, and it represents the analogue of Orowan's equation along the perturbation parameter.
This hypothesis is written as $\mvf(\prs{a}{ \delta \chi} ,\prs{a}{ \boldsymbol{\vel}} ,\prs{a}{ \boldsymbol{\lv}})=0$, and will be justified in~\S\ref{Constraints}.
Note that, under the glide assumption $\langle\prs{a}{\boldsymbol\pi},\prs{a}{ \boldsymbol{\vel}}\rangle=0$, the condition $\mvf(\prs{a}{ \delta \chi} ,\prs{a}{ \boldsymbol{\vel}} ,\prs{a}{ \boldsymbol{\lv}})=0$ implies $\mvf( \prs{a}{\delta \chi} , \prs{a}{\boldsymbol{\bv}} ,\prs{a}{ \boldsymbol{\lv}})=0$, which means isochoric variations, cf. Remark~\ref{rem:isochoric}.
As in Lemma~\ref{Lem:2-fam-comm}, the dislocation index $\indx{a}$ will be omitted for the sake of simplicity.

%---------------------
%---------------------
\begin{lem} \label{Lem:Orowan-variation}
Let $\{\dd^{\nu}\}$ be a closed decomposable dislocation field convected by a two-parameter family of material motions $\chi$ such that $\delta \chi$ satisfies $\iota_{\delta\chi } \iota_{  \boldsymbol{\vel} } \df=0$, i.e., $\mvf( \delta \chi , \boldsymbol{\vel} , \boldsymbol{\lv})=0$.
Then, Orowan's equation implies that the variation of the lattice coframe is written as
%---------------------
\begin{equation} \label{oro-variation}
	\delta \boldsymbol \vartheta^{\nu} = -\iota_{\delta \chi } \dd^{\nu} \,.
\end{equation}
%---------------------

\end{lem}
%---------------------
\begin{proof}
Since variations of a single dislocation motion are considered, from~\eqref{orowan-single} one can write the mixed derivative of a lattice form as
%---------------------
\begin{equation} 
	\partial_{\epsilon} \partial_t \boldsymbol \vartheta^{\nu} =
	- \partial_{\epsilon} \iota_{ \boldsymbol{\vel}} \dd^{\nu} =
	- \iota_{ \partial_{\epsilon} \boldsymbol{\vel}} \dd^{\nu}
	- \iota_{ \boldsymbol{\vel} } \partial_{\epsilon} \dd^{\nu} \,,
\end{equation}
%---------------------
where the Leibniz rule has been used to obtain the second equality.
Using Lemma~\ref{Lem:2-fam-comm} for the first term of the above identity, and Eq.~\eqref{evol-variation} for the second term, the mixed derivative of a lattice form is written as
%---------------------
\begin{equation} 
	\partial_{\epsilon} \partial_t  \boldsymbol \vartheta^{\nu} =
	- \iota_{\partial_{t} \boldsymbol{\var} } \dd^{\nu}
	- \iota_{[ \boldsymbol{\vel} , \boldsymbol{\var} ]} \dd^{\nu}
	+ \iota_{ \boldsymbol{\vel}} \mathrm d \iota_{ \boldsymbol{\var} }  \dd^{\nu}\,.
\end{equation}
%---------------------
Cartan's formula~\eqref{Cartan} implies that the operator
$\iota_{\boldsymbol U} \mathrm d \iota_{\boldsymbol W} - \iota_{\boldsymbol W} \mathrm d \iota_{\boldsymbol U} $
on the space of closed $2$-forms is equal to the operator
$- \mathrm d \iota_{\boldsymbol U} \iota_{\boldsymbol W} +\iota_{ [\boldsymbol U,\boldsymbol W] }$.\footnote{%
This can is obtained as follows
%---------------------
\begin{equation} 
\begin{split} 	
	2 \left( \iota_{\boldsymbol V} \mathrm d \iota_{\boldsymbol W} - \iota_{\boldsymbol W} \mathrm d \iota_{\boldsymbol V} \right) \boldsymbol\omega &=
	\left( \iota_{\boldsymbol V} \mathrm d \iota_{\boldsymbol W} \boldsymbol\omega - \iota_{\boldsymbol W} \mathrm d \iota_{\boldsymbol V} \boldsymbol\omega\right) +
	\left( \iota_{\boldsymbol V} \mathrm d \iota_{\boldsymbol W} \boldsymbol\omega- \iota_{\boldsymbol W} \mathrm d \iota_{\boldsymbol V} \boldsymbol\omega\right) \\
	&=
	\left(  \mathfrak L_{\boldsymbol V} \iota_{\boldsymbol W} \boldsymbol\omega - \mathrm d \iota_{\boldsymbol V} \iota_{\boldsymbol V} \boldsymbol\omega
	- \mathfrak L_{\boldsymbol W} \iota_{\boldsymbol V} \boldsymbol\omega + \mathrm d \iota_{\boldsymbol W} \iota_{\boldsymbol V} \boldsymbol\omega \right) 
	+ \left( \iota_{\boldsymbol V} \mathfrak L_{\boldsymbol W}  \boldsymbol\omega - \iota_{\boldsymbol W} \mathfrak L_{\boldsymbol V} \boldsymbol\omega  \right) \\
	&= \left( \iota_{[\boldsymbol V,\boldsymbol W]} \boldsymbol\omega +  \iota_{\boldsymbol W} \mathfrak L_{\boldsymbol V} \boldsymbol\omega - \mathrm d \iota_{\boldsymbol V} \iota_{\boldsymbol W} \boldsymbol\omega
	- \iota_{[\boldsymbol W,\boldsymbol V]} \boldsymbol\omega - \iota_{\boldsymbol V} \mathfrak L_{\boldsymbol W} \boldsymbol\omega + \mathrm d \iota_{\boldsymbol W} \iota_{\boldsymbol V} \boldsymbol\omega \right)
	+ \left( \iota_{\boldsymbol V} \mathfrak L_{\boldsymbol W}  \boldsymbol\omega - \iota_{\boldsymbol W} \mathfrak L_{\boldsymbol V} \boldsymbol\omega  \right) \\
	&= 2 \iota_{ [\boldsymbol V,\boldsymbol W] } \boldsymbol\omega - 2 \mathrm d \iota_{\boldsymbol V} \iota_{\boldsymbol W} \boldsymbol\omega
	\,,
\end{split}
\end{equation}
%---------------------
where use was made of Cartan's formula and $\mathrm d \boldsymbol\omega = 0$ in the second equality, and of the Leibniz rule in the third.
}
Moreover, one has $\partial_{\epsilon}  \partial_t  \boldsymbol \vartheta^{\nu}= \partial_t  \partial_{\epsilon} \boldsymbol \vartheta^{\nu}$, and hence
%---------------------
\begin{equation}
\begin{split}
	\partial_t \partial_{\epsilon} \boldsymbol \vartheta^{\nu} &=
	- \iota_{\partial_{t} \boldsymbol{\var} } \dd^{\nu}
	+ \iota_{ \boldsymbol{\var} } \mathrm d \iota_{ \boldsymbol{\vel} } \dd^{\nu}
	+\mathrm d \iota_{ \boldsymbol{\var} }  \iota_{ \boldsymbol{\vel} } \dd^{\nu} \\
	&= - \iota_{\partial_{t} \boldsymbol{\var} } \dd^{\nu}
	- \iota_{ \boldsymbol{\var} } \partial_{t} \dd^{\nu}
	+\mathrm d \iota_{ \boldsymbol{\var} }  \iota_{ \boldsymbol{\vel} }  \dd^{\nu} \\
	&= - \partial_{t} \left(\iota_{ \boldsymbol{\var} } \dd^{\nu}\right)
	+\mathrm d \iota_{ \boldsymbol{\var} }  \iota_{ \boldsymbol{\vel} }  \dd^{\nu} \,.
\end{split}
\end{equation}
%---------------------
Integrating over time between $t_1$ and $t$ and evaluating at $\epsilon=0$ one obtains
%---------------------
\begin{equation} \label{oro-variation-gen}
	\delta \boldsymbol \vartheta^{\nu} =
	-\iota_{\delta \chi } \dd^{\nu}
	+ \int_{t_1}^{t} \mathrm d \iota_{\delta \chi } \iota_{ \boldsymbol{\vel} }  \dd^{\nu} \mathrm \,dt \,,
\end{equation}
%---------------------
as all variations vanish at $t_1$.
The hypothesis $\iota_{\delta \chi} \iota_{ \boldsymbol{\vel}} \df=0$ implies $\iota_{\delta \chi} \iota_{ \boldsymbol{\vel}} \dd^{\nu}=0$ for all $\nu$, and hence, one obtains~\eqref{oro-variation}.
\end{proof}
%---------------------
%---------------------

Next we take variations of the action~\eqref{action}.
Since the kinetic energy does not depend on the material motions, one has $\delta\mathscr{L} = \delta \mathscr{T}_M - \delta\mathscr{F}$.
Then, we define the following generalized material forces:
%---------------------
\begin{equation} \label{generalized-forces}
	\boldsymbol Y_{\nu} = - \massd\, \frac{\partial \mathscr{L}}{\partial \boldsymbol \vartheta^{\nu} }
	= \massd \,\frac{\partial (\mathscr{F}-\mathscr{T}_M)}{\partial \boldsymbol \vartheta^{\nu} }
	\,,\quad
	\star^{\sharp}  \prs{b}{\boldsymbol{\xi}}_{\nu} = - \massd \,\frac{\partial \mathscr{L}}{\partial \prs{b}{\dd}^{\nu} }
	= \massd\, \frac{\partial (\mathscr{F}-\mathscr{T}_M)}{\partial \prs{b}{\dd}^{\nu} }
	\,,\quad
	\prs{b}{\boldsymbol \psi} = \massd \,\frac{\partial \mathscr{L}}{\partial \prs{b}{\boldsymbol{\vel}} } 
	= \massd \,\frac{\partial \mathscr{T}_M}{\partial \prs{b}{\boldsymbol{\vel}} }
	\,.
\end{equation}
%---------------------
In the case of negligible inertial effects associated with the dislocation motions, one can set $\mathscr{T}_M = 0$, so that the generalized material forces read
%---------------------
\begin{equation} \label{generalized-forces-no-inertia}
	\boldsymbol Y_{\nu} = \massd\, \frac{\partial \mathscr{F}}{\partial \boldsymbol \vartheta^{\nu} }
	\,,\qquad
	\star^{\sharp}  \prs{b}{\boldsymbol{\xi}}_{\nu} = \massd\, \frac{\partial \mathscr{F}}{\partial \prs{b}{\dd}^{\nu} }
	\,,\qquad
	\prs{b}{\boldsymbol\psi} = 0
	\,.
\end{equation}
%---------------------
Recall that we are looking at variations of the $\indx{a}$-th dislocation material motion while keeping all the others unperturbed, meaning that $\delta\prs{b}{\chi}$, $\delta\prs{b}{\dd}^{\nu}$, and $\delta\prs{b}{\boldsymbol{\vel}}$ vanish for $\indx{b}\neq\indx{a}$.
Therefore, using the definition of the generalized material forces~\eqref{generalized-forces}, and recalling that the mass and the volume forms are related through the mass density as $\boldsymbol{m}= \massd \mvf$, one can express the variation of the total Lagrangian as
%---------------------
\begin{equation} \label{mat-var1}
	\int_{\mathcal B} \delta \mathscr{L} \,\boldsymbol{m} =
	- \int_{\mathcal B} \left\langle \delta  \boldsymbol \vartheta^{\nu} , \boldsymbol Y_{\nu}  \right\rangle \mvf
	- \int_{\mathcal B}  \prs{a}{\boldsymbol{\xi}}_{\nu} \wedge \delta \prs{a}{\dd}^{\nu} 
	+ \int_{\mathcal B} \langle \prs{a}{\boldsymbol\psi} , \delta \prs{a}{\boldsymbol{\vel}} \rangle \, \mvf \,,
\end{equation}
%---------------------
where~\eqref{wedge-hodge} was used to write the pairing in the second term using the wedge product.
The generalized material forces have the following interpretations.
%---------------------
\begin{itemize}  \setlength\itemsep{0em}
	\item The microforces $\boldsymbol Y_{\nu}$ are a triplet of vectors that are work-conjugate to changes of the lattice frame.
	\item The microstresses $\prs{a}{\boldsymbol{\xi}}_{\nu}$ are triplets of $1$-forms that are work-conjugate to changes of the $\indx{a}$-th dislocation field, and have been defined so that conjugacy is expressed through the exterior product to give a $3$-form that is ready to be integrated on $\mathcal B$.
	\item The micro-inertial force $\prs{a}{\boldsymbol\psi}$ is a $1$-form that is work-conjugate to changes of the $\indx{a}$-th dislocation velocity.
\end{itemize}
%---------------------
%
Note that the action functional and the Lagrangian density are related through~\eqref{action}, and hence, integrating~\eqref{mat-var1} over the time interval $[t_1,t_2]$ one obtains $\delta \mathsf{A}$.
Thus, since the spatial variation $\delta\varphi$ is set to zero, the Lagrange-d'Alembert principle~\eqref{lag-dal} can be written as
%---------------------
\begin{equation} \label{mat-lag-dal}
	\int_{t_1}^{t_2} \int_{\mathcal B} \left\lbrace
	- \left\langle \delta  \boldsymbol \vartheta^{\nu} , \boldsymbol Y_{\nu}  \right\rangle \mvf 
	- \prs{a}{\boldsymbol{\xi}}_{\nu} \wedge \delta \prs{a}{\dd}^{\nu} 
	+ \left\langle \prs{a}{\boldsymbol\psi} , \delta \prs{a}{\boldsymbol{\vel}} \right\rangle \mvf
	+ \langle \prs{a}{\boldsymbol\tau} , \delta  \prs{a}{\chi} \rangle \, \mvf
	\right\rbrace \mathrm dt
	=0\,.
\end{equation}
%---------------------
Similar to spatial variations, the integral~\eqref{mat-lag-dal} can be written as a local functional of $\delta \prs{a}{\chi}$, which allows one to use the fundamental lemma of the calculus of variations and obtain the Euler-Lagrange equations associated with the internal degrees of freedom.
From the calculations given in~\S\ref{App:Calculations} that make use of the identities~\eqref{vel-variation},~\eqref{evol-variation}, and~\eqref{oro-variation} as well as of Stokes' theorem~\eqref{stokes}, one obtains:
%---------------------
\begin{equation} \label{mat-var2}
\begin{split}
	\int_{t_1}^{t_2}
	\int_{\partial \mathcal B} &
	\Big\langle
	\langle \prs{a}{\boldsymbol{\xi}} ,\prs{a}{\boldsymbol{\lv}} \rangle \, \boldsymbol \nu-
	\langle \boldsymbol \nu,\prs{a}{\boldsymbol{\lv}} \rangle \, \prs{a}{\boldsymbol{\xi}}  +
	\langle \boldsymbol \nu, \prs{a}{\boldsymbol{\vel}} \rangle \, \prs{a}{\boldsymbol\psi}
	~,\, \delta \prs{a}{\chi}
	\Big\rangle \boldsymbol \, \boldsymbol\varsigma \, \mathrm dt  \\
	&+ \int_{t_1}^{t_2} \int_{\mathcal B} \left\langle
	-\iota_{ \boldsymbol Y_{\nu} } \prs{a}{\dd}^{\nu}
	-\iota_{\star^{\sharp} \mathrm d \prs{a}{\boldsymbol{\xi}}_{\nu} } \prs{a}{\dd}^{\nu}
	-\mathsf T_{\prs{a}{\boldsymbol{\vel}}}   \prs{a}{\boldsymbol\psi}
	+\prs{a}{\boldsymbol\tau}
	~,\, \delta \prs{a}{\chi} \right\rangle \mvf \, \mathrm dt
	 =0	\,,
\end{split}
\end{equation}
%---------------------
with $\prs{a}{\boldsymbol{\xi}}=\prs{a}{\dens}\,\prs{a}{\bv}^{\nu}\,\prs{a}{\boldsymbol{\xi}}_{\nu}$, and where $\mathsf T_{\prs{a}{\boldsymbol{\vel}}}   \prs{a}{\boldsymbol\psi}$ denotes the Truesdell derivative along $\prs{a}{\boldsymbol{\vel}}$, see~\S\ref{App:Flows}.
Thus, the arbitrariness of $\delta \prs{a}{\chi}$ gives the following field equations:
%---------------------
\begin{equation} \label{EL-material}
	\iota_{ \boldsymbol Y_{\nu} } \prs{a}{\dd}^{\nu} +
	\iota_{\star^{\sharp} \mathrm d \prs{a}{\boldsymbol{\xi}}_{\nu} } \prs{a}{\dd}^{\nu}
	+
	\mathsf T_{\prs{a}{\boldsymbol{\vel}}}   \prs{a}{\boldsymbol\psi}
	= \prs{a}{\boldsymbol\tau}
	\,,\qquad\indx{a} = 1,2,\hdots,N \,.
\end{equation}
%---------------------
Eq.~\eqref{EL-material} constitutes the kinetic equations for the dislocation material motion, which we assumed to be conservative (i.e., inducing volume-preserving plastic deformation) in Lemma~\ref{Lem:Orowan-variation}.
Kinetic equations govern the evolution of the internal variables, and were introduced by~\citet{coleman1967thermodynamics} and \citet{rice1971inelastic} in the context of continuum thermodynamics and of inelastic solids.
The rate of change of the internal variables is usually given as an expression of the thermodynamic forces associated with them.
In the present setting, these equations are written in the following form
%---------------------
\begin{equation} \label{kinetic-rice}
	\prs{a}{\boldsymbol{\vel}} =  \prs{\,a}{\mathscr U}
	( \boldsymbol\vartheta^{\nu},\prs{a}{\dd}^{\nu},\boldsymbol Y_{\nu} ,\prs{a}{\boldsymbol{\xi}}_{\nu} )\,
	,\qquad\indx{a} = 1,2,\hdots,N \,.
\end{equation}
%---------------------
It should be noticed that, in general, the kinetic equations~\eqref{EL-material} we just derived cannot be reduced to the form~\eqref{kinetic-rice}.
However, this is possible under the assumption of negligible micro-inertial effects.
As a matter of fact, since $\prs{a}{\boldsymbol{\vel}}$ and $\prs{a}{\boldsymbol\tau}$ are in one-to-one correspondence by virtue of the convexity of the dissipation potential $\mathscr{D}$ with respect to $\prs{a}{\boldsymbol{\vel}}$, one can invert~\eqref{EL-material} to write
%---------------------
\begin{equation} \label{kinetic-rice2}
	\prs{a}{\boldsymbol{\vel}} =
	\tau^{-1} \Big(
	\iota_{ \boldsymbol Y_{\nu} } \prs{a}{\dd}^{\nu} +
	\iota_{\star^{\sharp} \mathrm d \prs{a}{\boldsymbol{\xi}}_{\nu} } \prs{a}{\dd}^{\nu}
	\Big)
	\,,\qquad\indx{a} = 1,2,\hdots,N \,.
\end{equation}
%---------------------
where we set $\prs{a}{\boldsymbol{\vel}}  = \tau^{-1}(\prs{a}{\boldsymbol\tau})$, and where the micro-inertial term $\mathsf T_{\prs{a}{\boldsymbol{\vel}}}   \prs{a}{\boldsymbol\psi}$ was neglected.
Lastly, it should be noticed that the kinetic equations~\eqref{EL-material} and~\eqref{kinetic-rice2} can also be interpreted as the equations of motion for the dislocation fields.

From~\eqref{mat-var2} one also obtains the following boundary conditions on $\partial \mathcal B$, viz.
%---------------------
\begin{equation} \label{EL-material-boundary}
	\langle \prs{a}{\boldsymbol{\xi}} ,\prs{a}{\boldsymbol{\lv}} \rangle \, \boldsymbol \nu
	-\langle \boldsymbol \nu,\prs{a}{\boldsymbol{\lv}} \rangle \,  \prs{a}{\boldsymbol{\xi}}
	+\langle \boldsymbol \nu, \prs{a}{\boldsymbol{\vel}} \rangle \,  \prs{a}{\boldsymbol\psi}=0
	\,,\qquad\indx{a} = 1,2,\hdots,N \,,
\end{equation}
%---------------------
where $\boldsymbol \nu$ is the unit normal $1$-form on $\partial \mathcal B$.
Eq.~\eqref{EL-material-boundary} represent natural boundary conditions, i.e., conditions that are not enforced on the material motion $\prs{a}{\chi}$ or on other internal variables, but instead on their thermodynamic-conjugate quantities.

It should be noted that the generalized stresses $\prs{a}{\boldsymbol{\xi}}_{\nu}$ induce a non-local effect,
in the sense that not only do they appear in the kinetic equations~\eqref{EL-material} in terms of their exterior derivative,
but they also appear in the boundary conditions~\eqref{EL-material-boundary}.
This happens because they are associated with changes of the dislocation fields, governed by Eq.~\eqref{evol-variation} and involving the derivatives of $\delta\prs{a}{\chi}$.
Other non-local contributions come from the micro-inertial forces $\prs{a}{\boldsymbol\psi}$,
as a consequence of~\eqref{vel-variation}, and hence of Lemma~\ref{Lem:2-fam-comm}.
These forces appear in the kinetic equations in terms of their Truesdell derivative,
as well as in the boundary conditions.
Note that this does not occur in the case of the macro-inertial forces, as no nonlocal expression of the acceleration is involved in the balance of linear momentum.
The reason for this difference is due to the way $\mathscr{T}_S$ and $\mathscr{T}_M$ depend on the respective velocities.
As a matter of fact, the spatial kinetic energy is defined in a Lagrangian way, i.e., by following the trajectories of the spatial motion.
On the other hand, since the dislocation motion does not involve the actual movement of material points, the kinetic energy $\mathscr{T}_M$ is defined in a Eulerian way, i.e., by keeping each point fixed.
This is also convenient as the other two arguments $\boldsymbol{C}^{\flat}$ and $\boldsymbol\vartheta^{\nu}$ are not convected quantities.
The details of the calculations are given in~\S\ref{App:Calculations}.

%----------------------------------------------------------------------
%----------------------------------------------------------------------
\subsection{Lattice constraints}   \label{Constraints}

Layered dislocation fields were introduced in~\S\ref{Layered} as those for which the Burgers director and the dislocation curves locally lie on a plane distribution.
Their glide motion was studied in~\S\ref{Glide} and is characterized by a velocity vector that lies on the plane distribution on which the dislocations are layered.
In particular, a decomposable layered dislocation field $ \prs{a}{\dd}^{\nu} = \prs{a}{\bv}^{\nu} \prs{a}{\df}$ gliding with velocity $ \prs{a}{\boldsymbol{\vel}}$ is subject to the conditions~\eqref{slip-constr}$_1$,~\eqref{slip-constr}$_2$ and~\eqref{glide}, viz.
%---------------------
\begin{equation} \label{constr-recap}
 	\prs{a}{\pi}_{\nu} \prs{a}{\bv}^{\nu} =0
	\,,\qquad
	\prs{a}{\boldsymbol\pi} \wedge \prs{a}{\df} =0
	\,,\qquad
	\langle \prs{a}{\boldsymbol\pi} , \prs{a}{\boldsymbol{\vel}} \rangle =0
	\,.
\end{equation}
%---------------------
These conditions can be seen as internal constraints that the lattice structure exerts on the dislocation field and its motion.
In particular, the layer constraints~\eqref{constr-recap}${}_1$ and~\eqref{constr-recap}${}_2$ are holonomic, while the glide constraint~\eqref{constr-recap}${}_3$ on the velocity is nonholonomic.\footnote{%
Here non-holonomic does not refer to the property of a moving frame on $\mathcal B$ of being induced by local charts and used to express the incompatibility of the lattice structure. It indicates that a constraint depends on the generalized velocities.}
It should be noted that in single crystals the scalars $\pi_{\nu}$ are uniform in space and constant in time, while the scalars $\bv^{\nu}$ evolve only through changes of the base point via~\eqref{evol-decomp}.
Therefore, if $\pi_{\nu}\bv^{\nu} =0$ at a given time, then $\pi_{\nu}\bv^{\nu} =0$ at all times, and hence the constraint~\eqref{constr-recap}${}_1$ is redundant.
Instead, as was shown in Lemma~\ref{Lem:glide}, the constraints~\eqref{constr-recap}${}_2$ and~\eqref{constr-recap}${}_3$ are independent, and hence they are both necessary.
Thus, we define the following lattice constraint functions:
%---------------------
\begin{equation} \label{l-g-constr}
	\prs{b}{f}_L (\boldsymbol\vartheta^{\nu}  , \prs{a}{\dd}^{\nu} ) =
	\prs{b}{\boldsymbol\pi} \wedge \prs{b}{\df} 
	\,,\qquad
	\prs{b}{f}_G  (\boldsymbol\vartheta^{\nu}  ,  \prs{a}{\boldsymbol{\vel}} ) =
	\langle \prs{b}{\boldsymbol \pi} , \prs{b}{\boldsymbol{\vel}} \rangle
	  \,.
\end{equation}
%---------------------
Note that these functions are independent of the spatial configuration, and hence the introduction of lattice constraints only affects material variations.
For this reason, we look at the Lagrange-d'Alembert principle corresponding to material variations in the form~\eqref{mat-var2}.

Denoting with $\prs{a}{\boldsymbol{\beta}}_{\partial}$ and $\prs{a}{\boldsymbol{\beta}}$ the $1$-forms in~\eqref{mat-var2} acting on $\delta \prs{a}{\chi}\vert_{\partial\mathcal{B}}$ and $\delta \prs{a}{\chi}$, respectively, the Lagrange-d'Alembert principle can be written in the form
%---------------------
\begin{equation} \label{mat-var-short}
	\int_{\partial  \mathcal B}
	\langle \prs{a}{\boldsymbol{\beta}}_{\partial} , \delta \prs{a}{\chi} \rangle
	\, \boldsymbol{\varsigma} 
	+
	\int_{\mathcal B}
	\langle \prs{a}{\boldsymbol{\beta}} , \delta \prs{a}{\chi} \rangle
	\, \mvf 
	= 0
	\,,\qquad\indx{a} =1,2,\hdots,N
	\,,
\end{equation}
%---------------------
so that the kinetic equations for the unconstrained problem read $\prs{a}{\boldsymbol{\beta}}=0$, and the associated boundary conditions are $\prs{a}{\boldsymbol{\beta}}_{\partial}=0$.
Because of the presence of the constraints $\prs{b}{f}_L$ and $\prs{b}{f}_G$, for $\indx{a} =1,2,\hdots,N$,~\eqref{mat-var-short} must hold only for all $N$-tuples of material variations $(\delta \prs{1}{\chi},\hdots,\delta \prsn{N}{\chi})$ in the space of virtual material displacements $\mathscr V$.
In the following we construct this space.

We start by looking at the holonomic layer constraint $\prs{b}{f}_L$.
Since we are considering variations of only the $\indx{a}$-th dislocation field, from~\eqref{evol-variation} and~\eqref{oro-variation} one can write
%---------------------
\begin{equation} \label{var-constr}
	\delta \prs{b}{f}_L = 
	\delta \prs{b}{\boldsymbol\pi} \wedge \prs{b}{\df} +  \prs{b}{\boldsymbol\pi} \wedge \delta \prs{b}{\df} =
	- \prs{b}{\pi}_{\nu} \, \iota_{\delta\prs{a}{\chi}} \prs{a}{\dd}^{\nu} \wedge  \prs{b}{\df}
	- \prs{a}{\boldsymbol\pi} \wedge \mathrm d  \iota_{\delta\prs{a}{\chi}} \prs{a}{\df}
	\,.
\end{equation}
%---------------------
Recalling the definition~\eqref{cross1} of slip coupling $1$-form, the first term in~\eqref{var-constr} can be written as $- \left\langle \prs{b\!a}{\boldsymbol\varpi}  , \delta\prs{a}{\chi} \right\rangle \mvf$.
Note also that $\delta\prs{a}{\chi}$ is assumed to lie on the plane distribution defined by $\prs{a}{\boldsymbol\pi}$ because of the glide constraint~\eqref{l-g-constr}$_2$, as will be explained later on.
Therefore, following the calculations given in Lemma~\ref{Lem:glide}, the second term in~\eqref{var-constr} can be written as $\mathrm d  \prs{a}{\boldsymbol\pi} \wedge (\prs{a}{\dens} \, \delta\prs{a}{\chi}^{\bot} \, \prs{a}{\boldsymbol\pi} )$, which does not involve the index $\indx{b}$.
Hence, we obtain
%---------------------
\begin{equation}
	\delta \prs{b}{f}_L = 
	- \left\langle \prs{b\!a}{\boldsymbol\varpi}  , \delta\prs{a}{\chi} \right\rangle \mvf 
	- \left\langle  \prs{a}{\dens} \, \mathfrak I_{\prs{a}{\Pi}} \, \prs{a}{\boldsymbol{\norm}}^{\flat}, \delta\prs{a}{\chi} \right\rangle \mvf
	\,,
\end{equation}
%---------------------
which implies that the variation $ \delta\prs{a}{\chi} $ must satisfy the following condition
%---------------------
\begin{equation} \label{virtual-displacements-layer}
	\int_{\mathcal B}
	\left\langle \prs{b\!a}{\boldsymbol\varpi} + \prs{a}{\dens} \, \mathfrak I_{\prs{a}{\Pi}} \, \prs{a}{\boldsymbol{\norm}}^{\flat} \,, \delta\prs{a}{\chi} \right\rangle
	\mvf
	=0
	\,,\qquad\indx{b}=1,2,\hdots,N \,.
\end{equation}
%---------------------
Eq.~\eqref{virtual-displacements-layer} for $\indx{a}=1,2,\hdots,N$ gives $N^2$ conditions for the material variations.
However, one should recall from~\S\ref{Glide} that the $1$-forms $\prs{a\!b}{\boldsymbol\varpi}$ and $\prs{b\!a}{\boldsymbol\varpi}$ only differ by a multiplicative factor,
and therefore, the conditions in~\eqref{virtual-displacements-layer} might not be independent, as in the case of Example~\ref{example-twoslips} for $N=2$. 

Next, we look at the nonholonomic glide constraint~\eqref{l-g-constr}$_2$.
The dynamics of mechanical systems with nonholonomic constraints can be formulated by following either the nonholonomic or the vakonomic approach \citep{lewis1995variational, cardin1996nonholonomic}.\footnote{%
In the nonholonomic approach, constraints are enforced after making the action functional stationary, and can be seen as the variational formulation of the principle of virtual work.
The vakonomic approach (from the acronym VAK, meaning variational axiomatic kind \citep{arnold1988dynamical}) consists of enforcing the constraints before minimization, and it is the equivalent of the method of Lagrange multipliers \citep{goldstein2002classical}.
These two approaches yield different equations although they coincide when the constraints are holonomic.
Given the Lagrange multipliers $\lambda_h$ associated with the constraints $f_h$, the Euler-Lagrange equations derived from the two approaches are
%---------------------
\begin{equation}
 	\frac{\mathrm d }{\mathrm d t}\frac{\partial \mathscr{L}}{\partial \dot q^i} -  \frac{\partial \mathscr{L}}{\partial q^i} = 
	\left\lbrace
	\begin{array}{lc}
		\sum_h \lambda_h \frac{\partial f_h}{\partial \dot q^i} & \text{(n.h.)} \\
		\sum_h \Big( \frac{\mathrm d \lambda_h}{\mathrm d t} \frac{\partial f_h}{\partial \dot q^i} + \lambda_h \frac{\mathrm d }
		{\mathrm d t}\frac{\partial f_h}{\partial \dot q^i} -  \lambda_h \frac{\partial f_h}{\partial q^i} \Big) & \text{(v.a.k.)}
	\end{array}
	\right. .
\end{equation}
%---------------------
}
We choose to follow the nonholonomic approach, as it allows one to work with variations that satisfy the hypotheses of Lemma~\ref{Lem:Orowan-variation}.
As a matter of fact, in the nonholonomic approach the material variation (material virtual displacement) $\delta \prs{a}{\chi}$ must satisfy
%---------------------
\begin{equation} \label{nh-constr-virtual}
	\bigg\langle \frac{\partial \prs{b}{f}_G }{\partial \prs{a}{\boldsymbol{\vel}}} , \delta \prs{a}{\chi} \bigg\rangle =
	\left\langle \prs{a}{\boldsymbol\pi} \,, \delta\prs{a}{\chi} \right\rangle = 
	0
	\,.
\end{equation}
%---------------------
Note that the previous condition does not involve the index $\indx{b}$ as the derivative of the $\indx{b}$-th constraint function with respect to the $\indx{a}$-th dislocation velocity vanishes for $\indx{a}\neq\indx{b}$.
Thus, we have obtained the following requirement for each $\delta \prs{a}{\chi}$:
%---------------------
\begin{equation} \label{virtual-displacements-glide}
	\int_{\mathcal B}
	\left\langle \prs{a}{\boldsymbol\pi} \,, \delta\prs{a}{\chi} \right\rangle
	\mvf
	=0  \,.
\end{equation}
%---------------------
One should note that the condition $ \langle  \prs{a}{\boldsymbol \pi} , \delta \prs{a}{\chi} \rangle =0 $ can also be written as $ \langle  \iota_{ \prs{a}{\boldsymbol{\vel}} }\prs{a}{\df} , \delta \prs{a}{\chi} \rangle =0 $ as both $\prs{a}{\df}$ and $\prs{a}{\boldsymbol{\vel}}$ are constrained to lie on $\prs{a}{\boldsymbol \pi}$. Hence, the hypotheses of Lemma~\ref{Lem:Orowan-variation} are satisfied.

\begin{figure}[tp]
\centering
\includegraphics[width=.95\textwidth]{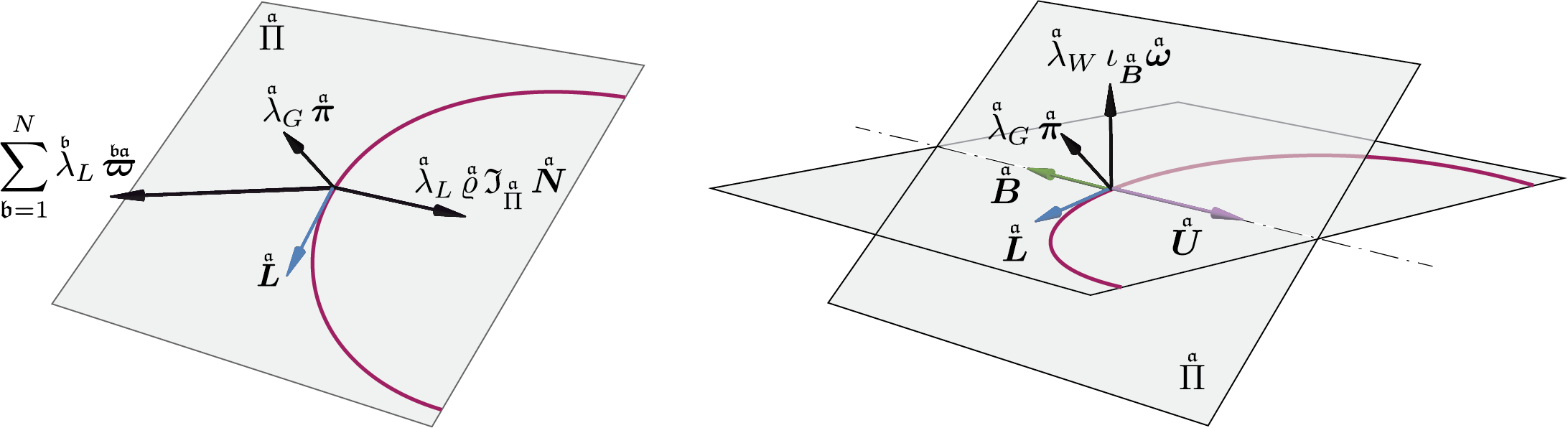}
\vskip 0.1in
\caption{Reaction forces exerted by the lattice on the dislocation fields. Left: Strongly layered dislocations. Right: Weakly layered dislocations.}
\label{fig:reactions}
\end{figure}

In the presence of the constraints $\prs{b}{f}_L$ and $\prs{b}{f}_G$, the space of virtual displacements $\mathscr V\subset T\mathcal B^N$ is made of those $N$-tuples $(\delta \prs{1}{\chi},\hdots,\delta \prsn{N}{\chi})$ that satisfy both~\eqref{virtual-displacements-layer} and~\eqref{virtual-displacements-glide}.
The orthogonal complement $\mathscr V^{\bot}\subset T^*\mathcal B^N$ is the space of nullifiers of $\mathscr V$, and is spanned by the $1$-forms $\prs{b\!a}{\boldsymbol\varpi}  + \prs{a}{\dens} \, \mathfrak I_{\prs{a}{\Pi}} \, \prs{a}{\boldsymbol{\norm}}^{\flat}$ and $\prs{a}{\boldsymbol\pi}$, for $\indx{a}, \indx{b}=1,2,\hdots,N$.
As was mentioned earlier, the integrals in~\eqref{mat-var-short} must vanish for all $\delta\prs{a}{\chi} \in \mathscr V$.
This implies that $\prs{a}{\boldsymbol{\beta}}$ must be in $\mathscr V^{\bot}$, and that it can be written as the following linear combination:
%---------------------
\begin{equation} \label{EL-constrained}
	\prs{a}{\boldsymbol{\beta}} =
	\prs{a}{\lambda}_L \, \prs{a}{\dens} \, \mathfrak I_{\prs{a}{\Pi}} \, \prs{a}{\boldsymbol{\norm}}^{\flat}
	+ \sum_{\indx{b}=1}^N \prs{b}{\lambda}_L \, \prs{b\!a}{\boldsymbol\varpi}
	+ \prs{a}{\lambda}_G \, \prs{a}{\boldsymbol\pi} \,.
\end{equation}
%---------------------
The $1$-form $\prs{a}{\boldsymbol{\beta}}_{\partial}$ associated with the boundary conditions is not affected.
The scalar fields $\prs{a}{\lambda}_L$ and $\prs{a}{\lambda}_G$ in~\eqref{EL-constrained} are the Lagrange multipliers associated with the $\indx{a}$-th layer and glide constraint, respectively.
Therefore, in the presence of lattice constraints, the kinetic equation~\eqref{EL-material} must be modified to read
%---------------------
\begin{equation} \label{EL-constrained2}
	\iota_{ \boldsymbol Y_{\nu} } \prs{a}{\dd}^{\nu} +
	\iota_{\star^{\sharp} \mathrm d \prs{a}{\boldsymbol{\xi}}_{\nu} } \prs{a}{\dd}^{\nu}
	+
	\mathsf T_{\prs{a}{\boldsymbol{\vel}}}   \prs{a}{\boldsymbol\psi}
	= \prs{a}{\boldsymbol\tau}
	+\prs{a}{\lambda}_L \, \prs{a}{\dens} \, \mathfrak I_{\prs{a}{\Pi}} \, \prs{a}{\boldsymbol{\norm}}^{\flat}
	+ \sum_{\indx{b}=1}^N \prs{b}{\lambda}_L \, \prs{b\!a}{\boldsymbol\varpi}
	+ \prs{a}{\lambda}_G \, \prs{a}{\boldsymbol\pi}
	\,,\qquad\indx{a}=1,2,\hdots,N 
	\,.
\end{equation}
%---------------------
The three terms added as a consequence of the constraints represent the reaction forces exerted by the lattice on the dislocation fields and are shown in Fig.~\ref{fig:reactions}.
They have the following interpretations.
%---------------------
%---------------------
%---------------------
%---------------------

%---------------------
\begin{itemize} \setlength\itemsep{0em}
\item
The $1$-form $\prs{a}{\lambda}_L \, \prs{a}{\dens} \, \mathfrak I_{\prs{a}{\Pi}} \, \prs{a}{\boldsymbol{\norm}}^{\flat}$ is the part of the reaction force that the lattice exerts on the $\indx{a}$-th dislocation field to keep its curves layered on its glide plane.
It is associated with the lack of integrability of the glide plane distribution.
It acts along the glide plane and orthogonal to the line director, see also Fig.~\ref{fig:coupling}.
\item
The $1$-form $\prs{b}{\lambda}_L \, \prs{b\!a}{\boldsymbol\varpi}$ is that part of the reaction force that the lattice exerts on the $\indx{a}$-th dislocation field to keep its curves layered on its glide plane.
It represents the effect of the $\indx{b}$-th slip system on the $\indx{a}$-th slip system, and hence it is a non-Schmid effect.\footnote{%
According to Schmid's law, a slip system is activated when the associated resolved shear stress reaches a critical value.
However, single crystals show deviations from it, as slip systems are activated by the effect of the respective resolved shear stress, as well as of the ones on different slip systems.
In our formulation, a non-Schmid response occurs every time the dynamics of a layered dislocation field, say the $\indx{a}$-th, is directly affected by the $\indx{b}$-th dislocation field, with $\indx{b}\neq\indx{a}$ and $\prs{b}{\Pi} \neq \prs{a}{\Pi}$
(note that $\prs{b\!a}{\boldsymbol\varpi}=0$ when $\prs{b}{\Pi}=\prs{a}{\Pi}$, see~\S\ref{Layered}).}
It is perpendicular to both the $\indx{a}$-th and $\indx{b}$-th line directors.
\item
The $1$-form $\prs{a}{\lambda}_G \, \prs{a}{\boldsymbol\pi}$ is the reaction force that keeps the glide velocity of the $\indx{a}$-th dislocation field to lie on its plane distribution, and is perpendicular to the glide plane.
\end{itemize}
%---------------------

One may ask what happens if the layer condition~\eqref{constr-recap}$_2$ is dropped while keeping the conditions~\eqref{constr-recap}$_1$ on $\prs{a}{\boldsymbol{\bv}}$ and~\eqref{constr-recap}$_3$ on $\prs{a}{\boldsymbol{\vel}}$.
As was pointed out earlier, the hypothesis $\iota_{\prs{a}{\boldsymbol{\vel}} }\iota_{\delta\prs{a}{\chi}}\prs{a}{\df} =0$ of Lemma~\ref{Lem:Orowan-variation}---that was used to obtain the kinetic equations---relies on the glide constraint $\prs{b}{f}_G$ as well as on the layer constraint $\prs{b}{f}_L$.
Hence, instead of simply discarding it, we replace the layer constraint with the weaker version~\eqref{weak-condition},
introducing the following constraint functions:
%---------------------
\begin{equation} \label{g-w-constr}
	\prs{b}{f}_W (\boldsymbol\vartheta^{\nu}  , \prs{a}{\dd}^{\nu}  ,  \prs{a}{\boldsymbol{\vel}} ) =
	\iota_{ \prs{b}{\boldsymbol{\vel}} } \iota_{ \prs{b}{\boldsymbol{\bv}} }  \prs{b}{\df} 
	\,,\qquad
	\prs{b}{f}_G  (\boldsymbol\vartheta^{\nu}  ,  \prs{a}{\boldsymbol{\vel}} ) =
	\langle \prs{b}{\boldsymbol \pi} , \prs{b}{\boldsymbol{\vel}} \rangle
	\,.
\end{equation}
%---------------------
Both functions represent a nonholonomic constraint, and are to be treated with the nonholonomic approach, giving conditions similar to~\eqref{nh-constr-virtual}, viz.
%---------------------
\begin{equation} \label{nh-constr-virtual-weak}
	\bigg\langle \frac{\partial \prs{b}{f}_W }{\partial \prs{a}{\boldsymbol{\vel}}} , \delta \prs{a}{\chi} \bigg\rangle =
	\big\langle \iota_{ \prs{a}{\boldsymbol{\bv}} }  \prs{a}{\df}  \,, \delta\prs{a}{\chi} \big\rangle = 0
	 \,,
	\,\qquad
	\bigg\langle \frac{\partial \prs{b}{f}_G }{\partial \prs{a}{\boldsymbol{\vel}}} , \delta \prs{a}{\chi} \bigg\rangle =
	\left\langle \prs{a}{\boldsymbol\pi} \,, \delta\prs{a}{\chi} \right\rangle = 0
	\,.
\end{equation}
%---------------------
As was mentioned earlier, under the glide assumption, the condition~\eqref{nh-constr-virtual-weak}$_1$ is equivalent to the hypothesis $\iota_{\prs{a}{\boldsymbol{\vel}} }\iota_{\delta\prs{a}{\chi}}\prs{a}{\df} =0$ of Lemma~\ref{Lem:Orowan-variation}, and hence, the calculations given in~\S\ref{Material}  are still valid.
Furthermore, the weakening of the layer constraint allows one to include a broader class of dislocations in the model.
Denoting with $\prs{a}{\lambda}_W$ the Lagrange multiplier associated with $\prs{a}{f}_W$, Eq.~\eqref{EL-constrained2} is modified to read
%---------------------
\begin{equation} \label{EL-constraint-weak}
	\iota_{ \boldsymbol Y_{\nu} } \prs{a}{\dd}^{\nu} +
	\iota_{\star^{\sharp} \mathrm d \prs{a}{\boldsymbol{\xi}}_{\nu} } \prs{a}{\dd}^{\nu} +
	\mathsf T_{\prs{a}{\boldsymbol{\vel}}}   \prs{a}{\boldsymbol\psi}
	=
	\prs{a}{\boldsymbol\tau} +
	\prs{a}{\lambda}_W \,  \iota_{ \prs{a}{\boldsymbol{\bv}} }  \prs{a}{\df} +
	\prs{a}{\lambda}_G \, \prs{a}{\boldsymbol\pi}
	\,,\qquad\indx{a}=1,2,\hdots,N \,.
\end{equation}
%---------------------
Each reaction force $\prs{a}{\lambda}_G \, \prs{a}{\boldsymbol\pi}$ and $\prs{a}{\lambda}_W \, \iota_{ \prs{a}{\boldsymbol{\bv}} } \prs{a}{\df}$ is perpendicular to the plane on which the respective constraint enforces glide, i.e., those defined by the $1$-forms $\prs{a}{\boldsymbol\pi}$ and $\iota_{ \prs{a}{\boldsymbol{\bv}} } \prs{a}{\df}$, as shown in Fig.~\ref{fig:reactions}.
Keeping in mind the characterization of weakly-layered dislocation fields given in \S\ref{Glide},
the force $\prs{a}{\lambda}_W \, \iota_{ \prs{a}{\boldsymbol{\bv}} } \prs{a}{\df}$ vanishes for dislocations that either have a screw character or are layered on $\prs{a}{\Pi}$, while in the case of non-screw dislocation fields that are not strongly layered, the two reaction forces are necessary to keep the velocity $\prs{a}{\boldsymbol{\vel}}$ parallel to the Burgers director~$\prs{a}{\boldsymbol{\bv}}$.

%------------------------------------------------------------------------------------------------
%------------------------------------------------------------------------------------------------
\subsection{Balance of energy}    \label{Entropy}

The governing equations were derived using a Lagrange-d'Alembert principle for non-conservative processes starting from a Lagrangian function.
Next we look at the rate of change of quantities such as free energy, entropy and total energy.
First, for the sake of simplicity, we assume negligible micro-inertial effects by setting $\mathscr{T}_M=0$, so that the generalized forces are given in~\eqref{generalized-forces-no-inertia}.
By doing so, both the microforces $\boldsymbol{Y}_{\nu}$ and microstresses $\prs{a}{\boldsymbol{\xi}}_{\nu}$ can be obtained by simply differentiating the free energy function $\mathscr{F}$.                                                                                                                                                                                                                                                                                                                                                                                                                                                                                                                                                                                                                                                                                                                                                                                                                                                                                                                                                                                                                                                                                                                                                                                                                                                                                                                                                                                                                                                                                                                                                                                                                                                                                                                                                                                                                                                         
Therefore, recalling~\eqref{rate-C-mandel}, one can write the rate of change of the free energy in terms of the generalized forces, viz.
%---------------------
\begin{equation} \label{rate-free-energy}
	\partial_t \mathscr{F} \boldsymbol{m} =
	\langle \boldsymbol{ M} , \mathbf{ L} \rangle \, \mvf + 
	 \left\langle \partial_t   \boldsymbol \vartheta^{\nu} , \boldsymbol Y_{\nu}  \right\rangle \mvf +
	\sum_{\indx a=1}^N  \prs{a}{\boldsymbol{\xi}}_{\nu} \wedge \partial_t  \prs{a}{\dd}^{\nu}
	\,.
\end{equation}
%---------------------
Following the calculations given in~\S\ref{App:Calculations}, one can use the kinetic equations~\eqref{EL-material} to write
%---------------------
\begin{equation} \label{first-law}
	\partial_t \mathscr{F}\, \boldsymbol{m} =
	\langle \op M , \mathbf{L} \rangle \, \mvf + 
	\sum_{\indx{a}=1}^N \langle \prs{a}{\boldsymbol\tau} , \prs{a}{\boldsymbol{\vel}} \rangle \,\mvf +
	\sum_{\indx{a}=1}^N \mathrm d \!\left( \prs{a}{\boldsymbol{\xi}}_{\nu} \wedge \iota_{ \prs{a}{\boldsymbol{\vel}} } \prs{a}{\dd}^{\nu}
	\right)
	\,.
\end{equation}
%---------------------
Note that the constraint reactions have not been included in~\eqref{first-law} as they do not do work on dislocation displacements.
This is a consequence of the fact that the holonomic constraint~\eqref{l-g-constr}$_1$ does not explicitly depend on time (it is scleronomous), while the nonholonomic constraint~\eqref{l-g-constr}$_2$ is linear in the dislocation velocities \citep{neimark2004dynamics,fasso2015conservation}.
Eq.~\eqref{first-law} constitutes the first law of thermodynamics for our system.
The term $\langle \op M , \mathbf{L} \rangle$ is the work done by the internal forces per unit material volume, while the other terms are associated with changes in entropy.
In particular, we define the entropy production density (per unit volume) as a scalar field $\Upsilon$, and the entropy flux as a vector field $\boldsymbol\Phi$ \citep{muller1967entropy}, viz.
%---------------------
\begin{equation} \label{prod-flux}
	\Upsilon =
	- \sum_{\indx{a}=1}^N \langle \prs{a}{\boldsymbol\tau} , \prs{a}{\boldsymbol{\vel}} \rangle
	\,,\qquad
	\boldsymbol\Phi =
	- \sum_{\indx{a}=1}^N
	 \star^{\sharp} \!\left(
	 \prs{a}{\boldsymbol{\xi}}_{\nu} \wedge \iota_{ \prs{a}{\boldsymbol{\vel}}} \prs{a}{\dd}^{\nu}
	 \right) \,.
\end{equation}
%---------------------
It should be noted that the entropy flux is generated by the microstresses, and hence comes from the explicit dependence of the free energy on the dislocation fields.
As constant thermodynamic temperature $\theta$ is assumed, the entropy density per unit volume $\mathscr{N}$ can be defined by
$\massd \,\theta\,\partial_t \mathscr{N}= \langle \op M , \mathbf{L} \rangle - \massd \,\partial_t \mathscr{F} $.
Thus, using~\eqref{ext-der-hodge} to express the non-local term in~\eqref{first-law} as the divergence of the entropy flux, the balance of entropy reads
%---------------------
\begin{equation}
	\massd \, \theta\,  \partial_t \mathscr{N} =	\Upsilon +	\operatorname{Div} \boldsymbol\Phi \,.
\end{equation}
%---------------------
Note that from the convexity of the dissipation potential $\mathscr{D}$, the entropy production is automatically non-negative, and hence the second law of thermodynamics is satisfied \citep{steigmann2020primer}.\footnote{%
This follows from convexity as well as from the assumption that $\mathscr{D}$ attains its minimum for vanishing $\prs{a}{\boldsymbol{\vel}}$.
The convexity of $\mathscr{D}$ in the variable $\prs{a}{\boldsymbol{\vel}}$ implies that
%---------------------
\begin{equation}
	\mathscr{D} (\prs{a}{\boldsymbol{\vel}})
	+\left\langle \frac{\partial \mathscr{D}}{\partial \prs{a}{\boldsymbol{\vel}} } (\prs{a}{\boldsymbol{\vel}}) , 
	\Delta\prs{a}{\boldsymbol{\vel}} \right\rangle 
	\leq \mathscr{D} (\prs{a}{\boldsymbol{\vel}}+\Delta\prs{a}{\boldsymbol{\vel}}) 
	\,.
\end{equation}
%---------------------
Choosing $\Delta\prs{a}{\boldsymbol{\vel}} =-\prs{a}{\boldsymbol{\vel}}$, one has
%---------------------
\begin{equation}
	\left\langle \frac{\partial \mathscr{D}}{\partial \prs{a}{\boldsymbol{\vel}} } (\prs{a}{\boldsymbol{\vel}}) , 
	\prs{a}{\boldsymbol{\vel}} \right\rangle \geq\mathscr{D} (\prs{a}{\boldsymbol{\vel}}) - \mathscr{D}(\boldsymbol 0)
	\geq 0 \,,
\end{equation}
%---------------------
as $\mathscr{D}$ attains its minimum for $\prs{a}{\boldsymbol{\vel}}=\boldsymbol 0$.
The entropy production $\Upsilon$ can be expressed by plugging~\eqref{drag} into~\eqref{prod-flux}$_1$, viz.
%---------------------
\begin{equation}
	\Upsilon =  \massd \sum_{\indx{a}=1}^N
	\left\langle \frac{\partial \mathscr{D}}{\partial \prs{a}{\boldsymbol{\vel}} } , \prs{a}{\boldsymbol{\vel}} \right\rangle \,.
\end{equation}
%---------------------
Therefore, we conclude that the entropy production $\Upsilon$ is a sum of non-negative terms, whence $\Upsilon \geq 0$.} 
In the case of a dissipation potential $\mathscr{D}$ that is quadratic in the dislocation velocities $\prs{a}{\boldsymbol{\vel}}$, the entropy production density can be written as $\Upsilon= 2\mathscr{ D}$ \citep{goldstein2002classical}.
Examples of this kind of dissipation potentials are given in~\S\ref{Dissipation}.

Under the assumption of vanishing $\mathscr{T}_M$, the total energy density is defined as the scalar field $\mathscr{E} = \mathscr{T}_S + \mathscr{F}$.
Hence, $\partial_t\mathscr{E} = \llangle \boldsymbol A, \boldsymbol V \rrangle + \partial_t\mathscr{F}$.
Note that the balance of linear momentum~\eqref{EL-spatial} allows one to write\footnote{%
For two-point tensors one can write $\nabla = \op F^{\star} \nabla^{\boldsymbol g}$, i.e., $\nabla_A = \mathsf F^a{}_A \nabla^{\boldsymbol g}_a$, see~\citep{marsden1983mathematical, sozio2020riemannian}.
Since $\mathbf{L}=\varphi^* \nabla^{\boldsymbol g} \boldsymbol v$, one has $ \langle \op M , \mathbf{L} \rangle =  \langle \op P , \nabla \boldsymbol v \rangle$.}
%---------------------
\begin{equation}
	\llangle \massd \boldsymbol A, \boldsymbol V \rrangle  + \langle \op M , \mathbf{L} \rangle =
	\langle \massd \mathbf{b}, \boldsymbol V \rangle 
	+ \langle \operatorname{Div}\op P, \boldsymbol V \rangle
	+ \langle \op M , \mathbf{L} \rangle =
	\langle \massd \mathbf{b}, \boldsymbol V \rangle + \operatorname{Div}(\op{P} \boldsymbol V) \,.
\end{equation}
%---------------------
Then, from the first law~\eqref{first-law} one obtains the local balance of total energy as
%---------------------
\begin{equation} \label{energy-rate}
	\partial_t\mathscr{E} \boldsymbol{m} =
	\langle \mathbf{b}, \boldsymbol V \rangle \, \boldsymbol{m} +
	\operatorname{Div}(\op{P} \boldsymbol V) \,\mvf +
	\sum_{\indx{a}=1}^N \langle \prs{a}{\boldsymbol\tau} , \prs{a}{\boldsymbol{\vel}} \rangle \,\mvf +
	\sum_{\indx{a}=1}^N \mathrm d\!\left(\prs{a}{\boldsymbol{\xi}}_{\nu}\wedge
	\iota_{\prs{a}{\boldsymbol{\vel}}}\prs{a}{\dd}^{\nu}\right)\,.
\end{equation}
%---------------------
The global balance of total energy can be obtained  by integrating \eqref{energy-rate} on $\mathcal B$.
Note that the integral of the entropy flux vanishes by virtue of the boundary conditions~\eqref{EL-material-boundary} with $\prs{a}{\boldsymbol\psi}=0$, while from $\operatorname{Div}(\op{P} \boldsymbol V)$ one recovers the tractions $\mathbf{t}$ through~\eqref{EL-spatial-boundary}.
Therefore, bringing the time derivative outside the integral, one obtains
%---------------------
\begin{equation} \label{balance-energy1}
	\frac{\mathrm{d}}{\mathrm{d}t} \int_{\mathcal B} \mathscr{E}\, \boldsymbol{m} =
	\int_{\partial \mathcal B} \langle \mathbf{t} , \boldsymbol V \rangle \, \boldsymbol{\varsigma} +
	\int_{\mathcal B} \langle \mathbf{b} , \boldsymbol V \rangle \, \boldsymbol{m} +
	\sum_{\indx{a}=1}^N \int_{\mathcal B} \langle \prs{a}{\boldsymbol\tau} ,
	 \prs{a}{\boldsymbol{\vel}} \rangle \, \mvf
	\,.
\end{equation}
%---------------------
The first two terms on the right-hand side of~\eqref{balance-energy1} represent the power supplied by the external forces, while the last term is the dissipated power by the effect of the dislocation motion.

When taking into account the contribution of the micro-inertial forces associated with the motion of dislocations, the rate of change of the free energy cannot be expressed in terms of the generalized forces alone, as the micro-kinetic energy $\mathscr{T}_M$ depends on the internal variables.
As a result, one would need to distinguish the contribution to microforces and microstresses that is due to $\mathscr{F}$ from the one coming from $-\mathscr{T}_M$, cf.~\eqref{generalized-forces}.
As for the total energy density in the presence of dislocation inertial forces, it is defined as $\mathscr{E} = \mathscr{T}_S + \mathscr{T}_M + \mathscr{F}$.
It should be noticed that the balance of the total energy~\eqref{balance-energy1} holds in this case as well.
In fact, by virtue of the boundary conditions~\eqref{EL-material-boundary}  the flux of micro-kinetic energy  across $\partial\mathcal B$ is counterbalanced by the nonlocal effects associated with microstresses.

%%%%%%%%%%%%%%%%%%%%%%%%%%%
%%%%%%%%%%%%%%%%%%%%%%%%%%%
\section{A Simplified Model for Nonlinear Dislocation Dynamics}   \label{Sec:Problem}

In the previous section we presented a variational formulation for the dynamics of solids with distributed dislocations, and derived the corresponding Euler-Lagrange equations.
The free energy was assumed to be a function of some external and internal variables.
In this section, we introduce a simplified model for gliding layered dislocations, in which we neglect the micro-inertial forces associated with the dislocation motions as well as other nonlocal effects.
We assume that the only contribution to the free energy comes from the elastic strains of the lattice.
As a consequence, elastic deformations induce a material force acting on dislocation fields known as the Peach-Koehler force \citep{PeachKoehler1950}.
We also discuss constitutive classes for the dissipation potential.

%---------------------------------------------------------------------
%---------------------------------------------------------------------
\subsection{The Peach-Koehler force}        \label{PK}

We start by assuming that free energy does not explicitly depend on the distribution of dislocations, i.e., $\mathscr{F}=\mathscr{F}( \boldsymbol{C}^{\flat}, \boldsymbol\vartheta^{\nu}  )$, and is entirely associated with the elastic response of the material.
This means that we assume the existence of an elastic energy $\mathscr{W} (\sube{\boldsymbol C}^{\flat})$ per unit mass,\footnote{%
In classical plasticity it is customary to work with an elastic energy density per unit volume, and defined in the intermediate configuration. Although for gliding dislocations the material volume is always conserved, that might not be the case for climbing dislocations, or for the evolution of other defects. Instead, mass is a conserved quantity for all types of deformations considered in plasticity. For this reason we choose to work with a density $\mathscr{W} (\sube{\boldsymbol C}^{\flat})$ per unit mass.} where the elastic pulled-back metric is defined as
%---------------------
\begin{equation} \label{elastic-pullback-metric}
	\sube{\boldsymbol C}^{\flat}(\boldsymbol V,\boldsymbol W)=
	\boldsymbol g (\sube{\op F}\boldsymbol V,\sube{\op F}\boldsymbol W)=
	\boldsymbol{C}^{\flat} (\subp{\op F}^{-1}\boldsymbol V,\subp{\op F}^{-1}\boldsymbol W) \,,
\end{equation}
%---------------------
for all vectors $\boldsymbol V,\boldsymbol W$, or in terms of the lattice coframe, as
%---------------------
\begin{equation} \label{elastic-pullback-metric2}
	\sube{\boldsymbol C}^{\flat} \left( \tfrac{\partial}{\partial Z^{\mu}} , \tfrac{\partial}{Z^{\nu}} \right) 
	= \boldsymbol{C}^{\flat}(\boldsymbol e_{\mu},\boldsymbol e_{\nu})	\,,
\end{equation}
%---------------------
with $Z^{\nu}$ indicating Cartesian coordinates on $\mathcal B$ defined in \S\ref{Sec:Lattice}.
In a more compact form, $\boldsymbol{C}^{\flat}=\subp{\op F}^{\star} \sube{\boldsymbol C}^{\flat} \subp{\op F}$.
The function $\mathscr{F}( \boldsymbol{C}^{\flat}, \boldsymbol\vartheta^{\nu}  )$ is then derived from the elastic energy $\mathscr{W} (\sube{\boldsymbol C}^{\flat})$ through the change of variables~\eqref{elastic-pullback-metric} or~\eqref{elastic-pullback-metric2}.
We also neglect the micro-kinetic energy $\mathscr{T}_M$, which would necessarily introduce a dependence on the dislocation fields in the Lagrangian, and hence non-vanishing microstresses $\prs{a}{\boldsymbol{\xi}}$.

\begin{figure}[tp]
\centering
\includegraphics[width=.55\textwidth]{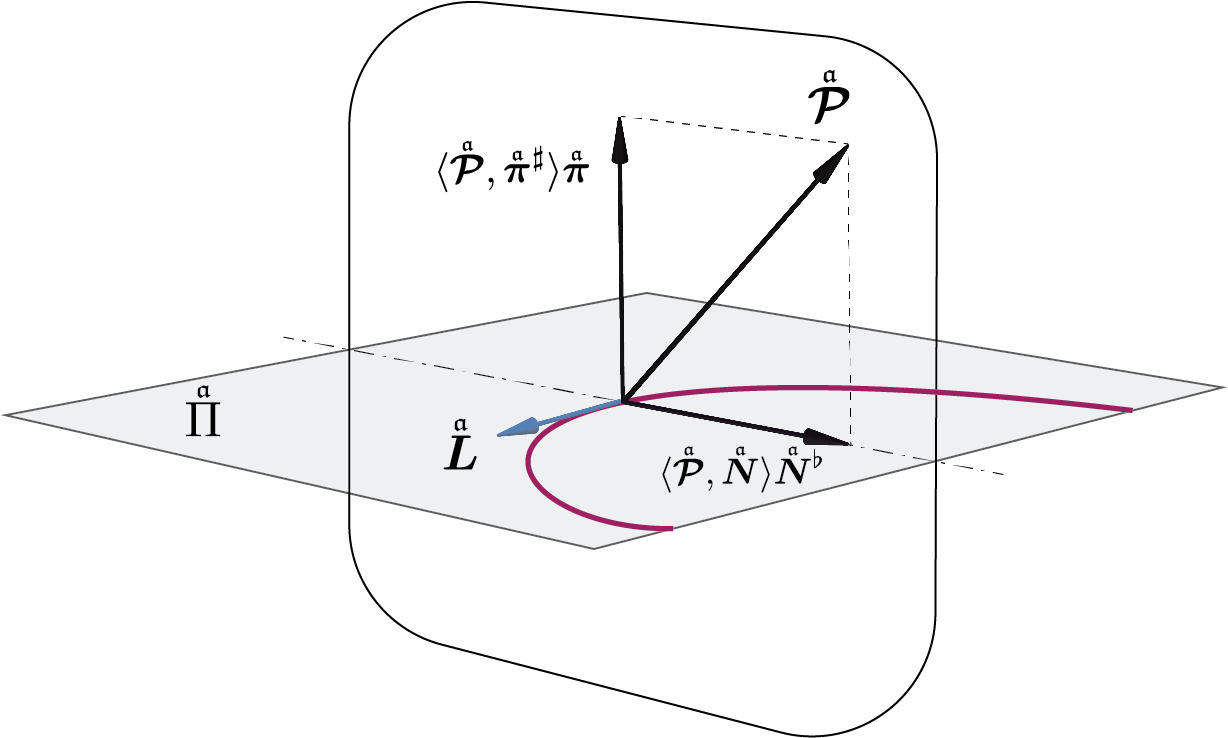}
\vskip 0.0 in
\caption{Components of the Peach-Koehler force with respect to a slip plane.}
\label{fig:pk}
\end{figure}

The microforces, defined in~\eqref{generalized-forces-no-inertia}${}_1$ as the derivative of $\mathscr{F}$ with respect to the lattice frame $\boldsymbol\vartheta^{\nu}$, are related to the second Piola-Kirchhoff stress $\boldsymbol S$, defined in~\eqref{second-pk} as the derivative with respect to $\boldsymbol{C}^{\flat}$.
In particular, one finds
%---------------------
\begin{equation} \label{Mandel-material}
	\boldsymbol Y_{\nu} = 
	\massd \frac{\partial \mathscr{F}}{\partial \boldsymbol\vartheta^{\nu} } =
	- \op M \boldsymbol e_{\nu} \,,
\end{equation}
%---------------------
where $\op M$ denotes the Mandel stress defined in~\S\ref{Spatial} as the lowering of $\boldsymbol S$ through $\boldsymbol C^{\flat}$.\footnote{%
Eq.~\eqref{Mandel-material} was derived in \citep{sozio2020riemannian} following a rather standard procedure in plasticity.
The calculations consist of applying the chain rule twice, viz.
%---------------------
\begin{equation} \label{force-mandel}
	\frac{\partial \mathscr{F}}{\partial \boldsymbol\vartheta^{\nu} } =
	\frac{\partial \mathscr{W}}{\partial \sube{\boldsymbol C}^{\flat} }  
	\frac{\partial \sube{\boldsymbol C}^{\flat} }{\partial \boldsymbol\vartheta^{\nu} } 
	=\frac{\partial \mathscr{F}}{\partial \boldsymbol C}^{\flat} 
	\frac{\partial \boldsymbol C^{\flat}}{\partial \sube{\boldsymbol C}^{\flat} }  
	\frac{\partial \sube{\boldsymbol C}^{\flat} }{\partial \boldsymbol\vartheta^{\nu} }
	 \,.
\end{equation}
%---------------------
}
Therefore, in the case of closed dislocation fields obeying Orowan's equation, from~\eqref{Mandel-material}, in addition to dissipation, the term in~\eqref{EL-material} coming from the microforces becomes
%---------------------
\begin{equation}
	\iota_{ \boldsymbol Y_{\nu} } \prs{a}{\dd}^{\nu}  =
	- \iota_{\op M \boldsymbol e_{\nu} } \prs{a}{\dd}^{\nu}  =
	- \iota_{\op M \boldsymbol e_{\nu} } \prs{a}{\bv}^{\nu}\prs{a}{\df} =
	- \iota_{\op M \prs{a}{\boldsymbol{\bv} } } \prs{a}{\df} \,.
\end{equation}
%---------------------
Then, we define the Peach-Koehler force associated with the $\indx{a}$-th dislocation field as the $1$-form
%---------------------
\begin{equation} \label{pk-force}
	\prs{a}{\boldsymbol\pk} = \iota_{\op M \prs{a}{\boldsymbol{\bv}} } \prs{a}{\df} \,,
\end{equation}
%---------------------
which is work-conjugate to the dislocation displacements, and constitutes the main driving force of dislocation dynamics.
Note that it can also be written as $\prs{a}{\boldsymbol\pk} = \prs{a}{\dens} \, \mvf( \prs{a}{\boldsymbol{\lv}} , \op M \prs{a}{\boldsymbol{\bv}})$.
Under these assumptions, the kinetic equation~\eqref{EL-constrained2} reads
%---------------------
\begin{equation} \label{EL-constrained3}
	\prs{a}{\boldsymbol\pk} +
	\prs{a}{\lambda}_L \, \mathfrak I_{\prs{a}{\Pi}} \prs{a}{\boldsymbol{\norm}}^{\flat}
	+ \sum_{\indx{b}=1}^N \prs{b}{\lambda}_L \, \prs{b\!a}{\boldsymbol\varpi}
	+ \prs{a}{\lambda}_G \, \prs{a}{\boldsymbol\pi}
	+ \prs{a}{\boldsymbol\tau}
	=0
	\,,\qquad\indx{a}=1,2,\hdots,N 
	\,,
\end{equation}
%---------------------
expressing the balance between the Peach-Koehler forces, the constraint reactions, and the dissipation forces on each slip system.
It should be noticed that the component of the Peach-Koehler force along the dislocation curves vanishes, as
%---------------------
\begin{equation}
	\langle \prs{a}{\boldsymbol\pk}  , \prs{a}{\boldsymbol{\lv}} \rangle =
	\langle \iota_{\op M \prs{a}{\boldsymbol{\bv}}  } \prs{a}{\df} ,  \prs{a}{\boldsymbol{\lv}} \rangle =
	-\langle \iota_{ \prs{a}{\boldsymbol{\lv}} } \prs{a}{\df} , \op M \prs{a}{ \boldsymbol{\bv} } \rangle =0 \,,
\end{equation}
%---------------------
where use was made of $\iota_{\prs{a}{\boldsymbol{\lv}} } \prs{a}{\df}=0$.
As for the component normal to the dislocation curves, since in~\S\ref{Layered} it was shown that $\iota_{ \prs{a}{\boldsymbol{\norm}} } \prs{a}{\df} = \prs{a}{\dens}  \prs{a}{\boldsymbol\pi}$, one has
%---------------------
\begin{equation} \label{rss}
	\langle \prs{a}{\boldsymbol\pk} , \prs{a}{ \boldsymbol{\norm}} \rangle =
	\langle \iota_{\op M  \prs{a}{ \boldsymbol{\bv} } } \prs{a}{\df} , \prs{a}{\boldsymbol{\norm}} \rangle =
	-\langle \iota_{ \prs{a}{\boldsymbol{\norm}} } \prs{a}{\df} , \op M \prs{a}{ \boldsymbol{\bv} } \rangle =
	-\prs{a}{\dens} \, \langle  \prs{a}{\boldsymbol\pi} , \op M \prs{a}{ \boldsymbol{\bv} } \rangle  \,,
\end{equation}
%---------------------
where $ \langle \prs{a}{\boldsymbol\pi} , \op M \prs{a}{ \boldsymbol{\bv} } \rangle$ is what is commonly called the resolved shear stress on the $\indx{a}$-th slip plane (see Fig.~\ref{fig:pk}).
This is usually defined as the component of stress on the slip plane in the direction of slip, i.e., in the direction of the Burgers director by the effect of~\eqref{orowan-slip}.
It should be emphasized that in the nonlinear setting the resolved shear stress is a Mandel stress.
Lastly, as the lattice reaction forces do not contribute to the balance of energy, from~\eqref{EL-constrained3} the entropy production is entirely given by the Peach-Koehler force, viz.
%---------------------
\begin{equation} \label{entropy-pk}
	\Upsilon_{\mathcal{N}} =
	-\sum_{\indx{a}=1}^N \langle \prs{a}{\boldsymbol\tau} , \prs{a}{\boldsymbol{\vel}} \rangle
	=\sum_{\indx{a}=1}^N \langle \prs{a}{\boldsymbol{\pk}} , \prs{a}{\boldsymbol{\vel}} \rangle	
	\,.
\end{equation}
%---------------------

%---------------------
%---------------------
%---------------------
%---------------------

%--------------------
%--------------------
\begin{remark} \label{Rem:Mandel-elastic}
As was mentioned in Remark~\ref{Rem:Mandel-int}, the Mandel stress is work-conjugate to $\mathbf{L}$.
Assuming that the free energy depends only on the elastic strain, from~\eqref{Mandel-material} one also obtains
%---------------------
\begin{equation} \label{mandel-Lp}
	\left\langle \boldsymbol Y_{\nu}, \partial_t  \boldsymbol \vartheta^{\nu}  \right\rangle =
	\left\langle \op M \boldsymbol e_{\nu} , \partial_t  \boldsymbol \vartheta^{\nu}  \right\rangle =
	 - \left\langle \op M , \subp{\mathbf{L}} \right\rangle \,,
\end{equation}
%---------------------
where use was made of $\subp{\mathbf{L}} =\boldsymbol e_{\nu} \otimes  \partial_t \boldsymbol\vartheta^{\nu}$.
This means that the Mandel stress is work-conjugate to $-\subp{\mathbf{L}}$ as well.
In particular, $ - \left\langle \op M , \subp{\mathbf{L}} \right\rangle$ represents the entropy production $\Pi_{\mathcal{N}}$ given in~\eqref{entropy-pk}.
Moreover, as $\mathbf{L}-\subp{\mathbf{L}} = \subp{\op F}^{-1} \sube{\mathbf{L}} \subp{\op F}$ from~\eqref{all-L}, one can write the rate of change of the elastic free energy as
%---------------------
\begin{equation}
	\dens \, \partial_t \mathscr{F}=
	\left\langle \op M , \mathbf{L} \right\rangle
	 - \left\langle \op M , \subp{\mathbf{L}} \right\rangle
	 =\left\langle \widetilde{\op{M}} , \sube{\mathbf{L}} \right\rangle \,,
\end{equation}
%---------------------
where $\widetilde{\op{M}} = \subp{\op F} \op M \subp{\op F}^{-1}$ is the Mandel stress brought to the ``intermediate configuration''.
This is due to the fact that the only deformation that the free energy sees is the elastic one.
It should be noted that the tensor $\widetilde{\op{M}}$ can be obtained by lowering with $\sube{\boldsymbol C}^{\flat}$ one index of the pushforward of $\boldsymbol S$ via $\subp{\op F}$, i.e.,
%---------------------
\begin{equation}
	\langle \widetilde{\op{M}}^* \boldsymbol\nu , \boldsymbol V \rangle =
	\llangle \subp{\op F} \boldsymbol S (\boldsymbol\nu) , \subp{\op F} \boldsymbol V \rrangle_{\sube{\boldsymbol C}^{\flat}} \,,
\end{equation}
%---------------------
cf.~\eqref{mixed-second-mandel},
and is sometimes defined as the primary variant of the Mandel stress \citep{mandel1971plasticite,cleja2000eshelby,lubliner2008plasticity}.
However, it should be noted that while the tensor $\op M$ is always work-conjugate to $\mathbf{L}$, the conjugacy between $\widetilde{\op{M}}$ and $\sube{\mathbf{L}}$, as well as the one between $-\op M$ and $\subp{\mathbf{L}}$, holds only under the assumption of an elastic free energy.
\end{remark}
%--------------------
%--------------------

%------------------------------------------------------
%------------------------------------------------------
\subsection{Dissipation} \label{Dissipation}

The dissipation potential $\mathscr{D}$ was introduced in \S\ref{Variational} as a convex function of the dislocation velocities.
The drag force $\prs{a}{\boldsymbol\tau}$ on the motion of the $\indx{a}$-th dislocation field was obtained by differentiating $\mathscr{D}$ with respect to $\prs{a}{\boldsymbol{\vel}}$ as in~\eqref{drag}.
In this subsection, we introduce some simplifications and provide explicit expressions for the function $\mathscr{D}$.
The first ansatz is to assume that each dislocation field dissipates energy independently from the others.
Therefore, we consider the following additive decomposition:
%---------------------
\begin{equation} \label{separate-dissipation}
	\mathscr{D} (\prs{1}{\boldsymbol{\vel}},\hdots,\prsn{N}{\boldsymbol{\vel}} )
	= \sum_{\indx{a}=1}^N  \prs{a}{\mathscr{D}} (\prs{a}{\boldsymbol{\vel}}) \,,
\end{equation}
%---------------------
where each function $\prs{a}{\mathscr{D}}$ depends only on the respective dislocation velocity---as well as on the other internal variables that we omit for the sake of simplicity---so that one obtains the dissipative force~\eqref{drag} associated with the material motion of the $\indx{a}$-th dislocation field as
%---------------------
\begin{equation} \label{dissip-form}
	\prs{a}{\boldsymbol\tau} = - \massd  \,  \frac{\partial \prs{a}{\mathscr{D}}}{\partial \prs{a}{\boldsymbol{\vel}}} \,.
\end{equation}
%---------------------
Next we assume that each function $\prs{a}{\mathscr{D}}$ depends only on the component of the dislocation velocity that is normal to the dislocation curves, i.e., such that $\prs{a}{\mathscr{D}} = \prs{a}{\mathscr{D}} (\prs{a}{\vel}^{\bot})$, and that it is convex in that argument, so that one obtains
%---------------------
\begin{equation} \label{dissip-norm-vel}
	\prs{a}{\boldsymbol\tau} = - \massd \, \frac{\partial \prs{a}{\mathscr{D}}}{\partial \prs{a}{\vel}^{\bot}} 
	\frac{\partial \langle \prs{a}{\boldsymbol{\norm}}^{\flat} , 
	\prs{a}{\boldsymbol\vel} \rangle }{\partial \prs{a}{\boldsymbol\vel} }
	= - \massd \, \frac{\partial \prs{a}{\mathscr{D}}}{\partial \prs{a}{\vel}^{\bot}} \prs{a}{\boldsymbol{\norm}}^{\flat} 	
	= - \massd \, \prs{a}{\mathscr{D}}'  (\prs{a}{\vel}^{\bot}) \, \prs{a}{\boldsymbol{\norm}}^{\flat}
	\,.
\end{equation}
%---------------------
Each scalar-valued dissipation potential $\prs{a}{\mathscr{D}}$ can be assumed quadratic in $\prs{a}{\boldsymbol\vel} $ and proportional to the scalar dislocation density $\prs{a}{\dens}$, i.e., $\prs{a}{\mathscr{D}} = \frac{1}{2} \prs{a}{\dens}\, \prs{a}{c}_o  \, (\prs{a}{\vel}^{\bot})^2$ for a constant drag coefficient  $\prs{a}{c}_o\geq 0$.
Hence, $\prs{a}{\boldsymbol\tau} = - \massd \,\prs{a}{\dens}\, \prs{a}{c}_o \, \prs{a}{\vel}^{\bot}  \prs{a}{\boldsymbol{\norm}}^{\flat}$.
In this case, the energy dissipated by the $\indx{a}$-th dislocation field is $\prs{a}{c}_o \, (\prs{a}{\vel}^{\bot})^2$.
Moreover, it is reasonable to assume that $\langle \prs{a}{\boldsymbol\tau} ,  \prs{a}{ \boldsymbol\pi}^{\sharp} \rangle =0$, i.e., that the dissipative forces do work only on dislocation displacements in the glide plane, which clearly holds for~\eqref{dissip-norm-vel}.
This also means that the $1$-forms $\prs{a}{\boldsymbol\tau} $ and $ \prs{a}{ \boldsymbol\pi}$ are linearly independent.

Viscous drag is not the only type of resistance that dislocations experience while they move across a crystal.
As a matter of fact, the motion of dislocations is only allowed when sufficiently large forces act on them, i.e., when the resolved shear stress reaches the value of the Peierls stress \citep{nabarro1967theory}.
We propose a penalty method that models this static friction via viscous resistance by incorporating it in the dissipation potential, with the goal of avoiding the introduction of discontinuities and singularities.
It should be noted that the resolved shear stress defined as $ \langle \prs{a}{\boldsymbol\pi} , \op M\prs{a}{\boldsymbol{\bv}} \rangle$ appears in the kinetic equations via the Peach-Koehler force, as was shown in~\eqref{rss}.
However, the Peach-Koehler force is not the only force at play in the dynamics of dislocations, as there might be other effects such as microstress and non-conservative forces considered in~\S\ref{Sec:Variational}, or reaction forces coming from the lattice constraints, see~\S\ref{Constraints}.
Hence, in the present theory the resolved shear (Mandel) stress does not fully determine whether the dislocation motion is allowed or not.
For this reason, we follow a strain-based approach, in the sense that the parameter that activates the motion is based on a measure of strain rather than stress.
The philosophy behind this choice lies in the fact that the energy required by the dislocations to overcome the potential barrier and move depends on the distances between lattice points \citep{nabarro1997fifty, joos1997peierls}, and hence on the elastic deformation.
In particular, we assume that this threshold mechanism is governed by a parameter $\prs{a}{r}$ denoting the shear strain on the slip plane in the direction of the Burgers director, viz.
%---------------------
\begin{equation}
	\prs{a}{r} =  \llangle \prs{a}{\boldsymbol\pi}^{\sharp},\prs{a}{\boldsymbol{\bv}} \rrangle_{\boldsymbol{C}^{\flat}} 
	=\langle \prs{a}{\boldsymbol\pi}, \op{C} \prs{a}{\boldsymbol{\bv}} \rangle
	= \prs{a}{\pi}_{\mu} \,  C^{\alpha}{}_{\mu} \, \prs{a}{\bv}^{\nu}\,.
\end{equation}
%---------------------
It should be noted that in our geometric approach, the pulled-back metric expressed in the lattice frame represents elastic strain, as shown in~\eqref{elastic-pullback-metric2}.
The motion of the $\indx{a}$-th dislocation field is activated when $\prs{a}{r}\geq r_o (\pi_{\nu},\bv^{\nu})$, for a smooth function $r_o$ of the slip plane orientation and of the Burgers vector direction. 
Therefore, we introduce a penalty function $p(\prs{a}{r})$ with the properties: $p(\prs{a}{r}) \gg 1$ for $\prs{a}{r} < s <  r_o (\pi_{\nu},\bv^{\nu})$, and $p(\prs{a}{r}) = 1$ for $\prs{a}{r} > S>  r_o (\pi_{\nu},\bv^{\nu})$, for $s,S>0$.
Based on the assumption~\eqref{separate-dissipation}, a modified dissipation potential is defined as
%---------------------
\begin{equation}
	\mathscr{D} = \sum_{\indx{a}=1}^N
	p(\prs{a}{r}) \,
	\prs{a}{\mathscr{D}} (\prs{a}{\boldsymbol{\vel}})
	\,,
\end{equation}
%---------------------
that models a static resistance by increasing the viscosity below the given threshold. The drag force can be written as
%---------------------
\begin{equation}
	\prs{a}{\boldsymbol\tau} =
	- \massd  \, p(\prs{a}{r}) \, \frac{\partial \prs{a}{\mathscr{D}}}{\partial \prs{a}{\boldsymbol{\vel}}} \,.
\end{equation}
%---------------------

%---------------------
%---------------------
\begin{remark}
In phenomenological plasticity, non-Schmid effects are usually encoded in a generalized stress measure that accounts for the resolved shear stress on other slip systems \citep{qin1992non, soare2014plasticity,salahshoor2018non}.
When this stress measure exceeds a certain threshold, the slip system is activated.
In our formulation, such an approach would require to define penalty functions $\prs{a}{p}$ that depend on the resolved strain associated with different dislocation fields, i.e., $\prs{a}{p}(\prs{1}{r}, \hdots , \prsn{N}{r} \,)$.
\end{remark}
%---------------------
%---------------------

%-------------------------------------------------------
%-------------------------------------------------------
\subsection{Summary of the governing equations} \label{ibvp}

In this section, we summarize the governing equations and formulate the initial boundary-value problem (IBVP) in the absence of micro-inertial forces and other nonlocal effects.
We take into account all the assumptions introduced so far; we consider closed dislocation fields obeying Orowan's equation, an elastic free energy, and the presence of lattice constraints that enforce the dislocation fields to glide and remain layered on the respective plane distributions.
The initial state of the crystal is given in terms of a lattice structure and of $N$ dislocation fields defined on $\mathcal B$.
Loads and boundary conditions at all times are considered as inputs as well.
The goal is to show how to use the governing equations derived so far in order to obtain the evolution of both the internal and external variables.
In particular, knowing the lattice structures at time $t$ allows one to obtain the geometry of the Riemannian manifold $(\mathcal B , \boldsymbol G_t )$,
and hence, to calculate the history of stress in the crystal during the plastic deformation process using the methods of nonlinear anelasticity.

The state of the system at time $t=0$ is defined
by a lattice structure in terms of the coframe $\{\boldsymbol\vartheta^{\nu}_0\}$,
and by $N$ closed decomposable dislocation fields $\prs{a}{\dd}^{\nu}_0$, each one layered on the plane distribution $\prs{a}{\Pi}_0$ described by $\prs{a}{\pi}_{\nu}$.
At time $t=0$ the incompatibility of the lattice forms must coincide with the sum of the dislocation fields as in~\eqref{gnd-sum}, viz.
%---------------------
\begin{equation} \label{incomp-eq-0}
	\mathrm d \boldsymbol \vartheta^{\nu}_0 = \sum_{\indx{a}=1}^{N} \prs{a}{\dd}^{\nu}_0\,.
\end{equation}
%---------------------
%
Energy is provided to the system through time-dependent body forces $\mathbf{b}(x,t)$ and tractions $\mathbf{t}(x,t)$, as well as through displacement boundary conditions $\overline{\varphi}_t$ on a subset $\partial_D\mathcal B$ of the boundary $\partial\mathcal B$ of the crystal.
It should be noted that contact forces are assigned on $\partial_N\mathcal B = \partial\mathcal B\setminus \partial_D\mathcal B$, while on $\partial_D\mathcal B$ they are reaction forces (Lagrange multipliers) associated with the displacement boundary conditions.
We denote the former with $\overline{\mathbf{t}}$ and the latter with $\mathbf{t}$.

As was mentioned earlier, for a given set of inputs---consisting of the initial internal state, the prescribed loads and the boundary conditions---our goal is to find the time-dependent configuration $\varphi_t$ and the material motions $\prs{a}{\chi}_t$ throughout the time interval of interest.
One should note that during this interval, the material motions might not be defined for all points in the body,
as was discussed in Remark~\ref{rem:motion-boundary}.
Therefore, it is convenient to express them in terms of their dislocation velocities $\prs{a}{\boldsymbol{\vel}}$, which are always well-defined.
Along with the material and spatial motions, we want to find the evolutions of the internal variables, i.e., the dislocation fields $\prs{a}{\dd}^{\nu}(X,t)$ and lattice forms $\boldsymbol\vartheta^{\nu}(X,t)$.
These evolve according to the evolution equation~\eqref{rate-Lie} and Orowan's equation~\eqref{orowan}, which can be integrated to give
%---------------------
\begin{equation} \label{evolution-final}
\begin{split}
	\boldsymbol\vartheta^{\nu} (X,t) &= \boldsymbol\vartheta^{\nu}(X,0) 
	- \sum_{\indx{a}=1}^{N} \int_0^t
	\iota_{\prs{a}{\boldsymbol{\vel}} (X,\overline t)} \prs{a}{\dd}^{\nu}(X,\overline t)
	\,\mathrm d \overline t
	\,,\\
	\prs{a}{\dd}^{\nu} (X,t)
	&= \prs{a}{\dd}^{\nu}(X,0) -  \int_0^t
	\mathrm d ( \iota_{ \prs{a}{\boldsymbol{\vel}} (X,\overline t) } \prs{a}{\dd}^{\nu} (X,\overline t) )
	\,\mathrm d \overline t
	\,,\qquad\indx{a}=1,2,\hdots,N \,.
\end{split}
\end{equation}
%---------------------
It should be noted that the evolution equations~\eqref{evolution-final} with the initial condition~\eqref{incomp-eq-0} guarantee that the incompatibility equation~\eqref{gnd-sum} is automatically satisfied at all times.
Decomposability and closedness of dislocation fields are preserved by~\eqref{evolution-final} as well.
At time $t$, $\prs{a}{\boldsymbol\pi}$, $\prs{a\!b}{\boldsymbol\varpi}$ and $\mathfrak I_{\prs{a}{\Pi}}$ and $\prs{a}{\boldsymbol{\norm}}$ can be expressed in terms of the $\indx{a}$-th dislocation field and of the lattice forms as was discussed in~\S\ref{Layered}.
The following layer and glide constraints hold:
%---------------------
\begin{equation}\label{constraints-final}
	\prs{a}{\boldsymbol \pi} \wedge \prs{a}{\df} = 0
	\quad\text{or}\qquad
	\iota_{\prs{a}{\boldsymbol{\vel}} } \iota_{\prs{a}{\boldsymbol{\bv}} } \prs{a}{\df} = 0
	\,,\qquad
	\langle \prs{a}{\boldsymbol \pi} , \prs{a}{\boldsymbol{\vel}} \rangle =0
	\,,\qquad\indx{a}=1,2,\hdots,N \,,
\end{equation}
%---------------------
together with the displacement boundary conditions $\varphi\vert_{\partial_D\mathcal B} (X,t) = \overline{\varphi}(X,t)$.
Recall that the constraint on the Burgers vector is redundant, as was discussed in~\S\ref{Constraints}.
All these unknowns, together with the Lagrange multipliers $\prs{a}{\lambda}_L (X,t)$, $\prs{a}{\lambda}_G (X,t)$, and $\mathbf{t}(x,t)$ associated with the constraints must satisfy the Euler-Lagrange equations associated with the constrained action principle, i.e., the kinetic equations~\eqref{EL-constrained3}, and the balance of linear momentum~\eqref{EL-spatial} with the associated boundary conditions~\eqref{EL-spatial-boundary}, viz.
%---------------------
\begin{align}
	\label{kinetic-final}
	&\prs{a}{\boldsymbol\pk} +
	\prs{a}{\lambda}_L \, \prs{a}{\dens} \, \mathfrak I_{\prs{a}{\Pi}} \, \prs{a}{\boldsymbol{\norm}}^{\flat}
	+ \sum_{\indx{b}=1}^N \prs{b}{\lambda}_L \, \prs{b\!a}{\boldsymbol\varpi}
	+ \prs{a}{\lambda}_G \, \prs{a}{\boldsymbol\pi}
	+ \prs{a}{\boldsymbol\tau}
	=0
	\,,\qquad\indx{a}=1,2,\hdots,N \,, \\
	\label{balance-final}
	&\operatorname{Div} \op P +  \massd \mathbf{b} = \massd \boldsymbol A^{\flat} \,,\qquad
	\op P^* \boldsymbol \nu \vert_{\partial_N\mathcal B} = \overline{\mathbf{t}}
	\,,\qquad
	\op P^* \boldsymbol \nu \vert_{\partial_D\mathcal B} = \mathbf{t}
	\,.
\end{align}
%---------------------
In the case of weakly layered dislocation distributions,~\eqref{kinetic-final} reads
%---------------------
\begin{equation}
	\prs{a}{\boldsymbol\pk} +
	\prs{a}{\lambda}_W \, \iota_{\prs{a}{\boldsymbol{\bv}} } \prs{a}{\df}
	+ \prs{a}{\lambda}_G \, \prs{a}{\boldsymbol\pi}
	+ \prs{a}{\boldsymbol\tau}
	=0
	\,,\qquad\indx{a}=1,2,\hdots,N \,.
\end{equation}
%---------------------
On should note that the kinetic boundary conditions~\eqref{EL-material-boundary} are trivially satisfied as in this simplified model we are not considering non-local effects.
The first Piola-Kirchhoff tensor $\op P$, the Mandel tensor $\op M$, the dissipation $1$-forms $\prs{a}{\boldsymbol\tau}$, and the Peach-Koehler force $\prs{a}{\boldsymbol\pk}$
defined in~\eqref{P-S},~\eqref{mixed-second-mandel},~\eqref{dissip-form}, and~\eqref{pk-force}, respectively,
and involved in the previous equations, are derived from an elastic free energy function $\mathscr{F}$ and a dissipation potential $\mathscr{D}$.
In components, one has
%---------------------
\begin{equation}
	P_a{}^A = 2 \massd \, g_{ab} F^b{}_B \frac{\partial \mathscr{F}}{\partial C_{AB} }
	\,,\quad
	M^A{}_B = 2 \massd \, G_{BH} \frac{\partial \mathscr{F}}{\partial C_{AH} }
	\,,\quad
	\prs{a}{\tau}_A = \massd \,  \frac{\partial \mathscr{D}}{\partial \prs{a}{\vel}^A }
	\,,\quad
	\prs{a}{\pk}_A =  M^K{}_H \bv^H \omega_{KA}
	\,.
\end{equation}
%---------------------

%---------------------
%---------------------
\begin{remark}
As was discussed in~\S\ref{Sec:Kinematics}, given some initial conditions, the material motions $\prs{a}{\chi}$ fully determine the internal variables via the evolution equations.
As a matter of fact, there are $9+9N$ evolution equations~\eqref{evolution-final} associated with the same number of unknown internal variables $\boldsymbol\vartheta^{\nu}$ and $\prs{a}{\dd}^{\nu}$.
In the case of uniform dislocation fields---those with uniform and constant Burgers director---a dislocation field $\{\prs{a}{\dd}^{\nu}\}$ is fully encoded in its dislocation $2$-form $\prs{a}{\df}$, and hence the number of the internal variables and the evolution equations is $9+3N$.
The $3+3N$ governing equations~\eqref{kinetic-final} and~\eqref{balance-final} were obtained from a variational principle associated with variations of the $3+3N$ unknowns $\varphi$ and $\prs{a}{\chi}$.
In addition to these, one has the $2N$ unknown Lagrange multipliers $\prs{a}{\lambda}_L $ and $\prs{a}{\lambda}_G$ that are associated with the $2N$ constraint equations~\eqref{constraints-final}$_2$ (whether strong or weak) and~\eqref{constraints-final}$_3$.
Hence, we have obtained $12+14N$ governing equations---$12+11N$ in the case of uniform dislocation fields---for the same number of unknowns.
\end{remark}
%---------------------
%---------------------

%%%%%%%%%%%%%%%%%%%%%%%%%%%%%%
%%%%%%%%%%%%%%%%%%%%%%%%%%%%%%
\section{Linearized theory}    \label{Sec:Linearization}

In the previous section a simplified nonlinear IBVP for dislocation dynamics was formulated.
The data for the problem consisted of the initial values for the internal variables (the initial lattice forms $\boldsymbol\vartheta^{\nu}(X,0)$ and the initial dislocation fields $\prs{a}{\dd}^{\nu}(X,0)$), the external loads (the body and contact forces $\mathbf{b}(x,t)$ and $\overline{\mathbf{t}}(x,t)$), and the displacement boundary condition $\overline{\varphi}$.
In this section we perturb these inputs and study the governing equations and their solutions in the linear approximation.
Similar to the notion of variations, a time-dependent tensor field $\boldsymbol A(X,t)$ on $\mathcal B$ is perturbed around $\boldsymbol A(X,t) \vert_o$ by considering one-parameter families $\boldsymbol A_{\epsilon}(X,t)$ such that $\boldsymbol A_{\epsilon=0}(X,t) = \boldsymbol A(X,t) \vert_o$, i.e., the zeroth-order term.
The $n$-th order term is defined as the $n$-th-order $\epsilon$-derivative evaluated at $\epsilon=0$.
We indicate the first-order term with $\delta \boldsymbol A$, the second-order term with $\delta^2 \boldsymbol A$, and so on.
Spatial quantities of the type $\boldsymbol f (X,t)= (\boldsymbol{\mathsf{f}} \circ \varphi_t )(X)$, are perturbed as $\boldsymbol f_{\epsilon} (X,t)=(\boldsymbol{\mathsf{f}} \circ \varphi_{t,\epsilon})(X)$.
Then, $\boldsymbol f(X,t)\vert_o =  (\boldsymbol{\mathsf{f}} \circ \varphi_t\vert_o )(X)$, while for the higher-order terms one must introduce a connection, as was discussed in~\S\ref{Rates} for the time derivatives. We simply use the ambient space connection $\nabla^{\boldsymbol g}$; the first-order term can be written as $\delta \boldsymbol f = \nabla^{\boldsymbol g}_{\delta\varphi} ( \boldsymbol{\mathsf{f}} \circ \varphi_{\epsilon}) \vert_{\epsilon=0}$.

Unlike what was done for the variations of the material dislocation motions, in the linearization process all the initial dislocation fields are perturbed at the same time. Moreover, since the lattice forms must satisfy~\eqref{incomp-eq-0}, they are perturbed together with the dislocation fields. Hence, all the initial internal variables are perturbed simultaneously, while the external loads are perturbed separately.

%-------------------------------------------------------------------------------------------------
%-------------------------------------------------------------------------------------------------
\subsection{Linearization of dislocation fields} \label{lin-disl}

First we focus on the perturbation of the dislocation fields.
Dislocation densities, and hence the overall incompatible content of the plastic deformation, are considered small compared to the inverse of a characteristic length $L_o$, i.e., $\prs{a}{\dens}\ll 1/L_o$.
Therefore, we perturb the initial dislocation fields $\prs{a}{\dd}^{\nu}_0$ around vanishing $2$-forms.
It should be noticed that in order for~\eqref{gnd-sum} to hold for $\epsilon=0$, one must perturb the lattice structure around a defect-free one.
However, in~\S\ref{lin-lat} we will introduce a slight modification of~\eqref{gnd-sum} that allows us to linearize around a distorted lattice while keeping the dislocation densities small.

We look at the zeroth-order evolution equations, i.e., at~\eqref{evolution-final} for $\epsilon=0$.
As the initial condition is $\prs{a}{\dd}^{\nu}_0\vert_o=0$, both $\partial_t \boldsymbol\vartheta^{\nu}\vert_{o}$ and $\partial_t\prs{a}{\dd}^{\nu}\vert_{o}$ vanish,
and hence $\prs{a}{\dd}^{\nu}\vert_{o}$ vanishes at all times (while $\boldsymbol\vartheta^{\nu}\vert_{o}$ remains constant).
As for the first-order evolution equations, since $\prs{a}{\dd}^{\nu}\vert_o=0$, one obtains
%---------------------
\begin{equation} \label{linear-evolution}
\begin{split}
	\delta \boldsymbol\vartheta^{\nu} (X,t) &= \delta \boldsymbol\vartheta^{\nu} (X,0)
	- \sum_{\indx{a}=1}^{N} \int_0^t
	\iota_{\prs{a}{\boldsymbol{\vel}}\vert_{o} (X,\overline t)} \delta \prs{a}{\dd}^{\nu} (X,\overline t)
	\,\mathrm d \overline t
	\,, \\
	\delta \prs{a}{\dd}^{\nu} (X,t) &= \delta \prs{a}{\dd}^{\nu} (X,0)
	- \int_0^t
	\mathrm d \iota_{\prs{a}{\boldsymbol{\vel}}\vert_{o} (X,\overline t)} \delta \prs{a}{\dd}^{\nu}(X,\overline t)
	\,\mathrm d \overline t
	\,,\qquad\indx{a}=1,2,\hdots,N \,.
\end{split}
\end{equation}
%---------------------
The last equation implies that the linearized dislocation fields $\delta\prs{a}{\dd}^{\nu}$ are convected by the zeroth-order dislocation velocities $\prs{a}{\boldsymbol{\vel}}\vert_{o}$.
We choose a decomposition $\dd^{\nu}=\bv^{\nu}\df$ where the only part that depends on $\epsilon$ is the $2$-form $\prs{a}{\df}$, with $\prs{a}{\df}\vert_o=0$.
Clearly one has $\delta\prs{a}{\dd}^{\nu}=\prs{a}{\bv}^{\nu} \,\delta\prs{a}{\df}$.
It should be emphasized that the Burgers director $\prs{a}{\boldsymbol{\bv}}=\prs{a}{\bv}^{\nu} \boldsymbol e_{\nu}$ depends on the lattice structure, and hence on $\epsilon$. Thus, one obtains $\prs{a}{\boldsymbol{\bv}}\vert_o=\prs{a}{\bv}^{\nu}\boldsymbol e _{\nu}\vert_o$, $\delta \prs{a}{\boldsymbol{\bv}}=\prs{a}{\bv}^{\nu} \delta\boldsymbol e _{\nu}$, and so on.
The linearization of~\eqref{gnd-sum-decomp-1}  reads
%---------------------
\begin{equation} \label{linearized-T-alpha}
	\mathrm d \delta \boldsymbol \vartheta^{\nu} = \sum_{\indx{a}=1}^{N} \prs{a}{\bv}^{\nu} \delta \prs{a}{\df}
	\,.
\end{equation}
%---------------------
Next we look at the consequences of the linearization of the dislocation fields on the kinetic equations~\eqref{kinetic-final}.
First one should note that $\prs{a}{ \df}\vert_o=0$ implies vanishing Peach-Koehler force $\prs{a}{\boldsymbol\pk}\vert_o$ and the $1$-forms $\prs{a\!b}{\boldsymbol\varpi}_0$ at $\epsilon=0$.
Hence, the the zeroth-order kinetic equation becomes
%---------------------
\begin{equation} \label{kin-zero}
	\prs{a}{\lambda}_G \vert_{o} \, \prs{a}{ \boldsymbol\pi} \vert_{o} \,+\, \prs{a}{\boldsymbol\tau} \vert_{o} \,=0 \,.
\end{equation}
%---------------------
As was discussed in~\S\ref{Dissipation}, $\prs{a}{\boldsymbol\tau}$ and $\prs{a}{ \boldsymbol\pi}$ are linearly independent.
Therefore, from~\eqref{kin-zero} one obtains $\prs{a}{\lambda}_G \vert_{o}=0$, and $\prs{a}{\boldsymbol\tau} \vert_{o}=0$.
Note that if one assumes a dissipation potential that is linear in the scalar dislocation densities (as in~\S\ref{Dissipation}), then $\prs{a}{\boldsymbol\tau} \vert_{o}=0$, which agrees with what was just obtained from the zeroth-order kinetic equations.
Instead, $\prs{a}{ \boldsymbol{\vel}} \vert_{o}$ and $\prs{a}{\lambda}_L \vert_{o}$ are unknowns and are to be determined using the rest of the first-order equations, which require more information and will be considered in the next subsection.

At $\epsilon=0$, the layer constraint equation is automatically satisfied as $\df\vert_o=0$.
The glide constraint equation gives
%---------------------
\begin{equation} \label{linear-constraint1}
	\langle \prs{a}{ \boldsymbol\pi} \vert_{o} , \prs{a}{ \boldsymbol{\vel}} \vert_{o} \rangle =0 \,,
\end{equation}
%---------------------
which means that the unknown dislocation velocities  $\prs{a}{ \boldsymbol{\vel}} \vert_{o}$ lie on the respective plane distributions.
At the first order, the layer constraint becomes
%---------------------
\begin{equation} \label{linear-constraint2}
	\boldsymbol\pi\vert_o \,\wedge\, \delta \df =0 \,,
\end{equation}
%---------------------
implying that the linearized dislocation fields are layered on the respective zeroth-order plane distributions, while we are not interested in the linearized glide constraint.
If one replaces the layer constraint with the weaker condition $\iota_{ \prs{a}{\boldsymbol{\vel}} } \iota_{ \prs{a}{\boldsymbol{\bv} }  }  \prs{a}{\df} = 0$, it is automatically satisfied at the zeroth order, while at the first order one has
%---------------------
\begin{equation} \label{linear-constraint3}
	\iota_{ \prs{a}{\boldsymbol{\vel}} \vert_o } \iota_{ \prs{a}{\boldsymbol{\bv} } \vert_o } \delta \prs{a}{\df} = 0 \,.
\end{equation}
%---------------------
By doing so it is possible to take into account a broader class of dislocation fields by allowing the dislocation curves to not be tangent to the glide planes.

With an abuse of notation, we define $\delta \prs{a}{\boldsymbol{\lv}}$ as the unit vector tangent to the dislocation curves associated with $\delta\prs{a}{\df}$, and $\delta \prs{a}{\boldsymbol{\norm}}$ as the unit vector that lies on $\prs{a}{\boldsymbol\pi} \vert_o$ and that is normal to the dislocation curves associated with $\delta\prs{a}{\df}$.
Note that one has $\delta \prs{a}{\dens} \, \delta \prs{a}{\boldsymbol{\lv}} = \star^{\sharp} \vert_o \delta \prs{a}{\df}$, where $\star^{\sharp} \vert_o$ is the raised Hodge operator induced by $\boldsymbol G\vert_{o}$.
From~\eqref{linear-constraint2}, similar to~\eqref{norm-dens}, one obtains $\iota_{\delta \prs{a}{\boldsymbol{\norm}}} \delta\prs{a}{\df} = \prs{a}{\dens} \,\prs{a}{\boldsymbol\pi} \vert_o$, and $\iota_{\delta \prs{a}{\boldsymbol{\norm}}} \, \iota_{\delta \prs{a}{\boldsymbol{\lv}}} \, \mvf\vert_o = \prs{a}{\boldsymbol\pi} \vert_o $.

%-------------------------------------------------------------------------------------------------
%-------------------------------------------------------------------------------------------------
\subsection{Linearization of the lattice structure}    \label{lin-lat}

First, we linearize the initial lattice structure $\boldsymbol\vartheta^{\nu}_0$ around a Cartesian coframe $\mathrm d Z^{\nu} = \mathrm d (z^{\nu} \circ \kappa)$, obtained by pulling back some Cartesian coordinates $z^{\nu}$ in the ambient space $\mathcal S$ via a reference map $\kappa:\mathcal B\to \mathcal S$,
as was shown in~\S\ref{Sec:Lattice}.\footnote{%
In the case in which we linearize around a compatible but non-Cartesian $\mathrm d Y^{\nu}$, one can simply reparametrize the material manifold to $(z^{-1}\circ Y )(\mathcal B) \subset \mathcal S$, with respect to which the coordinates $(Y^{\nu})$ are Cartesian.}
The dislocation fields are kept small as in~\S\ref{lin-disl}.
As was mentioned in~\S\ref{lin-disl}, the lattice structure at $\epsilon=0$ is time independent, and hence $\boldsymbol\vartheta^{\nu}\vert_{o} = \mathrm d Z^{\nu}$ at all times.
Therefore, one also has  $\boldsymbol T \vert_o=\boldsymbol 0$ and $\boldsymbol\alpha\vert_o=\boldsymbol 0$,
while the zeroth-order material metric is Euclidean and is given by $\boldsymbol G\vert_{o} = \kappa^* \boldsymbol g$.
As for plane distributions, one has $\prs{a}{\boldsymbol\pi}\vert_o = \prs{a}{\pi}_{\nu} \mathrm d Z^{\nu}$, and hence $\mathrm d \prs{a}{\boldsymbol\pi}\vert_o=0$.
This means that all plane distributions are integrable, i.e., $\mathfrak{I}_{\prs{a}{\Pi}}\vert_o = 0$ for all $\indx{a}$; in particular they define flat glide surfaces in $(\mathcal B, \boldsymbol G\vert_{o})$.
Considering the flat geometry induced by $\boldsymbol G\vert_{o}$, since from~\eqref{linear-evolution} and~\eqref{linear-constraint1} the first-order dislocation fields $\delta\prs{a}{\dd}^{\nu}$ glide on a flat surface given by $\prs{a}{ \boldsymbol\pi} \vert_{o} $, the classical dislocation kinematics is valid
\citep{sedlavcek2003importance,sedlavcek2007continuum}.
Linearizing~\eqref{gnd-sum-decomp-2} around the undistorted lattice structure, one has
%---------------------
\begin{equation}
	\delta \boldsymbol T = \sum_{\indx{a}=1}^N  \prs{a}{\boldsymbol{\bv}}\vert_o \,\otimes\, \delta\prs{a}{\df}
	\,,\qquad
	\delta \boldsymbol\alpha^{\nu} = \sum_{\indx{a}=1}^N \delta \prs{a}{\dens}\, \prs{a}{\bv}^{\nu} \, \delta\prs{a}{\boldsymbol{\lv}}
	\,,\qquad
	\delta \boldsymbol\alpha = \sum_{\indx{a}=1}^N  
	\delta \prs{a}{\dens}\, \prs{a}{\boldsymbol{\bv}}\vert_o \, \otimes \, \delta\prs{a}{\boldsymbol{\lv}}
	\,.
\end{equation}
%---------------------
We derive the analogue of Lemma~\ref{Lem:glide} in the linear approximation around a dislocation-free lattice,
which establishes the sufficiency of the zeroth-order glide constraint~\eqref{linear-constraint1} for the first-order layer constraint~\eqref{linear-constraint2}.
This also means that the glide of layered dislocation fields is always allowed.

%---------------------
%---------------------
\begin{lem}\label{lem:linear-int}
In the case of an integrable plane distribution, the glide condition~\eqref{linear-constraint1} implies the linearized layer condition~\eqref{linear-constraint2}.
\end{lem}
%---------------------
\begin{proof}
We apply Lemma~\ref{Lem:glide} to $\prs{a}{\df}$, $\prs{a}{ \boldsymbol\pi} \vert_{o}$, and $\prs{a}{ \boldsymbol{\vel}} \vert_{o}$.
Then, if $\langle \prs{a}{ \boldsymbol\pi} \vert_{o} , \prs{a}{ \boldsymbol{\vel}} \vert_{o} \rangle =0$, the first-order expansion of~\eqref{integr-bal} is a necessary and sufficient condition for $\prs{a}{ \boldsymbol\pi}\vert_o \,\wedge\,  \delta\prs{a}{\df} =0$.
Note that the zeroth-order expansion of~\eqref{sliprate} gives $\prs{a}{\gamma} \vert_o \,\prs{a}{ \boldsymbol\pi} \vert_{o} =0$, whence $\prs{a}{\gamma}\vert_o=0$.
Therefore, as $\partial_t \prs{a}{ \boldsymbol\pi} \vert_o=0$, and $\mathfrak{I}_{\prs{a}{\Pi}}\vert_o = 0$, the first-order expansion of~\eqref{integr-bal} is always satisfied.
Hence, assuming $\langle \prs{a}{ \boldsymbol\pi} \vert_{o} , \prs{a}{ \boldsymbol{\vel}} \vert_{o} \rangle =0$ one necessarily has $\prs{a}{ \boldsymbol\pi}\vert_o \wedge  \delta\prs{a}{\df} =0$.
\end{proof}
%---------------------
%---------------------

Next we consider distributions of low densities of dislocations that are layered on a lattice structure that is already dislocated.\footnote{%
See also \citep{SadikYavari2016}, in which $\boldsymbol G$ is perturbed around a non-Euclidean metric.}
Hence, we linearize the lattice structure around a $\boldsymbol\theta^{\nu}$ such that $\mathrm d \boldsymbol\theta^{\nu} \neq 0$.
As was mentioned earlier in this section, in order to allow this we need to introduce a slight modification to~\eqref{gnd-sum}, and assume the following:
%---------------------
\begin{equation} \label{sum-mod}
	\mathrm d \boldsymbol \vartheta^{\nu} \vert_o 
	= \mathrm d  \boldsymbol\theta^{\nu} + \sum_{\indx{a}=1}^{N} \prs{a}{\dd}^{\nu} \vert_o 
	= \mathrm d  \boldsymbol\theta^{\nu} \,,
\end{equation}
%---------------------
as all $ \prs{a}{\dd}^{\nu} \vert_o$ vanish.
By doing so, one is able to account for small dislocation densities distributed in a lattice with a large dislocation content.\footnote{The high dislocation content may be associated with dislocation fields on different slip systems, as well as with other defects such as disclinations, grain boundaries, etc.}
As for plane distributions, one has $\prs{a}{\boldsymbol\pi}\vert_o = \prs{a}{\pi}_{\nu} \boldsymbol\theta^{\nu}$, and therefore, since $\mathrm d  \boldsymbol\theta^{\nu} \neq 0$, integrability is not guaranteed. Hence, $\mathfrak I_{\prs{a}{\Pi}}\big\vert_o \neq 0$, in general.
It should be noted that under the modification~\eqref{sum-mod} the condition~\eqref{integr-bal6} does not hold anymore.
However, one can still rely on~\eqref{integr-bal} or~\eqref{integr-bal7}.
Therefore, one has the following analogue of Lemma~\ref{lem:linear-int}.

%---------------------
%---------------------
\begin{lem}\label{lem:linear-int2}
It is not possible to satisfy both the glide condition~\eqref{linear-constraint1} and the linearized layer condition~\eqref{linear-constraint2} on a non-integrable plane distribution $\Pi\vert_o$.
\end{lem}
%---------------------
\begin{proof}
The proof is similar to that of Lemma~\ref{lem:linear-int} as we apply Lemma~\ref{Lem:glide} to $\prs{a}{\df}$, $\prs{a}{ \boldsymbol\pi} \vert_{o}$, and $\prs{a}{ \boldsymbol{\vel}} \vert_{o}$.
In this case one still has $\prs{a}{\gamma}\vert_o=0$, and $\partial_t \prs{a}{ \boldsymbol\pi} \vert_o=0$, but the integrability object does not vanish.
Therefore,~\eqref{integr-bal} is simplified to read
%---------------------
\begin{equation}
	\boldsymbol{\vel}\vert_o^{\bot } \, \mathfrak I_{\prs{a}{\Pi}} \vert_o \, \delta\prs{a}{\boldsymbol{\norm}}^{\flat} =0
	\,,
\end{equation}
%---------------------
where $\delta\prs{a}{\boldsymbol{\norm}}^{\flat}$ was defined in~\S\ref{lin-disl}, and indices are lowered using $\boldsymbol G\vert_o$.
As $\mathfrak{I}_{\prs{a}{\Pi}}\vert_o \neq 0$, the right-hand side vanishes only for $\delta\prs{a}{\gamma}=0$, i.e., for vanishing dislocation velocities (modulo vector fields that are tangent to the dislocation lines, as was discussed in~\S\ref{Orowan}).
\end{proof}
%---------------------
%---------------------

The previous result can be explained by noticing that in the linear approximation dislocations are not dense enough to change a finitely distorted lattice in order to accommodate their motion.
In other words, linearized dislocations cannot glide on highly distorted lattices.
This also means that the lack of integrability of a plane distribution anchors the motion of low dislocation densities, and the possibility of glide on a non-integrable plane distribution discussed in~\S\ref{Glide} is a nonlinear effect.\footnote{It is straightforward to see that satisfying the layer condition $\langle \boldsymbol\pi\vert_o,\boldsymbol{\vel}\vert_o \rangle=0$ at all times does not imply the validity of the glide condition $\boldsymbol\pi\vert_o\,\wedge\,\delta\df=0$.}

%----------------
%----------------
\begin{table}[tp!]
%\setstretch{1.5}
\centering
\begin{tabular}{rlccccc}
\toprule
\multicolumn{2}{c}{}  & \multicolumn{2}{c}{Glide} & & \multicolumn{2}{c}{Reactions~~~~~~~~~~~~}\\
%\hline
\multicolumn{2}{c}{} & \multicolumn{1}{c}{Defect-free} & \multicolumn{1}{c}{Dislocated} & & \multicolumn{1}{c}{Defect-free} & \multicolumn{1}{c}{Dislocated} \\ 
\midrule 
\multirow{2}*{Constraint} &  $f_L~~~$   & In plane  & $\boldsymbol\times$  & & $\delta\prs{a}{\lambda}_G \, \prs{a}{\boldsymbol\pi} \vert_o$  & $\prs{a}{\lambda}_L\vert_o  \, \delta\prs{a}{\dens} \, \mathfrak I_{\prs{a}{\Pi}} \Big\vert_o \, \delta\prs{a}{\boldsymbol{\norm}}^{\flat}  \,+\, \delta\prs{a}{\lambda}_G \, \prs{a}{\boldsymbol\pi} \vert_o$  \\
                                          & $f_W$   &  In plane  & Out of plane  & &  $\delta\prs{a}{\lambda}_G \, \prs{a}{\boldsymbol\pi} \vert_o$  & $ \prs{a}{\lambda}_W \vert_o \,\, \iota_{ \prs{a}{\boldsymbol{\bv}} \vert_o  } \delta\prs{a}{\df} \,+\, \delta\prs{a}{\lambda}_G \, \prs{a}{\boldsymbol\pi} \vert_o $    \\
\bottomrule
\end{tabular}
%\setstretch{1}
\vskip 0in
\caption{For linearized dislocation densities, the defect content of the initial lattice structure can predict the occurrence of dislocation motion.
In the first row, layered dislocation fields are considered, while in the second row we assume the weaker constraint.
From Lemma~\ref{lem:linear-int} in the integrable case the glide constraint implies that initially layered dislocation fields stay layered. Hence, in the defect-free case the glide motion is on the glide surfaces.
From Lemma~\ref{lem:linear-int2}, the strong layer constraint obstructs the glide motion, while in the weaker case dislocations are allowed to leave the glide plane, as long as they move with a velocity that is parallel to the Burgers director.}
\label{Tab:linear}
\end{table}
%----------------
%----------------

Next we look at the first-order kinetic equations.
From~\eqref{pk-force} one obtains the linearized Peach-Koehler force as
%---------------------
\begin{equation} 
	\delta \prs{a}{\boldsymbol\pk} = \iota_{ \op M \vert_o \prs{a}{\boldsymbol{\bv}} \vert_o  } \delta\prs{a}{\df} \,,
\end{equation}
%---------------------
where the linearized Mandel stress can be calculated starting from the linearized first Piola-Kirchhoff stress as $\op M\vert_o=\op P \vert_o \, \op F \vert_o$, with $\op K=T\kappa=\op F\vert_o$ being the tangent map of the zeroth-order configuration mapping.
The tensor $\op P\vert_{o}$ appears in the zeroth-order balance of linear momentum, that reads
%---------------------
\begin{equation} \label{linear-balance}
	\operatorname{Div}\vert_{o} \op P\vert_{o} +  \massd\vert_{o} \,  \boldsymbol \beta = 0 \,,\qquad
	\op P\vert_{o}^{\star} \, \boldsymbol \nu \vert_{\partial_N\mathcal B} = \overline{\boldsymbol \tau}  
	\,,\qquad
	\delta\op P \vert_{o}^{\star} \, \boldsymbol \nu \vert_{\partial_D\mathcal B} =  \boldsymbol \tau    \,,
\end{equation}
%---------------------
with the constraint $\varphi\vert_{\partial_D\mathcal B} = \overline\varphi$.
It should be noticed that the linearization is defined with respect to a parameter related to the internal variables and not with respect to the stiffness of the crystal, so nothing can be said about the magnitude of stresses and strains, see~\S\ref{lin-loads}.
Moreover, from the proof of Lemma~\ref{lem:linear-int2} one can calculate the linearization of the second term in the kinetic equations~\eqref{kinetic-final}, i.e., of lattice reaction that depends on the integrability object associated with a plane distribution.
In particular, one has
 %---------------------
\begin{equation} \label{non-integrability-force-lin}
	\delta \left(  \prs{a}{\lambda}_L  \, \prs{a}{\dens} \, \mathfrak I_{\prs{a}{\Pi}} \, \prs{a}{\boldsymbol{\norm}}^{\flat}  \right) =
	\prs{a}{\lambda}_L\vert_o  \, \delta\prs{a}{\dens} \, \mathfrak I_{\prs{a}{\Pi}} \Big\vert_o \, 
	\delta\prs{a}{\boldsymbol{\norm}}^{\flat}
	\,.
\end{equation}
%---------------------
This term represents the reaction force that prevents the motion of dislocations, as prescribed by~\eqref{lem:linear-int2}.
It should be noticed that in the linearization around a defect-free lattice, the dislocation motion is allowed because of Lemma~\eqref{lem:linear-int}, and hence~\eqref{non-integrability-force-lin} vanishes as $ \mathfrak I_{\prs{a}{\Pi}} \big\vert_o=0$.
Since the $1$-forms $\prs{a\!b}{\boldsymbol\varpi}$ are quadratic in the dislocation densities, their linearizations vanish and so does the third term in the kinetic equations~\eqref{kinetic-final}.
Recalling $\prs{a}{\lambda}_G\vert_o=0$ from~\S\ref{lin-disl}, the linearized kinetic equations~\eqref{kinetic-final} can now be written as
%---------------------
\begin{equation} \label{lin-kin-stress}
	\delta\prs{a}{\boldsymbol\pk}
	+\prs{a}{\lambda}_L\vert_o  \, \delta\prs{a}{\dens} \, \mathfrak I_{\prs{a}{\Pi}} \Big\vert_o \, \delta\prs{a}{\boldsymbol{\norm}}^{\flat}
	+ \delta \prs{a}{\lambda}_G \, \prs{a}{\boldsymbol\pi} \vert_o
	+ \delta \prs{a}{\boldsymbol\tau} =0 \,.
\end{equation}
%---------------------
Eq.~\eqref{lin-kin-stress} is the equation of motion of the dislocation fields in the linearized theory.
The term $\delta\prs{a}{\boldsymbol\pk}$ is the driving force for the dislocation motion, whereas $\delta \prs{a}{\lambda}_G \, \prs{a}{\boldsymbol\pi} \vert_o$ is the force that the lattice exerts on the dislocations in order to keep them gliding and layered on the slip surface.
In short, when the zeroth-order lattice structure is dislocation-free, the motion is allowed and is driven by the linearized Peach-Koehler force.
The linearized dissipation force $\delta \prs{a}{\boldsymbol\tau} $ has the following simplified expression
%---------------------
\begin{equation}
	\delta \prs{a}{\boldsymbol\tau}  = - \massd  \,  \frac{\partial^2 \prs{a}{\mathscr{D}}}{\partial \prs{a}{\boldsymbol{\vel}} 
	\partial\prs{a}{\dens} } \,\delta \prs{a}{\dens} \,,
\end{equation}
%---------------------
that can be simplified further to read $- \massd \, \delta \prs{a}{\dens}\, \prs{a}{c}_o \, \prs{a}{\vel}^{\bot}  \delta \prs{a}{\boldsymbol{\norm}}^{\flat}$ when $\prs{a}{\mathscr{D}} = \frac{1}{2} \prs{a}{\dens}\, \prs{a}{c}_o  \, (\prs{a}{\vel}^{\bot})^2$ as in~\S\ref{Dissipation}.
It should be noted that by the effect of the Peierls stress barrier (that can be embedded in the dissipative force $\delta \prs{a}{\boldsymbol\tau} $ through a penalty function as was discussed in~\S\ref{Dissipation}) the glide motion is obstructed if the stresses---and hence the Peach-Koehler force---are not large enough, see Table~\ref{Tab:linear} and~\S\ref{lin-loads}.

\begin{figure}[tp!]
\centering
\vskip 0.1in
\includegraphics[width=.8\textwidth]{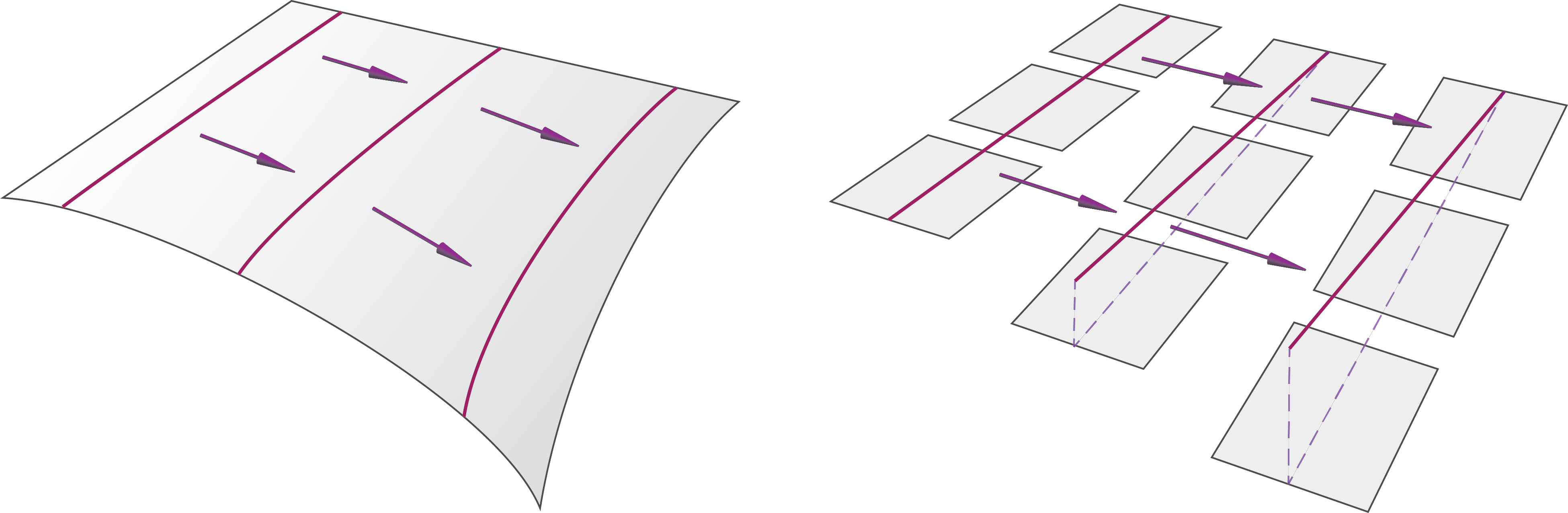}
\vskip 0.1in
\caption{Glide of layered dislocation fields. Left: When the slip plane distribution is integrable, the glide motion keeps the dislocations on the prescribed slip surface. Right: In the non-integrable case the glide motion causes the dislocations to leave the slip plane.}
\label{fig:glide-linear}
\end{figure}

For weakly layered dislocation fields obeying~\eqref{linear-constraint3}, Lemma~\ref{lem:linear-int2} does not hold, and dislocations are now allowed to glide on non-integrable plane distributions by leaving the slip plane.
In the integrable case, since the glide constraint~\eqref{linear-constraint1} implies that an initially layered dislocation field stays layered (Lemma~\ref{lem:linear-int}), the only way to go out of the glide plane is by the effect of the non-integrability of the plane distribution itself, as shown in Fig.~\ref{fig:glide-linear}.
Linearizing the modified constrained Euler-Lagrange equations~\eqref{EL-constraint-weak}, Eq.~\eqref{lin-kin-stress} must be modified to read
%---------------------
\begin{equation} \label{linear-kine-weak}
	\delta\prs{a}{\boldsymbol\pk}
	+ \prs{a}{\lambda}_W \vert_o \,\, \iota_{ \prs{a}{\boldsymbol{\bv}} \vert_o  } \delta\prs{a}{\df} 
	+ \delta \prs{a}{\lambda}_G \, \prs{a}{\boldsymbol\pi} \vert_o
	+ \delta \prs{a}{\boldsymbol\tau} =0 \,.
\end{equation}
%---------------------
It should be noticed that the reaction force $\prs{a}{\lambda}_W \vert_o \,\, \iota_{ \prs{a}{\boldsymbol{\bv}} \vert_o  } \delta\prs{a}{\df} $ vanishes for screw dislocations, as both $\prs{a}{\boldsymbol{\bv}} \vert_o $ and $\delta\prs{a}{\boldsymbol{\lv}}$ belong to $\prs{a}{\Pi}\vert_o$.
In the integrable case, one of the two reaction forces in~\eqref{linear-kine-weak} is redundant as they act perpendicular to the same plane.
Instead, when $\mathfrak I_{\prs{a}{\Pi}} \neq 0$ it acts in such a way to keep the dislocation velocity $\prs{a}{\boldsymbol{\vel}} \vert_o $ parallel to the Burgers director $\prs{a}{\boldsymbol{\bv}} \vert_o $, together with the reaction force $\delta \prs{a}{\lambda}_G \, \prs{a}{\boldsymbol\pi} \vert_o$, see also Table~\ref{Tab:linear}.

%---------------------
%---------------------
%---------------------
%---------------------

%------------------------------------------------------
%------------------------------------------------------
\subsection{Examples of linearized glide}

As was mentioned earlier, the linear approximation allows one to consider the motion of dislocations on a time-independent lattice structure.
We consider the kinematics of a linearized dislocation field $\delta\dd^{\nu}$ in a distorted lattice structure defined by the triplet $\{\boldsymbol\theta^{\nu}\}$, with $\mathrm d \boldsymbol\theta^{\nu}\neq 0$.
By the effect of the pre-existing dislocations, the slip plane distribution $\Pi\vert_o$ assigned to $\delta\dd^{\nu}$ is non-integrable.
Hence, glide can only occur by relaxing the strong layer condition, i.e., by allowing the dislocation curves to leave the slip plane while satisfying~\eqref{linear-constraint3}.
This implies that the dislocation velocity $\boldsymbol\vel\vert_o$ is in the direction of the Burgers director.
We also assume that the dislocation velocity is time independent---and hence induces an autonomous material flow---but is non-uniform.
Setting Cartesian coordinates $(X,Y,Z)$, we consider the cube $\mathcal{Q} =[0,L]^3$, with the goal of studying the effects of both the non-integrability of $\Pi\vert_o$  and the non-uniformity of $\boldsymbol\vel\vert_o$ on the induced linearized plastic slip $\delta\boldsymbol\vartheta^{\nu}$.
We consider an initial lattice structure given by
%---------------------
\begin{equation} \label{initial-lat-str}
	\boldsymbol\theta^1 =  \mathrm d X \,,\qquad
	\boldsymbol\theta^2 = \mathrm d Y \,,\qquad
	\boldsymbol\theta^3= K Y \,\mathrm d X + \mathrm d Z \,.
\end{equation}
%---------------------
The material metric induced by~\eqref{initial-lat-str} is non-Euclidean, with the associated material volume form $\mvf\vert_o = \mathrm d Z^1\wedge\mathrm d Y\wedge\mathrm d Z^3$.
The fact that the material volume form is the same as the Cartesian volume form $\mathrm d Z^1\wedge\mathrm d Y\wedge\mathrm d Z^3$ indicates that~\eqref{initial-lat-str} represents a simple shear.
Since $\mathrm d\boldsymbol\theta^3= - K \,\mathrm d X \wedge \mathrm d Y$, the initial lattice structure is associated with a dislocation content,
that can be achieved by the dislocation form $K \,\mathrm d X \wedge \mathrm d Y$ with Burgers director $-\boldsymbol e_3 = -\frac{\partial}{\partial Z}$.
From~\eqref{gnd-sum-decomp-2}$_3$, the initial dislocation content can also be expressed by 
$\boldsymbol\alpha \vert_o = - \boldsymbol e_3 \otimes \, K \frac{\partial}{\partial Z}  = -K \frac{\partial}{\partial Z} \otimes \frac{\partial}{\partial Z}$, implying that~\eqref{initial-lat-str} is equivalent to a field of uniformly distributed screw dislocations oriented along the $Z$-axis.

%---------------------
\begin{figure}[t]
\centering
\includegraphics[width=\textwidth]{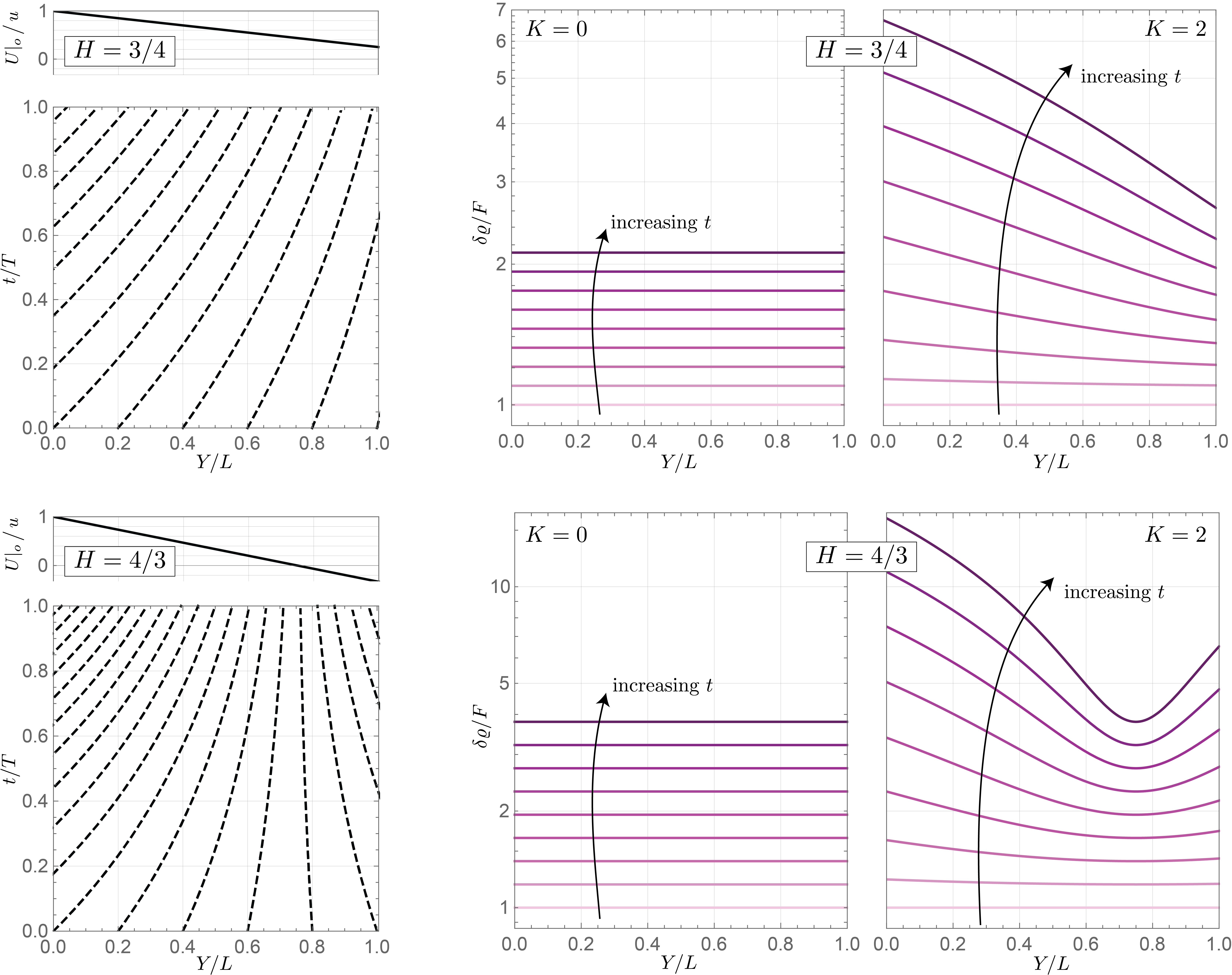}
\vskip 0.1in
\caption{Time-evolution of a field of edge dislocations over a distorted lattice structure.
Top: $H=3/4$, the dislocation velocity reverses outside the cube $\mathcal{Q} = [0,L]^3$ at $Y=4/3$.
Bottom: $H=4/3$, the dislocation velocity reverses inside the cube $\mathcal{Q} $ at $Y=3/4$.
Left: The solution to the equation of motion for the linearized dislocation fields is represented by points $(\chi(Y,t),t)$ for different initial positions $Y$.
Center: Scalar dislocation density along $Y$ at different times for $K=0$.
Right: Scalar dislocation density along $Y$ at different times for $K=2$.}
\label{fig:ex-nonint-1}
\end{figure}
%---------------------

Next we consider the slip plane distribution $\Pi\vert_o$ defined by $\boldsymbol\pi\vert_o = \boldsymbol\theta^3$, and a linearized dislocation field $\delta\dd^{\nu} = \bv^{\nu} \delta\df$ associated with it.
We assume uniform and constant $\bv^{\nu}=\{0,-1,0\}$, i.e., $\boldsymbol\bv\vert_o = -\boldsymbol e_2 = -\frac{\partial}{\partial Y}$.
Note that the plane distribution $\Pi\vert_o$ is non-integrable, as $\mathfrak I_{\Pi\vert_o} = \langle \boldsymbol\theta^3\otimes\boldsymbol\theta^3 , \boldsymbol\alpha \vert_o \rangle = -K$.
We look at the weakly-layered glide of $\delta\df$ on $\Pi\vert_o$, meaning that it occurs along the Burgers director.
By the effect of both the distorted lattice and the external loads the crystal is subject to a non-vanishing zeroth-order Mandel stress field $\op M \vert_o$, that we assume drives the motion of $\delta\dd^{\nu}$ along a time-independent dislocation velocity of the type $\boldsymbol\vel\vert_o=\vel\vert_o(Y) \frac{\partial}{\partial Y}$.
In particular, we take a linear function, viz.
%---------------------
\begin{equation} \label{ex-velocity1}
	\boldsymbol U \vert_o = u\, ( 1-HY ) \tfrac{\partial}{\partial Y} \,,
\end{equation}
%---------------------
where $u$ and $H$ are two constants.
More specifically, $1/H$ is the position on the $Y$-axis at which the dislocation velocity changes sign.
When $H<0$ or $1/H<L$ the velocity reverses outside $\mathcal{Q}$, while for $1/H<L$ that occurs inside $\mathcal{Q}$, see Fig.~\ref{fig:ex-nonint-1}.
With an abuse of notation we denote with $\chi$ the component representation of the material motion along $Y$.
Hence, for a fixed $Y\in[0,L]$, from~\eqref{ex-velocity1} the material motion satisfies the following ODE:
%---------------------
\begin{equation} \label{ex-ode}
	\frac{d \chi}{\mathrm dt}  = u\, ( 1- H \chi ) \,,
\end{equation}
%---------------------
with initial condition $\chi(Y,0)=0$. Eq.~\eqref{ex-ode} admits the following solution:
%---------------------
\begin{equation} \label{ex-velocity2}
	\chi (Y,t) =  \frac{e^{-uHt} }{H} \left( HY + e^{uHt}-1 \right) \,.
\end{equation}
%---------------------
Next we look at the evolution of the dislocation form $\delta\df$.
We take the initial condition $\delta\df (Y,0) = F  [ \mathrm d Y \wedge \mathrm d Z  - K Y\,\mathrm d X \wedge \mathrm d Y ]$, where the constant $F$ is a reciprocal length representing a characteristic dislocation density.
It is straightforward to prove that $\delta\df (Y,0)$ is strongly layered on $\Pi\vert_o$.
This means that we start with some distributed dislocations that are strongly layered and look at them leave the slip plane as a consequence of glide on a non-integrable slip distribution.
Since the initial lattice structure, the dislocation velocity, and the initial dislocation field depend only on $Y$, we consider solutions of the evolution equation~\eqref{linear-evolution} of the type $\delta\df = a(Y,t) \, \mathrm d Y \wedge \mathrm d Z + b(Y,t) \, \mathrm d X \wedge \mathrm d Y$.
Therefore, recalling the calculations given in Example~\ref{example-oneslip}, one obtains the following system of PDEs for $a(Y,t)$ and $b (Y,t)$:
%---------------------
\begin{equation} \label{ex-pde}
	\frac{\partial a}{\partial t}+ \frac{\partial (U\vert_o \,a)}{\partial Y} = 0	\,,\qquad
	\frac{\partial b}{\partial t} + \frac{\partial (U\vert_o \,b)}{\partial Y}=0 \,,
\end{equation}
%---------------------
where $U\vert_o(Y)$ has the expression given in~\eqref{ex-velocity1}.
Together with the initial condition,~\eqref{ex-pde} admits the following solution:
%---------------------
\begin{equation} \label{ex-disl-form}
	\delta\df (Y,t)=
	F e^{uHt} \left[
	\mathrm d Y \wedge \mathrm d Z -
	\frac{K}{H} e^{uHt} \left(1- e^{uHt} (1-HY)\right)
	\,\mathrm d X \wedge \mathrm d Y \right] \,.
\end{equation}
%---------------------
It is straightforward to show that $\llangle\delta\boldsymbol\lv , \boldsymbol\bv\vert_o \rrangle_{\boldsymbol G} = 0$, and hence~\eqref{ex-disl-form} represents a field of edge dislocations with respect to the metric $\boldsymbol G$.
Note that the left-hand side of~\eqref{linear-constraint2} reads
%---------------------
\begin{equation}
	\delta\df\wedge\boldsymbol\pi\vert_o = \frac{F K }{H} e^{uHt} (e^{uHt}-1 ) ( 1-HY )  \,\mvf\vert_o \,.
\end{equation}
%---------------------
Therefore, for $K=0$ (integrable $\Pi\vert_o$) the strong layer condition is always satisfied.
For $H\to 0$ (uniform $\vel\vert_o(Y)\equiv u$) one obtains $\delta\df\wedge\boldsymbol\pi\vert_o \to FKut  \,\mvf\vert_o$, showing that in the non-integrable case the dislocation curves leave the slip plane regardless of the uniformity of the dislocation velocity.
From~\eqref{ex-disl-form} it is possible to calculate the scalar dislocation density as $\delta\dens = \Vert\, \star^{\sharp}\vert_o \delta\df \, \Vert_{\boldsymbol G\vert_o}$, viz.
%---------------------
\begin{equation}
	\delta\dens (Y,t)=\frac{F}{H} e^{uH t}  \sqrt{  K^2  \left(e^{uH t}-1\right)^2  \left( 1 -HY \right)^2 + H^2} \,,
\end{equation}
%---------------------
which we show in Fig.~\ref{fig:ex-nonint-1}.
The fact that an initially-uniform scalar dislocation density becomes non-uniform is due to the combined effects of the non-uniform velocity and of the non-integrable slip plane distribution, as both $H$ and $K$ must be non-vanishing in order for $\delta\dens$ to be $Y$-dependent.

%---------------------
\begin{figure}[tp]
\centering
\includegraphics[width=\textwidth]{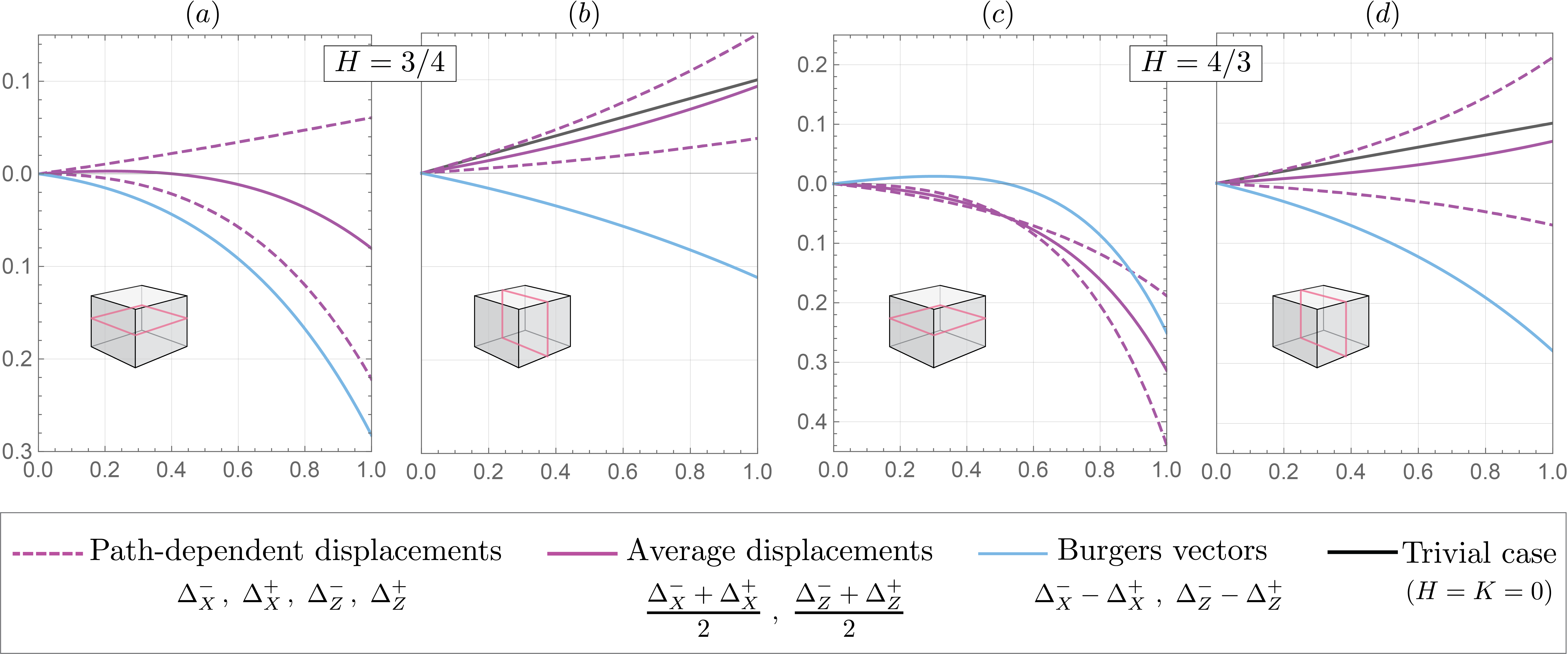}
\vskip 0.1in
\caption{Plastic displacement in the $2$-direction (the $Y$ direction), representing the plastic shear of the cube $\mathcal{Q}=[0,L]^3$.
Since $\delta\boldsymbol\vartheta^2$ is incompatible, these displacements are path dependent.
Each plot shows the displacements obtained by following two opposite sequences of paths, their average, and their difference (the Burgers vector associated with the loops obtained by joining opposite paths).
Displacements are normalized with respect to $L$ and expressed in percentage, e.g., $100 \, \Delta_X^-/L \,\%$.
In (a) and (b) $H=\frac{3}{4}$, while in (c) and (d) $H=\frac{4}{3}$.
In (a) and (c) the shear deformations of the type $Y$-$X$ are shown, while in (b) and (d) those of the type $Y$-$Z$ are shown.
In all the plots $K=2$. When $K=0$ all quantities in (a) and (c) vanish, while (b) and (d) are not affected by $K$.
}
\label{fig:ex-nonint-2}
\end{figure}
%---------------------

Lastly, we look at the linearized plastic slip induced by the motion of $\delta\df$.
The first-order Orowan's equation gives
%---------------------
\begin{equation}
	\partial_t\delta\boldsymbol\vartheta^2 =
	Fu e^{uHt} (1-HY)
	\left[  -\frac{ K }{H} \left( 1-e^{uHt}(1-HY ) \right)  \mathrm d X
	+    \mathrm d Z \right] \,,
\end{equation}
%---------------------
while $\partial_t\delta\boldsymbol\vartheta^1=\partial_t\delta\boldsymbol\vartheta^3=0$.
The rate of change of the lattice forms can be integrated with respect to time to give $\delta\boldsymbol\vartheta^{\nu} $, $\nu=1,2,3$.
These can be written in terms of the linearized plastic deformation gradient $\delta \subp{\op F}$, viz.
%---------------------
\begin{equation} \label{ex-def}
    \delta \subp{\op F} =
    \begin{bmatrix}
    	1&0&0 \\
	\frac{FK \left(e^{uHt}-1\right) (1- HY)  \left[2H- \left(e^{uHt}-1 \right) (1-HY) \right]}{2 H^2} & 1 
	& \frac{F (1-HY) \left(e^{uHt}-1\right)}{H} \\
    	0&0&1
    \end{bmatrix} \,,
\end{equation}
%---------------------
where the initial condition $\delta \subp{\op F} (t=0)=\op I$ was assumed.
Taking the exterior derivative of the second row of~\eqref{ex-def}, one obtains
%---------------------
\begin{equation}
    \mathrm d \delta \boldsymbol\vartheta^2 =
    F ( e^{uHt} -1)
    \left[
    - \mathrm d Y \wedge \mathrm d Z
     +\frac{K}{H} \left( HY - e^{uHt} (1-HY)  \right)  \mathrm d X \wedge \mathrm d Y
     \right]
    \,,
\end{equation}
%---------------------
from which it can be concluded that the plastic deformation is incompatible.
It should be noticed that the incompatibility of the linearized plastic deformation is due to the non-uniformity of the dislocation velocity rather than the non-integrability of the slip plane distribution, although it plays a role.
As a matter of fact, one has $\mathrm d \delta \boldsymbol\vartheta^2 \to 0$ for $H\to 0$, while for $K=0$ one has $\mathrm d \delta \boldsymbol\vartheta^2 = - F(e^{uHt}-1) \, \mathrm d Y \wedge \mathrm d Z$.
As a consequence of this incompatibility, there exists no unique plastic displacement associated with the plastic deformation~\eqref{ex-def}.
For this reason, we define the following quantities representing the shear deformations of the cube $\mathcal{Q}$:
%---------------------
\begin{equation}
\begin{aligned}
	&\Delta_X^- (t)= \int_0^L \delta  \boldsymbol\vartheta^2{}_X (0,t) \,\mathrm dX 
	+ \int_0^L \delta  \boldsymbol\vartheta^2{}_Y (Y,t) \,\mathrm dY \,, &
	\Delta_X^+ (t)= \int_0^L \delta  \boldsymbol\vartheta^2{}_Y (Y,t) \,\mathrm dY 
	+  \int_0^L \delta  \boldsymbol\vartheta^2{}_X (L,t) \,\mathrm dX  \,, \\
	&\Delta_Z^-(t) = \int_0^L \delta  \boldsymbol\vartheta^2{}_Z (0,t) \,\mathrm dX 
	+ \int_0^L \delta  \boldsymbol\vartheta^2{}_Y (Y,t) \,\mathrm dY \,, &
	\Delta_Z^+ (t)= \int_0^L \delta  \boldsymbol\vartheta^2{}_Y (Y,t) \,\mathrm dY 
	+  \int_0^L \delta  \boldsymbol\vartheta^2{}_Z (L,t) \,\mathrm dX  \,.
\end{aligned}
\end{equation}
%---------------------
$\Delta_X^- $ and $\Delta_X^+$ are displacements corresponding to shear deformations of the type $Y$-$X$, and obtained by moving in the plane $X$-$Y$ following opposite sequences, as shown in Fig.~\ref{fig:ex-nonint-2}.
Similarly, $\Delta_Z^- $ and $\Delta_Z^+$ correspond to shear deformations of the type $Y$-$Z$, obtained by moving in the plane $Y$-$Z$ following opposite sequences.
Moreover we define the differences
%---------------------
\begin{equation}
	\mathscr B_X = \Delta_X^+ - \Delta_X^-  \,,\qquad	\mathscr B_Z = \Delta_Z^- -\Delta_Z^+ \,,
\end{equation}
%---------------------
representing the Burgers vector associated with the loops obtained by joining opposite paths.
Fig.~\ref{fig:ex-nonint-2} shows the dependence of all these quantities on time, $K$ and $H$.

%-------------------------------------------------------------------------------------
%-------------------------------------------------------------------------------------
\subsection{Linearization of the external loads}  \label{lin-loads}

Next we consider perturbations of the external loads and boundary conditions.
It should be emphasized that, unlike what was done for the internal variables in~\S\ref{lin-disl} and~\S\ref{lin-lat}, this perturbation parameter is different from the one that was introduced for the internal variables.
As a matter of fact, external loads are considered small compared to a measure of stiffness, such as any elastic modulus $E_o$, regardless of the initial internal state.
In particular, we are assuming that the tractions are small compared to $E_o$, while the body forces are small compared to $E_o/L_o$, where $L_o$ is a characteristic length.
As for the displacement boundary condition, one assumes that $\overline\varphi$ is close to $\kappa\vert_{\partial_D\mathcal B}$ (the reference configuration restricted to the displacement boundary) if compared to $L_o$.
These assumptions imply small stresses compared to $E_0$, and hence small elastic deformations.
However, displacements might be large by the effect of stress-free plastic deformations, i.e., those inducing a Euclidean material metric, see Remark~\ref{rem:contorted-aeolotropy}.
It should be noticed that since the perturbations of the internal and external variables are independent, the linearized kinetic and evolution equations can be coupled with the balance of linear momentum considered at both the zeroth and the first order, depending on the ratio between the Peierls stress and $E_o$ \citep{kamimura2018peierls}.
Recalling the definition of $r_o(\pi_{\nu},\bv^{\nu})$ given in~\S\ref{Dissipation} to account for the Peierls barrier through the dissipation potential, if a crystal has $r_o \ll  1$, then the motion of dislocations is unlocked while the solid is still in the linear elastic regime (small displacements if one assumes zero initial plastic deformation).
Conversely, if $r_o \sim 1$ the motion of dislocations is unlocked after the solid has entered the nonlinear elastic regime.

We denote the load parameter with $\widetilde\epsilon$, while the zeroth and first-order expansions with respect to it are denoted with $\vert_{\widetilde o} $ and $\widetilde\delta$.
We linearize the displacement boundary condition around $\overline\varphi\vert_{\tilde o} = \kappa\vert_{\partial_D\mathcal B}$, while the external forces are linearized around vanishing $\mathbf{b}\vert_{\widetilde o}$ and $\overline{\mathbf{t}}\vert_{\widetilde o}$ at all times $t$.
It should be noticed that keeping the non-Euclidean part of the plastic deformations small is a necessary condition for small elastic deformations.
However, as was mentioned earlier, one can allow large plastic deformations that induce a Euclidean material metric.
Under these assumptions, the configuration at order zero is simply the reference configuration $\kappa$ at all times, which implies that $\op P\vert_{\widetilde o}\,=\boldsymbol{0}$ and $\op M\vert_{\widetilde o}\,=\boldsymbol{0}$, whence the validity of the zeroth-order balance of linear momentum.
The first-order balance of linear momentum reads
%---------------------
\begin{equation} \label{linear-balance}
	\operatorname{Div}\vert_{\widetilde o}  \,\widetilde\delta \op P +  \massd\vert_{\widetilde o} \, 
	\widetilde\delta \mathbf{b} 
	= \massd\vert_{\widetilde o} \, \delta \ddot{\varphi} \,,\qquad
	\widetilde\delta\op P^* \boldsymbol \nu \vert_{\partial_N\mathcal B} 
	= \widetilde\delta \,\overline{\mathbf{t}}
	\,,\qquad
	\widetilde\delta\op P^* \boldsymbol \nu \vert_{\partial_D\mathcal B} =\widetilde \delta\, \mathbf{t} \,,
\end{equation}
%---------------------
constrained by the displacement boundary condition $\widetilde\delta\varphi\vert_{\partial_D\mathcal B}=\boldsymbol 0$.
Recalling the relation $ \op M=\op P \op F$ given in~\S\ref{Spatial}, the linearized Mandel stress can be obtained from the linearized first Piola-Kirchhoff stress as $\widetilde\delta \op M=\widetilde\delta\op P \, \op K$, with $\op K=T\kappa=\op F \vert_{\widetilde o}$ being the tangent map of the reference configuration.
Since the initial internal variables and the external loads are perturbed by different parameters,
one can $\widetilde\epsilon$-linearize the $\epsilon$-linearized equations (and vice versa) without changing their structure (although, technically, one obtains second-order equations).
In particular, the two-time linearized Peach-Koehler force depends on $\widetilde\delta \op M \vert_o$, i.e., the Mandel stress of order zero with respect to $\epsilon$ and of order one with respect to $\widetilde\epsilon$.

%%%%%%%%%%%%%%
%%%%%%%%%%%%%%
\section{Conclusions} \label{Sec:Conclusions}

In this paper we formulated a geometric theory of dislocation mechanics and finite plasticity in single crystals.
This may be considered a mesoscale theory, in which dislocations are single-valued smooth fields inducing the defect content of a lattice in a deterministic way.
In particular, dislocation fields were defined as $N$ triplets of differential $2$-forms $\prs{a}{\dd}^{\nu}$ related to the lattice $1$-forms $\boldsymbol \vartheta^{\nu}$ through~\eqref{gnd-sum}.
Both the lattice and the dislocation forms constitute the internal variables of the theory.
Inside the lattice, dislocation fields are only allowed to move and migrate through the boundary of the crystal, which is the only source and sink of dislocations.

Decomposable dislocation fields were defined as those that can be written as the product of a Burgers director and a $2$-form that carries the information about the dislocation curves.
Each decomposable dislocation field is assumed to be convected by a material dislocation motion in the crystal.
We saw that this induces a class of evolution equations for the lattice forms.
An example is Orowan's equation~\eqref{orowan}, which holds under the assumption of closed decomposable dislocation fields.
This implies that at any time the $3+3N$ internal variables $\prs{a}{\dd}^{\nu}$ and $\boldsymbol\vartheta^{\nu}$ are completely determined by the history of the $N$ material motions $\prs{a}{\chi}$ (or of the material velocities $\prs{a}{\boldsymbol{\vel} }$) with assigned initial conditions.
Hence, the configuration mapping $\varphi$ and the material motions  $\prs{a}{\chi}$ (or the material velocities $\prs{a}{\boldsymbol{\vel} }$) constitute the kinematical degrees of freedom of the system.
Starting from a variational principle of the Lagrange-d'Alembert type based on a two-potential approach, we derived the kinetic equations~\eqref{EL-constrained2} as the Euler-Lagrange equations associated with variations of the material motions.
In the setting of a simplified model, an expression for the Peach-Koehler force in the nonlinear setting was derived by assuming a hyperelastic free energy, and a penalty approach was proposed to include the effect of the Peierls stress in the dissipation potential.

We defined layered decomposable dislocation fields as those whose Burgers and line directors locally lie on a plane distribution.
Note that a plane distribution $\prs{a}{\Pi}$ may not be integrable, i.e., surfaces that are tangent to it may not exist.
We showed that this property is encoded in what we called the integrability object $\mathfrak I_{\prs{a}{\Pi}}$, which carries information on the total dislocation content.
Lemma~\ref{Lemma:int} states that the integrability of a plane distribution is controlled only by those decomposable dislocation fields whose Burgers director densities do not belong to the plane distribution.
Moreover, we saw that each decomposable dislocation field participates in the integrability of a plane distribution proportionally to the normal component of the Burgers director with respect to the plane distribution.
Hence, layered decomposable dislocation fields do not affect the integrability of the plane distribution they are layered on.
Then, we assumed that layered decomposable dislocation fields can only glide on the plane distribution they are layered on.
Using the geometric framework, we studied how the lack of integrability of a plane distribution affects the glide motion.
According to Lemma~\ref{Lem:glide}, the glide condition does not imply that a dislocation field remains layered, as the condition~\eqref{integr-bal} is also needed. This equation involves the integrability object $\mathfrak I_{\prs{a}{\Pi}}$, and prescribes how the lattice structure must evolve in order to accommodate the glide of dislocations on a non-integrable slip plane distribution.
In order to force the decomposable dislocation fields to be layered and to glide on a prescribed plane distribution 
we introduced some internal lattice constraints.
The Lagrange multipliers associated with these constraints represent the lattice forces that keep the gliding dislocations layered on the respective plane distribution.
The reason that constraints are necessary, and it is not possible to formulate the dynamics of each layered decomposable dislocation field directly on its glide surface, is due to the fact that in the nonlinear setting the lattice structure evolves in time.
Moreover, as a consequence of Lemma~\ref{Lem:glide}, in the nonlinear setting the layer and the glide conditions are distinct.
The introduction of these constraints generates new terms in the kinetic equation~\eqref{EL-constrained2}, and they represent the reaction forces exerted by the lattice on dislocation fields.
In particular, for the $\indx{a}$-th equation we obtained two terms that come from the layer condition and depend on the integrability of the $\indx{a}$-th plane distribution, as well as on the other slip systems (hence representing a non-Schmid effect), plus a reaction force normal to the $\indx{a}$-th plane distribution coming from the glide condition.
We also considered a weak version of the layer constraint, and obtained the corresponding kinetic equations~\eqref{EL-constraint-weak}. By doing so, we allowed the dislocation fields to leave the prescribed slip plane as long as the motion remains conservative.

We derived a linear theory under the assumption of small dislocation densities.
We perturbed the initial lattice structure and studied its influence on the governing equations in the linearized setting.
In particular, Lemma~\ref{lem:linear-int} shows that if the slip plane distribution is integrable, then the glide condition implies the layer condition.
This means that the glide motion is always allowed (at least for large enough values of the Peach-Koehler force), and that the reaction forces associated with the lattice constraint do not appear in the linearized kinetic equations~\eqref{lin-kin-stress}.
When the slip plane distribution is non-integrable, according to Lemma~\ref{lem:linear-int2} the layer and the glide conditions cannot be satisfied simultaneously.
Therefore, glide is only allowed in the case of weakly layered dislocation fields, for which we derived the linearized kinetic equations~\eqref{linear-kine-weak}.
The derivation of the linearized theory also allowed us to identify the purely nonlinear effects, such as the interaction between different slip planes by the effect of the layer constraint, which is a second-order force.

The following are a few directions for future research. Exploring the existence of solutions, their regularity and stability will be crucial for numerical implementations of this geometric formulation. 
Another extension of the present theory is to consider the potential effects of distributed sources of dislocations such as Frank-Read sources, and the possibility of changes in the topology of the dislocation curves. One should also extend the model to include different types of dislocation interactions, such as annihilation and cross slip, as well as interactions with other defects, such as vacancies and grain boundaries. It would also be interesting to consider the thermal effects and their influence on climb.
Another potential extension of the present theory is to develop a geometric formulation for statistical dislocation dynamics in the nonlinear setting.
The present work is a good starting point for formulating such a theory, although it would require a change of perspective; instead of considering the superposition of $N$ dislocation fields, one would need to look at random collections of dislocations---potentially infinitely many---with an assigned probability distribution.

%---------------------------
%------------------------------
\section*{Acknowledgement}

We benefited from discussions with David L. McDowell. 
This work was partially supported by NSF -- Grant No. CMMI 1561578, 1939901, and  ARO Grant No. W911NF-18-1-0003.

%-----------------------------------------
%-----------------------------------------
\setlength{\bibsep}{6pt plus 0.3ex}
\bibliographystyle{plainnat}
\bibliography{Refs}

\appendix

%-------------------------------------------------------------
%-------------------------------------------------------------
\section{Differential Forms} \label{App:Differential}

The following facts about differential forms can be found in \citep{spivak1970comprehensive} and \citep{marsden1983mathematical}.
We consider an $n$-dimensional manifold $\mathcal B$.
The interior product (or contraction) between a differential $k$-form $\boldsymbol\omega$ and a vector field $\boldsymbol V$ is denoted $\iota_{\boldsymbol V} \boldsymbol\omega$,
and the exterior derivative of a differential $k$-form $\boldsymbol\omega$ is a $(k+1)$ form that is denoted $\mathrm d\boldsymbol\omega$.
In the case of a $1$-form $\boldsymbol\gamma$, the following identities hold for all vector fields $\boldsymbol V$ and $\boldsymbol W$:
%---------------------
\begin{equation}\label{der-1-form}
	\iota_{\boldsymbol V} \boldsymbol\gamma =\langle \boldsymbol\gamma,  \boldsymbol V \rangle
	\,,\qquad
	\mathrm d \boldsymbol\gamma (\boldsymbol V, \boldsymbol W) =
	\langle \mathrm d \langle \boldsymbol\gamma , \boldsymbol W \rangle    , \boldsymbol V \rangle
	-\langle \mathrm d \langle \boldsymbol\gamma, \boldsymbol V \rangle   , \boldsymbol W \rangle
	-\langle \boldsymbol\gamma , [\boldsymbol V, \boldsymbol W]\rangle \,,
\end{equation}
%---------------------
where the Lie bracket is defined as $[\boldsymbol V, \boldsymbol W]=\boldsymbol V \boldsymbol W-\boldsymbol W \boldsymbol V$.

Both the interior product and the exterior derivative satisfy the following properties for all differential forms $\boldsymbol{\omega}_1$ and $\boldsymbol{\omega}_2$, and vector fields $\boldsymbol V$:
%---------------------
\begin{align}
	\label{contr}
	\iota_{\boldsymbol V} (\boldsymbol{\omega}_1 \wedge \boldsymbol{\omega}_2) &=
	\iota_{\boldsymbol V} \boldsymbol{\omega}_1 \wedge \boldsymbol{\omega}_2 +
	(-1)^{\operatorname{deg}\boldsymbol{\omega}_1} \,\boldsymbol{\omega}_1 \wedge 
	\iota_{\boldsymbol V} \boldsymbol{\omega}_2  \,,\\
	\label{ext-der}
	\mathrm d (\boldsymbol{\omega}_1 \wedge \boldsymbol{\omega}_2) &=
	\mathrm d \boldsymbol{\omega}_1 \wedge \boldsymbol{\omega}_2 +
	(-1)^{\operatorname{deg}\boldsymbol{\omega}_1} \, \boldsymbol{\omega}_1 \wedge 
	\mathrm d \boldsymbol{\omega}_2 \,,
\end{align}
%---------------------
where $\operatorname{deg}$ gives the degree of a form.
In particular, given a scalar field $f$ and a differential form $\boldsymbol{\omega}$, one has $\mathrm d (f\boldsymbol{\omega})= \mathrm df \wedge \boldsymbol{\omega} + (-1)^{\operatorname{deg}\boldsymbol{\omega}}  f \mathrm d \boldsymbol{\omega}$.
Given a vector field $\boldsymbol V$, Cartan's formula relates the Lie derivative $\mathfrak L_{\boldsymbol V}$, the interior product $\iota_{\boldsymbol V}$, and the exterior derivative $\mathrm d$ of a differential form $\boldsymbol\omega$ as
%---------------------
\begin{equation}\label{Cartan}
	\mathfrak L_{\boldsymbol V} \boldsymbol\omega =
	\mathrm d\iota_{\boldsymbol V} \boldsymbol\omega +
	\iota_{\boldsymbol V} \mathrm d \boldsymbol\omega \,.
\end{equation}
%---------------------
%
%
Given a metric tensor $\boldsymbol G$ on $\mathcal B$, the Hodge operator assigns to a $k$-form $\boldsymbol\omega$ the $(n-k)$-form $\star\boldsymbol\omega$ such that
%---------------------
\begin{equation} \label{hodge-star}
	(\star \boldsymbol \omega)( \boldsymbol V_1, \hdots,  \boldsymbol V_k) =
	\boldsymbol \omega ( \boldsymbol V_{k+1}, \hdots,  \boldsymbol V_n) \,,
\end{equation}
%---------------------
for any $\boldsymbol G$-orthonormal frame $\{\boldsymbol V_1,\hdots,\boldsymbol V_n\}$.
Note that $\star\star\df=(-1)^{k(n-k)}\df$.
In particular, for the volume $n$-form $\mvf$ associated with the metric $\boldsymbol G$ one has $\star \mvf=1$.

The raised Hodge operator is defined by raising all the indices of the Hodge star operator, i.e., $\star^{\sharp} \boldsymbol\omega= (\star \boldsymbol\omega)^{\sharp}$.
The result is an alternating contravariant tensor.
For any $k$-form $\boldsymbol{\omega}_1$ and any $(n-k)$-form $\boldsymbol{\omega}_2$ the following identities hold:
%---------------------
\begin{equation} \label{wedge-hodge}
	\boldsymbol{\omega}_1 \wedge \boldsymbol{\omega}_2=
	\langle \boldsymbol{\omega}_1  , \star^{\sharp}  \boldsymbol{\omega}_2 \rangle \,\mvf=
	\langle \boldsymbol{\omega}_2 , \star^{\sharp}  \boldsymbol{\omega}_1 \rangle \,\mvf \,,
\end{equation}
%---------------------
where the pairing $\langle,\rangle$ of forms with alternating multivectors is defined by tensor contraction of increasing ordered indices.\footnote{Instead of summing on only those permutations with increasingly-ordered indices, one can take into account all the index permutations, so that the pairing form-multivector becomes a simple contraction of tensors.
In this case, the factors $\frac{1}{(\operatorname{deg}\boldsymbol{\omega}_1) !}=\frac{1}{k!}$ and $\frac{1}{ (\operatorname{deg}\boldsymbol{\omega}_2) !}=\frac{1}{(n-k)!}$ show up in the second and third terms of~\eqref{wedge-hodge}, respectively.}
In particular, the raised Hodge operator of an $(n-1)$-form $\boldsymbol{\omega}$ gives a vector $\star^{\sharp}  \boldsymbol{\omega}$ such that
%---------------------
\begin{equation} \label{hodge-n-1}
	\boldsymbol{\omega} = \iota_{\star^{\sharp}  \boldsymbol{\omega}} \, \mvf \,,
\end{equation}
%---------------------
where $\mvf$ is the volume form associated to the metric tensor $\boldsymbol G$.
By virtue of~\eqref{wedge-hodge}, the exterior product between a $1$-form $\boldsymbol\gamma$ and a $2$-form $\boldsymbol\omega$ is the contraction $\langle \boldsymbol\gamma,\star^{\sharp} \boldsymbol\omega \rangle$ (in vector calculus this becomes a scalar product).
Also, from~\eqref{hodge-n-1} the interior product between a vector $\boldsymbol V$ and a $2$-form $\boldsymbol\omega$ works as a cross product, i.e., $\iota_{\boldsymbol V} \boldsymbol\omega = \mvf( \star^{\sharp} \boldsymbol\omega , \boldsymbol V)$.

A $k$-form $\boldsymbol{\omega}$ on $\mathcal B$ is closed if $\mathrm d \boldsymbol{\omega}= 0$, and is exact if there exists a $(k-1)$-form $\boldsymbol\chi$ such that $\boldsymbol{\omega}=\mathrm d \boldsymbol\chi$.
An exact $k$-form is necessarily closed, while the converse holds only when the $k$-th de Rham cohomology group is trivial.
Since closedness can be seen as the local version of exactness, holonomicity becomes quite clear: the existence of local coordinates $( Y^{\nu} )$ such that $\boldsymbol \vartheta^{\nu}=\mathrm dY^{\nu} $ is guaranteed whenever the lattice forms are closed.
It should be noticed from~\eqref{der-1-form} that the exterior derivative of a differential $1$-form is written similarly to a curl in Cartesian coordinates, i.e., $(\mathrm d\gamma)_{AB} = \gamma_{B,A}-\gamma_{A,B}$.
More precisely, in the geometric setting the curl of a vector field $\boldsymbol V$ is defined as $\operatorname{Curl} \boldsymbol V =\star^{\sharp} \mathrm d (\boldsymbol V^{\flat})$.
For this reason, a closed (exact) differential $1$-form is the analogue of an irrotational (conservative) vector field, while an exact $2$-form can be thought as the curl of a vector potential field.

Given a volume form $\mvf$, the divergence of a vector field $\boldsymbol V$ is defined as
$(\operatorname{Div} \boldsymbol V) \mvf=\mathfrak L_{\boldsymbol V}\mvf$, and by Cartan's formula $(\operatorname{Div} \boldsymbol V) \mvf= \mathrm d \iota_{ \boldsymbol V} \mvf$.
From~\eqref{hodge-n-1} the exterior derivative of a $2$-form $\boldsymbol \omega$ can be expressed using the Hodge operator as:\footnote{If the connection is induced from the same metric that induces $\mvf$, the component form of this identity is $(\mathrm d \omega)_{ABC}= \nabla_I\left(\mu^{IJK}\omega_{JK}\right) \mu_{ABC}$.}
%---------------------
\begin{equation} \label{ext-der-hodge}
	\mathrm d \boldsymbol \omega = \operatorname{Div} (\star^{\sharp} \boldsymbol\omega ) \, \mvf \,.
\end{equation}
%---------------------
A vector field $\boldsymbol V$ is said to be solenoidal if $\operatorname{Div} \boldsymbol V=0$.
Note that if $\mvf$ is induced from $\boldsymbol G$ and $\nabla$ is the Levi-Civita connection associated with $\boldsymbol G$, then $\operatorname{Div} \boldsymbol V = \operatorname{tr} \nabla \boldsymbol V$.
Furthermore, from~\eqref{ext-der-hodge}, we conclude that the exterior derivative of a differential $2$-form is analogous to the divergence of its axial vector. This implies that a closed $2$-form can be thought of as a solenoidal vector field.

Let $\incl :\mathcal N \to \mathcal B$ be a surface in $(\mathcal B,\boldsymbol G)$.
Let us denote with $\boldsymbol \nu$ the associated $\boldsymbol G$-normal $1$-form, i.e.,
a field $\mathcal N \to T^*\mathcal B $ such that i) $\langle \boldsymbol \nu, \incl_* \boldsymbol Y \rangle=0$ for any $\boldsymbol Y\in T\mathcal N$ , and ii) $\Vert\boldsymbol \nu^{\sharp}\Vert =1$.
Then, $\star\boldsymbol \nu$ is a $2$-form.
We set $\boldsymbol\varsigma=\incl^*\star\boldsymbol \nu$, which can be proved to be the area form on $\mathcal N$ associated with the metric $\incl^* \boldsymbol G$ inherited from $\mathcal B$.
Then, the inclusion map satisfies the following identity
%---------------------
\begin{equation} \label{pullback-forms}
	\incl^*( \iota_{\boldsymbol V}\mvf )=
	\langle \boldsymbol\nu , \boldsymbol V\rangle \,\boldsymbol\varsigma
	\,,
\end{equation}
%---------------------
for all $\boldsymbol V\in T\mathcal B$.
For a $2$-form $\boldsymbol{\omega}$ one has $\incl^*\boldsymbol{\omega} = \incl^* (\iota_{\star^{\sharp}\boldsymbol{\omega}} \mvf) = \langle \boldsymbol\nu,  \star^{\sharp}\boldsymbol{\omega}  \rangle \, \boldsymbol\varsigma$,
i.e., the inclusion pullback filters out $2$-forms whose raised Hodge transformations are tangent to the surface.
In the geometric framework, integrals on an $n$-manifold are associated with $n$-forms, hence volume integrals require a $3$-form, while surface integrals such as boundary integrals are associated with $2$-forms.
Stokes' theorem and its metric-dependent variant using \eqref{pullback-forms}---the divergence theorem---are, respectively, written as
%---------------------
\begin{equation} \label{stokes}
	\int_{\mathcal B} \mathrm d \boldsymbol\omega = \int_{\partial\mathcal B} \incl^* \boldsymbol\omega
	\,,\qquad
	\int_{\mathcal B} (\operatorname{Div} \boldsymbol V) \, \mvf =
	\int_{\partial\mathcal B} \langle \boldsymbol\nu , \boldsymbol V \rangle \,\boldsymbol\varsigma
	\,.
\end{equation}
%---------------------

%-------------------------------
%-------------------------------
\section{Distributions} \label{App:Frobenius}

Given a differentiable manifold $\mathcal B$, a smooth collection $\Pi^{(k)}$ of $k$-dimensional subspaces $\Pi^{(k)}_X$ of $T_X\mathcal B$, $X\in\mathcal B$, is called a $k$-dimensional distribution \citep{spivak1970comprehensive, bryant2013exterior}.
A $k$-dimensional distribution is integrable if there exists a family of $k$-dimensional submanifolds (constituting a foliation) that are tangent to the distribution, i.e., given an integral submanifold $i: \mathcal S \hookrightarrow \mathcal B$, one has $i_*(T_X\mathcal S) = \Pi^{(k)}_X$.
Note that the integrability of a distribution is a local property.
A $k$-dimensional distribution is involutive if for any two vector fields $\boldsymbol V,\boldsymbol W \in \Pi^{(k)}$ one has $[\boldsymbol V,\boldsymbol W]\in\Pi^{(k)}$.
The Frobenius theorem establishes the equivalence between integrability and involutivity (or simply the sufficiency of involutivity as necessity is trivial).

The Frobenius theorem can also be formulated in terms of differential forms.
We denote with $\mathscr I \Pi^{(k)}$ the set of forms that vanish when they are paired with vectors in $\Pi^{(k)}$, constituting an ideal.
Then, $\Pi^{(k)}$ is integrable if and only if $\mathscr I \Pi^{(k)}$ is closed under exterior differentiation.
Consider the $n-k$ $1$-forms $\{\boldsymbol\gamma^J\}_{J=1,2,\hdots,N-k}$ that generate $\mathscr I\Pi^{(k)}$.
The ideal $\mathscr I\Pi^{(k)}$ is closed under exterior differentiation if and only if there exist $(n-k)^2$ $1$-forms $\boldsymbol\beta^J{}_K$ (not necessarily in $\mathscr I\Pi^{(k)}$) such that
%---------------------
\begin{equation}
	\mathrm d \boldsymbol\gamma^J = \sum_K\boldsymbol\beta^J{}_K \wedge \boldsymbol\gamma^K \,,
	\qquad J=1,2,\hdots,N-k \,.
\end{equation}
%---------------------
Equivalently, one can require
%---------------------
\begin{equation} \label{Frob-forms-2}
	\mathrm d \boldsymbol\gamma^J \wedge \boldsymbol\gamma^1 \wedge \hdots \wedge \boldsymbol\gamma^{n-k} =0 \,,
	\qquad J=1,2,\hdots,N-k \,.
\end{equation}
%---------------------
The fact that a $1$-distribution is always integrable follows trivially from~\eqref{Frob-forms-2}.
In other words, given a vector field on $\mathcal B$, one can always find a parametrization that guarantees the existence of integral curves.

Next we consider the case $n=3$, $k=2$.
The distribution of planes $\Pi^{(2)}$ can be defined by a $1$-form $\boldsymbol \pi$ that vanishes on $\Pi^{(2)}$.
Then, from~\eqref{Frob-forms-2} $\Pi^{(2)}$ is integrable if and only if $\mathrm d \boldsymbol \pi \wedge \boldsymbol \pi = 0$.
Note that $\Pi^{(2)}$ can be expressed by any $1$-form $\tilde{\boldsymbol\pi} = f \boldsymbol \pi$, where $f$ is a non-vanishing scalar field.
As a matter of fact, the integrability condition for $\boldsymbol \pi$ is the same as that for $\tilde{\boldsymbol\pi}$, because
%---------------------
\begin{equation}
	\mathrm d (f\boldsymbol \pi) \wedge (f \boldsymbol \pi )
	= f \mathrm d f \wedge \boldsymbol \pi \wedge \boldsymbol \pi + f^2 \mathrm d \boldsymbol \pi \wedge \boldsymbol \pi
	= 0 \,.
\end{equation}
%---------------------
The integrability of $\Pi^{(2)}$ implies the existence of a $2$-dimensional foliation of $\mathcal B$.
This allows one to define foliation coordinates $(Y^1,Y^2,h)$ such that the leaves are the level surfaces of $h$.
Moreover, $\mathrm d h=f \boldsymbol \pi$ for some scalar field $f$.

%-----------------------------
%-----------------------------
\section{Flows} \label{App:Flows}

We look at a smooth map $\chi:\mathcal B \times \mathbb R\to\mathcal B$, with $\chi(X,0)=X$.
Its generator is the vector field $\boldsymbol{\vel}(X,t)$ defined as the velocity of the trajectories $t\mapsto \chi(X,t)$, with components $U^A=\partial\chi^A/\partial t$. We refer to the variable $t$ as time.
Fixing $t$, we denote with $\chi_t$ the diffeomorphism $X\mapsto \chi(X,t)$.
We also set $\chi^t_s = \chi_s \circ \chi_t^{-1}:\chi_t(\mathcal{B})\rightarrow \chi_s(\mathcal{B})$, so that $\chi_s = \chi^t_s \circ \chi_t$, and of course $\chi^s_s=\operatorname{id}_{\mathcal B}$.

The non-autonomous Lie derivative of a time-dependent tensor field is defined as
%---------------------
\begin{equation} \label{non-lie-def}
	\mathsf L_{\boldsymbol{\vel}_t} \boldsymbol A_t =
	\partial_s  \left[\left(\chi^t_s\right)^* \boldsymbol A_s \right] \Big\vert_{s=t}  \,.
\end{equation}
%---------------------
The autonomous Lie derivative of a time-dependent tensor field $\boldsymbol A$ along $\boldsymbol{\vel}$ is defined as
%---------------------
\begin{equation} \label{lie-def}
	\mathfrak L_{\boldsymbol{\vel}_t} \boldsymbol A_t =
	\partial_s  \left[\left(\chi^t_s\right)^* \boldsymbol A_t \right] \Big\vert_{s=t}  \,.
\end{equation}
%---------------------
Invoking the chain rule, one obtains
%---------------------
\begin{equation} \label{non-aut-lie}
	\mathsf L_{\boldsymbol{\vel}_t} \boldsymbol A_t =
	\partial_t \boldsymbol A_t + \mathfrak L_{\boldsymbol{\vel}_t} \boldsymbol A_t \,.
\end{equation}
%---------------------
A time-dependent field $\boldsymbol A(X,t)$ is said to be convected by the flow $\chi$ if one has $\boldsymbol A_t = \chi_{t*} \boldsymbol A_0$, or $\boldsymbol A_s = (\chi^t_s)_{*}\,\boldsymbol A_t$ for any $s,t$.
Clearly, the non-autonomous Lie derivative of a convected field vanishes, as $\partial_s  \left[\left(\chi^t_s\right)^* \boldsymbol A_s \right]=\partial_s\boldsymbol A_t=\boldsymbol 0 $. Therefore, one has
%---------------------
\begin{equation} \label{evolution}
	\partial_t \boldsymbol A_t = - \mathfrak L_{\boldsymbol{\vel}_t } \boldsymbol A_t \,.
\end{equation}
%---------------------

In addition to the Lie derivative, a similar operation, called the Truesdell derivative, is defined.
It relies on the definition of volume form $\mvf$, that for now we assume to be time-independent.
Then, we define the scalar field $J^s_t$ as $(\chi^t_s)^*\mvf=J^t_s\mvf$ expressing the change of volume associated with the flow.
Trivially, $J^t_t = 1$.
Moreover,
%---------------------
\begin{equation} \label{truesd-vol}
	\partial_s J^t_s  \Big\vert_{s=t}
	= \operatorname{Div}\boldsymbol{\vel}_t
	\,.
\end{equation}
%---------------------
The Truesdell derivative of a time-dependent tensor $\boldsymbol A_t$ is defined as
%---------------------
\begin{equation} \label{Truesdell-auto1}
	\mathfrak T_{\boldsymbol{\vel}_t }  \boldsymbol A_t =
	\partial_s
	\left[
	J^t_s \,\, (\chi^t_s)^*   \boldsymbol A_t
	\right] \Big\vert_{s=t}
	\,.
\end{equation}
%---------------------
From~\eqref{truesd-vol}, one obtains the following expression for the Truesdell derivative:
%---------------------
\begin{equation} \label{Truesdell-auto2}
	\mathfrak T_{\boldsymbol{\vel}} \boldsymbol A =
	\mathfrak L_{\boldsymbol{\vel}}  \boldsymbol A +
	(\operatorname{Div} \boldsymbol{\vel}) \boldsymbol A
	\,.
\end{equation}
%---------------------
As for the Lie derivative, one defines the non-autonomous Truesdell derivative as
%---------------------
\begin{equation} \label{Truesdell-non1}
	\mathsf T_{\boldsymbol{\vel}_t }  \boldsymbol A_t =
	\partial_s
	\left[
	J^t_s \,\, (\chi^t_s)^*   \boldsymbol A_s
	\right] \Big\vert_{s=t}
	\,,
\end{equation}
%---------------------
for which one has
%---------------------
\begin{equation} \label{Truesdell-non2}
	\mathsf T_{\boldsymbol{\vel}} \boldsymbol A =
	\mathsf L_{\boldsymbol{\vel}}  \boldsymbol A +
	(\operatorname{Div} \boldsymbol{\vel}) \boldsymbol A
	\,.
\end{equation}
%---------------------
The Truesdell derivatives can also be defined as $\mathfrak T_{\boldsymbol{\vel}}  \boldsymbol A \otimes \mvf = \mathfrak L_{\boldsymbol{\vel}}  (\boldsymbol A \otimes \mvf)$, and $\mathsf T_{\boldsymbol{\vel}}  \boldsymbol A \otimes \mvf = \mathsf L_{\boldsymbol{\vel}}  (\boldsymbol A \otimes \mvf)$, as in~\citep{marsden1983mathematical}.

%---------------------------------------------------------------------------------------------------
%---------------------------------------------------------------------------------------------------
%---------------------------------------------------------------------------------------------------
\section{The Euler-Lagrange equations corresponding to the material variations}    \label{App:Calculations}

In this appendix we obtain the Euler-Lagrange equations from the Lagrange-d'Alembert principle~\eqref{mat-lag-dal}, viz.
%---------------------
\begin{equation} \label{mat-lag-dal-app}
	\int_{t_1}^{t_2} \int_{\mathcal B} \left\lbrace
	- \left\langle \delta  \boldsymbol \vartheta^{\nu} , \boldsymbol Y_{\nu}  \right\rangle \mvf 
	- \prs{a}{\boldsymbol{\xi}}_{\nu} \wedge \delta \prs{a}{\dd}^{\nu} 
	+ \left\langle \prs{a}{\boldsymbol\psi} , \delta \prs{a}{\boldsymbol{\vel}} \right\rangle \mvf
	+ \langle \prs{a}{\boldsymbol\tau} , \delta  \prs{a}{\chi} \rangle \, \mvf
	\right\rbrace \mathrm dt
	=0\,.
\end{equation}
%---------------------
Similar to what is done for spatial variations, the integral~\eqref{mat-lag-dal-app} can be written as a local functional of $\delta \prs{a}{\chi}$, which allows one to use the fundamental lemma of the calculus of variations and obtain the Euler-Lagrange equations.
First, we look at the term $\langle  \delta  \boldsymbol \vartheta^{\nu} ,\boldsymbol Y_{\nu} \rangle$ coming from the explicit dependence of the free energy on the lattice coframe.
We assume the hypotheses of Lemma~\ref{Lem:Orowan-variation}: the dislocation fields obey Orowan's equation, and the variation of their motions is coplanar with both the dislocation curves and the dislocation velocities.
Under these premises, one can write the variation of the energy due to changes of the lattice frame by invoking~\eqref{oro-variation} as
%---------------------
\begin{equation} \label{latt-var}
	\left\langle \delta  \boldsymbol \vartheta^{\nu} , \boldsymbol Y_{\nu} \right\rangle =
	- \left\langle \boldsymbol Y_{\nu},  \iota_{\delta \prs{a}{\chi}} \prs{a}{\dd}^{\nu} \right\rangle =
	\left\langle \iota_{ \boldsymbol Y_{\nu} } \prs{a}{\dd}^{\nu} , \delta \prs{a}{\chi}\right\rangle \,,
\end{equation}
%---------------------
where we used the identity $\left\langle \iota_{\boldsymbol X} \boldsymbol \nu ,\boldsymbol Y  \right\rangle =- \left\langle \iota_{\boldsymbol Y} \boldsymbol \nu ,\boldsymbol X  \right\rangle$ for all $2$-forms $\boldsymbol \nu$ and vectors $\boldsymbol X,\boldsymbol Y$.
Next, we consider the second term in~\eqref{mat-lag-dal-app}, coming from the explicit dependence of the free energy on the distribution of dislocations in the crystal.
From~\eqref{evol-variation}, the variation of the free energy reads
%---------------------
\begin{equation}
	\prs{a}{\boldsymbol{\xi}}_{\nu} \wedge \delta \prs{a}{\dd}^{\nu} =
	- \prs{a}{\boldsymbol{\xi}}_{\nu} \wedge \mathfrak L_{\delta \prs{a}{\chi} } \prs{a}{\dd}^{\nu} =
	- \prs{a}{\boldsymbol{\xi}}_{\nu} \wedge\mathrm d \iota_{\delta \prs{a}{\chi} } \prs{a}{\dd}^{\nu} \,,
\end{equation}
%---------------------
as the dislocation fields are closed.
Note that both the Lie derivative along $\delta \prs{a}{\chi}$ and the exterior derivative are nonlocal functionals of $\delta \prs{a}{\chi}$.
We use Stokes' theorem~\eqref{stokes} in order to obtain a local expression.
Invoking~\eqref{ext-der}, and recalling that $\boldsymbol{\xi}_{\nu}$ is a $1$-form for $\nu=1,2,3$, one obtains
%---------------------
\begin{equation}
	- \prs{a}{\boldsymbol{\xi}}_{\nu} \wedge\mathrm d \iota_{\delta \prs{a}{\chi} } \prs{a}{\dd}^{\nu} =
	\mathrm d ( \prs{a}{\boldsymbol{\xi}}_{\nu} \wedge \iota_{\delta \prs{a}{\chi}} \prs{a}{\dd}^{\nu} )
	- \mathrm d \prs{a}{\boldsymbol{\xi}}_{\nu} \wedge  \iota_{\delta \prs{a}{\chi}} \prs{a}{\dd}^{\nu} \,.
\end{equation}
%---------------------
The $3$-form in the last term can be written as a $1$-form acting on $\delta \prs{a}{\chi}$ times the volume form $\mvf$ by using the raised Hodge operator~\eqref{wedge-hodge}, viz.
%---------------------
\begin{equation}
	- \mathrm d \prs{a}{\boldsymbol{\xi}}_{\nu} \wedge  \iota_{\delta \prs{a}{\chi}} \prs{a}{\dd}^{\nu}
	=
	- \langle \iota_{\delta \prs{a}{\chi}} \prs{a}{\dd}^{\nu} , \star^{\sharp}\mathrm d \prs{a}{\boldsymbol{\xi}}_{\nu} \rangle
	\, \mvf
	=
	\langle \iota_{\star^{\sharp} \mathrm d\prs{a}{\boldsymbol{\xi}}_{\nu} } \prs{a}{\dd}^{\nu}
	, \delta \prs{a}{\chi} \rangle
	\, \mvf	\,.
\end{equation}
%---------------------
Then, one can use Stokes' theorem to write the second term of~\eqref{mat-lag-dal-app} as
%---------------------
\begin{equation} \label{dd-var2}
	\int_{\mathcal B}  \prs{a}{\boldsymbol{\xi}}_{\nu} \wedge \delta \prs{a}{\dd}^{\nu}  =
	\int_{\partial \mathcal B} \incl^* \left(\prs{a}{\boldsymbol{\xi}}_{\nu} \wedge \iota_{\delta \prs{a}{\chi}} \prs{a}{\dd}^{\nu} \right)
	+ \int_{\mathcal B} \left\langle
	\iota_{\star^{\sharp} \mathrm d\prs{a}{\boldsymbol{\xi}}_{\nu} } \prs{a}{\dd}^{\nu}
	, \delta \prs{a}{\chi} \right\rangle
	\mvf \,,
\end{equation}
%---------------------
where $\incl$ is the inclusion map $\partial\mathcal B\hookrightarrow\mathcal B$.
Lastly, we look at the third term in~\eqref{mat-lag-dal-app}, coming from the micro-kinetic energy.
From \eqref{vel-variation} one has
%---------------------
\begin{equation} \label{vel-var}
	\left\langle \prs{a}{\boldsymbol\psi}, \delta \prs{a}{\boldsymbol{\vel}} \right\rangle =
	\left\langle \prs{a}{\boldsymbol\psi}, \partial_t \delta \prs{a}{\chi} \, \right\rangle +
	\left\langle \prs{a}{\boldsymbol\psi}, [ \prs{a}{\boldsymbol{\vel}}, \delta \prs{a}{\chi} ] \, \right\rangle \,.
\end{equation}
%---------------------
For the first term in~\eqref{vel-var} one has
%---------------------
\begin{equation} \label{vel-var-2}
	\left\langle \prs{a}{\boldsymbol\psi}, \partial_t \delta \prs{a}{\chi} \, \right\rangle  =
	\partial_t  \langle \prs{a}{\boldsymbol\psi},  \delta \prs{a}{\chi}  \rangle -
	\left\langle \partial_t \prs{a}{\boldsymbol\psi}, \delta \prs{a}{\chi} \right\rangle \,,
\end{equation}
%---------------------
where the first term vanishes after time integration as $\delta\prs{a}{\chi}(t_1)=\delta\prs{a}{\chi}(t_2)=0$, and hence it will not be considered in the following developments.
As for the second term in~\eqref{vel-var}, it can written as
%---------------------
\begin{equation} \label{vel-var-3}
	\langle \prs{a}{\boldsymbol\psi}, [ \prs{a}{\boldsymbol{\vel}} , \delta \prs{a}{\chi} ] \rangle =
	\langle \prs{a}{\boldsymbol\psi} , \mathfrak L_{\prs{a}{\boldsymbol{\vel}}}  \delta \prs{a}{\chi} \rangle =
	\mathfrak L_{\prs{a}{\boldsymbol{\vel}}} \langle \prs{a}{\boldsymbol\psi} ,  \delta \prs{a}{\chi} \rangle -
	\langle \mathfrak L_{\prs{a}{\boldsymbol{\vel}}}  \prs{a}{\boldsymbol\psi} ,  \delta \prs{a}{\chi} \rangle \,,
\end{equation}
%---------------------
where the Leibniz rule was used.
Note that the Lie derivative of a scalar is a simple derivative, and so, after multiplying by $\mvf$, one can write
%---------------------
\begin{equation} \label{vel-var-4}
	\mathfrak L_{\prs{a}{\boldsymbol{\vel}}} \langle \prs{a}{\boldsymbol\psi} ,  \delta \prs{a}{\chi} \rangle \,\mvf=
	\iota_{\prs{a}{\boldsymbol{\vel}}}  \mathrm d \langle \prs{a}{\boldsymbol\psi} ,  \delta \prs{a}{\chi} \rangle \,\mvf =
	\iota_{\prs{a}{\boldsymbol{\vel}}} \mvf \wedge\mathrm d \langle \prs{a}{\boldsymbol\psi} ,  \delta \prs{a}{\chi} \rangle =
	\mathrm d \! \left(  \langle \prs{a}{\boldsymbol\psi} ,  \delta \prs{a}{\chi} \rangle \, \iota_{\prs{a}{\boldsymbol{\vel}}} \mvf \right) -
	\mathrm d (\iota_{\prs{a}{\boldsymbol{\vel}}} \mvf ) \, \langle \prs{a}{\boldsymbol\psi} ,  \delta \prs{a}{\chi} \rangle \,,
\end{equation}
%---------------------
where use was made of~\eqref{contr} and~\eqref{ext-der}.
Note that, by definition of divergence, one also has $\mathrm d (\iota_{\prs{a}{\boldsymbol{\vel}}} \mvf )= (\operatorname{Div}\prs{a}{\boldsymbol{\vel}}) \,\mvf$.
Therefore, gathering~\eqref{vel-var-2},~\eqref{vel-var-3} and~~\eqref{vel-var-4}, and invoking the notion of non-autonomous Truesdell derivative~\eqref{Truesdell-non1}, i.e.,
%---------------------
\begin{equation}
	\mathsf T_{\prs{a}{\boldsymbol{\vel}}}   \prs{a}{\boldsymbol\psi} =
	\partial_t \prs{a}{\boldsymbol\psi} +
	\mathfrak L_{\prs{a}{\boldsymbol{\vel}}}  \prs{a}{\boldsymbol\psi} +
	(\operatorname{Div}\prs{a}{\boldsymbol{\vel}}) \prs{a}{\boldsymbol\psi}
	\,,
\end{equation}
%---------------------
and using Stokes' theorem on the non-local term, the last term in~\eqref{mat-lag-dal} can be written as
%---------------------
\begin{equation} \label{vel-var-5}
	\int_{\mathcal B}   \left\langle \prs{a}{\boldsymbol\psi} , \delta \prs{a}{\boldsymbol{\vel}} \right\rangle \mvf
	=
	\int_{\partial \mathcal B} \incl^* \left( \langle \prs{a}{\boldsymbol\psi} , \delta \prs{a}{\chi} \rangle \, 
	\iota_{\prs{a}{\boldsymbol{\vel}}} \mvf \right)
	-
	\int_{\mathcal B} \langle \mathsf T_{\prs{a}{\boldsymbol{\vel}}}  \prs{a}{\boldsymbol\psi} \,, \,\delta \prs{a}{\chi}\, \rangle \mvf
	\,.
\end{equation}
%---------------------
Hence, from~\eqref{latt-var},~\eqref{dd-var2} and~\eqref{vel-var-5} one can write~\eqref{mat-lag-dal-app} as
%---------------------
\begin{equation} \label{mat-var3}
\begin{split}
	\int_{t_1}^{t_2}  \Bigg\lbrace
	- \int_{\mathcal B} \left\langle \iota_{ \boldsymbol Y_{\nu} } \prs{a}{\dd}^{\nu} , \delta \prs{a}{\chi}\right\rangle \mvf
	- \int_{\mathcal B} \left\langle \iota_{\star^{\sharp} \mathrm d\prs{a}{\boldsymbol{\xi}}_{\nu} } \prs{a}{\dd}^{\nu} , 
	\delta \prs{a}{\chi} \right\rangle \mvf
	-\int_{\partial \mathcal B} \incl^* \left(\prs{a}{\boldsymbol{\xi}}_{\nu} \wedge \iota_{\delta \prs{a}{\chi}} \prs{a}{\dd}^{\nu} \right)
	& \\
	+ \int_{\partial \mathcal B} \incl^* \left( \langle \prs{a}{\boldsymbol\psi} , 
	\delta \prs{a}{\chi} \rangle \, \iota_{\prs{a}{\boldsymbol{\vel}}} \mvf \right)
	-\int_{\mathcal B} \langle \mathsf T_{\prs{a}{\boldsymbol{\vel}}}  \prs{a}{\boldsymbol\psi} \,, \,\delta \prs{a}{\chi}\, \rangle \mvf
	&+
	\langle \prs{a}{\boldsymbol\tau} , \delta  \prs{a}{\chi} \rangle \, \mvf
	\Bigg\rbrace \mathrm dt = 0 \,.
\end{split}
\end{equation}
%---------------------
%
%
%
Next we look for a more convenient expression for the boundary terms.
First we rewrite the integrand of the boundary term in~\eqref{dd-var2} using~\eqref{contr}, viz.
%---------------------
\begin{equation}
	\prs{a}{\boldsymbol{\xi}}_{\nu} \wedge \iota_{\delta \prs{a}{\chi}} \prs{a}{\dd}^{\nu} =
	-  \iota_{\delta \prs{a}{\chi}} (\prs{a}{\boldsymbol{\xi}}_{\nu} \wedge \prs{a}{\dd}^{\nu})
	+ \langle \prs{a}{\boldsymbol{\xi}}_{\nu},\delta \prs{a}{\chi} \rangle  \prs{a}{\dd}^{\nu} \, .
\end{equation}
%---------------------
Then,
still denoting with $\boldsymbol\varsigma$ the area form and with $\boldsymbol\nu$ the unit normal $1$-form on $\partial\mathcal B$,
we make use of the identity~\eqref{pullback-forms} for the pullback of $2$-forms via the inclusion map.
Hence, setting $\prs{a}{\boldsymbol{\xi}}=\prs{a}{\dens}\,\prs{a}{\bv}^{\nu}\,\prs{a}{\boldsymbol{\xi}}_{\nu}$ for the sake of simplicity,
one can write
%---------------------
\begin{equation}
	\incl^* \left( \prs{a}{\boldsymbol{\xi}}_{\nu} \wedge \iota_{\delta \prs{a}{\chi}} \prs{a}{\dd}^{\nu} \right)
	=  - \langle \prs{a}{\boldsymbol{\xi}} ,\prs{a}{\boldsymbol{\lv}} \rangle \langle \boldsymbol \nu,\delta \prs{a}{\chi} \rangle \, \boldsymbol\varsigma
	+ \langle \boldsymbol \nu,\prs{a}{\boldsymbol{\lv}} \rangle \langle  \prs{a}{\boldsymbol{\xi}} , \delta \prs{a}{\chi} \rangle \, \boldsymbol\varsigma
	\,.
\end{equation}
%---------------------
Similarly, the boundary term in~\eqref{vel-var-5} can be written as
%---------------------
\begin{equation}
	\incl^* \left( \langle \prs{a}{\boldsymbol\psi} , \delta \prs{a}{\chi} \rangle \, \iota_{\prs{a}{\boldsymbol{\vel}}} \mvf \right) =
	\langle \boldsymbol \nu, \prs{a}{\boldsymbol{\vel}} \rangle \langle \prs{a}{\boldsymbol\psi} , 
	\delta \prs{a}{\chi} \rangle \, \boldsymbol\varsigma
	\,.
\end{equation}
%---------------------
%
%
Therefore, the Lagrange-d'Alembert principle~\eqref{mat-lag-dal-app} for material variations gives
%---------------------
\begin{equation} 
\begin{split}
	\int_{t_1}^{t_2}
	\int_{\partial \mathcal B} &
	\Big\langle
	\langle \prs{a}{\boldsymbol{\xi}} ,\prs{a}{\boldsymbol{\lv}} \rangle \, \boldsymbol \nu-
	\langle \boldsymbol \nu,\prs{a}{\boldsymbol{\lv}} \rangle \, \prs{a}{\boldsymbol{\xi}}  +
	\langle \boldsymbol \nu, \prs{a}{\boldsymbol{\vel}} \rangle \, \prs{a}{\boldsymbol\psi}
	~,\, \delta \prs{a}{\chi}
	\Big\rangle \boldsymbol \, \boldsymbol\varsigma \, \mathrm dt  \\
	&+ \int_{t_1}^{t_2} \int_{\mathcal B} \left\langle
	-\iota_{ \boldsymbol Y_{\nu} } \prs{a}{\dd}^{\nu}
	-\iota_{\star^{\sharp} \mathrm d \prs{a}{\boldsymbol{\xi}}_{\nu} } \prs{a}{\dd}^{\nu}
	-\mathsf T_{\prs{a}{\boldsymbol{\vel}}}   \prs{a}{\boldsymbol\psi}
	+\prs{a}{\boldsymbol\tau}
	~,\, \delta \prs{a}{\chi} \right\rangle \mvf \, \mathrm dt
	 =0	\,.
\end{split}
\end{equation}
%---------------------

\end{document}